\documentclass[twocolumn]{aastex63}
\pdfoutput=1 
\usepackage{amsmath,amstext}
\usepackage[T1]{fontenc}
\usepackage{apjfonts} 
\usepackage[figure,figure*]{hypcap}
\usepackage{longtable}
\usepackage{threeparttable}

\usepackage{amssymb}
\usepackage{amsmath}
\usepackage{fontawesome}
\usepackage{gensymb}
\usepackage{mathrsfs}
\usepackage{upgreek}
\usepackage{fontenc}
\usepackage{hyperref}
\usepackage{float}
\usepackage{tabularx}

\newcommand{\ergs}{\text{erg s}\ensuremath{^{-1}}}

\newcommand{\kms}{\text{km s}\ensuremath{^{-1}}}
\newcommand{\voff}{\ensuremath{v_{\rm off}}}

\renewcommand{\d}{\ensuremath{\rm d}}

\newcommand{\rev}{}
\newcommand{\rrev}{}

\shorttitle{CLASS Survey Description}
\shortauthors{M. Reefe et al.}

\graphicspath{{./}{figures/}}

\begin{document}

\title{CLASS Survey Description: Coronal Line Needles in the SDSS Haystack}

\author[0000-0003-4701-8497]{Michael Reefe}
\altaffiliation{National Science Foundation, Graduate Research Fellow}
\affiliation{George Mason University, Department of Physics and Astronomy, MS3F3, 4400 University Drive, Fairfax, VA 22030, USA}
\affiliation{Massachusetts Institute of Technology, Kavli Institute for Astrophysics and Space Research, 77 Massachusetts Avenue, Cambridge, MA 02139}
\author[0000-0003-3432-2094]{Remington O. Sexton}
\affiliation{George Mason University, Department of Physics and Astronomy, MS3F3, 4400 University Drive, Fairfax, VA 22030, USA}
\affiliation{U.S. Naval Observatory, 3450 Massachusetts Avenue NW, Washington, DC 20392-5420, USA}
\author[0000-0003-3152-4328]{Sara M. Doan}
\affiliation{George Mason University, Department of Physics and Astronomy, MS3F3, 4400 University Drive, Fairfax, VA 22030, USA}
\author[0000-0003-2277-2354]{Shobita Satyapal}
\affiliation{George Mason University, Department of Physics and Astronomy, MS3F3, 4400 University Drive, Fairfax, VA 22030, USA}
\author[0000-0002-4902-8077]{Nathan J. Secrest}
\affiliation{U.S. Naval Observatory, 3450 Massachusetts Avenue NW, Washington, DC 20392-5420, USA}
\author[0000-0003-1051-6564]{Jenna M. Cann}
\altaffiliation{NASA Postdoctoral Program}
\affiliation{NASA Goddard Space Fligh Center, 8800 Greenbelt Rd, Greenbelt, Maryland 20771 USA}

\begin{abstract}

Coronal lines are a powerful, yet poorly understood, tool to identify and characterize Active Galactic Nuclei (AGNs). 
There have been few large scale surveys of coronal lines in the general galaxy population in the literature so far. Using a novel pre-selection technique with a flux-to-RMS ratio $\mathcal{F}$, followed by Markov-Chain Monte Carlo (MCMC) fitting, we searched for the full suite of 20 coronal lines in the optical spectra of almost 1 million galaxies from the Sloan Digital Sky Survey (SDSS) Data Release 8.  We present a catalog of the emission line parameters for the resulting 258 galaxies with detections. The Coronal Line Activity Spectroscopic Survey (CLASS) includes
line properties, host galaxy properties, and selection criteria for all galaxies in which at least one line is detected.  This comprehensive study reveals that a significant fraction of coronal line activity is missed in past surveys based on a more limited set of coronal lines; $\sim$60\% of our sample do not display the more widely surveyed [\ion{Fe}{10}] $\lambda$6374.  In addition, we discover a strong correlation between coronal line and WISE W2 luminosities, suggesting that the mid-infrared flux can be used to predict coronal line fluxes.  For each line we also provide a confidence level that the line is present, generated by a novel neural network, trained on fully simulated data. We find that after training the network to detect individual lines using 100,000 simulated spectra, we achieve an overall true positive rate of 75.49\% and a false positive rate of only 3.96\%.



\end{abstract}


\section{Introduction}

Since the advent of large-scale ground-based spectroscopic campaigns, detailed studies of the brightest emission lines in the optical regime have been carried out on millions of galaxies. These studies have provided unequivocal evidence for the ubiquity of supermassive black holes (SMBHs) in massive galaxies and have revealed that when accreting, SMBHs can have a profound impact on the host galaxies in which they reside. However, there have not yet been any systematic studies of the much fainter \rev{optical} coronal line spectrum in a large sample of galaxies. This is a significant deficiency because these \rev{coronal} emission lines, which arise from collisionally excited forbidden fine-structure transitions from highly ionized species, are a powerful diagnostic in uncovering active galactic nuclei (AGNs) in galaxy populations in which all other available tools are often ineffective \citep[e.g.,][]{2009MNRAS.398.1165G,2015ApJ...811...26T, 2018ApJ...858...38S, 2021ApJ...906...35S}. \rev{Near-infrared coronal line studies in particular have further traced nuclear activity and ionized gas outflows in highly obscured AGNs \citep{2002ApJ...579..214R,2006A&A...457...61R,2017MNRAS.467..540L,2018ApJ...858...48M}.} Moreover, coronal line emission can potentially be used to constrain accretion disk models \citep[e.g.,][]{1996A&A...315L.109M,2000ApJ...536..710A}, study the properties of the dense gas in close proximity to the SMBH, and study outflows of the highly ionized gas near their launch origin \citep[e.g.,][]{1997A&A...323..707E, 2006ApJ...653.1098R,2021ApJ...911...70B,2021MNRAS.506.3831F}, which is of crucial importance in understanding AGN feedback in the host. The coronal line region samples gas with a wide range of densities, and it can extend several hundred parsecs from the nucleus \citep[e.g.,][]{2011ApJ...739...69M, 2021MNRAS.506.3831F}, with some of the emission originating between the broad line region and the narrow line region, as demonstrated by recent VLT/GRAVITY observations \citep{2021A&A...648A.117G}. Because the spectral energy distribution (SED) of the accretion disk hardens with decreasing black hole mass, coronal line ratios may even have the potential to constrain the black hole mass and reveal accreting intermediate mass black holes \citep[IMBHs, e.g.,][]{2018ApJ...861..142C, 2021ApJ...912L...2C, 2021ApJ...910....5M, 2021ApJ...922..155M}. 

\rev{Studying the optical coronal line spectrum in particular is scientifically advantageous due to the many large-scale optical ground-based surveys (e.g. the Sloan Digital Sky Survey, or SDSS) that have public data and cover wide regions of the sky.  This is in contrast to other wavelength regimes, e.g. the infrared, where the coronal lines are intrinsically orders of magnitude brighter and hence easier to detect than in the optical, but where data is limited due to the IR opacity of the atmosphere, a problem that will soon be remedied by JWST.}

In order to explore the host galaxy demographics and coronal line spectrum of a large sample of galaxies, we searched for the full suite of 20 optical coronal lines in the spectra of roughly 1 million galaxies in the SDSS DR8. Unlike previous studies that often target a small set of coronal lines \citep[e.g.][\rev{which only examine \lbrack \ion{Fe}{10}\rbrack $\lambda$6374}]{10.1111/j.1365-2966.2009.14961.x,2021ApJ...922..155M}, we provide a comprehensive study of all optical coronal lines. The detection statistics and host galaxy demographics of these coronal line emitters are presented in \citet{Reefe_2022}. \rev{In this previous work, we identified the most commonly detected lines in the survey to be [\ion{Fe}{10}] $\lambda$6374, [\ion{Fe}{7}] $\lambda\lambda$6087,5720, and [\ion{Ne}{5}] $\lambda\lambda$3426,3346, the latter of which were also the brightest. Among the rarest were [\ion{S}{12}] $\lambda$7609 and [\ion{Fe}{5}] $\lambda\lambda$3891,3839. The luminosities of these lines are typically 1--3 orders of magnitude larger than what would be expected from purely stellar processes (i.e. Wolf-Rayet stars, supernovae, planetary nebulae, or shocks from starburst-driven winds, see \citet{1999A&A...345L..17S,2008ApJ...678..686A,2008ApJS..178...20A,2012ApJ...754...28Z,2012MNRAS.427.1229I,2021MNRAS.508.2556I}), which seems to suggest that they are overwhelmingly produced by accreting SMBHs or TDEs. However, examining the Baldwin-Phillips-Terlevich (BPT) ratios \citep{1981PASP...93....5B} and WISE colors of these coronal line emitters reveals a large population outside the AGN cutoffs of both metrics. The measured star formation rates (SFR) of these galaxies tend to lie more than 1$\sigma$ above the $M_*$-SFR main-sequence relationship from \citet{2017ApJ...851...22M}, suggesting that enhanced accretion activity onto the SMBH is being falsely identified as an enhanced SFR and causing these emitters to not be identified as AGNs by the typical metrics. Even more strikingly, we found coronal line-emitters to have preferentially lower stellar masses ($\lesssim 10^{10}$ M$_\odot$) than traditionally identified BPT or WISE AGNs, hinting that a lower mass allows for a hotter accretion disk and therefore an enhanced high-ionization line spectrum \citep{2018ApJ...861..142C}.}

In this work, we present and make publicly available the Coronal Line Activity Spectroscopic Survey (CLASS) catalog. \rev{In contrast to our previous work, we do not focus on host galaxy demographics or properties and instead further examine} detection statistics and physical properties (i.e. luminosities, widths, velocity offsets, \&c.), distributions, and correlations between coronal line occurrences \rev{in order to provide context, guidelines, and caveats for anyone wishing to utilize the survey in future work}.  The results from this study may \rev{also} prove useful in predicting coronal emission line strengths in high redshift galaxies observable by the James Webb Space Telescope (JWST). Throughout this work, we adopt a flat $\Lambda$CDM cosmological model with $H_0 = 70$ \kms~Mpc$^{-1}$, $\Omega_m = 0.3$, and $\Omega_\Lambda = 0.7$.

\par

This paper is organized as follows. In ${\S}$\ref{sect:sample}, we describe our sample pre-selection and fitting techniques. Then, in ${\S}$\ref{sect:results}, we discuss properties and statistics of the CLASS survey catalog. In ${\S}$\ref{sect:description} we describe the format of the catalog and how the data may be accessed.

\section{Sample Selection}
\label{sect:sample}

We searched for coronal line emission in the optical ($\sim$3300--8000 \AA) spectra of the 952,138 galaxies in the  SDSS MPA/JHU Data Release 8 catalog \citep{2011AJ....142...72E,2011ApJS..193...29A,2006AJ....131.2332G,2013AJ....146...32S}\footnote{\href{https://wwwmpa.mpa-garching.mpg.de/SDSS/DR7/}{https://wwwmpa.mpa-garching.mpg.de/SDSS/DR7/}}. Contrary to several other works, we do not impose any initial pre-selection cuts to the spectra based on redshift, AGN classification, or the strength or presence of specific lines (e.g. [\ion{O}{1}] $\lambda$6302 or [\ion{Fe}{10}] $\lambda$6374). Instead, our goal was to conduct an unbiased search for the 20 optical coronal lines accessible in the SDSS spectrum in the general galaxy population, and to study the host galaxy demographics of the resulting population. We searched systematically for a set of 20 optical coronal lines listed in Table \ref{tab:lines}, ranging from [\ion{Ne}{5}] $\lambda\lambda$3346,3426 to [\ion{Fe}{11}] $\lambda$7892, providing for the first time a catalog of the emission line properties in a large sample of galaxies drawn from SDSS.

\subsection{Pre-selection filtering}
In order to reduce computation time, we begin with a filtering algorithm  to identify candidate spectra with potential robust detections. For each coronal line, we first subtract a linear slope from the continuum around the line, which is calculated by taking the average flux in 20 \AA-wide regions located $\pm$30 \AA\ from the line's rest wavelength. If another emission or absorption line happens to fall within either of these regions, we adjust the regions accordingly such that they encompass a flat and featureless part of the spectrum.  We then take the average continuum-subtracted flux within $\pm$10 \AA\ of the line's rest wavelength ($\bar{f_\lambda}$) and divide by the root-mean-square (RMS) deviation of the flux within the same 20 \AA-wide reference windows at $\pm$30 \AA\ from the line ($\sigma_\lambda$):
\begin{equation}
    \mathcal{F} = \frac{\bar{f_\lambda}}{\sigma_\lambda}
\end{equation}


We select all galaxies where $\mathcal{F} \geqslant 4$ for at least one coronal line, \rev{which totaled 2,079 galaxies}\footnote{there was a slight error in the number reported in our previous paper, \citet{Reefe_2022}, caused by counting an additional non-coronal line}. We find that this method, on its own, is highly sensitive to noise spikes and improper sky line subtractions.  To mitigate these effects, we impose the additional criteria that at least 3 continuous pixels within $\pm 10$ \AA\ from the line must be $\geqslant 3\sigma_\lambda$, which filters out noise spikes that are 1 or 2 pixels wide; and the emission line must not be within $\pm$20 \AA\ (or roughly $\pm$1000 \kms) of the four most prominent sky lines in the observed frame ($\lambda = $ 5578.5, 5894.6, 6301.7, and 7246.0 \AA), which filters out poor subtractions.  Our filtering procedure is carried out with a custom publicly available Python package called BIFR\"OST\footnote{\href{https://github.com/Michael-Reefe/bifrost}{https://github.com/Michael-Reefe/bifrost}}.   We note that this pre-selection algorithm is designed to identify the most robust detections in the entire sample, and will likely exclude marginal detections.  Our resulting sample of candidate coronal line emitters, \rev{after applying these two filters,  contains a total of 434 galaxies}.  We note that we use the $\mathcal{F} \geqslant 4$ requirement only as a method of filtering the SDSS catalog for follow-up analysis;  we do not impose this criterion to consider a line a detection within this subsample of \rev{434}.

\startlongtable
\begin{deluxetable*}{lccccccccccc}
\tabletypesize{\footnotesize}
\tablecaption{Optical coronal emission lines from 3346--7892 \AA. $\lambda_{\rm rest}$ is the rest wavelength, $\rho_{\rm crit}$ is the critical density, \rev{IP is the ionization potential, $L$ Threshold is the average $1\sigma$ luminosity detection threshold across all spectra}, $L$ is the luminosity, FWHM is the full-width at half-maximum of the line profile, $v_{\rm off}$ is the velocity offset relative to the stellar velocity, and $N$ is the number of detections. For $L$, FWHM, and $v_{\rm off}$, the values shown are (mean) $\pm$ (standard deviation).}
\tablehead{\colhead{Line} & \colhead{$\lambda_{\rm rest}$}\textsuperscript{1} & \colhead{$\rho_{\rm crit}$} & \colhead{IP}\textsuperscript{2} & \colhead{Transition} & \colhead{$L_{\rm thresh}$} & \colhead{$L$} & \colhead{FWHM} & \colhead{$v_{\rm off}$} & \colhead{Detections}\textsuperscript{3} & \colhead{Detections}\textsuperscript{3} & \colhead{$N$} \\
\colhead{} & \colhead{(\AA)} & \colhead{(cm$^{-3}$)} & \colhead{(eV)} & \colhead{} & \colhead{$\log L/$cgs} & \colhead{$\log L/$cgs} & \colhead{(km s$^{-1}$)} & \colhead{(km s$^{-1}$)} & \colhead{(\%)} & \colhead{(\%)} & \colhead{}}
\decimals
\startdata
& & & & & & & & & \rev{(CL avg)} & \rev{(CL max)} & \\
\lbrack \ion{Fe}{11}\rbrack & 7891.800 & $6.39 \times 10^8$ & 262.10 & $^3$P$_2 - ^3$P$_1$ & 39.18 & $38.92 \pm 1.27$ & $409 \pm 234$ & $-77 \pm 137$ & ${0.0064}^{+0.0029}_{-0.0021}$ & ${0.00136}^{+0.00062}_{-0.00044}$ & 9 \\ 
\lbrack \ion{S}{12}\rbrack & 7611.000 & $7.09 \times 10^9$ & 504.78 & $^2$P$^0_{1/2} - ^2$P$^0_{3/2}$ & 39.25 & $39.40 \pm 0.28$ & $394 \pm 32$ & $-191 \pm 67$ & ${0.00058}^{+0.00076}_{-0.00037}$ & ${0.00040}^{+0.00053}_{-0.00026}$ & 2 \\ 
\lbrack \ion{Fe}{10}\rbrack & 6374.510 & $4.45 \times 10^8$ & 235.04 & $^2$P$^0_{3/2} - ^2$P$^0_{1/2}$ & 39.45 & $39.00 \pm 1.05$ & $699 \pm 876$ & $-91 \pm 216$ & ${0.0407}^{+0.0043}_{-0.0039}$ & ${0.0125}^{+0.0013}_{-0.0012}$ & 107 \\ 
\lbrack \ion{Fe}{7}\rbrack & 6087.000 & $4.46 \times 10^7$ & 99.00 & $^3$F$_3 - ^1$D$_2$ & 39.47 & $39.96 \pm 0.59$ & $577 \pm 409$ & $-27 \pm 128$ & ${0.0146}^{+0.0015}_{-0.0014}$ & ${0.0124}^{+0.0013}_{-0.0012}$ & 111 \\ 
\lbrack \ion{Fe}{7}\rbrack & 5720.700 & $3.72 \times 10^7$ & 99.00 & $^3$F$_2 - ^1$D$_2$ & 39.41 & $39.77 \pm 0.59$ & $615 \pm 563$ & $-6 \pm 177$ & ${0.0158}^{+0.0016}_{-0.0015}$ & ${0.0125}^{+0.0013}_{-0.0012}$ & 114 \\ 
\lbrack \ion{Ar}{10}\rbrack & 5533.265 & $2.36 \times 10^9$ & 422.60 & $^2$P$^0_{3/2} - ^2$P$^0_{1/2}$ & 39.38 & $38.83 \pm 2.83$ & $380 \pm 346$ & $-387 \pm 225$ & ${0.0028}^{+0.0019}_{-0.0012}$ & ${0.00054}^{+0.00037}_{-0.00023}$ & 5 \\ 
\lbrack \ion{Fe}{6}\rbrack & 5335.180 & $6.32 \times 10^6$ & 75.00 & $^4$F$_{3/2} - ^4$P$_{1/2}$ & 39.40 & $39.74 \pm 2.14$ & $25 \pm 42$ & $-311 \pm 303$ & ${0.00131}^{+0.00064}_{-0.00045}$ & ${0.00087}^{+0.00043}_{-0.00030}$ & 8 \\ 
\lbrack \ion{Ca}{5}\rbrack & 5309.110 & $6.63 \times 10^7$ & 67.10 & $^3$P$_2 - ^1$D$_2$ & 39.39 & $40.11 \pm 0.58$ & $404 \pm 333$ & $20 \pm 171$ & ${0.00137}^{+0.00055}_{-0.00040}$ & ${0.00120}^{+0.00048}_{-0.00035}$ & 11 \\ 
\lbrack \ion{Fe}{14}\rbrack & 5302.860 & $3.99 \times 10^8$ & 361.00 & $^2$P$^0_{1/2} - ^2$P$^0_{3/2}$ & 39.40 & $39.94 \pm 0.66$ & $450 \pm 237$ & $-148 \pm 210$ & ${0.00095}^{+0.00051}_{-0.00035}$ & ${0.00076}^{+0.00041}_{-0.00028}$ & 7 \\ 
\lbrack \ion{Fe}{7}\rbrack & 5276.380 & $2.98 \times 10^6$ & 99.00 & $^3$F$_4 - ^3$P$_2$ & 39.40 & $37.08 \pm 1.76$ & $118 \pm 74$ & $-234 \pm 322$ & ${0.0057}^{+0.0026}_{-0.0018}$ & ${0.00155}^{+0.00071}_{-0.00051}$ & 9 \\ 
\lbrack \ion{Fe}{6}\rbrack & 5176.040 & $3.29 \times 10^7$ & 75.00 & $^4$F$_{9/2} - ^2$G$_{9/2}$ & 39.35 & $39.47 \pm 2.55$ & $81 \pm 62$ & $-110 \pm 184$ & ${0.00070}^{+0.00068}_{-0.00038}$ & ${0.00033}^{+0.00032}_{-0.00018}$ & 3 \\ 
\lbrack \ion{Fe}{7}\rbrack & 5158.890 & $3.44 \times 10^6$ & 99.00 & $^3$F$_3 - ^3$P$_1$ & 39.40 & $37.70 \pm 1.45$ & $224 \pm 150$ & $-205 \pm 108$ & ${0.0025}^{+0.0020}_{-0.0012}$ & ${0.00132}^{+0.00104}_{-0.00063}$ & 4 \\ 
\lbrack \ion{Fe}{6}\rbrack & 5145.750 & $2.29 \times 10^7$ & 75.00 & $^4$F$_{7/2} - ^2$G$_{7/2}$ & 39.47 & $40.65 \pm 0.48$ & $51 \pm 124$ & $-302 \pm 160$ & ${0.00077}^{+0.00041}_{-0.00028}$ & ${0.00076}^{+0.00041}_{-0.00028}$ & 7 \\ 
\lbrack \ion{Fe}{7}\rbrack & 4893.370 & $3.09 \times 10^6$ & 99.00 & $^3$F$_2 - ^3$P$_1$ & 39.51 & $38.94 \pm 2.21$ & $188 \pm 311$ & $15 \pm 204$ & ${0.0054}^{+0.0036}_{-0.0023}$ & ${0.00054}^{+0.00037}_{-0.00023}$ & 5 \\ 
\lbrack \ion{Fe}{5}\rbrack & 4180.600 & $1.86 \times 10^8$ & 54.80 & $^5$D$_1 - ^3$P2$_0$ & 39.42 & $39.45 \pm 2.15$ & $66 \pm 131$ & $-57 \pm 387$ & ${0.00117}^{+0.00079}_{-0.00051}$ & ${0.00054}^{+0.00037}_{-0.00023}$ & 5 \\ 
\lbrack \ion{Fe}{5}\rbrack & 3891.280 & $1.61 \times 10^8$ & 54.80 & $^5$D$_4 - ^3$F2$_4$ & 39.45 & $40.62 \pm 1.06$ & $1 \pm 0$ & $7 \pm 73$ & ${0.00022}^{+0.00029}_{-0.00014}$ & ${0.00022}^{+0.00029}_{-0.00014}$ & 2 \\ 
\lbrack \ion{Fe}{5}\rbrack & 3839.270 & $1.00 \times 10^8$ & 54.80 & $^5$D$_3 - ^3$F2$_3$ & 39.43 & $41.20 \pm 0.60$ & $16 \pm 33$ & $-387 \pm 211$ & ${0.00054}^{+0.00037}_{-0.00023}$ & ${0.00054}^{+0.00037}_{-0.00023}$ & 5 \\ 
\lbrack \ion{Fe}{7}\rbrack & 3758.920 & $4.02 \times 10^7$ & 99.00 & $^3$F$_4 - ^1$G$_4$ & 39.65 & $40.38 \pm 0.90$ & $34 \pm 89$ & $-164 \pm 570$ & ${0.00097}^{+0.00048}_{-0.00033}$ & ${0.00088}^{+0.00043}_{-0.00030}$ & 8 \\ 
\lbrack \ion{Ne}{5}\rbrack & 3425.881 & $1.90 \times 10^7$ & 97.11 & $^3$P$_2 - ^1$D$_2$ & 39.72 & $41.21 \pm 0.29$ & $439 \pm 172$ & $-111 \pm 91$ & ${0.0123}^{+0.0018}_{-0.0016}$ & ${0.0123}^{+0.0018}_{-0.0016}$ & 61 \\ 
\lbrack \ion{Ne}{5}\rbrack & 3345.821 & $1.14 \times 10^7$ & 97.11 & $^3$P$_1 - ^1$D$_2$ & 39.83 & $40.91 \pm 0.24$ & $418 \pm 131$ & $-135 \pm 81$ & ${0.0085}^{+0.0018}_{-0.0015}$ & ${0.0085}^{+0.0018}_{-0.0015}$ & 31 \\ 
\enddata
\begin{tablenotes}
    \item[1] \textsuperscript{1}Wavelengths taken from: \url{https://physics.nist.gov/PhysRefData/ASD/lines_form.html}.
    \item[2] \textsuperscript{2}Ionization potential taken from: \url{https://physics.nist.gov/PhysRefData/ASD/ionEnergy.html}.
    \rev{\item[3] \textsuperscript{3}The detection percentage columns are calculated as the number $N$ over the total number of spectra that could have detected the line. The denominator of the left detection column is determined by the number of galaxies whose luminosity threshold exceeds 3 times the {\it average} CL luminosity and the right is determined with respect to 3 times the {\it maximum} CL luminosity. In both cases, the errors are 68\% confidence intervals calculated with binomial statistics.}
    \item[4] \textsuperscript{$\dagger$}Lines previously observed from ground- and space-based observatories.
\end{tablenotes}
\label{tab:lines}
\end{deluxetable*}

\subsection{MCMC Fitting}
After generating our subsample of \rev{434} galaxies, we model each one by fitting the flux first with a likelihood maximization, followed by a Markov-Chain Monte Carlo (MCMC) simulation, with the Bayesian AGN Decomposition Analysis for SDSS Spectra (BADASS) software \citep{sexton_2020}, version 8.0.12. BADASS uses basin-hopping with the sequential least-squares programming (SLSQP) method implemented with SciPy to perform the maximum likelihood fit. These parameters are then used as the starting point for the MCMC, which uses the affine-invariant Emcee sampler \citep{2013PASP..125..306F} to obtain robust parameter uncertainties and covariances.  BADASS deconvolves the spectrum into its base components, fitting an AGN power-law, stellar line-of-sight velocity distribution \citep[LOSVD; which heavily utilizes the pPXF method from][]{2017MNRAS.466..798C}, \ion{Fe}{2} emission, and a standardized set of permitted and forbidden transition lines (with narrow, broad, and outflow components), upon which we append the 20 coronal lines from Table \ref{tab:lines}.  The AGN continuum is fit as a simple power law, stellar kinematics are fit using the Indo-US template library, and the \ion{Fe}{2} emission uses the spectrum of I Zw 1 from \citet{2004A&A...417..515V} as a base.  Optionally, BADASS can fit directly with a simple stellar population (SSP) host galaxy template instead of the stellar kinematics. Each emission line is fitted with a Gaussian profile with a free amplitude, offset, and width (except doublets, in which case the amplitudes, widths, and offsets are tied), in other words two free moments.  Broad lines (e.g. H$\alpha$, [\ion{O}{3}] $\lambda$5007) are instead fit with pseudo-Voigt profiles with a free mixing parameter (which determines the weight given to the Gaussian and Lorentzian components).   For a few select prominent coronal lines ([\ion{Fe}{10}] $\lambda$6374, [\ion{Fe}{7}] $\lambda\lambda$6087,5720, and [\ion{Ne}{5}] $\lambda\lambda$3426,3346) we rerun our analysis allowing free third and fourth moments (skewness and kurtosis) to analyze any potential correlations with lower order moments (velocity offset and width) and to search for outflows, but we do not find any significant correlations.

We decompose each spectrum into 8 regions and fit regions 2--7 separately to decrease the simulation times and allow for a more robust power-law fit.  Region 1 covers wavelengths $> 8000$ \AA\ and region 8 covers wavelengths $< 2000$ \AA, neither of which we consider.  The other regions are defined as:
\begin{itemize}
    \item Region 2 (coronal lines): 6800--8000 \AA
    \item Region 3 (H$\alpha$ region): 6200--6800 \AA 
    \item Region 4 (coronal lines): 5500--6200 \AA
    \item Region 5 (H$\beta$/[\ion{O}{3}] region): 4400--5500 \AA
    \item Region 6 (Balmer region): 3500--4400 \AA
    \item Region 7 (\ion{Mg}{2} region): 2000--3500 \AA
\end{itemize}
\rev{Regions 5 and 6 both contain strong absorption features (Ca K+H and \ion{Mg}{2}, respectively) that make it preferable to fit using stellar kinematics using the Indo-US stellar template libraries with the pPXF technique, whereas regions 2--4 contain no similarly strong absorption, so we instead fit them using the SSP host galaxy templates as placeholders for the galaxy's stellar contribution.  It is technically possible to recover the stellar kinematics with both of these methods, but it is debatable whether one is better than another. As such, to prevent any biases in our velocity offsets, we always measure the stellar velocity relative to the region 5 fit with the Indo-US templates.}  Each MCMC is run for 1,500 burn-in steps and 2,500 steps, which we find is generally long enough to reach convergence.  We do not check for auto-correlation of the chains to end the MCMC early.  The average luminosity $L$, FWHM, and velocity offset $v_{\rm off}$ of each line from Table \ref{tab:lines} are all calculated from our MCMC posteriors, and \voff\ is measured with respect to the stellar velocity in region 5: $v_{\rm off} = v_{\rm off, SDSS} - v_{\rm stel}$.

Finally, after fitting each coronal line, we impose the constraint that the fitted flux must be significant to at least $\geqslant 3\sigma$. We then perform a final round of visual inspection to confirm each detection.  A summary of the characteristics and detection statistics of each line we analyzed is presented in Table \ref{tab:lines}.


\section{Catalog Statistics}
\label{sect:results}

\subsection{\rev{Line Detection Statistics}}
\begin{figure}
    \centering
    \includegraphics[width=\columnwidth]{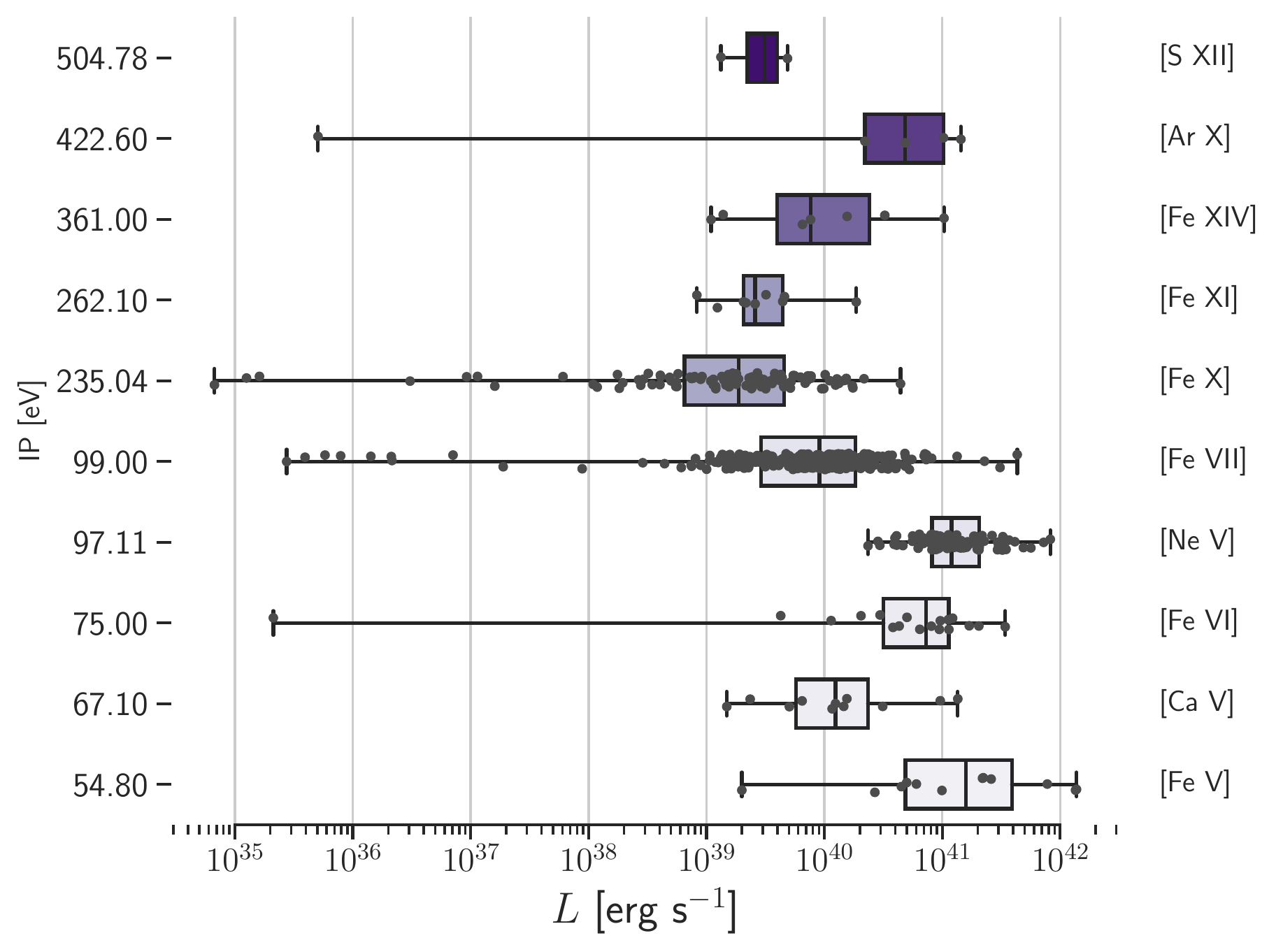}
    \includegraphics[width=\columnwidth]{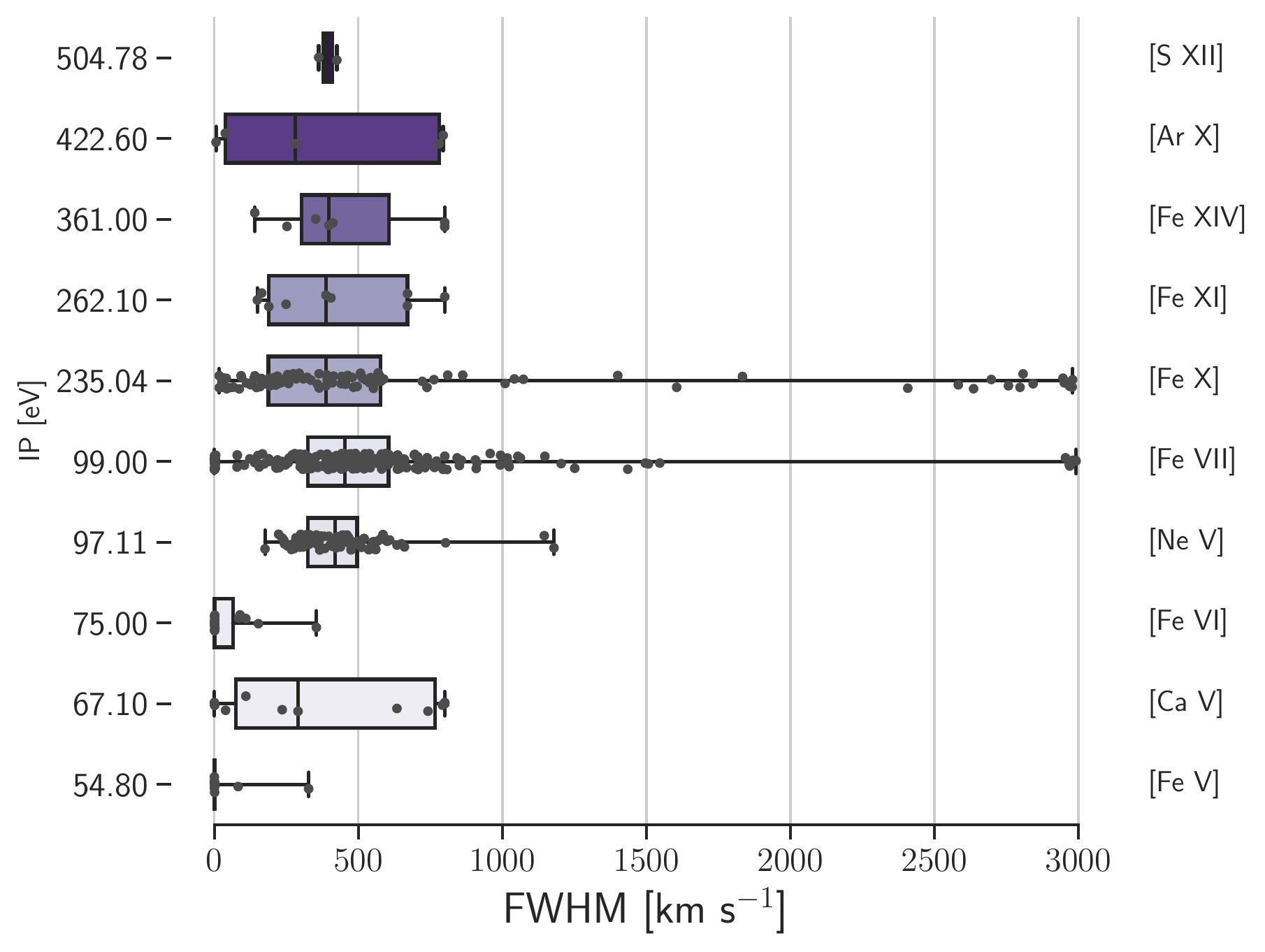}
    \includegraphics[width=\columnwidth]{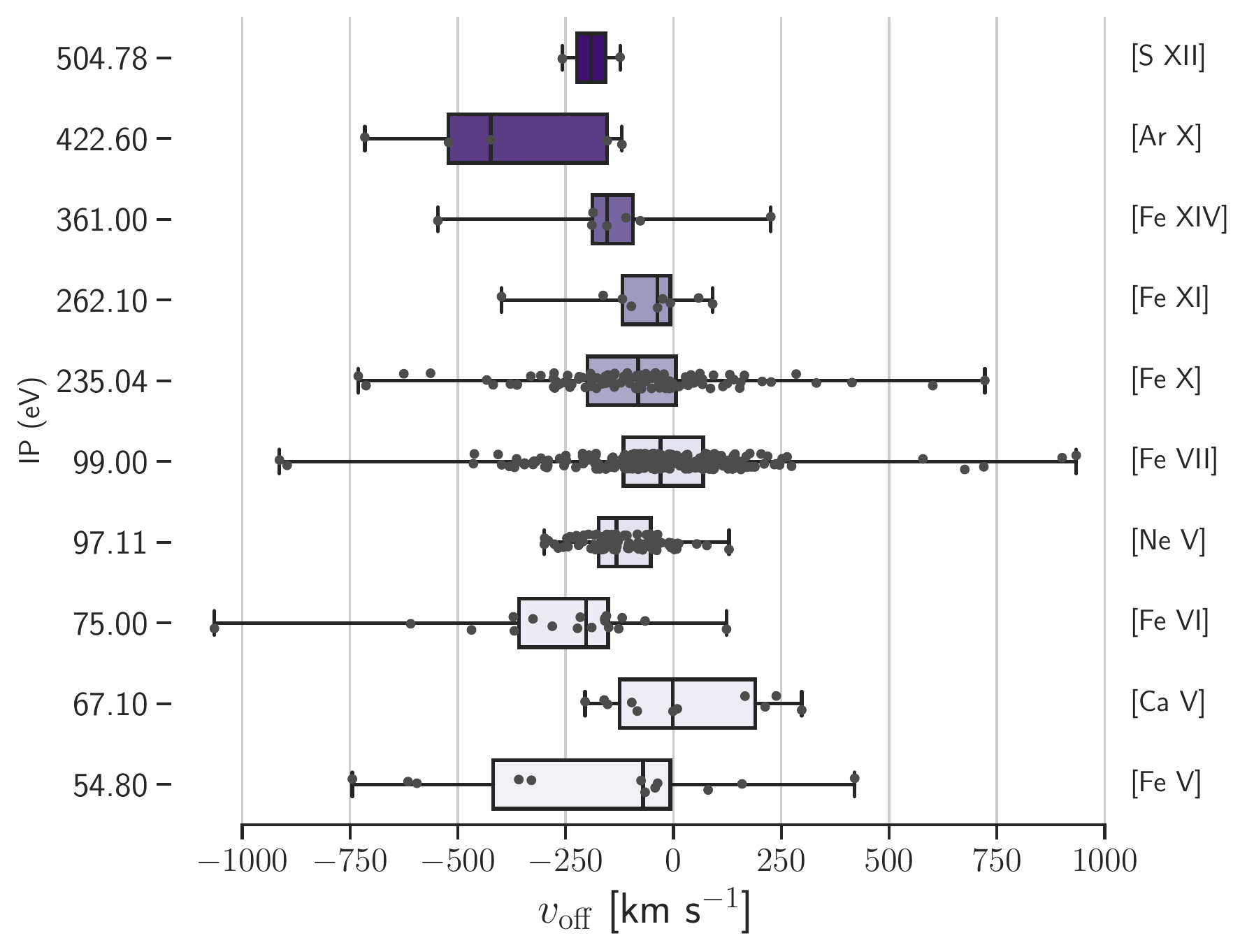}
    \caption{Distributions of each coronal line over the luminosity ($L$), FWHM, and velocity shift (\voff) organized by the IP in eV.  The luminosity is in \ergs, and FWHM and \voff\ are in \kms. Each line is annotated on the right side of the corresponding IP bin.  The colors also indicate the ionization potential, with more saturated purples corresponding to higher IPs. Note that the vertical axes are not scaled linearly; each ionization potential is shown categorically.}
    \label{fig:lum_dist_ip}
\end{figure}

\begin{figure}
    \centering
    \includegraphics[width=\columnwidth]{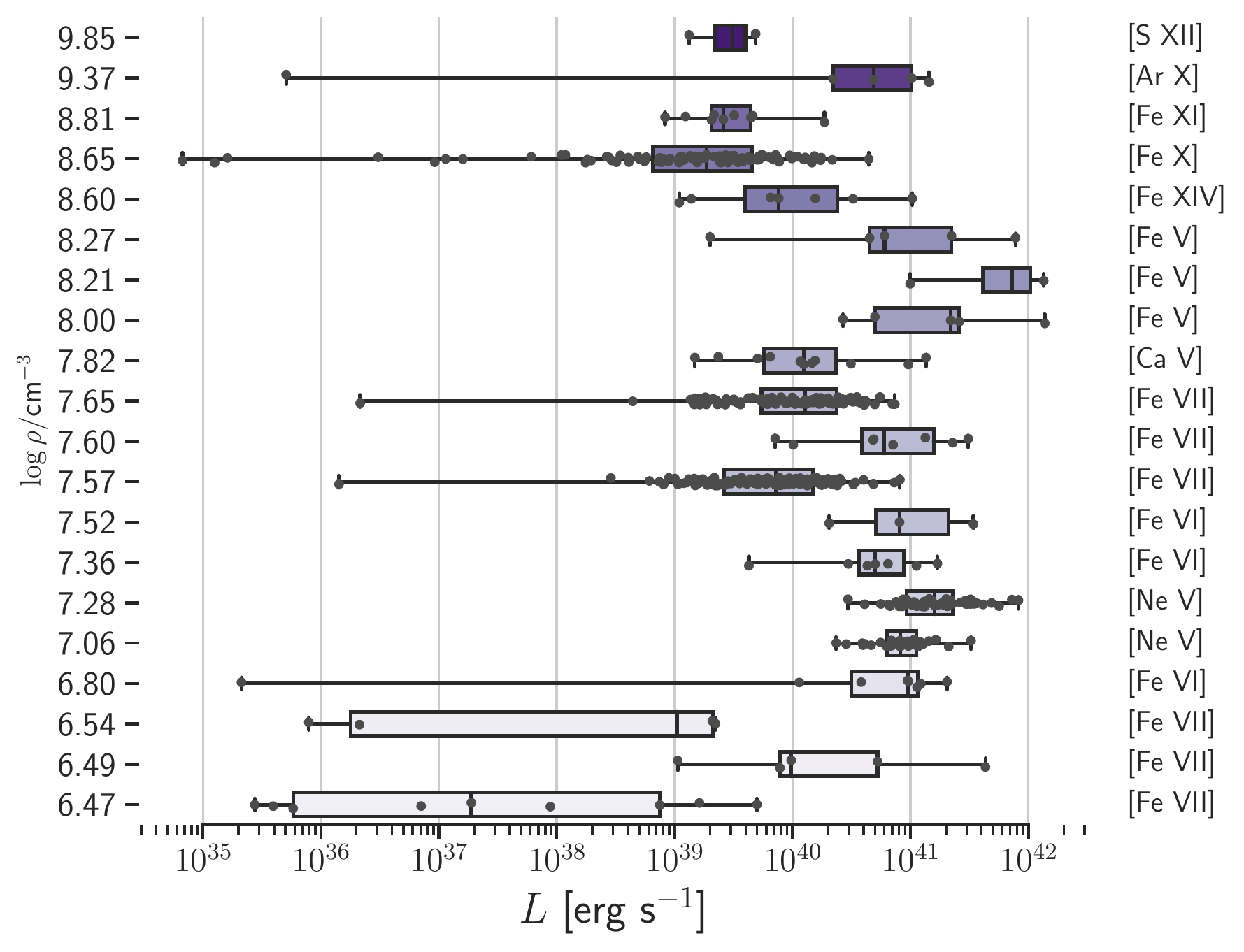}
    \includegraphics[width=\columnwidth]{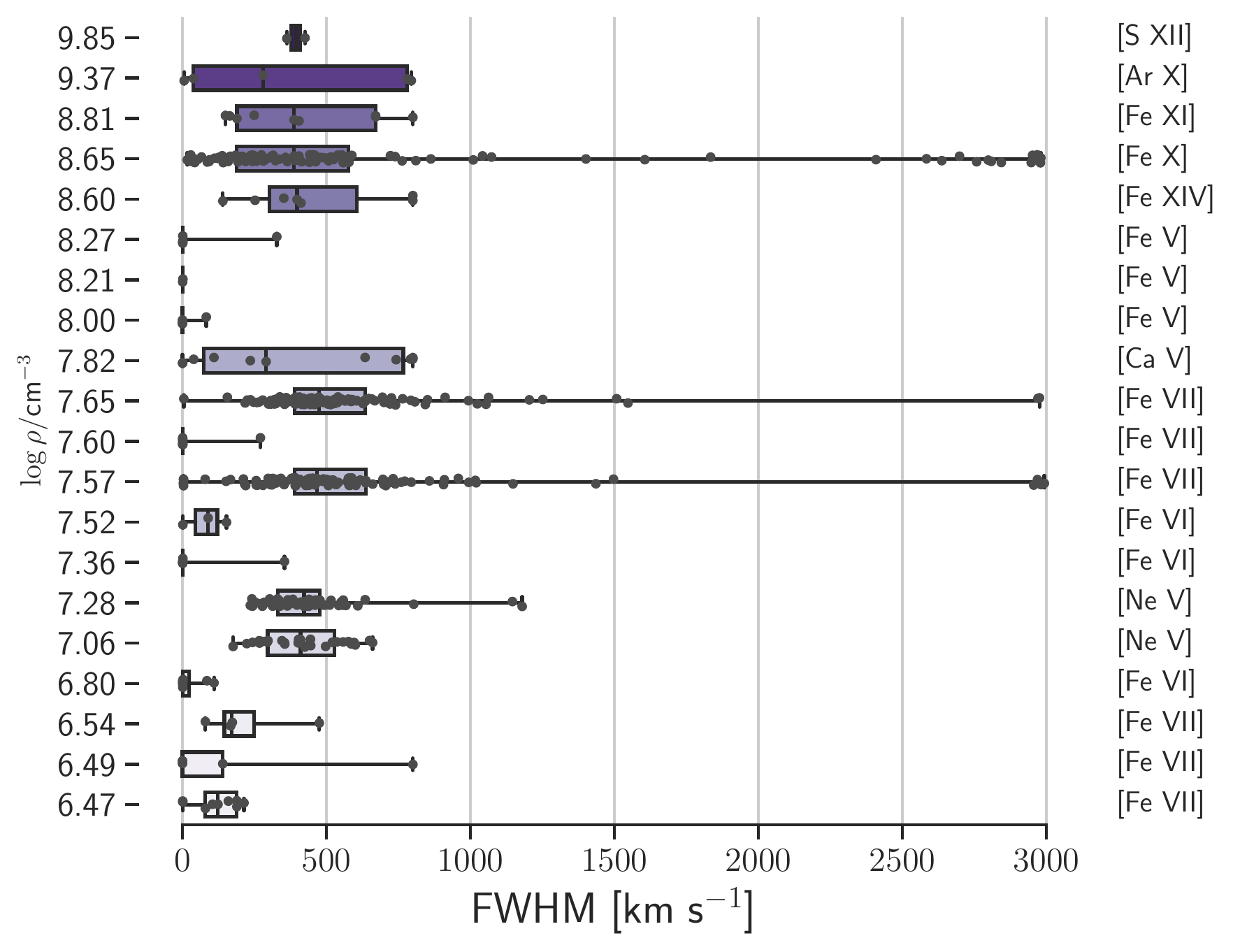}
    \includegraphics[width=\columnwidth]{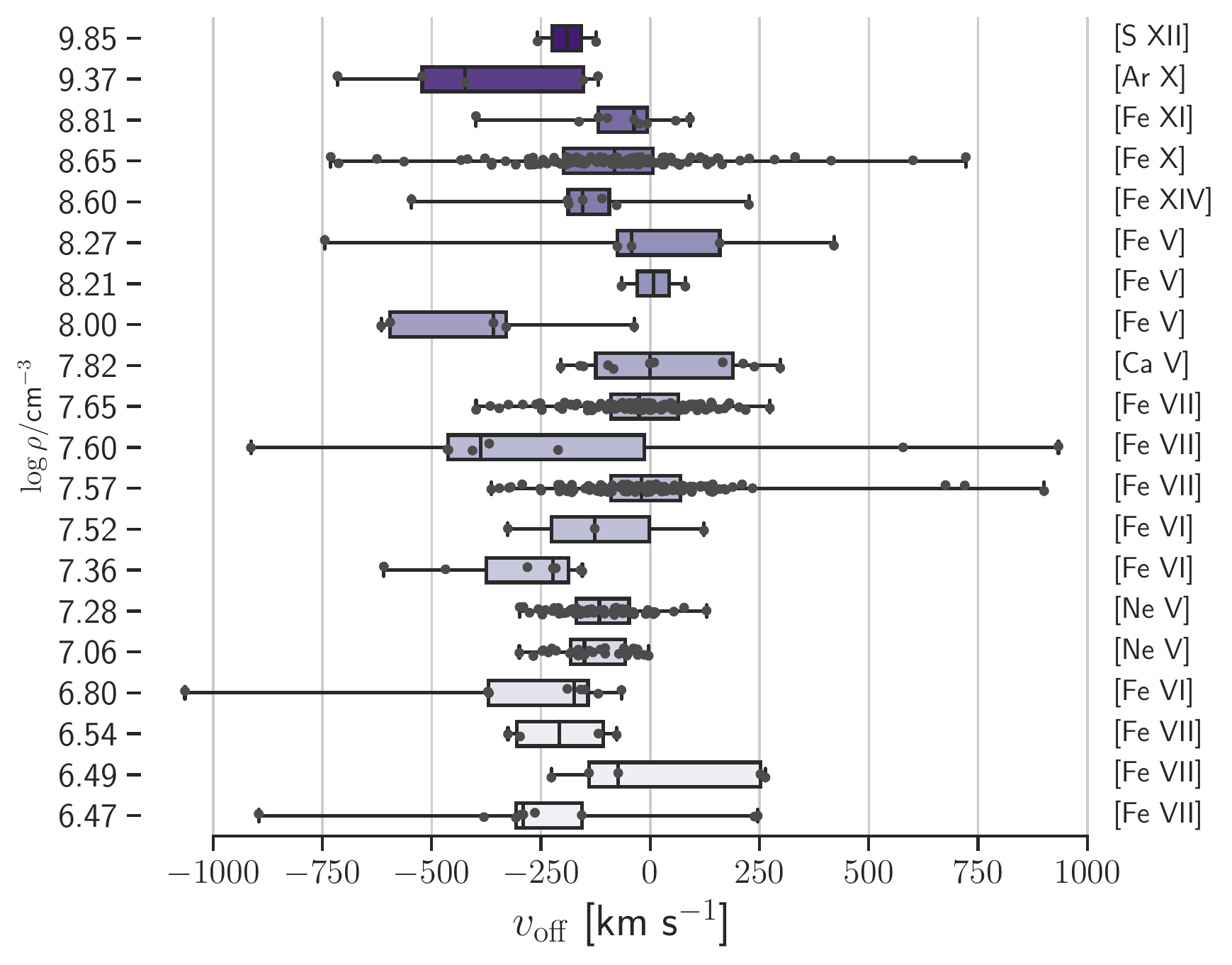}
    \caption{Distributions of each coronal line over the luminosity ($L$), FWHM, and velocity shift (\voff) organized by the critical density in cm$^{-3}$.  The luminosity is in \ergs, and FWHM and \voff\ are in \kms. Each line is annotated on the right side of the corresponding $\log\rho$ bin.  The colors also indicate the critical density, with more saturated purples corresponding to higher $\rho$. Note that the vertical axes are not scaled logarithmically; each critical density is shown categorically.}
    \label{fig:lum_dist_cd}
\end{figure}

Our finalized survey contains 258 spectra with coronal line detections, \rev{a recovery of $\sim$60\% of our pre-selection sample.  The other $\sim$40\% of the pre-selection sample that did not pass our rigorous fitting procedures is likely due to a combination of 1) $\mathcal{F}$ ratios from [\ion{Fe}{10}] $\lambda$6374 being biased by its proximity to the [\ion{O}{1}] $\lambda$6365 line, and 2) spectra with missing data at wavelengths near the lines in question}. As shown in Table \ref{tab:lines}, our detection statistics \rev{for individual lines} range from 0.00058\% at the low end to 0.0407\% at the high end, with the percentages being out of the overall MPA catalog.  Coronal line emission is thus extremely rare in all cases given the sensitivity of SDSS.  These detection percentages take into consideration the flux threshold of each spectrum---we eliminate any spectrum with a flux threshold greater than 1/3 of the average flux of the coronal line in question, which removes any spectra that are too noisy to detect the line in question up to 3$\sigma$ significance.  This accounts for biases in spectra $S/N$. \rev{We also report alternative, more conservative estimates of the detection percentages that use the maximum coronal line flux instead of the average coronal line flux when comparing to the flux threshold of each spectrum. The error bounds on both of these percentage columns are 68\% confidence intervals calculated using binomial statistics.} The flux thresholds themselves are calculated by integrating a Gaussian distribution with amplitude equal to the RMS of the spectrum around the line and width equal to the instrumental FWHM dispersion at the line. \rev{These are then converted to luminosities using the standard luminosity distance $d_L$, and the averages are reported in Table \ref{tab:lines} along with the detection percentages.}  

Comparing our 258 coronal line detections to the [\ion{Fe}{10}] $\lambda$6374-emitting dwarfs found by \citet{2021ApJ...922..155M} and \citet{10.1111/j.1365-2966.2009.14961.x}, we find that 16 and 59 of their detections, respectively, were also listed in the MPA/JHU catalog. Despite this, we only detected 1 of these galaxies from \citet{2021ApJ...922..155M} and 3 from \citet{10.1111/j.1365-2966.2009.14961.x} as [\ion{Fe}{10}] emitters using our methods. This is likely due to our stricter pre-selection filtering methods, which would eliminate borderline detections that would have been identified by a more lenient algorithm.  However, we also find many coronal line emitting spectra that were not found by \citet{2021ApJ...922..155M} or \citet{10.1111/j.1365-2966.2009.14961.x}, not only because of our larger starting sample, but also by virtue of the 19 other coronal lines that we searched for.  In fact, $\sim 60$\% of our coronal line detections do not exhibit any [\ion{Fe}{10}] emission whatsoever, demonstrating that coronal line searches based on a single or limited set of coronal lines will be highly incomplete. 

\rev{Many of our final coronal line detections with $S/N \geqslant 3$ do not have $\mathcal{F} \geqslant 4$. This is because we fit all coronal lines in any spectrum that had \textit{at least one} line with an $\mathcal{F} \geqslant 4$, since the detection of one coronal line increases the likelihood that more than one coronal line is present in the spectrum. However, these $\mathcal{F}$ ratios demonstrate that with our filtering scheme of $\mathcal{F} \geqslant 4$, we may be excluding additional coronal line detections that were not identified in our screening procedure.}

\subsection{\rev{Redshift Distribution}}

\begin{figure}
    \centering
    \includegraphics[width=\columnwidth]{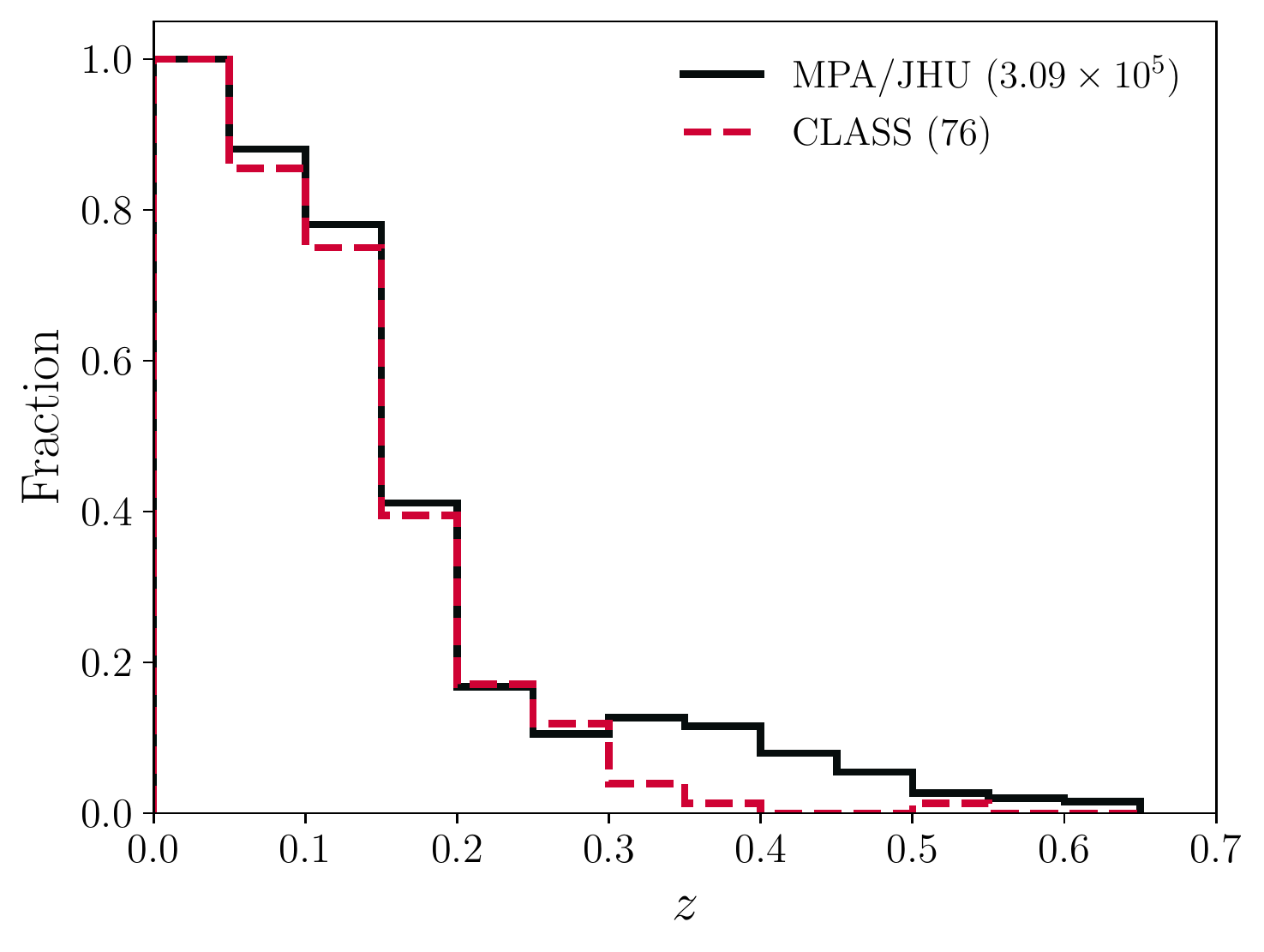}
    \caption{\rev{Redshift distribution of the MPA/JHU catalog and CLASS, normalized such that the maximum is 1. The normalization constants for both distributions are shown in the legend.}}
    \label{fig:redshift}
\end{figure}

\rev{Shown in Figure \ref{fig:redshift} is the redshift distribution of CLASS with respect to the overall MPA/JHU catalog, both of which have been rescaled. We see that generally the two samples follow the same shape, but the CLASS survey \rrev{largely drops off by around} $z = 0.3$, \rrev{with only a handful (6/258) of sources having $z > 0.3$,} while the MPA/JHU catalog steadily falls off until upwards of $z \sim 0.65$. This is to be expected considering the increased difficulty of measuring emission lines with significant $S/N$ at higher redshifts, and the reddest optical coronal lines will begin being redshifted out of the SDSS wavelength range at as low as $z = 0.166$.}

\subsection{\rev{Line Properties}}

As can be seen from Table \ref{tab:lines}, average coronal line luminosities range from $10^{38}$--$10^{41}$ \ergs, with FWHMs typically $>$100 \kms, and \voff\, typically $< 0$ \kms.  Histograms of these properties are shown in Figure \ref{fig:lum_dist_ip} and \ref{fig:lum_dist_cd}, organized vertically by the IP and critical density ($\rho_{\rm crit}$), respectively. As can be seen, the [\ion{Ne}{5}] $\lambda\lambda$3346,3426 doublet displays the largest luminosities with little dynamic range seen in the sample, which explains its prominence.  [\ion{Fe}{10}] $\lambda$6374 and [\ion{Fe}{7}] $\lambda\lambda$5720,6087 are luminous and commonly detected as well. 


Most lines show a wider distribution in FWHMs from 0--800 \kms, while a few clustered near 0 hint at being near the lower FWHM resolution threshold for SDSS. Unfortunately our limited sample size precludes us from investigating any trends in line width with IP or critical density. However, an emergent, robust trend seen in the \voff\ is that all line distributions are at least slightly blueshifted relative to the stellar velocity.  This is in agreement with smaller coronal line surveys in the optical and NIR \citep[such as][]{2006ApJ...653.1098R,2011ApJ...743..100R, 2011ApJ...739...69M}.  Aside from this, however, there are no obvious correlations in $L$, FWHM, or \voff\ with respect to the IP or $\rho_{\rm crit}$ of each line. 

\begin{figure*}
    \centering
    \includegraphics[width=\columnwidth]{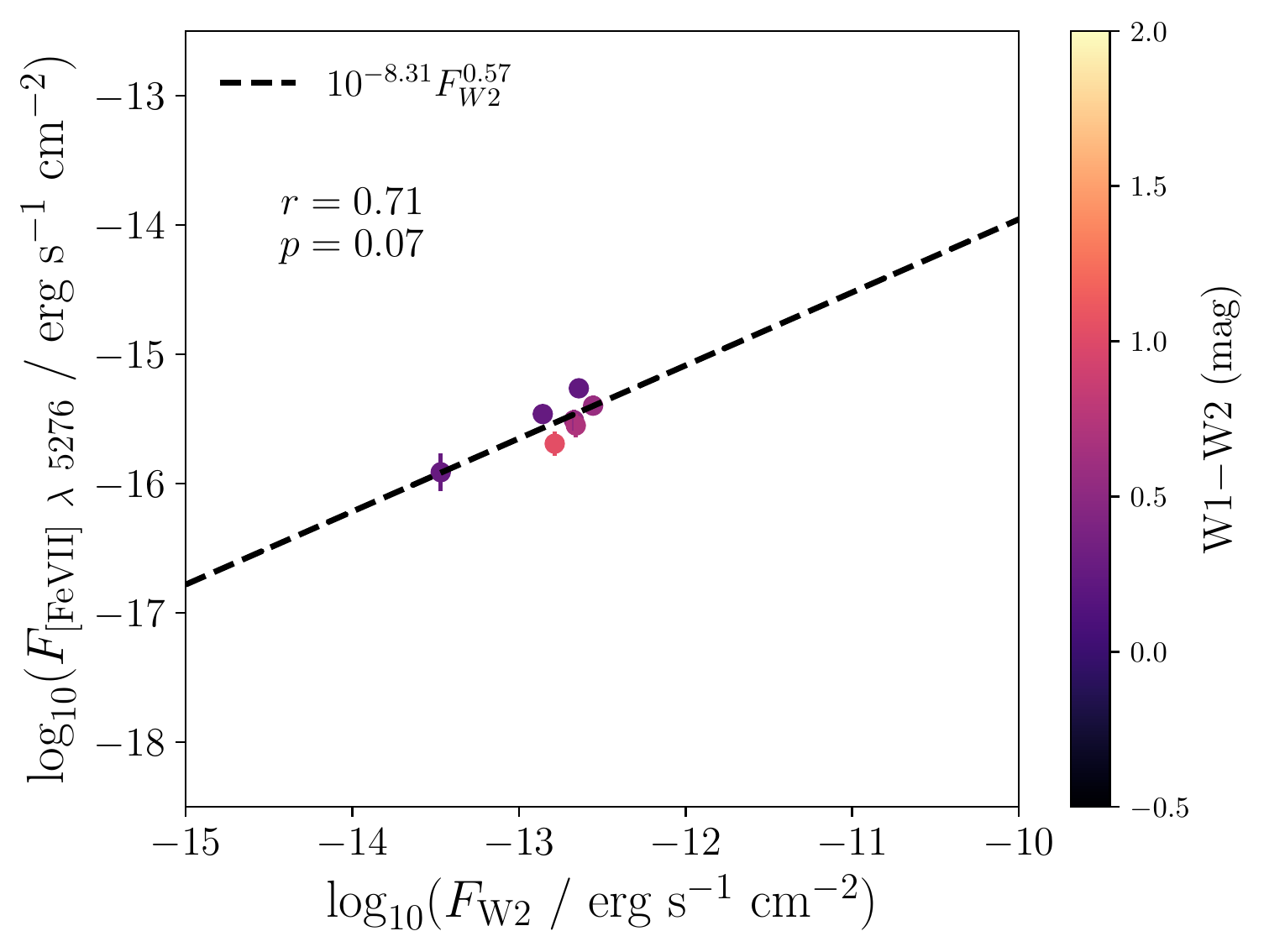}
    \includegraphics[width=\columnwidth]{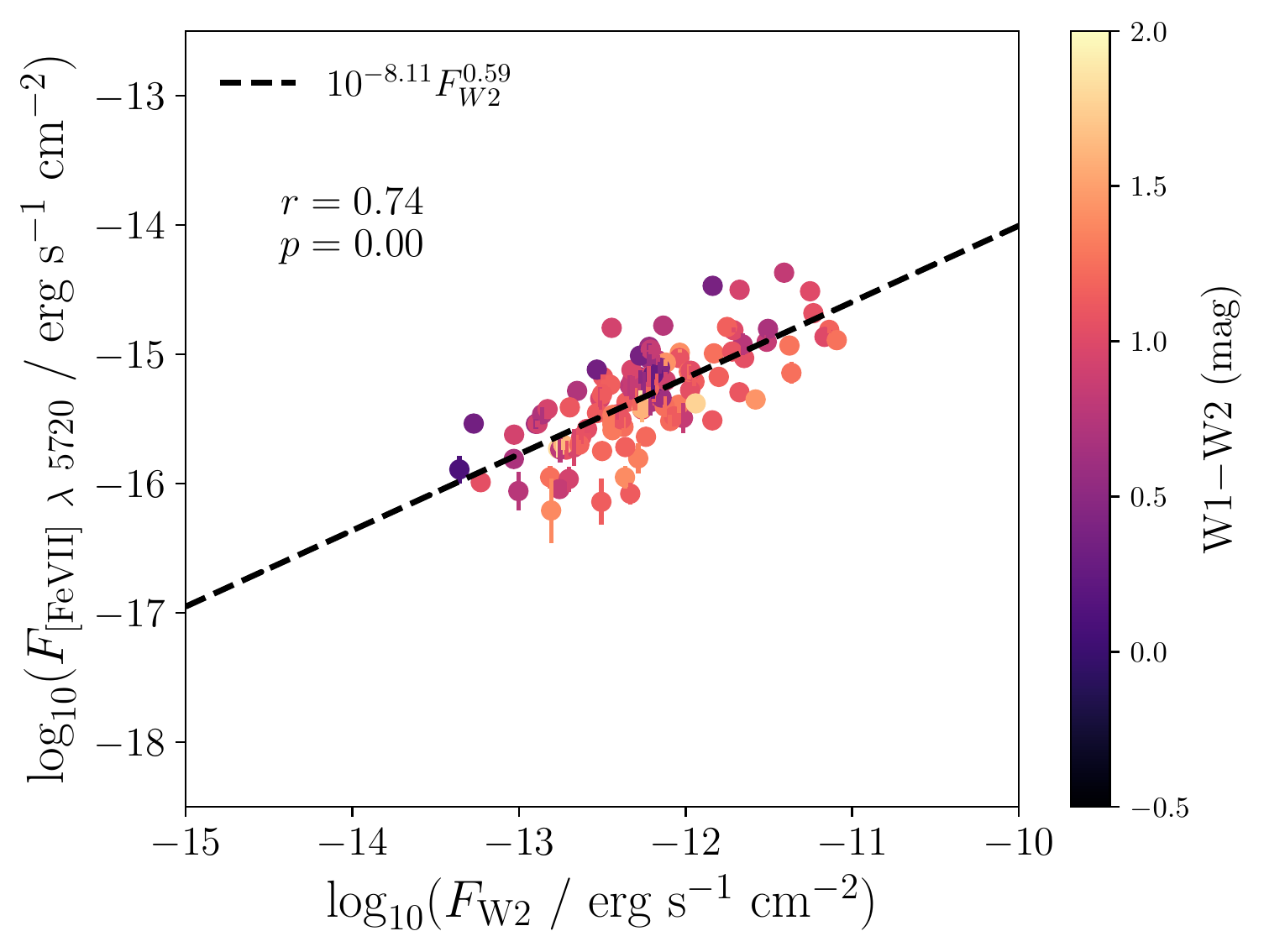}
    \includegraphics[width=\columnwidth]{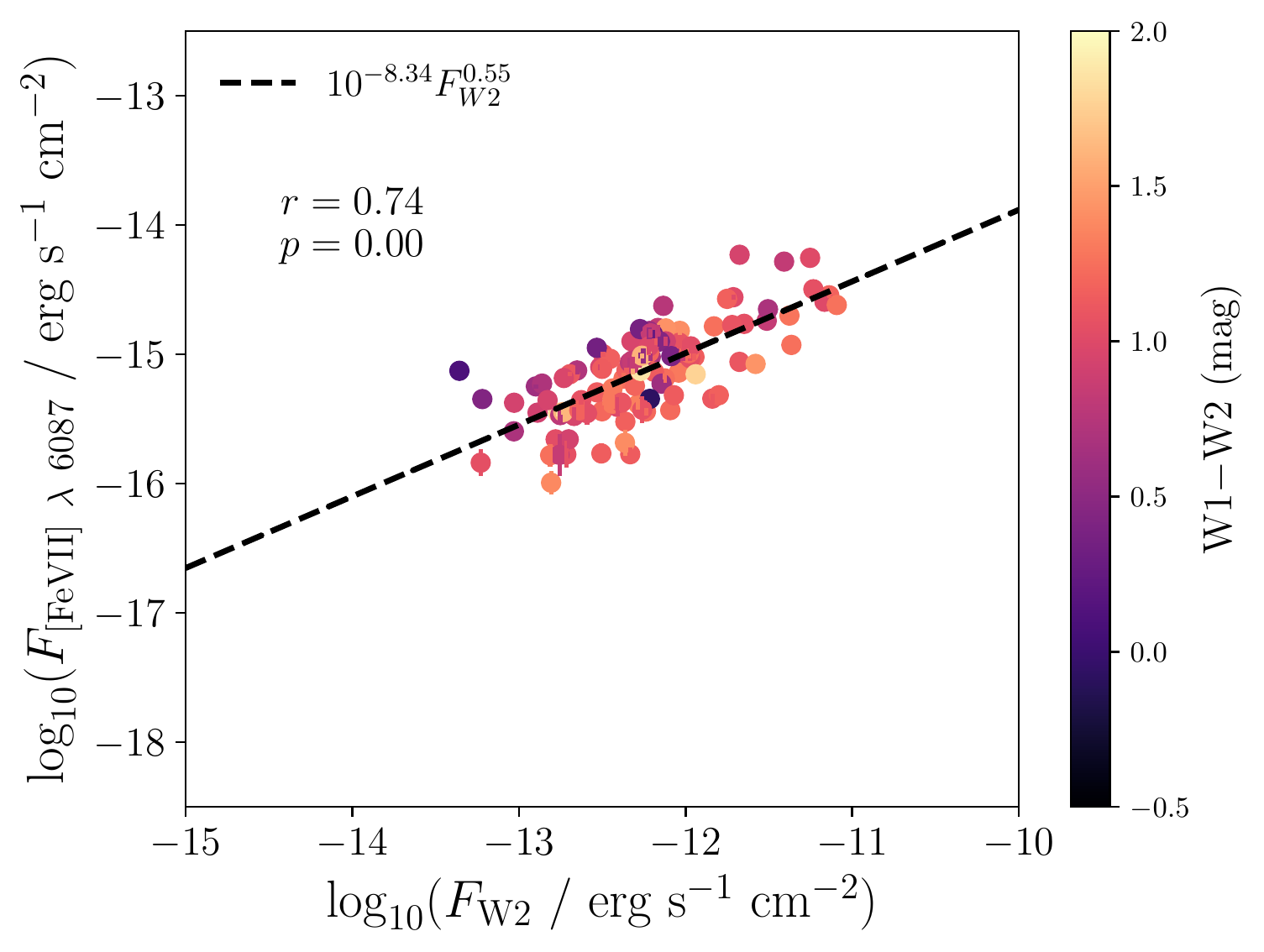}
    \includegraphics[width=\columnwidth]{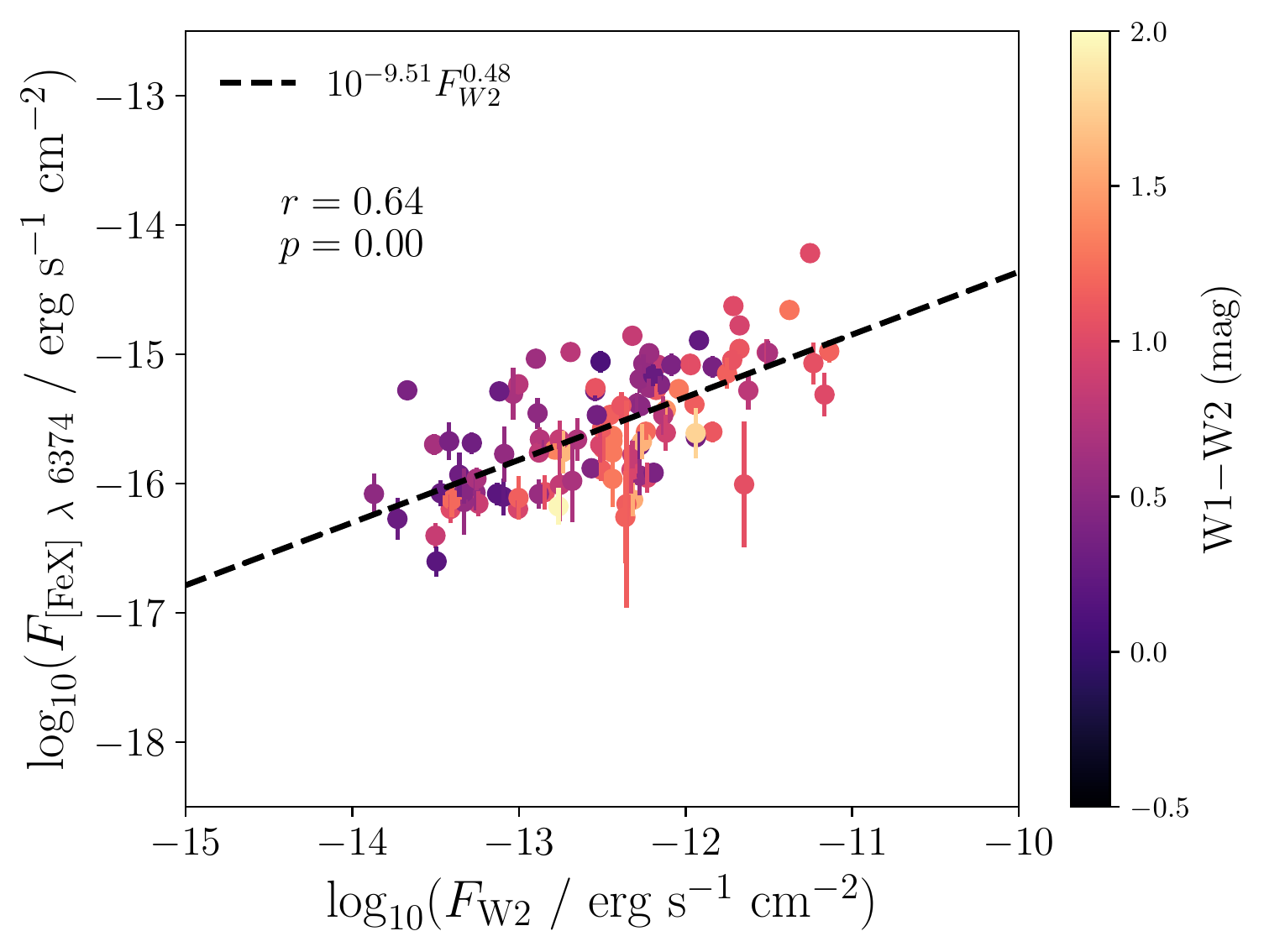}
    \includegraphics[width=\columnwidth]{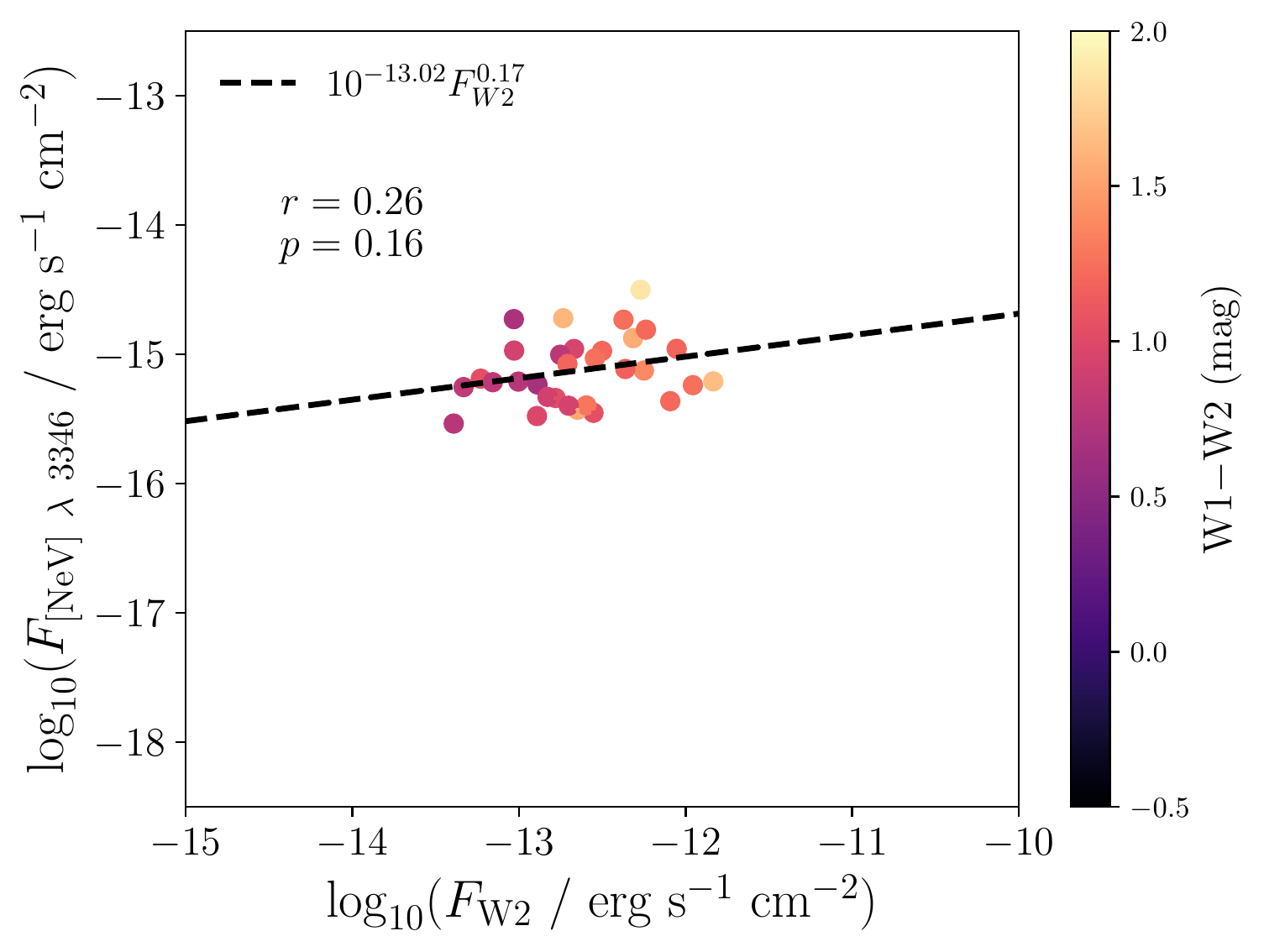}
    \includegraphics[width=\columnwidth]{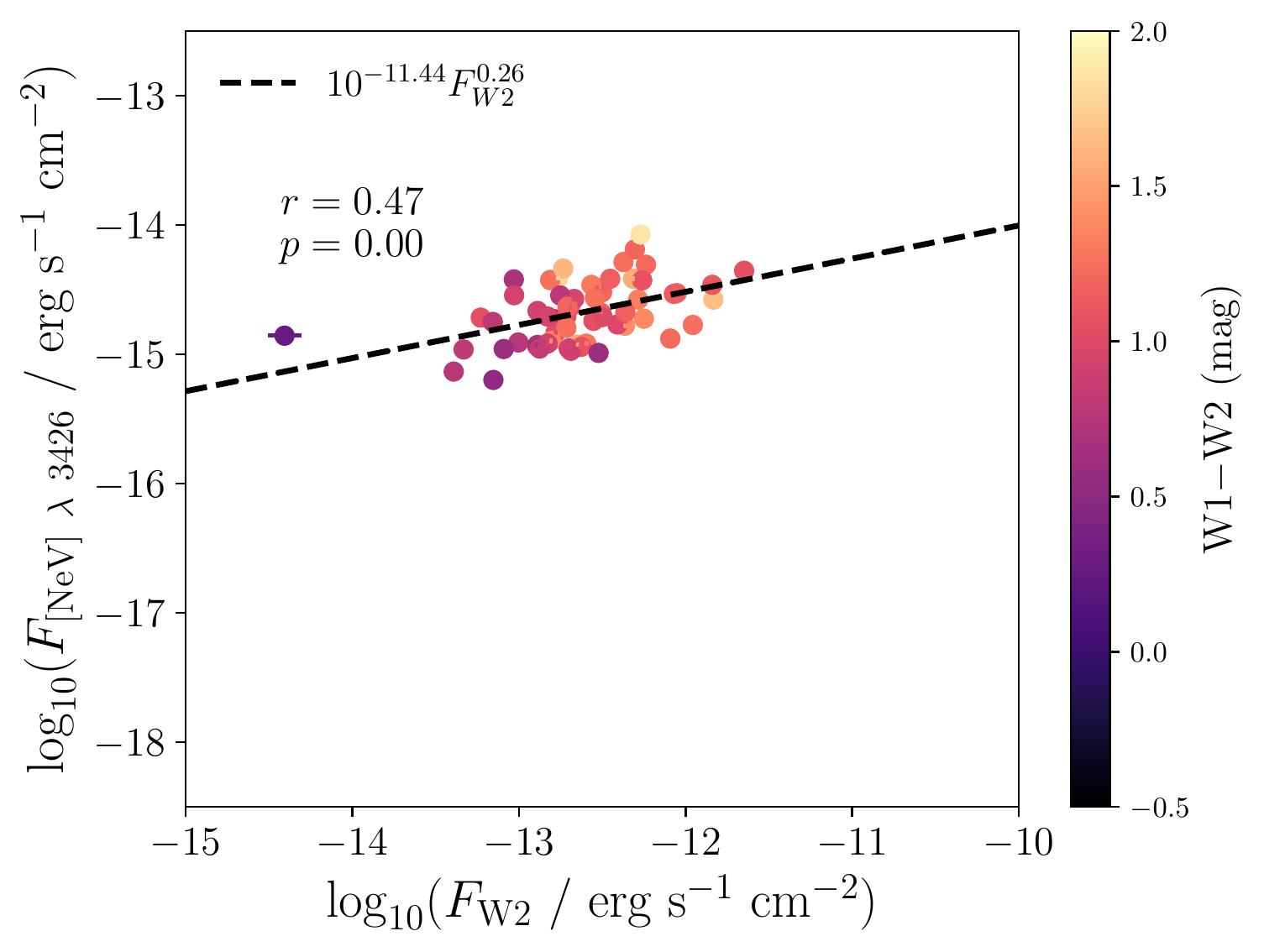}
    \caption{A series of logarithmic scatterplots showing correlations between individual coronal line \rev{fluxes} and the WISE 2 band \rev{fluxes} of the host galaxy. Each panel has identical vertical and horizontal scales. Best-fit power laws are plotted with dashed lines over the data, and the best-fit parameters are shown in the legend of each plot as a function of $F_{\rm W2}$, the W2 \rev{flux}. The points are colored according to the WISE 1 $-$ WISE 2 color (in magnitudes). Each panel lists the Spearman correlation coefficient $r$ and corresponding $p$-value of the distribution.}
    \label{fig:w2_lum}
\end{figure*}

In Figure \ref{fig:w2_lum}, we examine a correlation between individual coronal line \rev{fluxes} and host galaxy \rev{fluxes} in the WISE 4.6 $\mu$m (W2) band.  The W2 flux is calculated from the W2 magnitude using
\begin{equation}
    F_{W2} (\text{erg s}^{-1}\text{cm}^{-2}) = 171.787 \times 10^{-m_{W2}/2.5 - 23} \Delta\nu
\end{equation}
\citep{2012wise.rept....1C}, where $m_{W2}$ is the magnitude in the W2 band and $\Delta\nu$ is the width of the W2 passband ($\sim 1.4653 \pm 0.0018 \times 10^{13}$ Hz). A best-fit power law is shown as a dashed line in each panel, fit using the least squares method with SciPy. The power-law slopes of each best-fit line shown fall relatively close to each other, all \rev{between 0--1}.  Most of the other lines not shown in Figure \ref{fig:w2_lum} either have too few data points, or their data's W2 \rev{fluxes} cover too narrow of a range to obtain robust power law fits.  
The Spearman correlation coefficients ($r$) and $p$-values annotated in each plot quantify the strength of the linearity of each flux-flux relationship. Thus, each coronal line appears to follow a similar power law relationship, the most robust of which are also enumerated in Table \ref{tab:w2_fits}.  These relationships also seem to be independent of W1$-$W2 color, as indicated by the colorbars in each panel, suggesting that the relationships presented here can potentially be used to predict the coronal line strengths in galaxies that are not identified as AGNs based on their mid-infrared color.
\begin{table}[]
    \centering
    \begin{tabular}{lcccc}
        \hline
        Line & $\alpha$ & $\beta$ & $r$ & $p$ \\
        \hline
        \lbrack \ion{Fe}{11}\rbrack $\lambda$7892 & 0.16 & 13.25 & 0.32 & 0.41 \\
        \lbrack \ion{Fe}{10}\rbrack $\lambda$6374 & 0.48 & 9.51 & 0.64 & 0.00 \\
        \lbrack \ion{Fe}{7}\rbrack $\lambda$6087 & 0.55 & 8.34 & 0.74 & 0.00 \\
        \lbrack \ion{Fe}{7}\rbrack $\lambda$5720 & 0.59 & 8.11 & 0.74 & 0.00 \\
        \lbrack \ion{Ca}{5}\rbrack $\lambda$5309 & 0.10 & 14.34 & 0.15 & 0.68 \\
        \lbrack \ion{Fe}{14}\rbrack $\lambda$5303 & 0.19 & 13.06 & 0.46 & 0.29 \\
        \lbrack \ion{Fe}{7}\rbrack $\lambda$5276 & 0.57 & 8.31 & 0.71 & 0.07 \\
        \lbrack \ion{Ne}{5}\rbrack $\lambda$3426 & 0.26 & 11.44 & 0.47 & 0.00 \\
        \lbrack \ion{Ne}{5}\rbrack $\lambda$3346 & 0.17 & 13.02 & 0.26 & 0.16 \\
        \hline
    \end{tabular}
    \caption{Power law fitting results for each coronal line, \rev{where $\log F_{CL} = \alpha \log F_{W2} - \beta$.}  The Spearman coefficients ($r$) and $p$-values for each line are also shown. We only show lines that have at least 5 datapoints.}
    \label{tab:w2_fits}
\end{table}

\begin{figure*}
    \centering
    \includegraphics[width=\columnwidth]{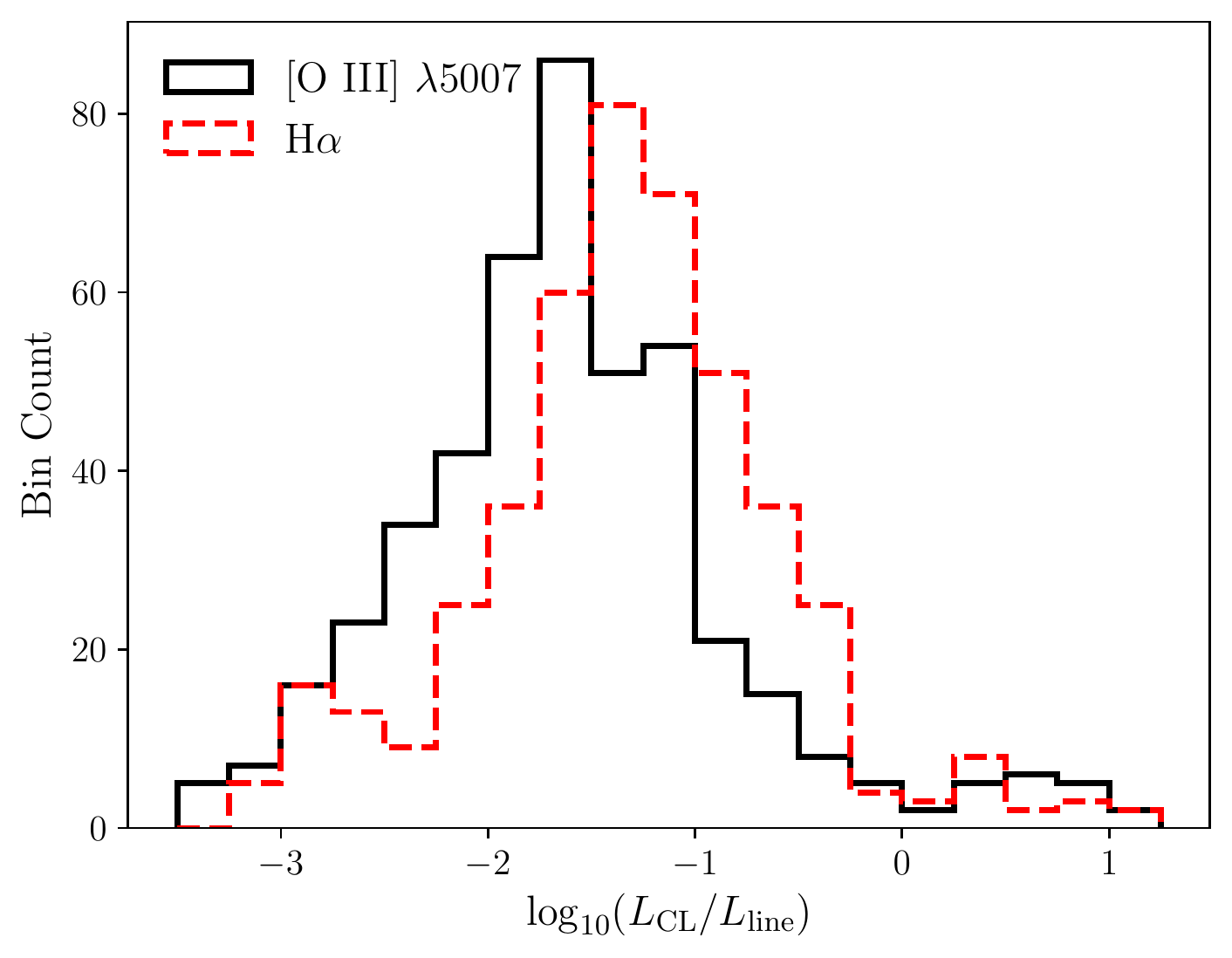}
    \includegraphics[width=\columnwidth]{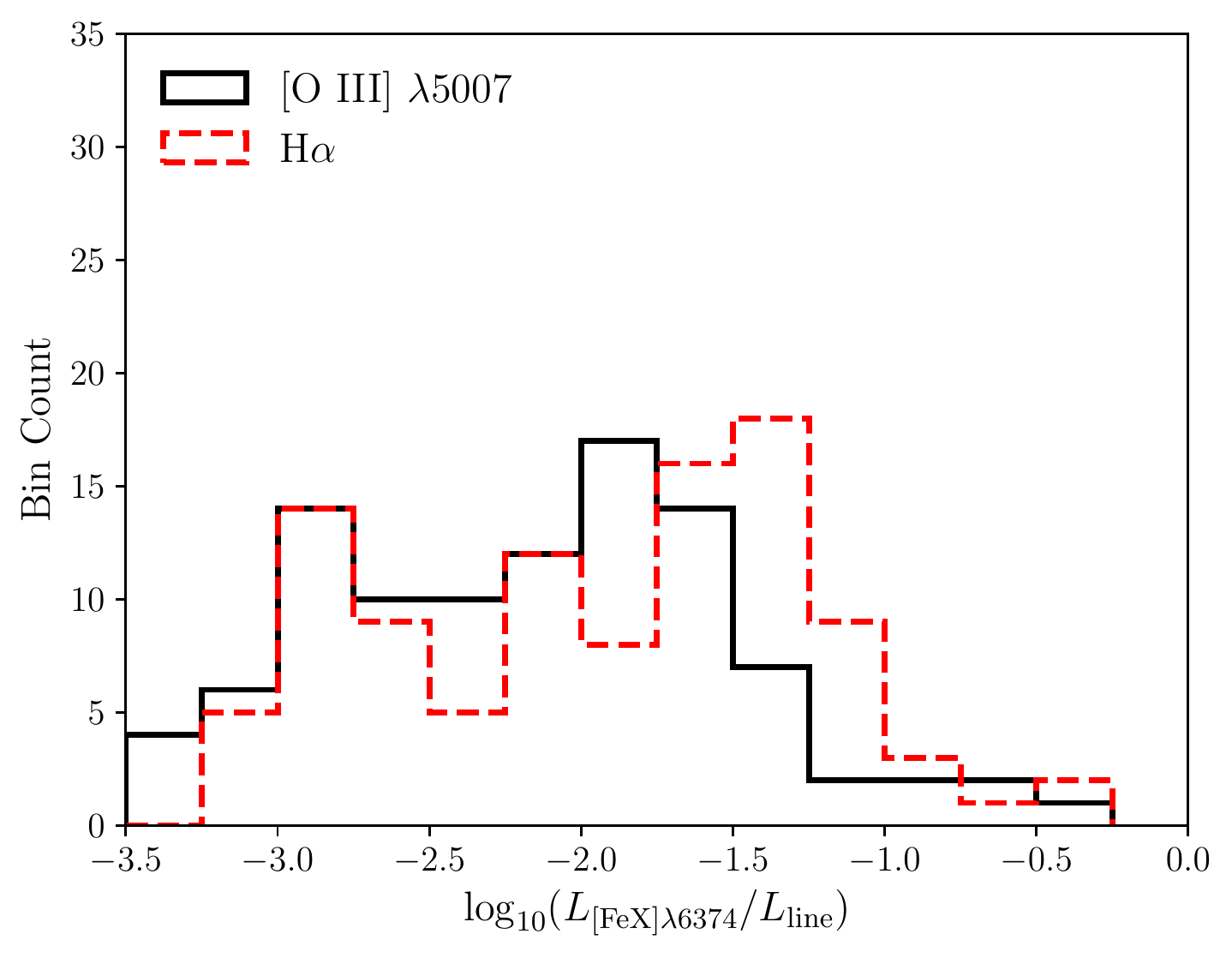}
    \includegraphics[width=\columnwidth]{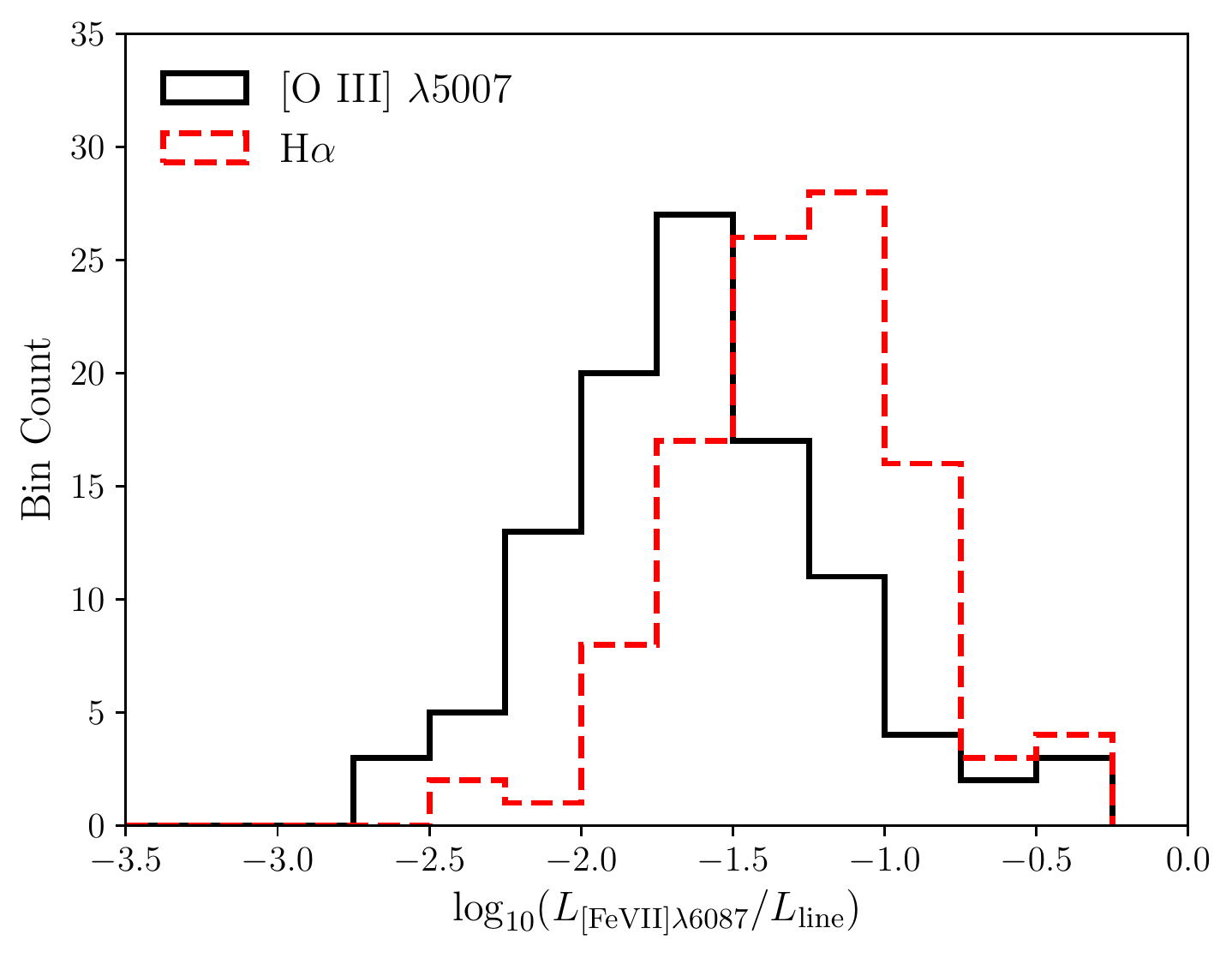}
    \includegraphics[width=\columnwidth]{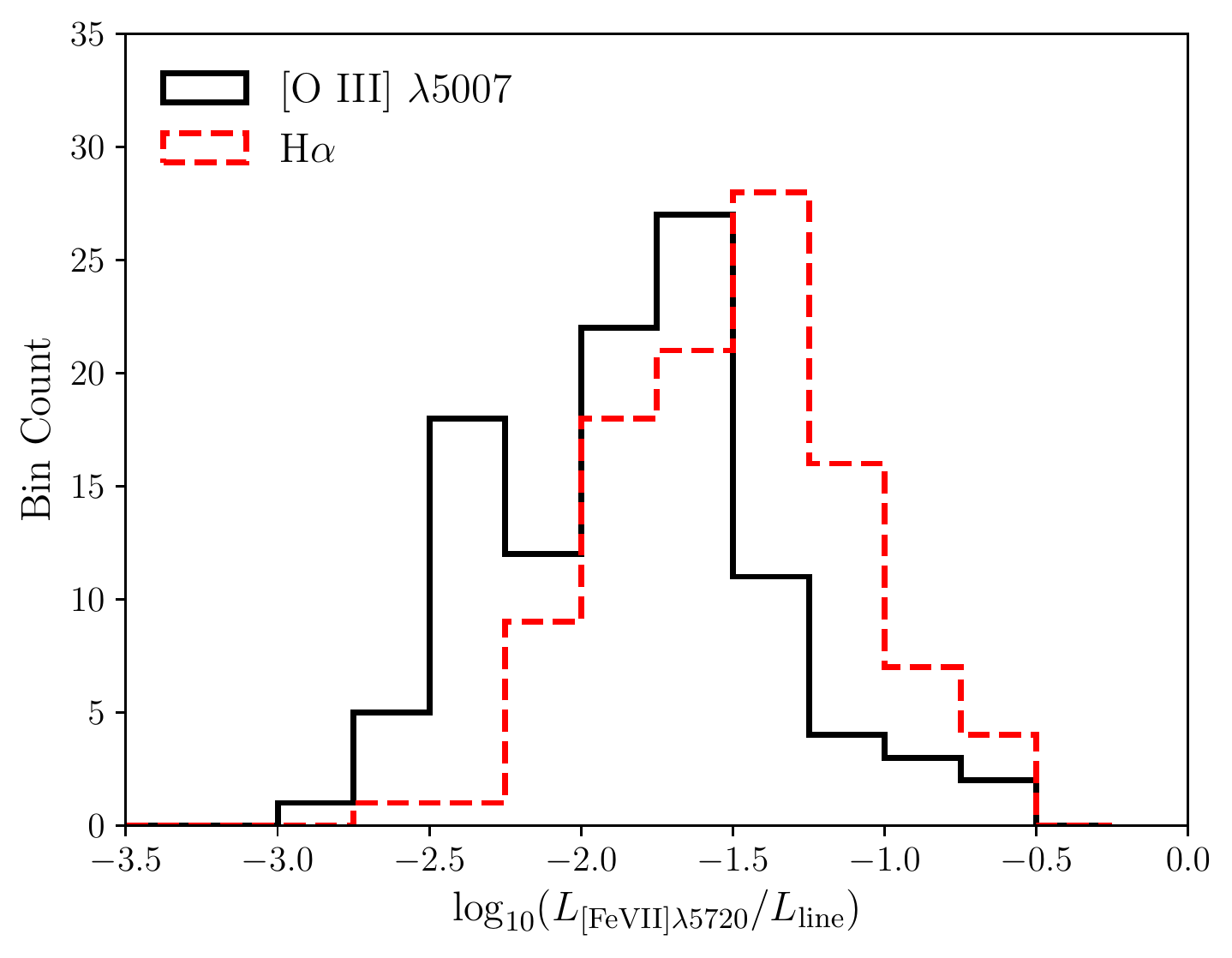}
    \includegraphics[width=\columnwidth]{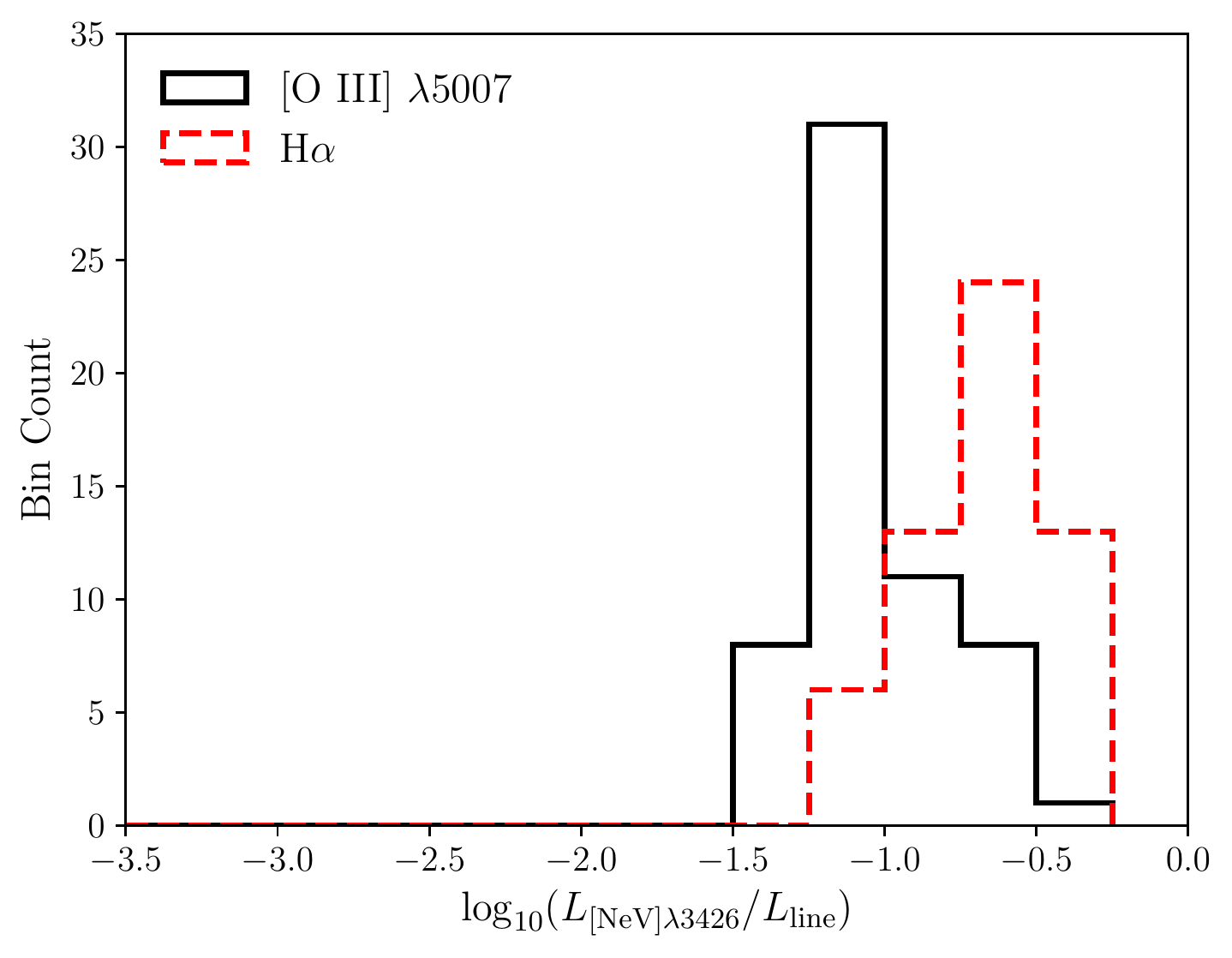}
    \includegraphics[width=\columnwidth]{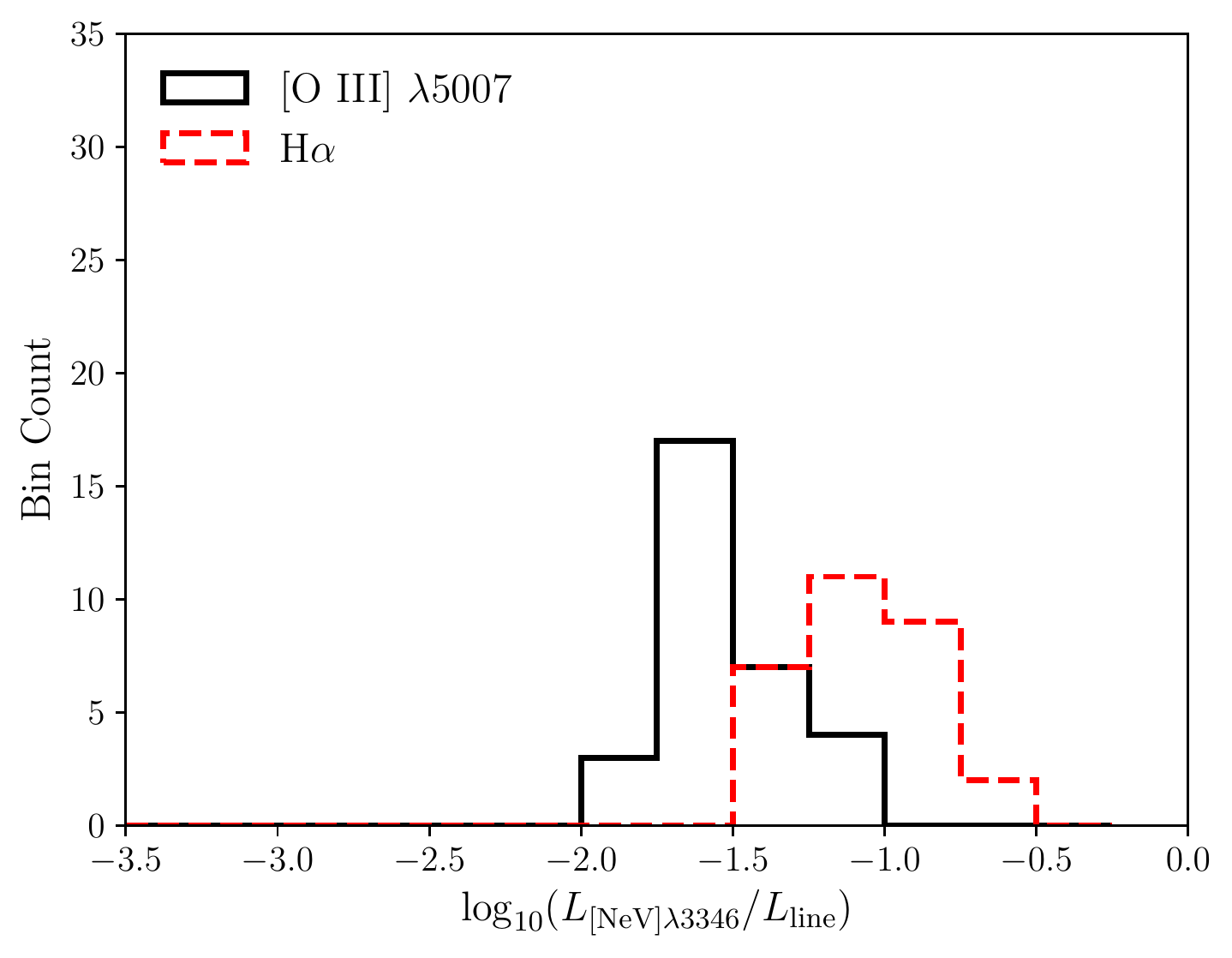}
    \caption{A series of histograms of various coronal line luminosity ratios, in logarithmic units. The solid black histograms show the luminosity ratios with [\ion{O}{3}] $\lambda$5007, while the red dashed histograms show H$\alpha$.  Each panel displays a different coronal line, labeled on the horizontal axes, with the exception of the first (top-left) panel, which shows a concatenation of all the coronal lines together.  The horizontal and vertical scales of each plot are identical, again with the exception of the top-left panel.}
    \label{fig:oiii_ha_ratios}
\end{figure*}

\begin{table}[]
    \centering
    \begin{tabular}{l|ccc|ccc}
        \hline
        & & [\ion{O}{3}] & & & H$\alpha$ & \\
        Line & Min & Med & Max & Min & Med & Max \\
        \hline
        \lbrack\ion{Fe}{11}\rbrack $\lambda$7892 & -2.09 & -1.10$^\dagger$ & 0.64$^\dagger$ & -2.07 & -1.01 & -0.47 \\
        \lbrack\ion{Fe}{10}\rbrack $\lambda$6374 & -10.17 & -2.08 & 0.96$^\dagger$ & -10.16 & -1.78 & -0.16 \\
        \lbrack\ion{Fe}{7}\rbrack $\lambda$6087 & -2.61 & -1.62 & 0.60$^\dagger$ & -2.37 & -1.26 & 0.09 \\
        \lbrack\ion{Fe}{7}\rbrack $\lambda$5720 & -2.78 & -1.83 & -0.16$^\dagger$ & -2.54 & -1.48 & -0.10 \\
        \lbrack\ion{Ca}{5}\rbrack $\lambda$5309 & -2.40 & -2.14 & -1.07$^\dagger$ & -2.06 & -1.76 & -1.68 \\
        \lbrack\ion{Fe}{14}\rbrack $\lambda$5303 & -1.60 & -0.58$^\dagger$ & 0.80$^\dagger$ & -1.69 & -1.09 & -0.32 \\
        \lbrack\ion{Fe}{7}\rbrack $\lambda$5276 & -3.34 & -2.74 & -2.66 & -2.81 & -2.63 & -2.50 \\
        \lbrack\ion{Ne}{5}\rbrack $\lambda$3426 & -1.43$^\dagger$ & -1.06$^\dagger$ & -0.46$^\dagger$ & -1.16 & -0.62 & -0.07 \\
        \lbrack\ion{Ne}{5}\rbrack $\lambda$3346 & -1.84 & -1.52$^\dagger$ & -1.06$^\dagger$ & -1.43 & -1.09 & -0.71 \\
        \hline
    \end{tabular}
    \caption{Minimum, median, and maximum flux ratios between each coronal line and [\ion{O}{3}] $\lambda$5007 and H$\alpha$, in log units. Only lines where we have at least 3 detections with an [\ion{O}{3}] or H$\alpha$ measurement at $S/N > 3$ are shown. A $\dagger$ signifies that the given [\ion{O}{3}] ratio would be detectable to an $S/N$ of 3 in the $z \sim 7.7$ spectrum obtained from JWST presented by \citet{2022arXiv220710034S} if the coronal line were present.}
    \label{tab:oiii_ha_ratios}
\end{table}

We next consider the flux ratios between each coronal line and the prominent H$\alpha$ and [\ion{O}{3}] $\lambda$5007 emission lines. In Figure \ref{fig:oiii_ha_ratios}, we show histograms of these various ratios, demonstrating that the flux ratios of the coronal lines with these optical strong lines span a wide dynamic range. Coronal line fluxes are typically a factor of $\sim$1000--2 times weaker than [\ion{O}{3}] and H$\alpha$ flux. The [\ion{Ne}{5}] $\lambda\lambda$3426,3346 doublet, shown in the bottom two panels, has notably higher flux ratios; the minimum flux ratio of the brighter 3426 line relative to the [\ion{O}{3}] line is $\sim$30.  Statistics on the minimum, median, and maximum fractions for each line, including some not shown in Figure \ref{fig:oiii_ha_ratios}, are given in Table \ref{tab:oiii_ha_ratios}.  The recent release and analysis of an extremely high redshift ($z \sim 7.7$) high signal-to-noise galaxy spectrum (06355) from JWST by \citet{2022arXiv220710034S} raises the question of how far we are still able to detect these coronal line signatures. \rev{The spectrum was gathered with the NIRSpec Micro-Shutter Assembly with the gratings and filters G235M/F170LP and G395M/F290LP, the latter of which we analyzed for the [\ion{O}{3}] line. The effective exposure time was 8,754 seconds.}  We estimate from this spectrum, using a simple Gaussian fit, an [\ion{O}{3}] $S/N \sim 103$, which would allow for coronal line detections up to $\sim34$ times weaker (or $-1.53$ in log units) to be detected with a $S/N$ of 3, given this exposure time .  Then, accordingly from our median CL/[\ion{O}{3}] ratios, we would expect to be able to detect [\ion{Fe}{11}] $\lambda$7892, [\ion{Fe}{14}] $\lambda$5303, or [\ion{Ne}{5}] $\lambda\lambda$3426,3346 a majority of the time when they are physically present, if the line ratios present in the local galaxies observed by SDSS are consistent with those present in high redshift galaxies, \rev{and assuming the same JWST configuration and sensitivity limit}.  Additionally, all of our maximum ratios exceed $-1.53$, except for [\ion{Fe}{7}] $\lambda$5276, meaning it would be possible to detect any of these lines in high redshift galaxies observed by JWST if they display line flux ratios relative to the [\ion{O}{3}] $\lambda$5007 line consistent with the maximum  observed in our local sample. \rev{The G395M/F290LP grating is one of the most sensitive configurations available for NIRSpec, therefore the detection of weak coronal lines in deep exposures of high redshift galaxies should be possible.}

\subsection{Line Detection Covariances}

\begin{figure*}
    \centering
    \includegraphics[width=\textwidth]{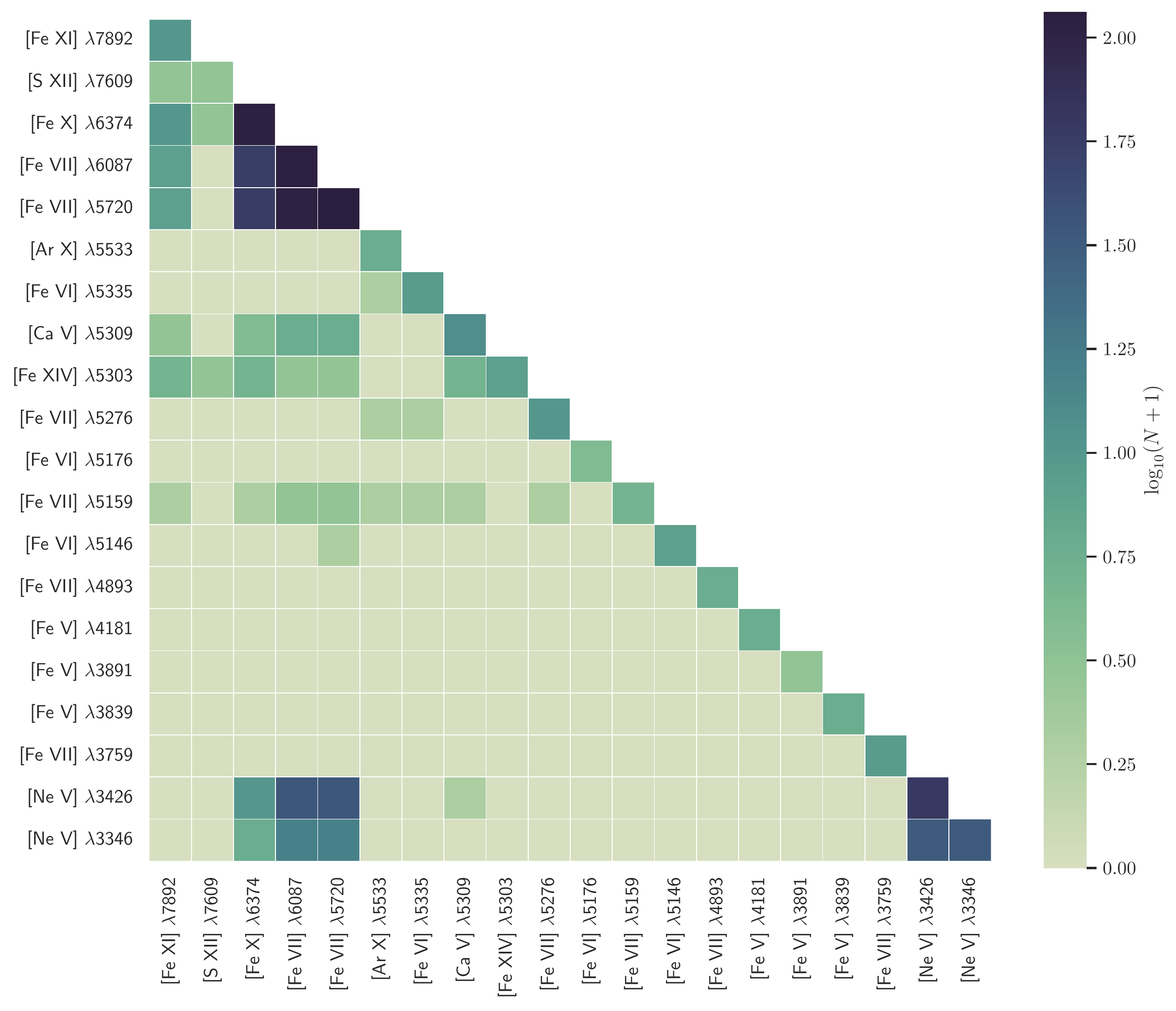}
    \caption{A 2D correlation plot between each pair of lines.  The color of each box represents the number of spectra in which both lines from the horizontal and vertical labels are found together.  Along the diagonal, the color instead represents the pure number of detections of that single line. The color bar is logarithmic, corresponding to $\log_{10}(N+1)$ where $N$ is the number of detections.}
    \label{fig:line_corr}
\end{figure*}

We next look at how often lines are detected together within the same spectrum.  Figure \ref{fig:line_corr} shows a correlation matrix for each pair of lines, where we notice three prominent peaks in the number of lines found together.  The first is around [\ion{Fe}{10}] $\lambda$6374 and [\ion{Fe}{7}] $\lambda\lambda$6087,5720.  The second is between the doublet [\ion{Ne}{5}] $\lambda\lambda$3426,3346.  And the third is between the two previous peaks, showing that the iron and neon lines are frequently found together. Worth noting is the fact that many high-IP lines are often \textit{not} found with other, lower IP lines from the same species.  A bigger sample size would be required to accurately quantify the correlations between the lines which currently have 0 detections.  A possible explanation for the relative lack of lines that are found together is the differences in critical densities. [\ion{Ar}{10}] for example is found in 5 galaxies, none of which show [\ion{Ca}{5}], [\ion{Fe}{6}], [\ion{Fe}{7}], or [\ion{Fe}{10}], all of which have lower IPs.  However, [\ion{Ar}{10}] also has one of the highest $\rho_{\rm crit}$ of any line, at $2.36 \times 10^9$ cm$^{-3}$.  It is possible that the absence of other lines may be due to an intrinsic difference in the physical properties of the gas or the ionizing radiation field.  This same argument applies to [\ion{S}{12}], [\ion{Fe}{11}], and to a lesser extent [\ion{Fe}{14}], which all have $\rho_{\rm crit}$ higher than most other lines.  By contrast, when detections are observed in [\ion{Fe}{7}] and/or [\ion{Ne}{5}], detections in other lines with similar $\rho_{\rm crit}$ tend to be more common, possibly indicating similar physical conditions in the line emitting gas in these galaxies.

A line not shown in Figure \ref{fig:line_corr} but nevertheless of interest is the [\ion{O}{1}] $\lambda\lambda$6302,6365 doublet found directly adjacent to [\ion{Fe}{10}] $\lambda$6374.  We find that only 2 of our [\ion{Fe}{10}] detections (1.87\%) do not also display [\ion{O}{1}], despite the fact that we did not impose any [\ion{O}{1}] pre-selection criteria in our search. This hints at a strong correlation between the presence of \rev{[\ion{Fe}{10}] and [\ion{O}{1}].  However, this relationship does not hold in the opposite direction: of the 244,806 spectra in the MPA/JHU catalog that detect [\ion{O}{1}] to an $S/N$ of $\geqslant 3$, 99.96\% (244,701) do not detect [\ion{Fe}{10}].}

\subsection{Neural Network Confidences}
We have created a framework for training and utilizing neural networks using the implementation of the Keras deep learning framework \citep{2018ascl.soft06022C} within Tensorflow \citep{tensorflow}.  Each coronal line has a separately trained neural network to allow for optimizations for each line (i.e. in the case of doublets or [\ion{Fe}{10}] with the surrounding [\ion{O}{1}] pair).  Each network is structured with an input layer where an item in the array corresponds to a pixel of the input spectrum. For uniformity of input, all spectra are resampled (while conserving flux) onto an identical wavelength grid using the Spectres package \citep{spectres}.  \rev{We then renormalize the flux by subtracting the mean (bringing the new mean to 0) and dividing by the standard deviation (bringing the new standard deviation to 1). For each line's network, we consider only a window of the spectrum around the line 100 \AA\ wide. For doublets, we extend the window such that both lines are encompassed, with a padding of 50 \AA\ on the left and right of each line.}  The network then consists of 3--6 dense layers with 1--100 neurons per layer using the Rectified Linear Unit (${\rm ReLU}(x) = {\rm max}(0, x)$) activation function.  The final dense layer has a single neuron with a sigmoid activation function, which gives the resulting confidence level, from 0--1, that a line is present in the input spectrum.  The weights of each neuron are trained on a set of 100,000 simulated spectra per coronal line.  The simulated spectra consist of a simple power law and \rrev{a line profile chosen randomly to be a Gaussian, Lorentzian, or Voigt distribution}, with added random noise at varying amplitudes.  Half of the spectra contain a line, and the other half do not. \rev{For the spectra that do contain a line, we draw the amplitude from a uniform distribution $\mathcal{U}(0,3)$, in normalized flux units. The FWHMs are similarly drawn from $\mathcal{U}(10,400)$ km/s, whereas the velocity shifts are drawn from a normal distribution centered at 0 km/s with a standard deviation of 100 km/s, $\mathcal{N}(0,100)$. If the line profile is Voigt, the mixing parameter is drawn from $\mathcal{U}(0,1)$.  For all simulated spectra, random noise is added using $\mathcal{N}(0,\sigma)$, where the amplitude of the noise is itself drawn from a uniform distribution, $\sigma = \mathcal{U}(0.01,3)$ in normalized flux units. This allows our spectra to vary from very high $S/N$ to very low $S/N$. Additionally, a shallow power law is added with a power slope drawn from $\mathcal{U}(0,3)$.} For doublets, we ensure that when one line is present, the other is as well. And for [\ion{Fe}{10}] in particular, we also simulate the [\ion{O}{1}] doublet since, as discussed, the vast majority of our visually inspected [\ion{Fe}{10}] detections also exhibit [\ion{O}{1}].

In preparation for the training process, each spectrum is assigned a truth value $t_i \in \{0,1\}$, corresponding to whether or not the line is present in the spectrum with an $S/N$ of at least 3.  We take 80,000 of the 100,000 simulated spectra to use in the training set.  Then, after outputting probabilities $0 \leqslant p_i \leqslant 1$, $p_i \in \mathbb{R}$ for each training spectrum, the loss function is computed as a metric of how well the network performed.  We choose to utilize a binary cross-entropy as our loss function:
\begin{equation}
    \log({\rm loss}) = -\frac{1}{N}\sum_{i=1}^N t_i\log(p_i) + (1-t_i)\log(1-p_i)
\end{equation}
The weights of each neuron are then adjusted to minimize this loss using the gradient descent method of backpropagation \citep{rumelhart1995backpropagation}.  This process is repeated for 11 epochs.

This entire process constitutes a single training iteration.  We wrap this within a Bayesian searching process, during which we allow the number of dense layers $\eta_d$, the neurons per layer $\eta_n$, and the learning rate $\eta_\ell$ of the network itself to vary as hyperparameters.  A total of 50 learning iterations are performed while varying these hyperparameters.  Uniform priors are placed on $\eta_d$ and $\eta_n$, with $\mathcal{U}(3,6)$ and $\mathcal{U}(0,100)$ respectively, and $\eta_\ell$ is given a log-uniform prior with bounds ($10^{-10}$, $10^{-1}$).  The Bayesian search process attempts to find the network configuration that minimizes the loss of the trained network by testing it on a validation set of 10,000 spectra, different from the 80,000 used to train the data. At the end, it selects the model hyperparmaters that minimized the loss the best before performing a final test with another test set of 10,000 spectra, different from the validation set and the training set.

After training the networks in this way, we compare the results of the generated neural network confidences to the coronal line detections that we have verified as a part of the CLASS survey selection.  Using a confidence level of 0.7 as the cutoff between a detection and a non-detection, this produces an overall true-positive (false-negative) rate of 75.49\% (24.51\%) and a false-positive (true-negative) rate of 3.96\% (96.04\%).  With further optimizations in the simulated spectra and the training process, this accuracy may be improved in the future and may allow this neural network to become a useful tool for probing new coronal line detections that have so far been missed.


\section{Catalog Availability}
\label{sect:description}

We show a sample of selected spectra for each coronal line in Figures \ref{fig:fexi_7892}--\ref{fig:nev_3426} in the Appendix.  Many of these lines are broad, with FWHMs $> 300$ \kms, and with velocity shifts sometimes approaching hundreds of \kms\ as well. It is also readily apparent how bright the [\ion{Ne}{5}] lines are in comparison to the other lines.  \rev{A few of these lines are very narrow and very bright (such as the three [\ion{Fe}{6}] $\lambda$5276 detections in Figure \ref{fig:misc_3}), and in these cases our Gaussian fitting tends to underestimate the amplitude of the line because the line shapes are not strictly Gaussian.  The $S/N$ ratios of these lines indicate that they are not noise spikes, and they have been vetted for sky lines, so their non-Gaussian shapes can most likely be attributed to their narrow widths not being properly sampled by the SDSS resolution limit of $\sigma \sim$70 \kms. We do not attempt to refit these lines with differently shaped profiles so as to maintain uniformity in CLASS.}

The CLASS catalog itself is available for download as a table in CSV or ASCII format. Two versions of the table are provided, with the full version including all 952,138 objects from the original MPA/JHU DR8 sample, and the small version including only our 258 coronal line detections. We provide the smaller table in the online journal, due to size limitations, and the full table will be available through Vizier. Descriptions of each column in the table, as well as the units, data type, and source, are provided in Table \ref{tab:column_descriptions} in the Appendix.  Note that the SDSS Spec Object ID column is a 64-bit integer, and must be read appropriately (either as a long integer, or a string) to avoid losing information. All velocity offsets for coronal lines are measured relative to the stellar velocity, and all equivalent widths use a sign convention where emission is positive and absorption is negative.



\section{Conclusions}
\label{sect:conclusion}
In summary, we have produced the first catalog of the emission line properties of the comprehensive set of 20 optical coronal lines detected in the SDSS MPA/JHU DR8 catalog. The results from the CLASS survey demonstrate the extreme rarity and large variance in parameter space (i.e. $L$, FWHM, $v_{\rm off}$, etc.) of optical coronal line detections.  No significant correlations were found between the IP or $\rho_{\rm crit}$ of a line and its luminosity, FWHM, or $v_{\rm off}$, but a general trend of coronal lines being slightly blueshifted relative to the stellar velocity is seen.  We also find strong power-law correlations between the WISE W2 luminosity of a galaxy and its corresponding coronal line luminosities for the [\ion{Fe}{7}] $\lambda$5276, [\ion{Fe}{7}] $\lambda\lambda$5720,6087, and [\ion{Fe}{10}] $\lambda$6374 lines, which is independent of the WISE W1-W2 color. The coronal line fluxes are significantly weaker than the widely detected optical strong lines. Coronal line fluxes are typically a factor of at least two to over three orders of magnitude weaker than the [\ion{O}{3}] $\lambda$5007 and H$\alpha$ flux. These flux ratios, as well as our W2 correlation relations, can be used to predict coronal line luminosities in future ground-based studies of local galaxies or high redshift studies with JWST. 
The development of the neural network for detecting these lines is another promising step towards more accurate and efficient detection algorithms, and it may be utilized in future surveys to quickly filter results, in much the same manner as we utilized the $\mathcal{F}$ ratio in this survey.

\section{Acknowledgements}

J.M.C. would like to acknowledge a NASA Postdoctoral
Program (NPP) fellowship at Goddard Space Flight Center,
administered by ORAU through contract with NASA.

The simulations carried out in this work were run on ARGO and HOPPER, research computing clusters provided by the Office of Research Computing at George Mason University, VA. (\url{ http://orc.gmu.edu})

This research made use of Astropy,\footnote{\url{http://www.astropy.org}} a community-developed core Python package for Astronomy \citep{2013A&A...558A..33A}, as well as \textsc{topcat} \citep{2005ASPC..347...29T}.  

Funding for SDSS-III has been provided by the Alfred P. Sloan Foundation, the Participating Institutions, the National Science Foundation, and the U.S. Department of Energy Office of Science. The SDSS-III web site is \href{http://www.sdss3.org/}{http://www.sdss3.org/}.

SDSS-III is managed by the Astrophysical Research Consortium for the Participating Institutions of the SDSS-III Collaboration including the University of Arizona, the Brazilian Participation Group, Brookhaven National Laboratory, Carnegie Mellon University, University of Florida, the French Participation Group, the German Participation Group, Harvard University, the Instituto de Astrofisica de Canarias, the Michigan State/Notre Dame/JINA Participation Group, Johns Hopkins University, Lawrence Berkeley National Laboratory, Max Planck Institute for Astrophysics, Max Planck Institute for Extraterrestrial Physics, New Mexico State University, New York University, Ohio State University, Pennsylvania State University, University of Portsmouth, Princeton University, the Spanish Participation Group, University of Tokyo, University of Utah, Vanderbilt University, University of Virginia, University of Washington, and Yale University. 

This publication makes use of data products from the Wide-field Infrared Survey Explorer, which is a joint project of the University of California, Los Angeles, and the Jet Propulsion Laboratory/California Institute of Technology, and NEOWISE, which is a project of the Jet Propulsion Laboratory/California Institute of Technology. WISE and NEOWISE are funded by the National Aeronautics and Space Administration.

The {\it JWST} spectrum referenced in this paper can be found in MAST: \dataset[10.17909/jb8d-8379]{https://dx.doi.org/10.17909/jb8d-8379}

\facilities{Sloan, WISE}

\software{
\texttt{astropy} \citep{2013A&A...558A..33A}, 
\texttt{numpy},\citep{2020Natur.585..357H}
\texttt{scipy},\citep{2020NatMe..17..261V}
\texttt{emcee} \citep{2013PASP..125..306F},
\texttt{pPXF} \citep{2017MNRAS.466..798C},
\textsc{topcat} \citep{2005ASPC..347...29T},
\textsc{badass} \citep{sexton_2020},
\texttt{bifrost} (\href{https://github.com/Michael-Reefe/bifrost}{https://github.com/Michael-Reefe/bifrost})
}

\nocite{1988AJ.....95...45A}
\nocite{2018ApJ...861..142C}

\bibliographystyle{yahapj}
\bibliography{ref}






\appendix

\begin{figure*}
    \centering
    \includegraphics[width=.48\textwidth]{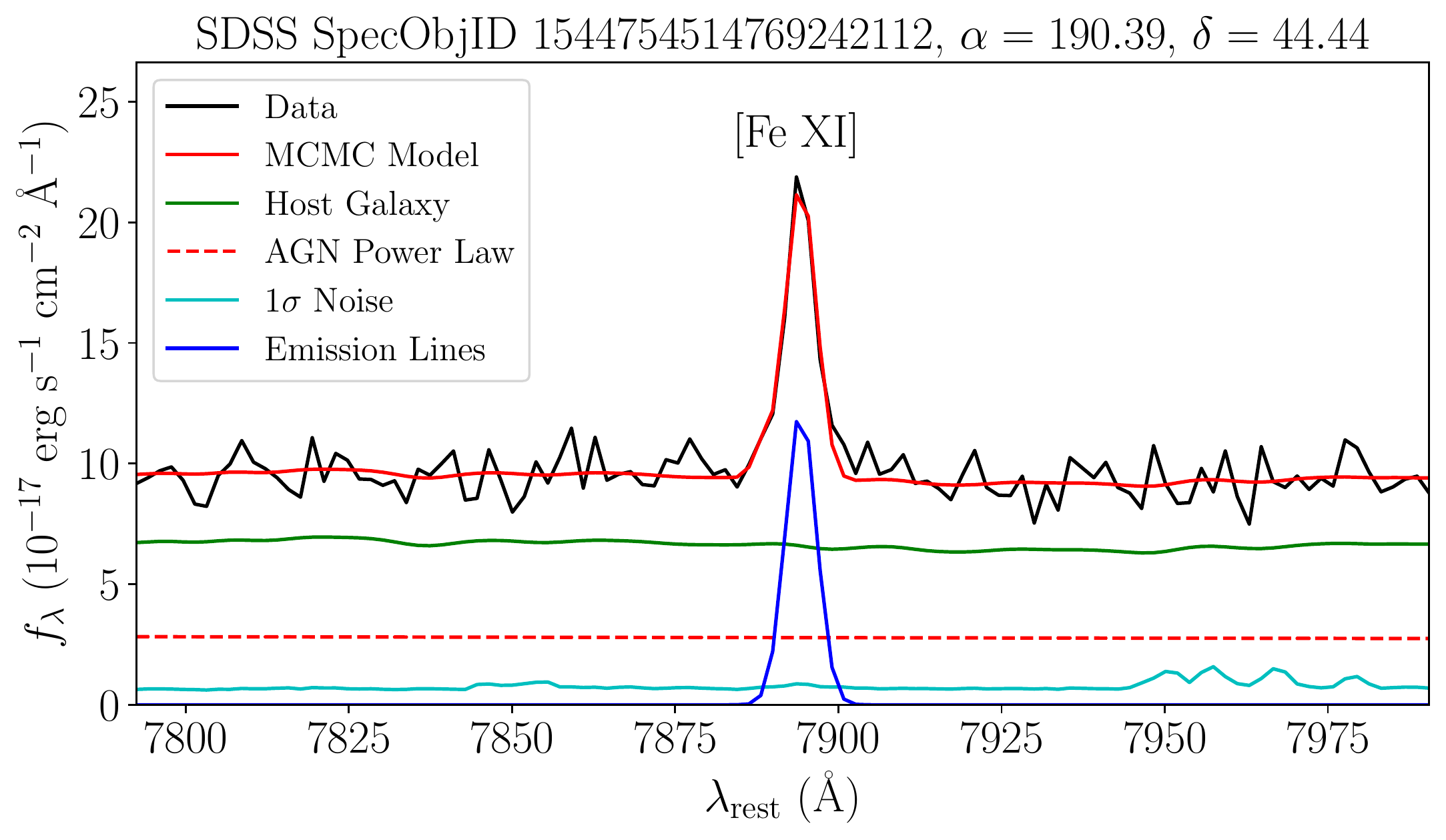}
    \includegraphics[width=.48\textwidth]{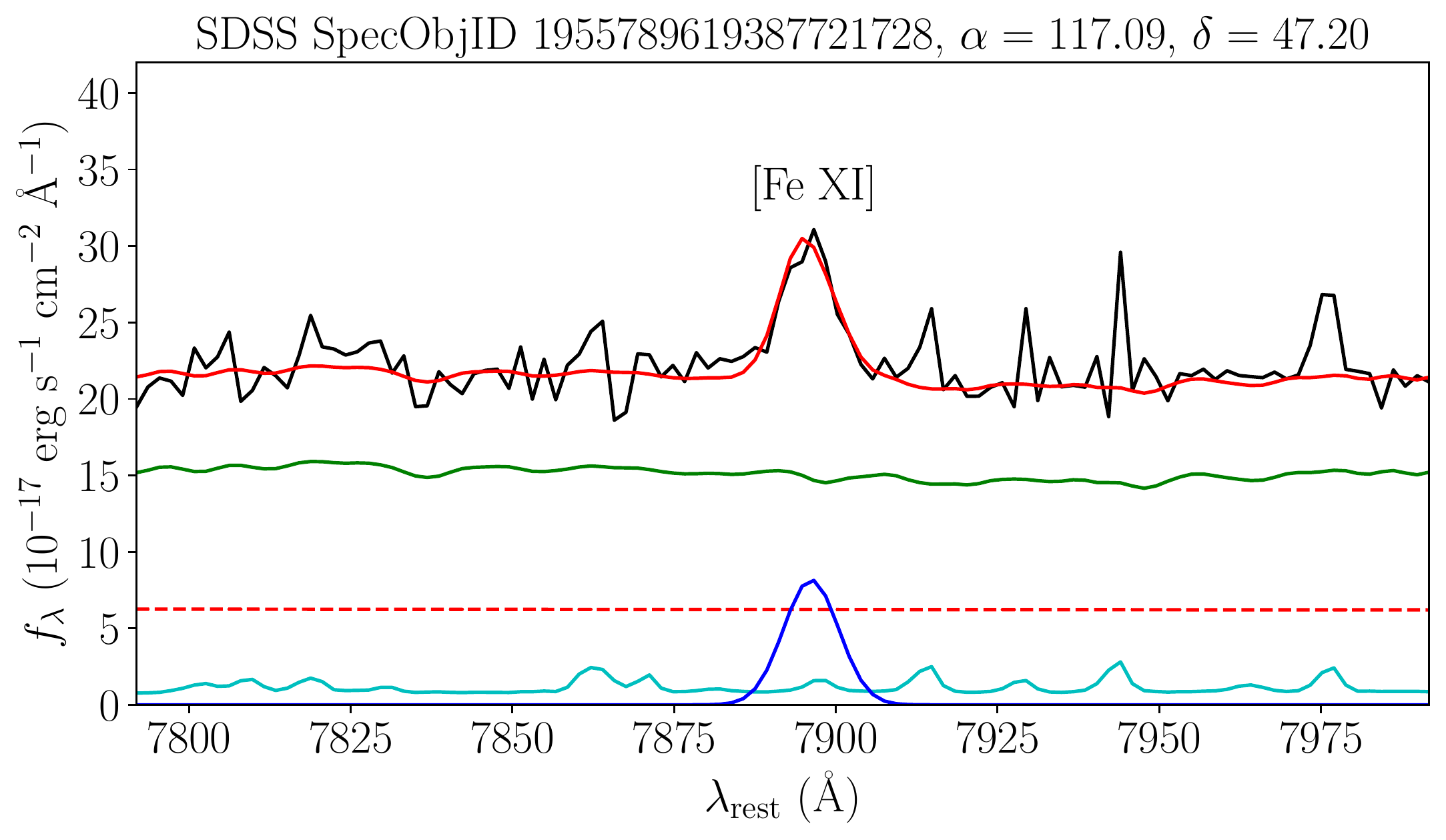}
    \includegraphics[width=.48\textwidth]{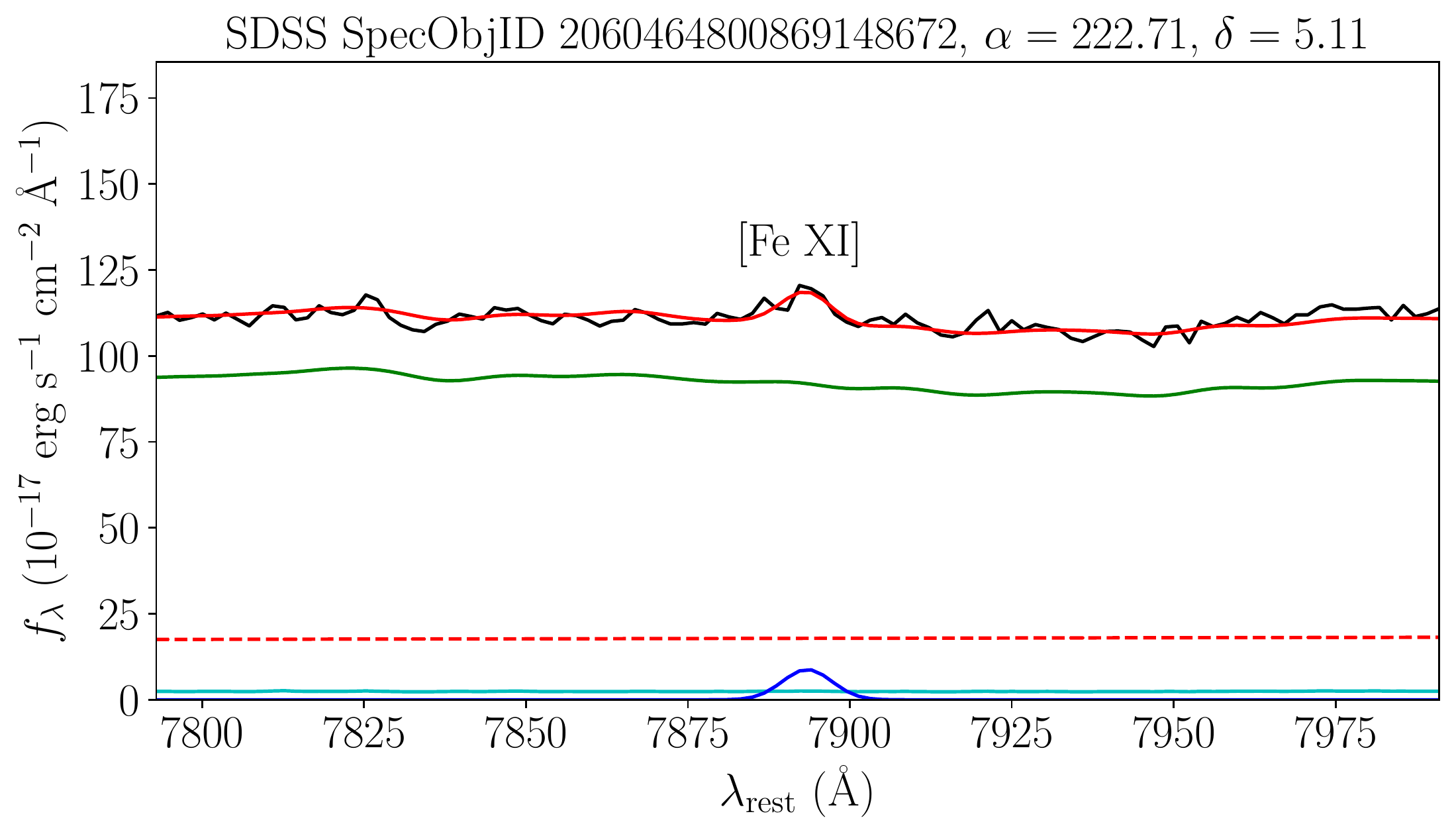}
    \includegraphics[width=.48\textwidth]{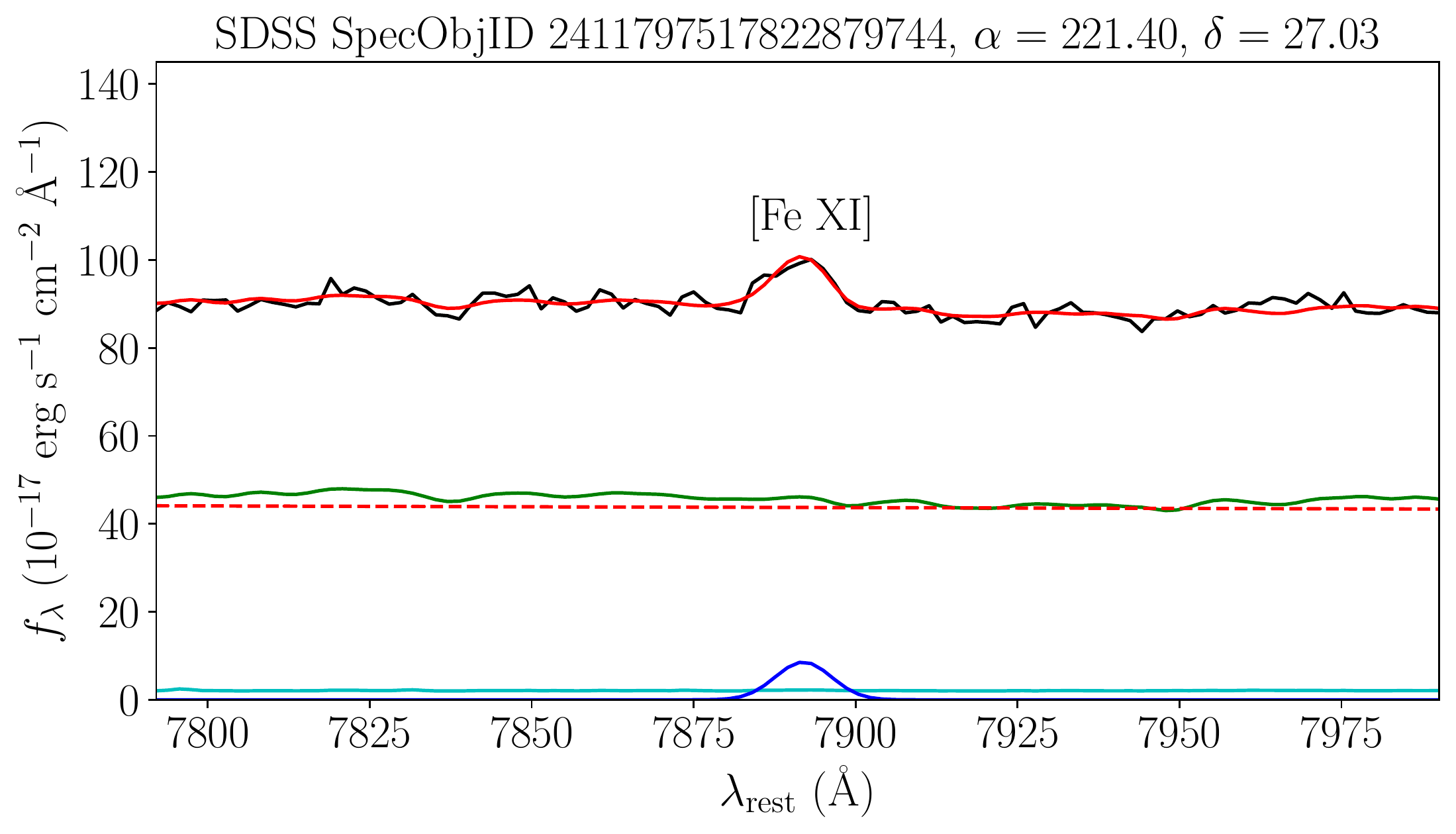}
    \includegraphics[width=.48\textwidth]{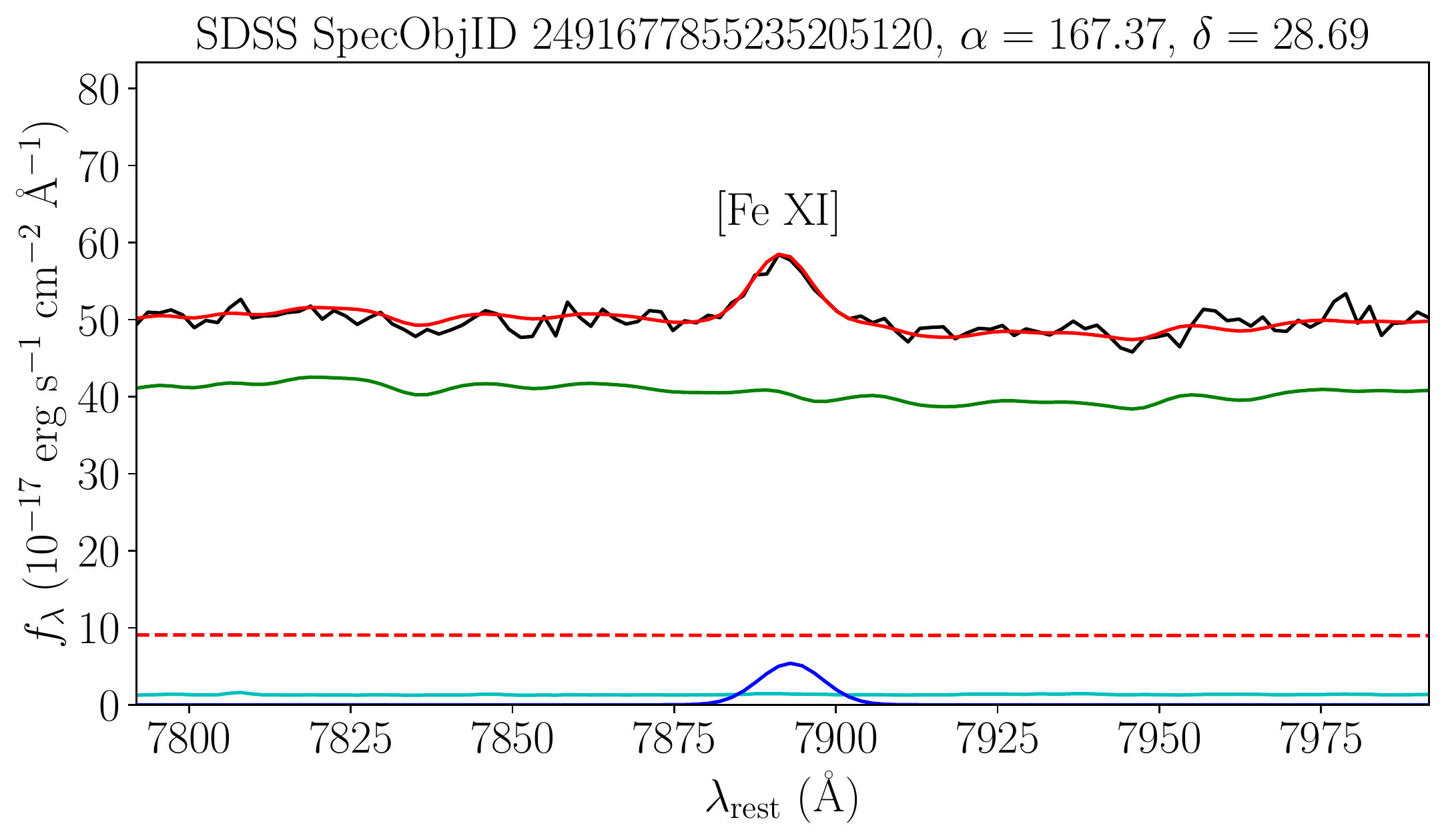}
    \includegraphics[width=.48\textwidth]{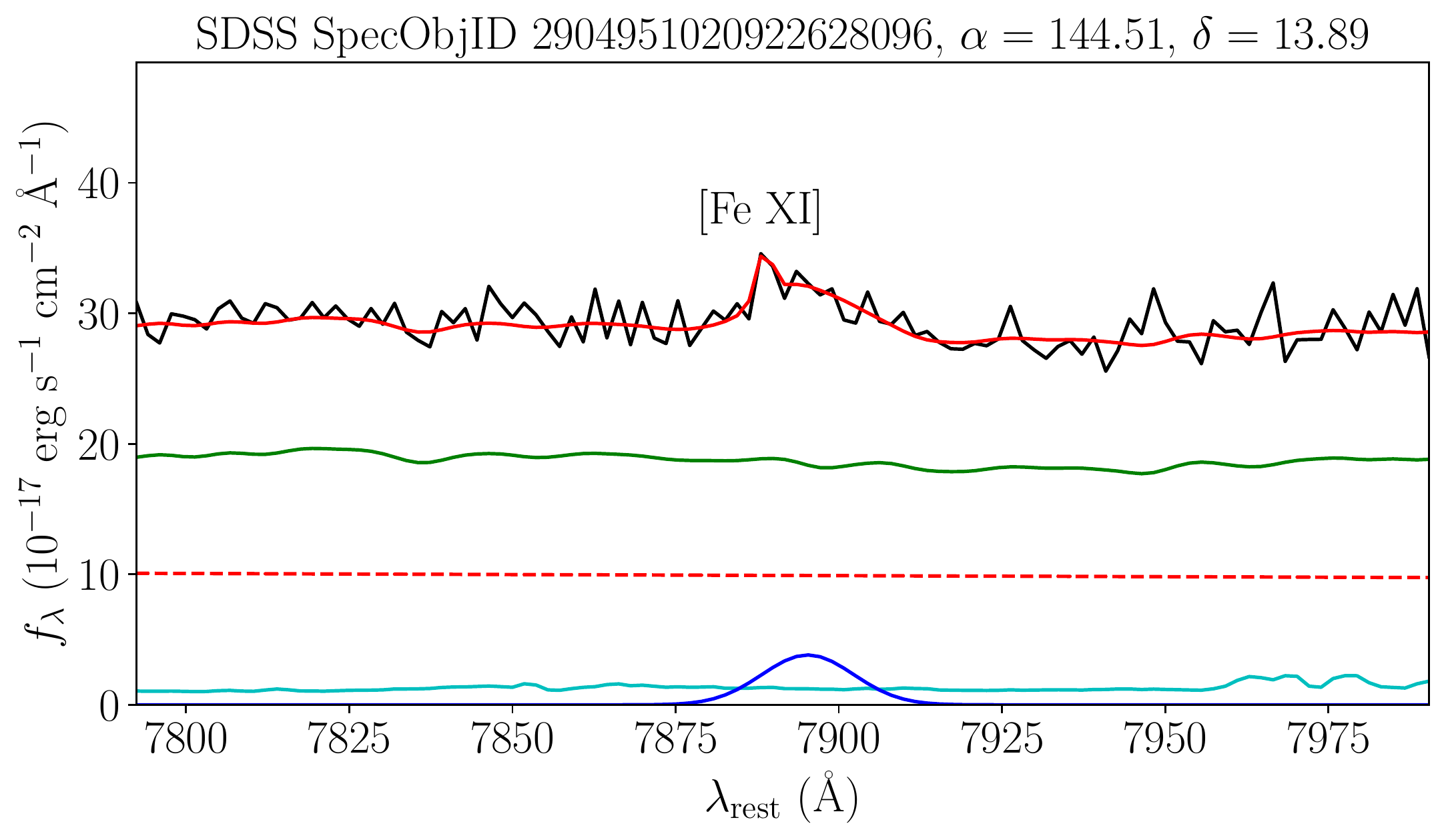}
    \includegraphics[width=.48\textwidth]{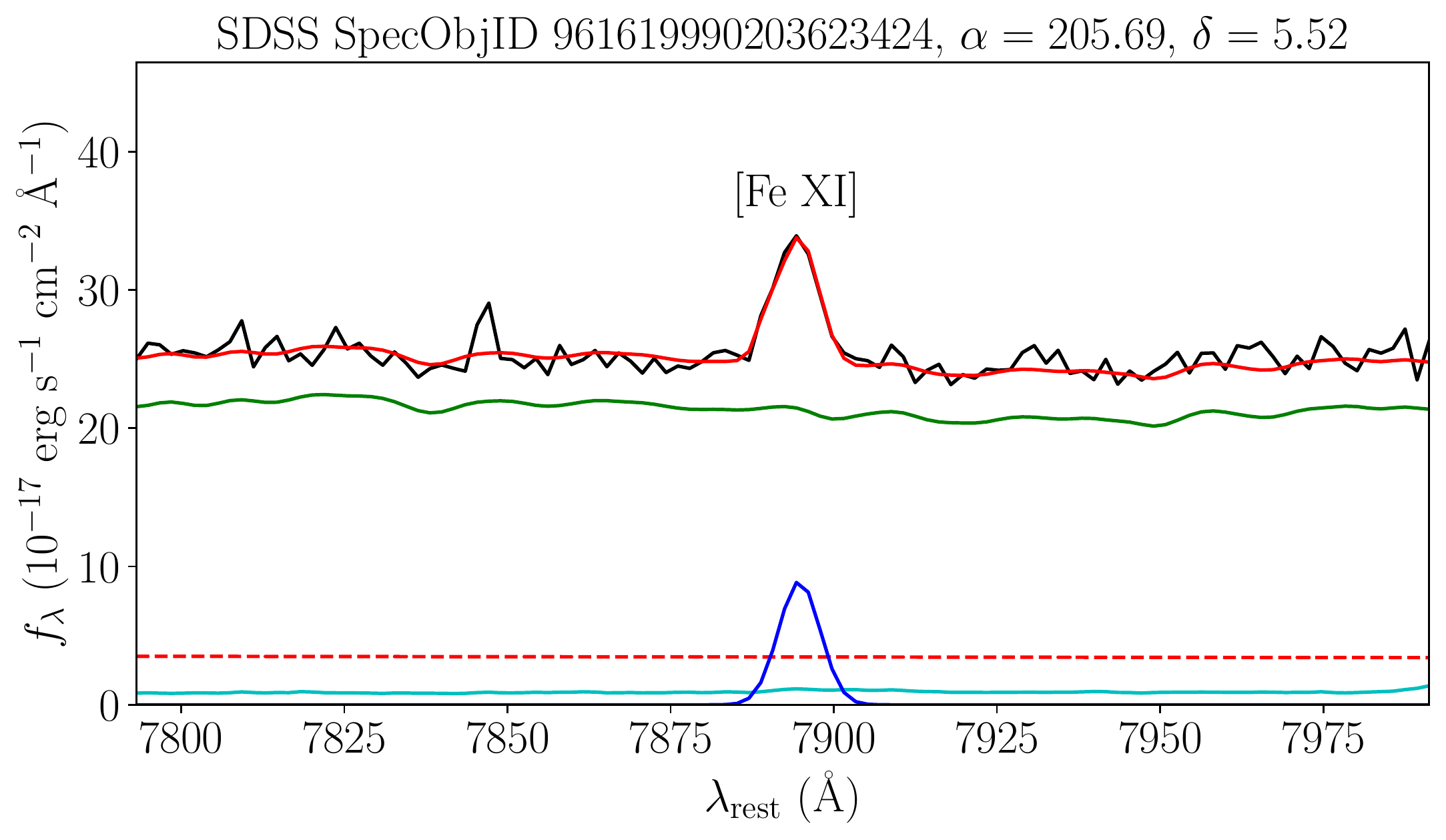}
    \includegraphics[width=.48\textwidth]{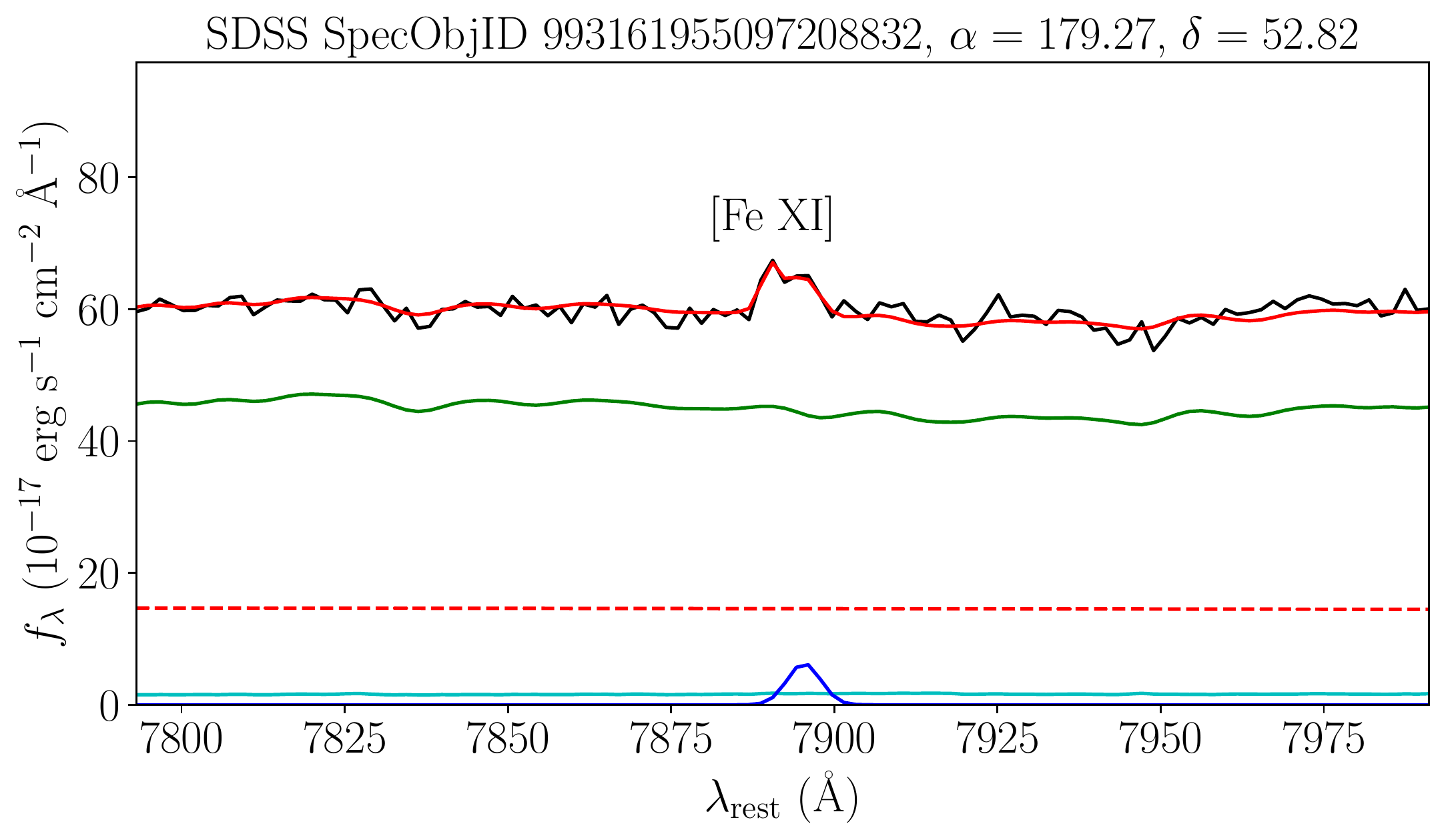}
    \caption{A selection of [\ion{Fe}{11}] $\lambda$7892 detections from our subsample. The raw flux, corrected for redshift and galactic extinction, is plotted in black.  The BADASS model and each of its components (emission lines, host galaxy, AGN power law) are plotted in colors indicated by the legend.  Each coronal line is labeled, and the spectra's coordinates and SDSS Spec Object ID are given in the plot titles.}
    \label{fig:fexi_7892}
\end{figure*}

\begin{figure*}
    \centering
    \includegraphics[width=.48\textwidth]{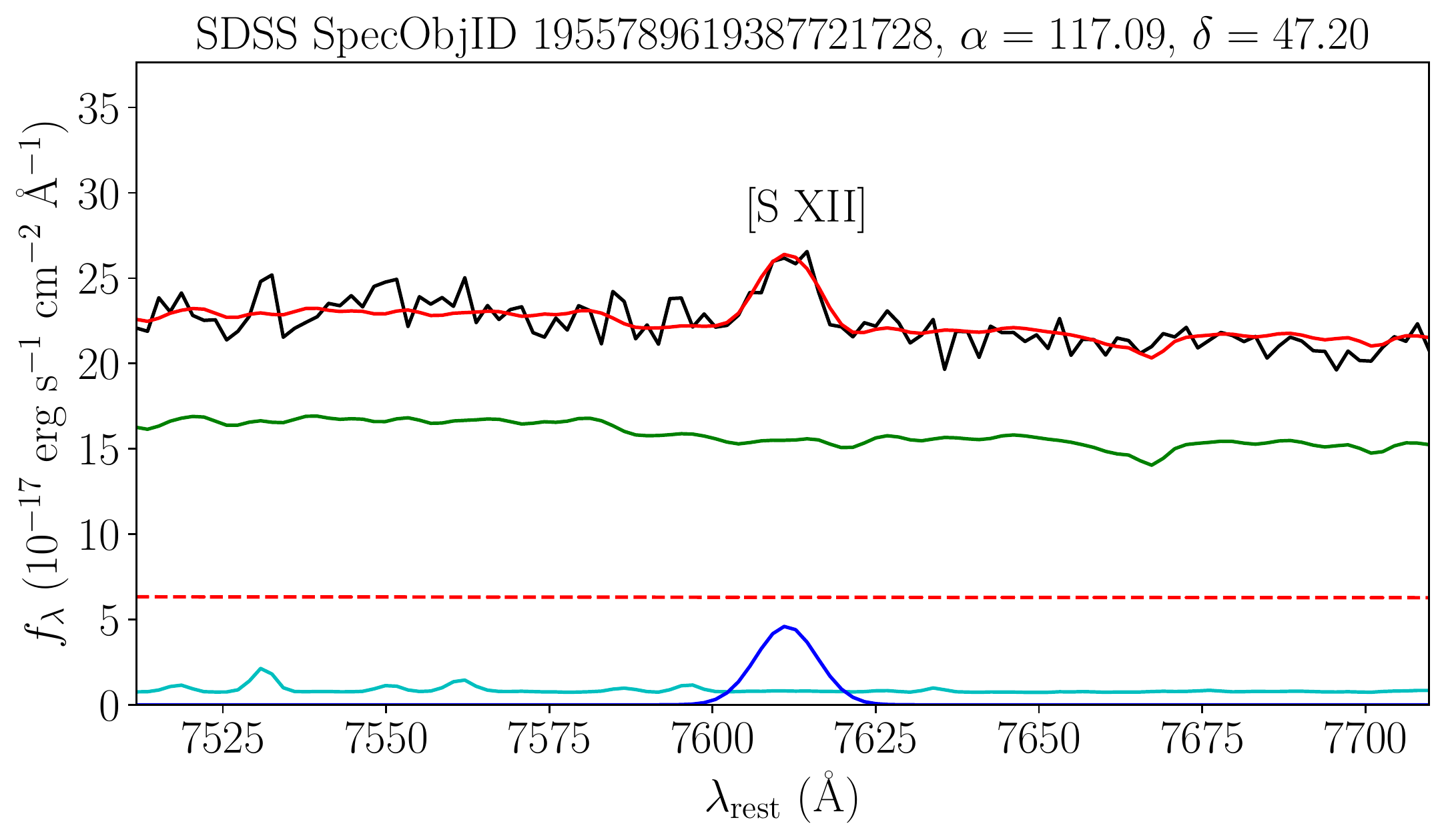}
    \includegraphics[width=.48\textwidth]{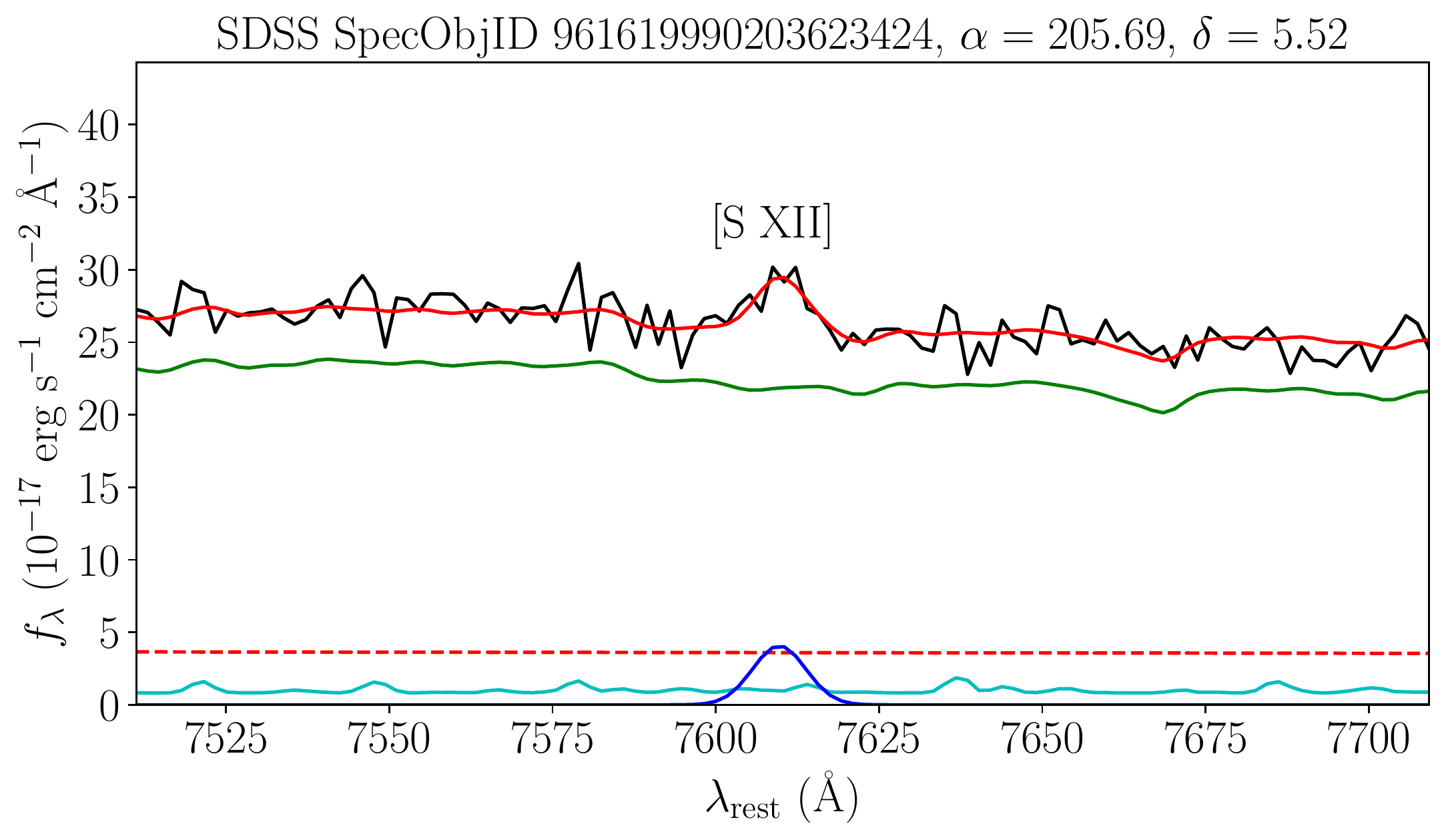}
    \includegraphics[width=.48\textwidth]{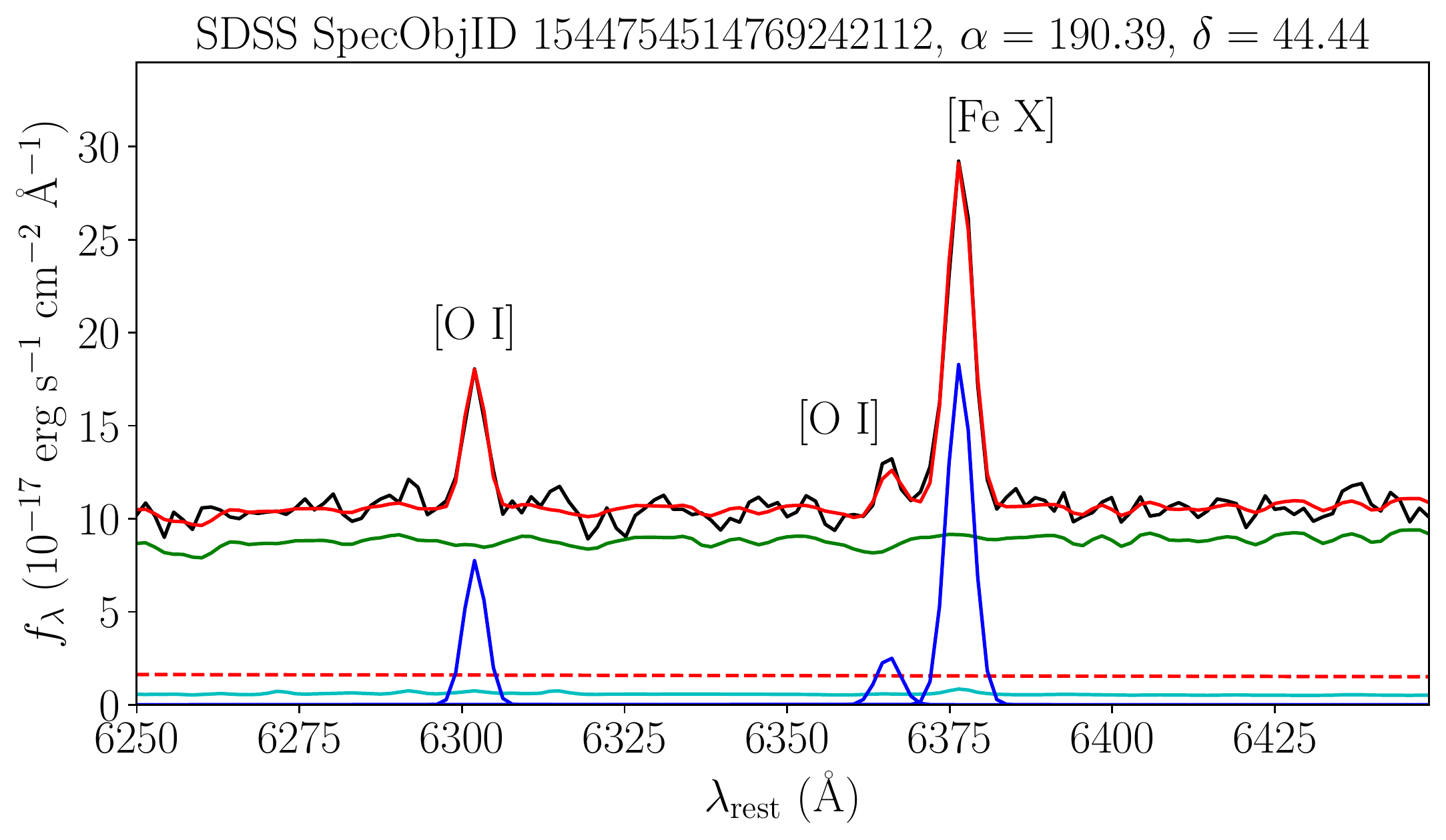}
    \includegraphics[width=.48\textwidth]{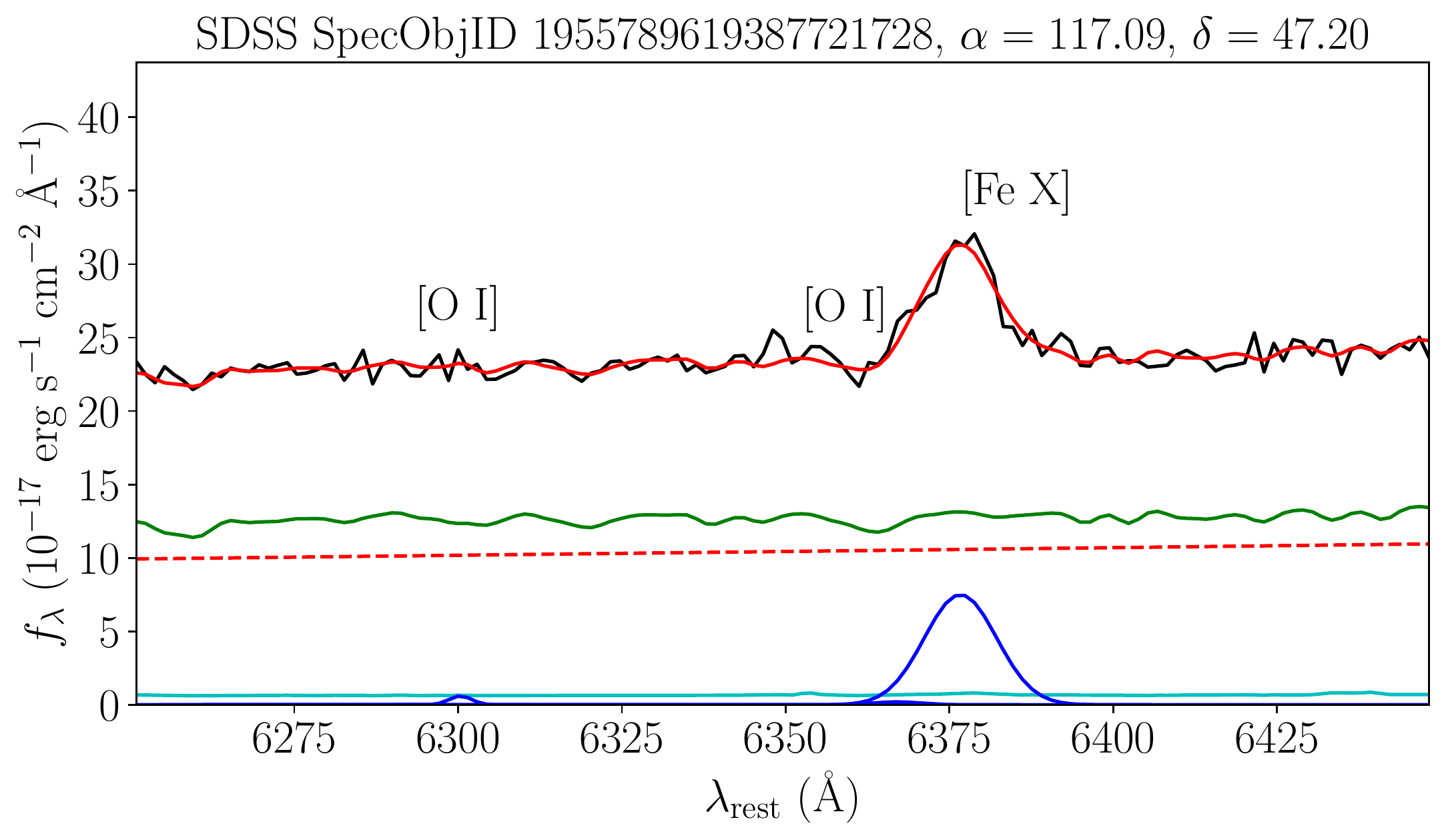}
    \includegraphics[width=.48\textwidth]{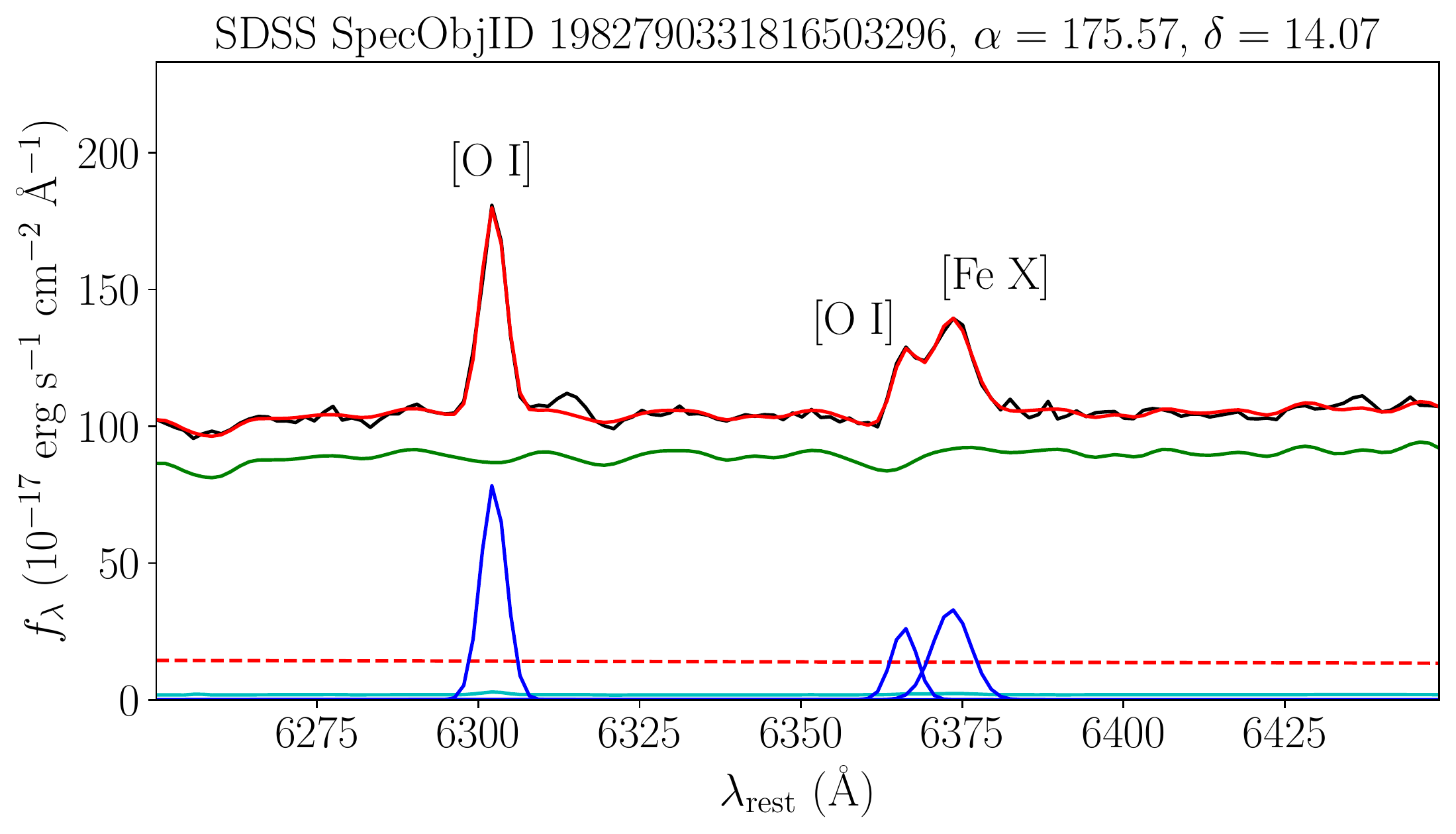}
    \includegraphics[width=.48\textwidth]{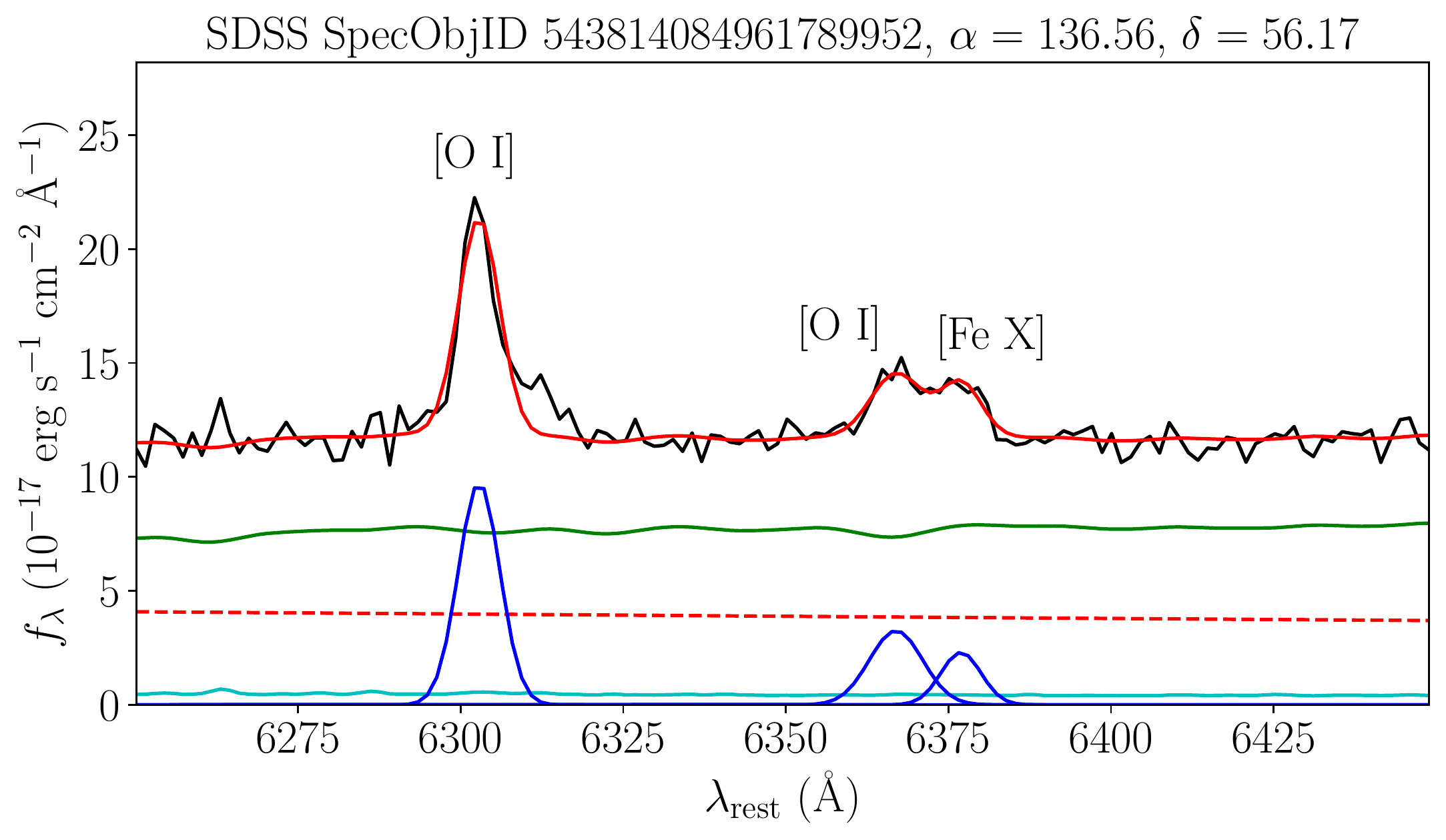}
    \includegraphics[width=.48\textwidth]{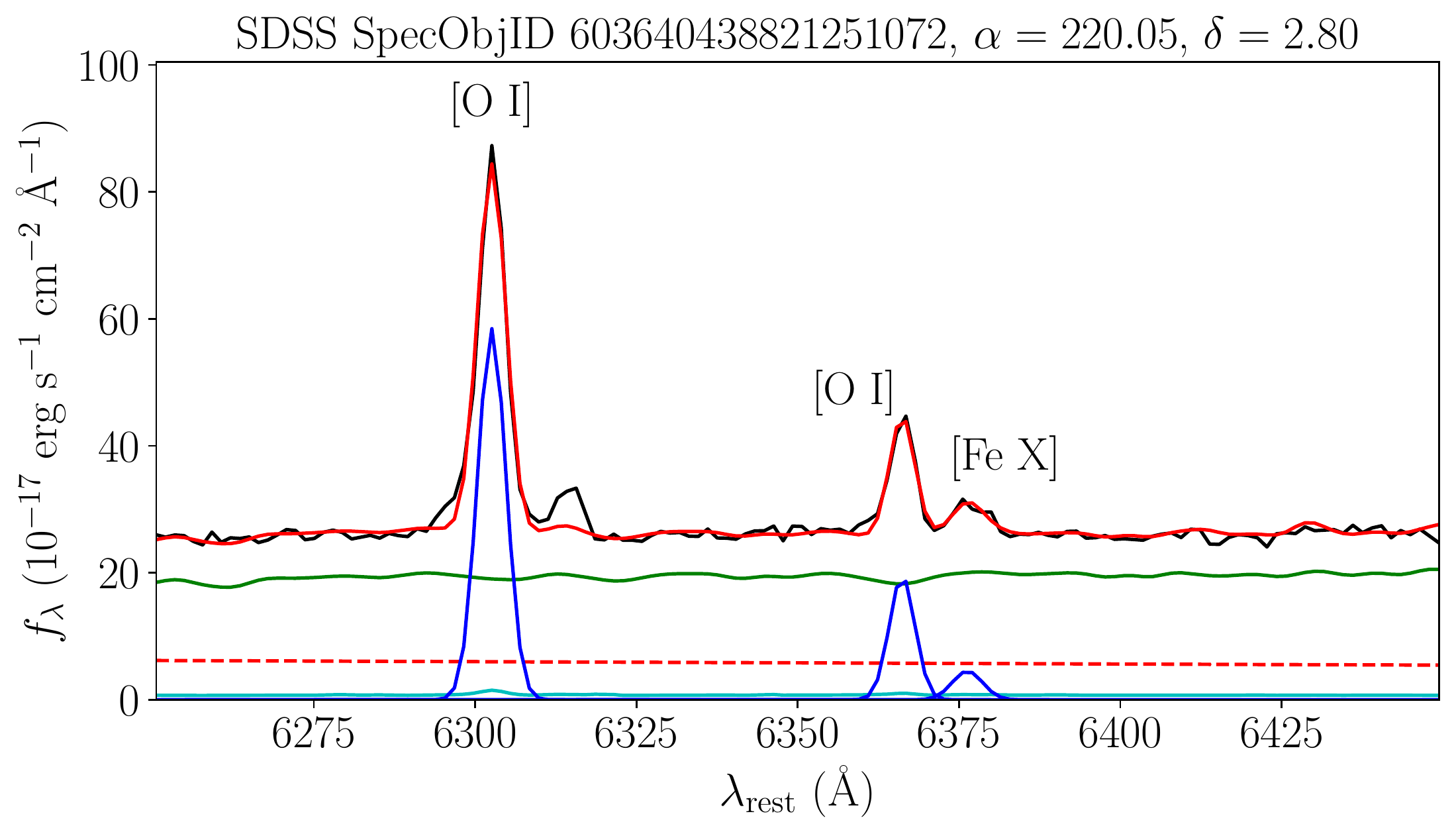}
    \includegraphics[width=.48\textwidth]{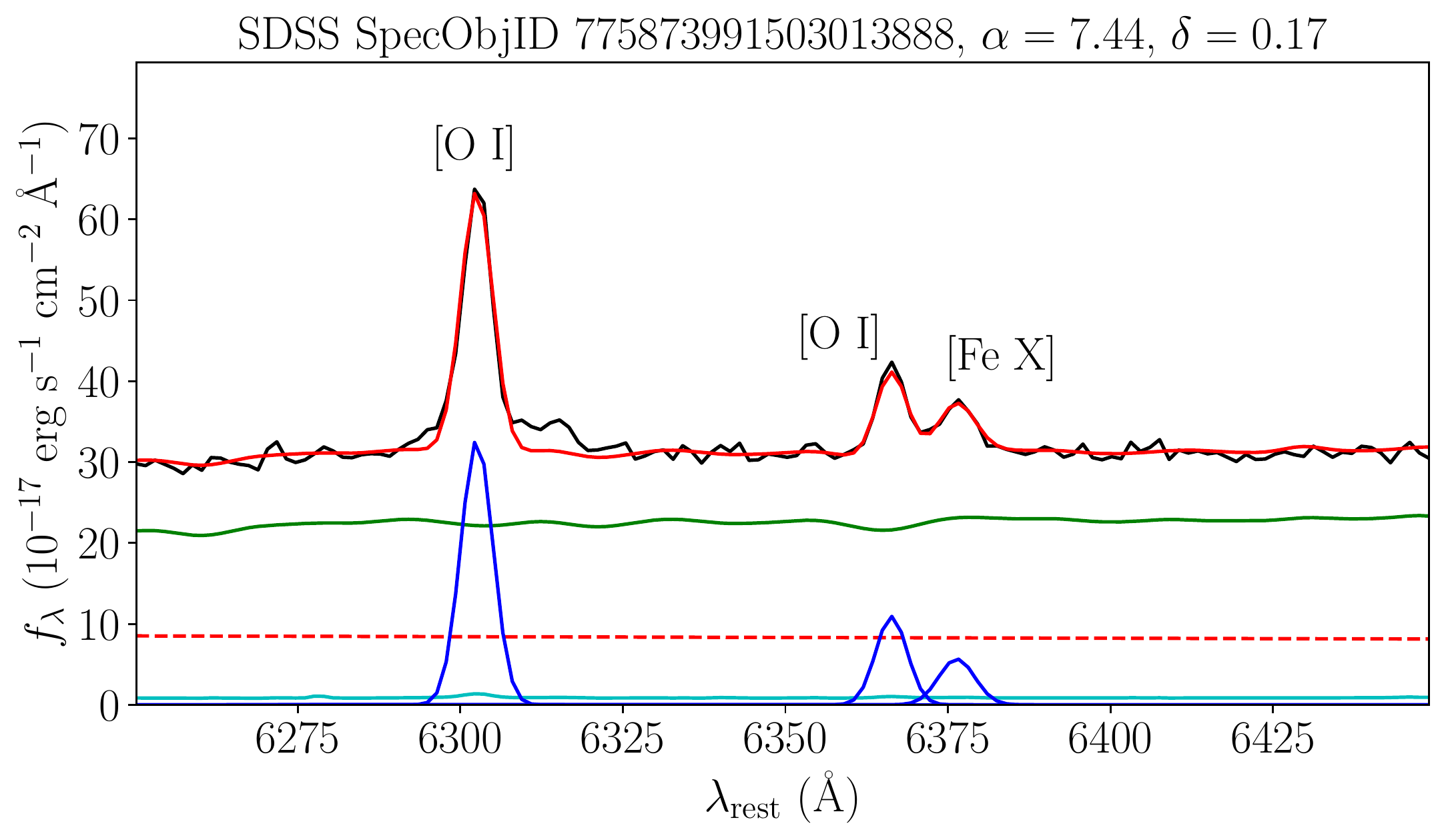}
    \caption{\rev{A selection of [\ion{S}{12}] $\lambda$7609 and} [\ion{Fe}{10}] $\lambda$6374 detections from our subsample. The raw flux, corrected for redshift and galactic extinction, is plotted in black.  The BADASS model and each of its components (emission lines, host galaxy, AGN power law) are plotted in colors indicated by the legend.  Each coronal line is labeled, and the spectra's coordinates and SDSS Spec Object ID are given in the plot titles.}
    \label{fig:fex_6374}
\end{figure*}

\begin{figure*}
    \centering
    \includegraphics[width=.48\textwidth]{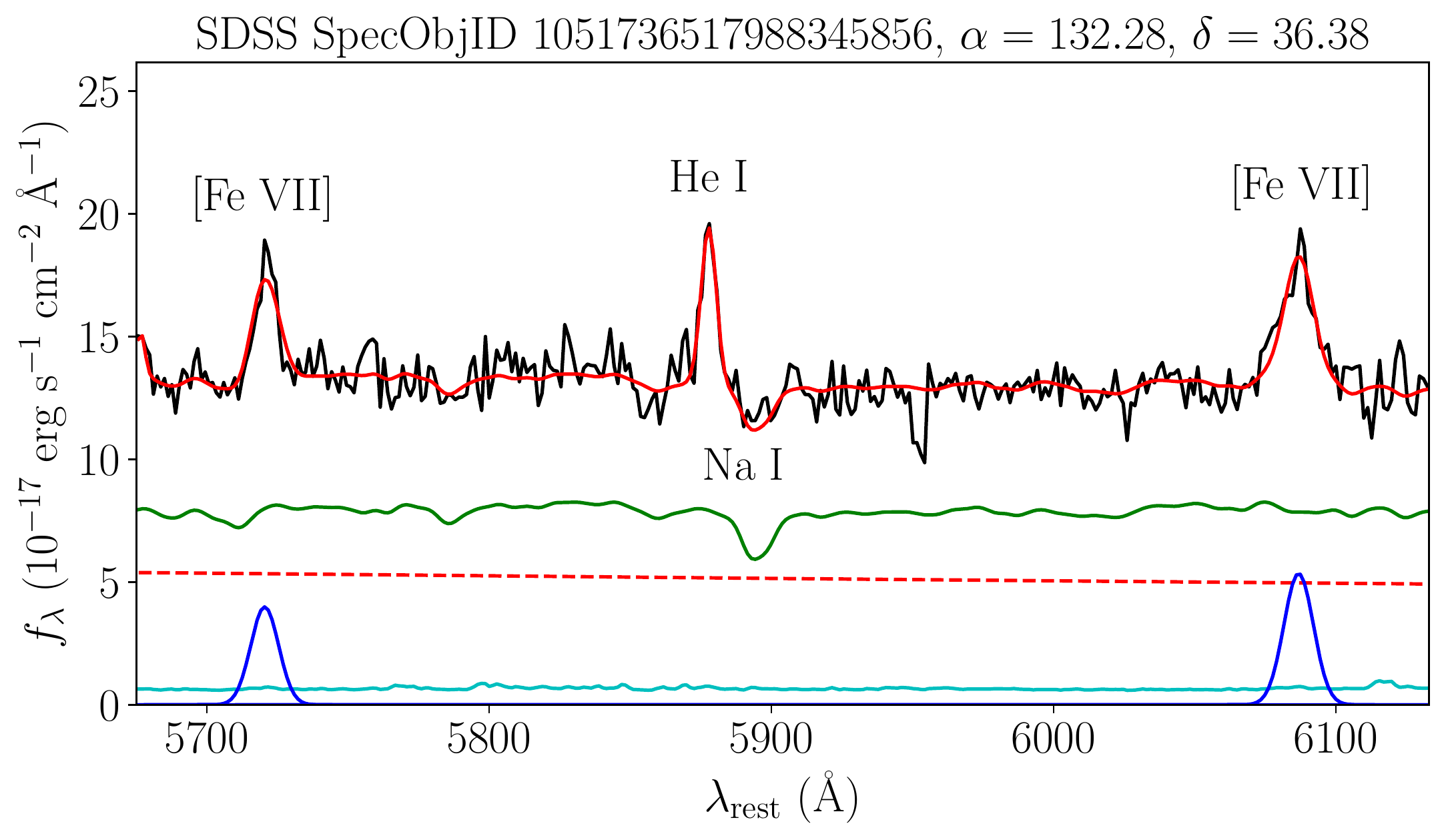}
    \includegraphics[width=.48\textwidth]{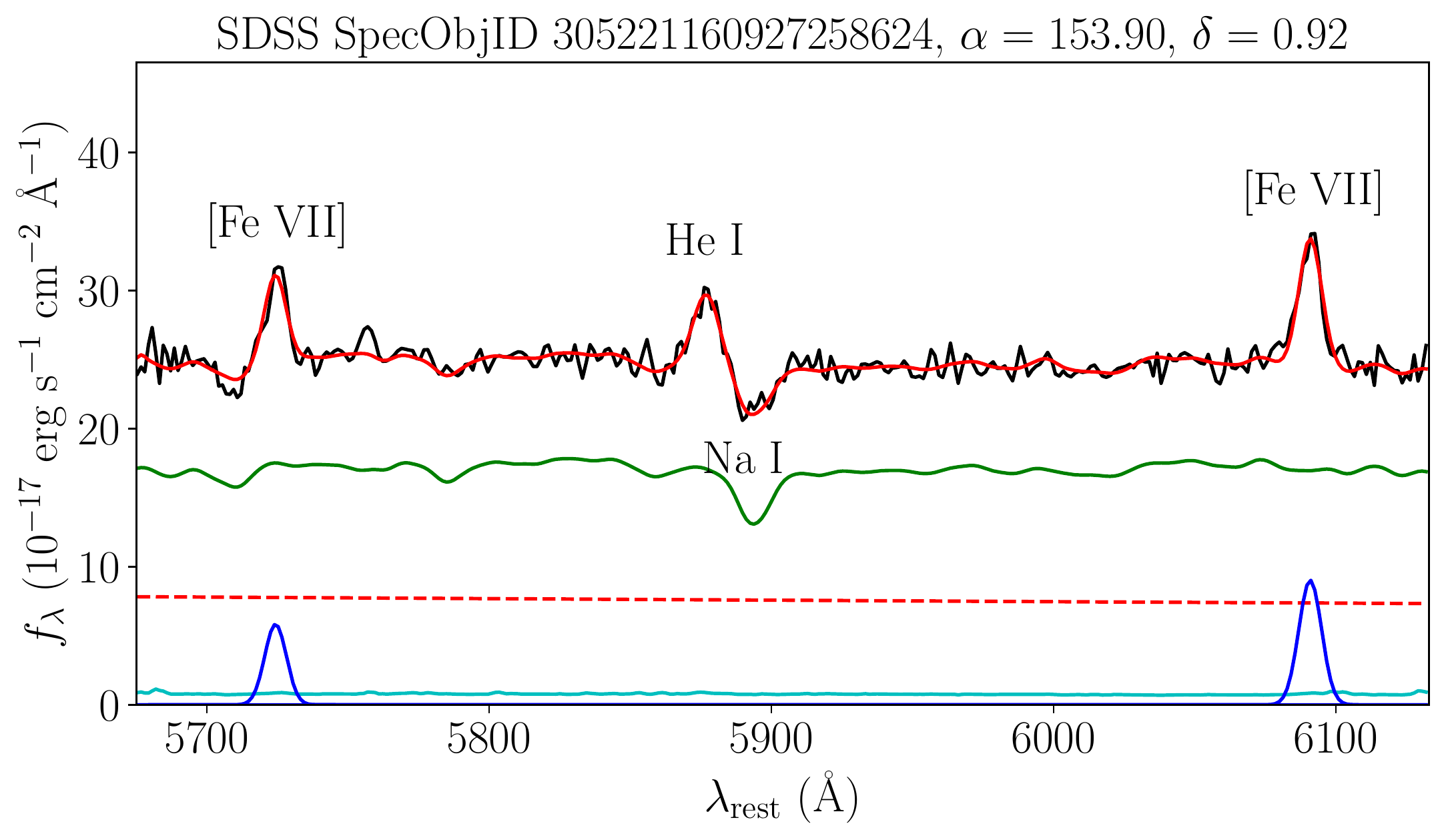}
    \includegraphics[width=.48\textwidth]{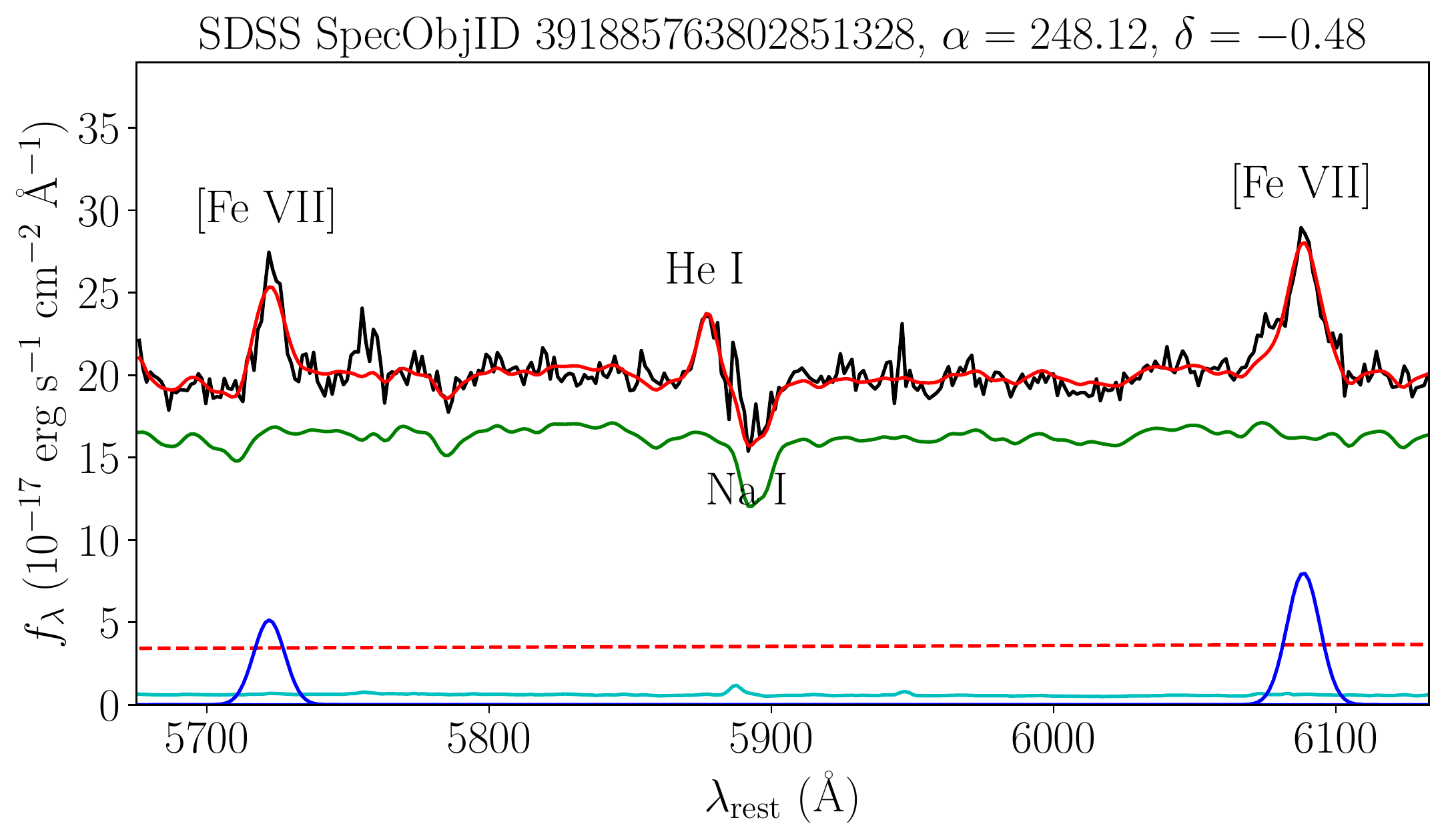}
    \includegraphics[width=.48\textwidth]{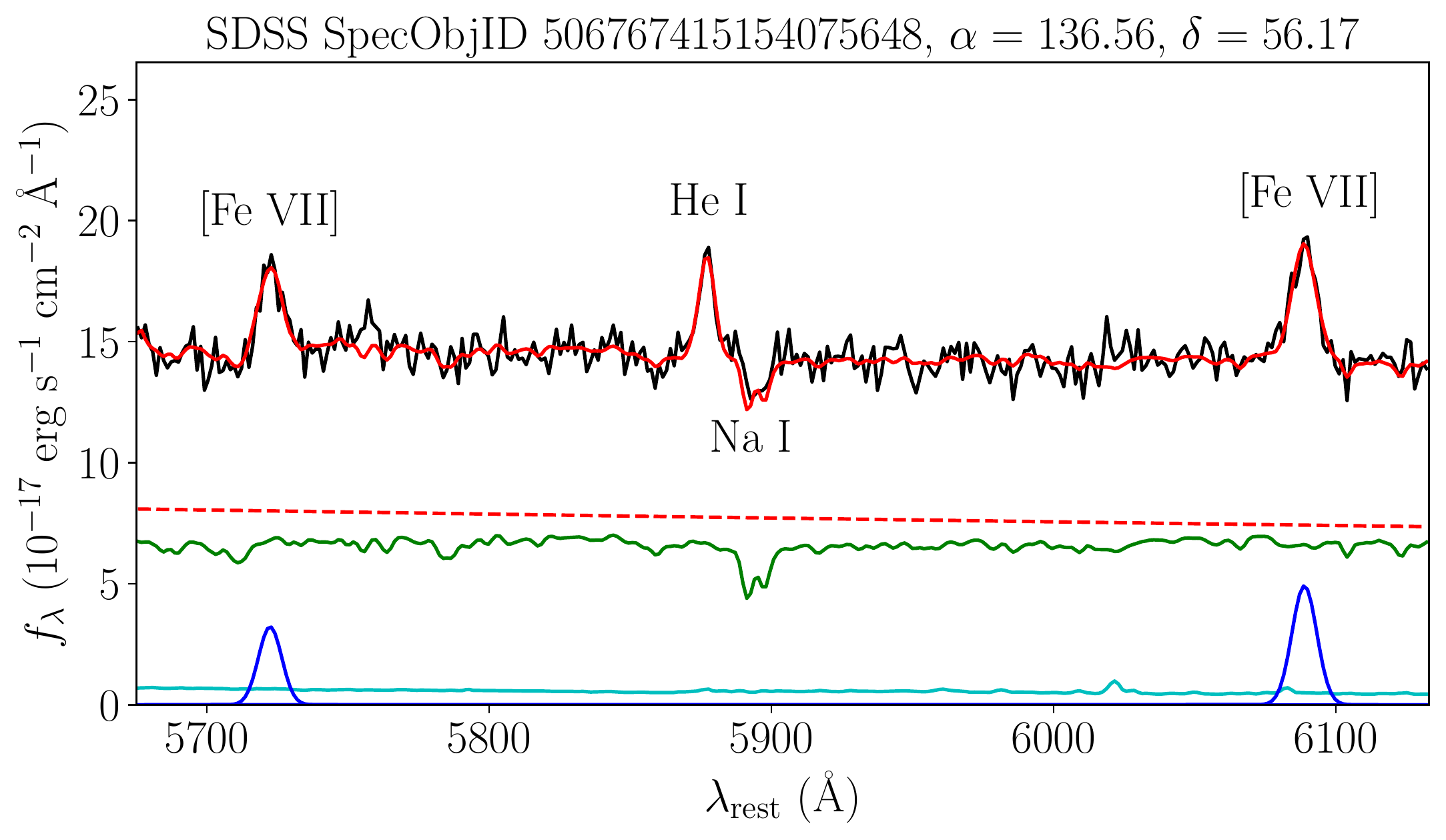}
    \includegraphics[width=.48\textwidth]{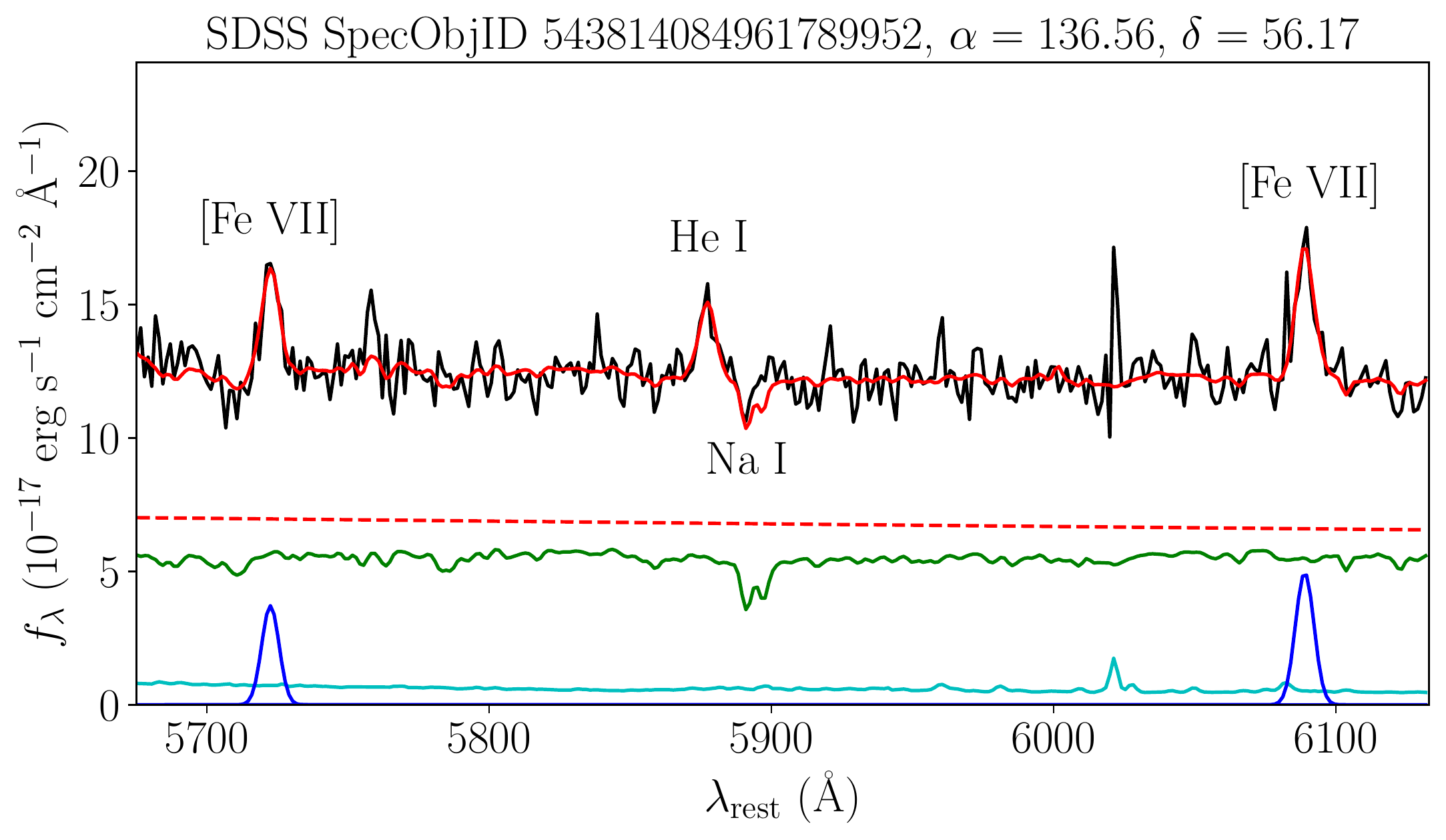}
    \includegraphics[width=.48\textwidth]{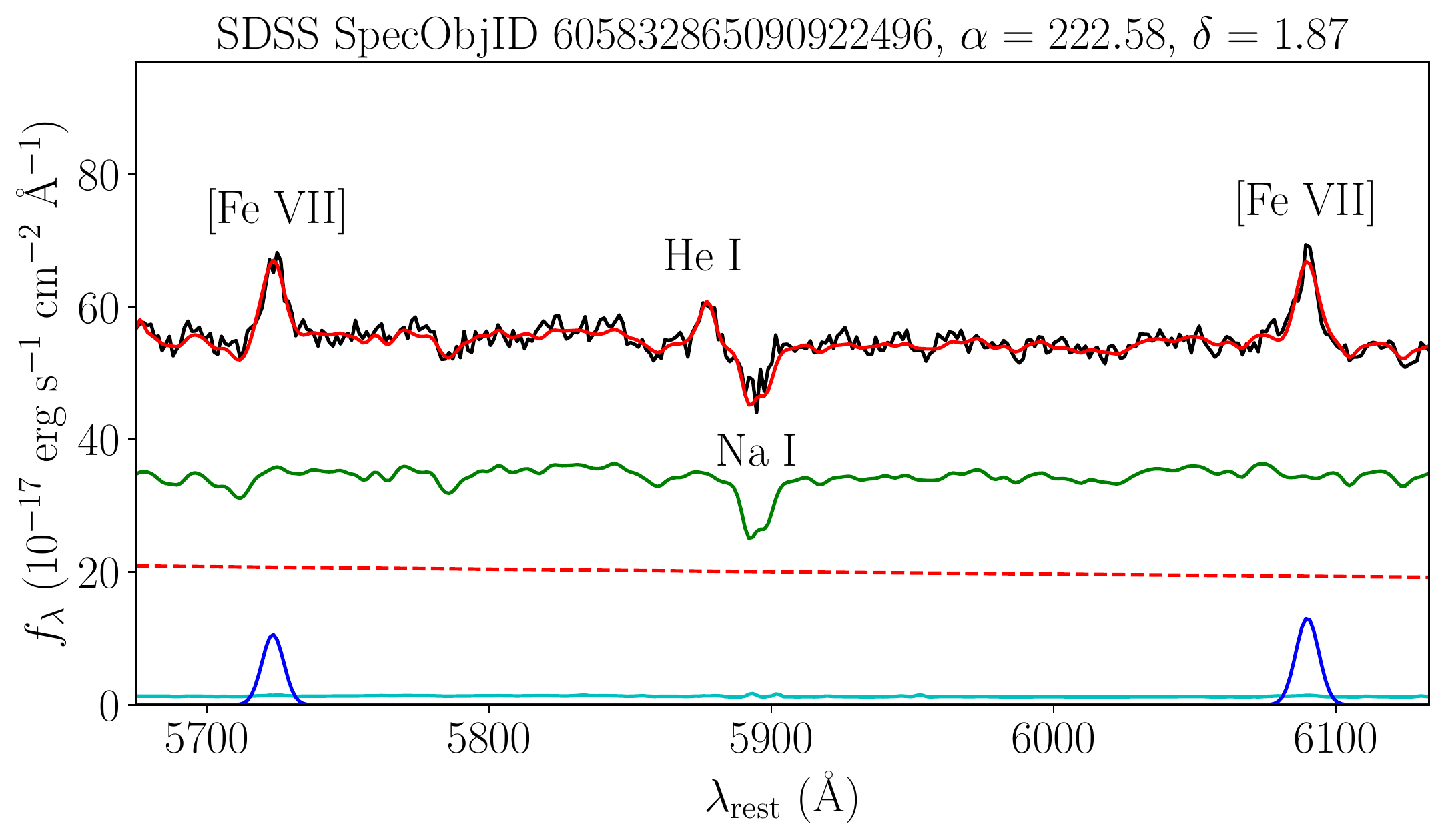}
    \includegraphics[width=.48\textwidth]{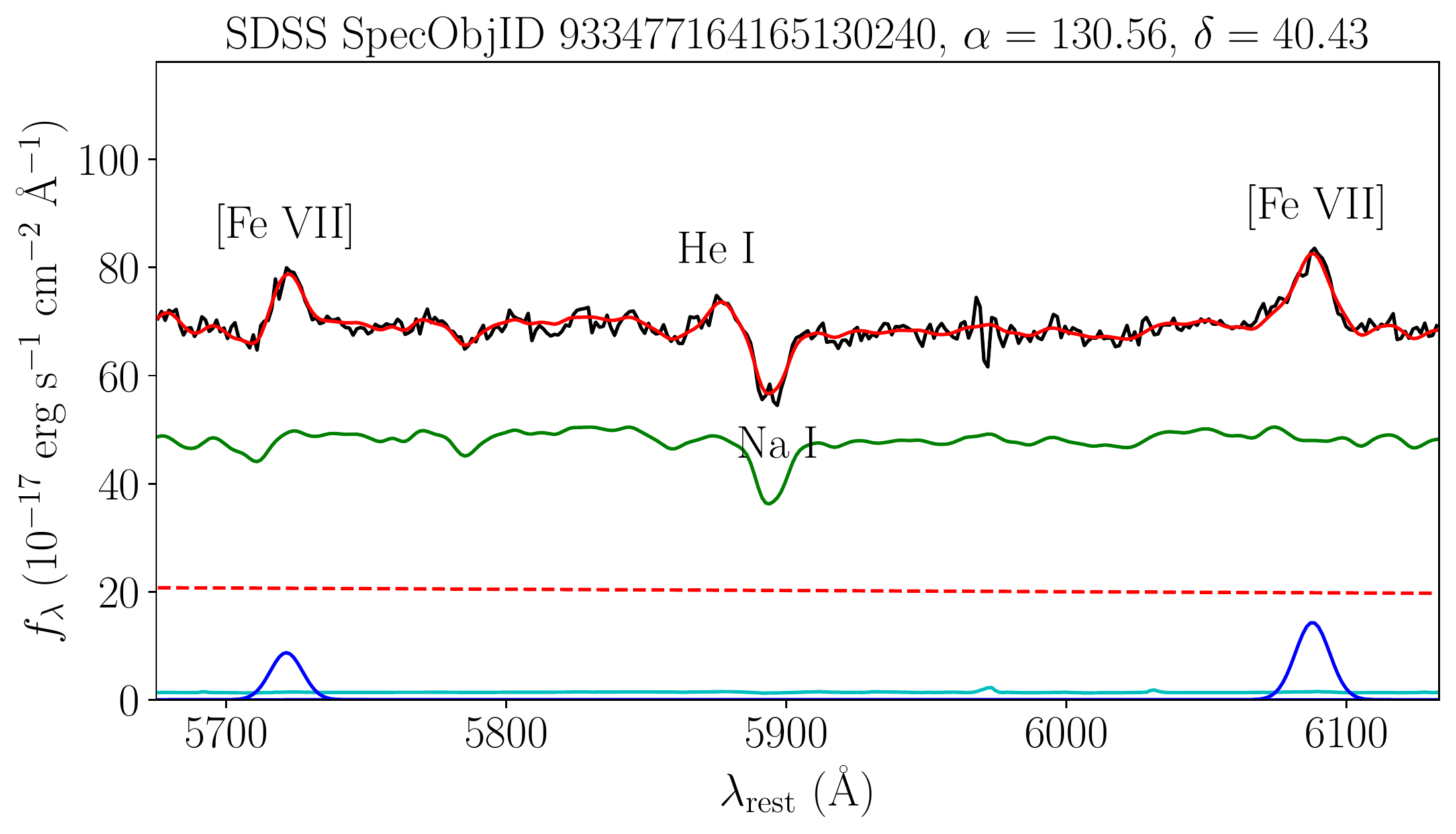}
    \includegraphics[width=.48\textwidth]{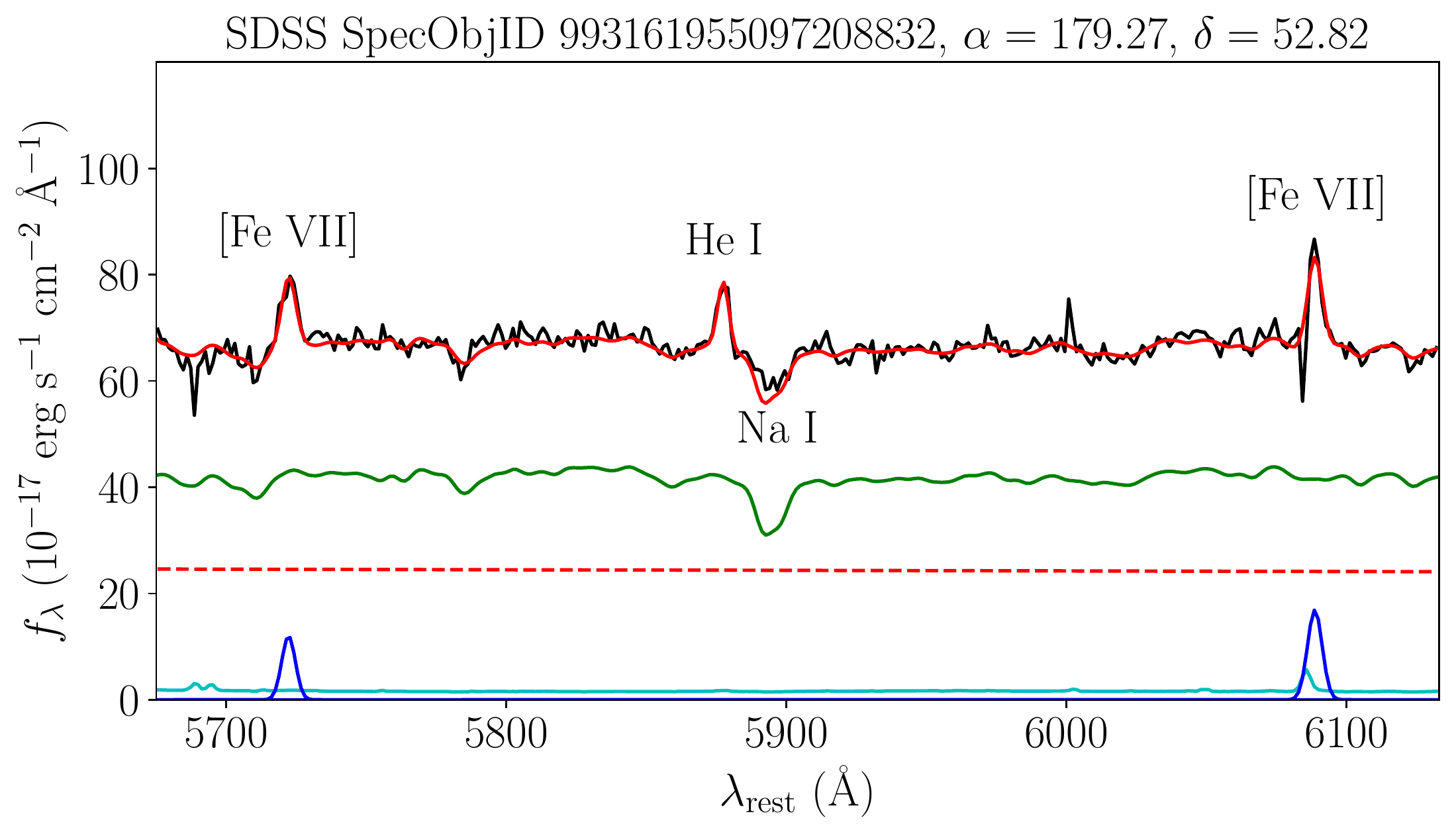}
    \caption{A selection of [\ion{Fe}{7}] $\lambda\lambda$5720,6087 detections from our subsample. The raw flux, corrected for redshift and galactic extinction, is plotted in black.  The BADASS model and each of its components (emission lines, host galaxy, AGN power law) are plotted in colors indicated by the legend.  Each coronal line is labeled, and the spectra's coordinates and SDSS Spec Object ID are given in the plot titles.}
    \label{fig:fevii_6087}
\end{figure*}

\begin{figure*}
    \centering
    \includegraphics[width=.48\textwidth]{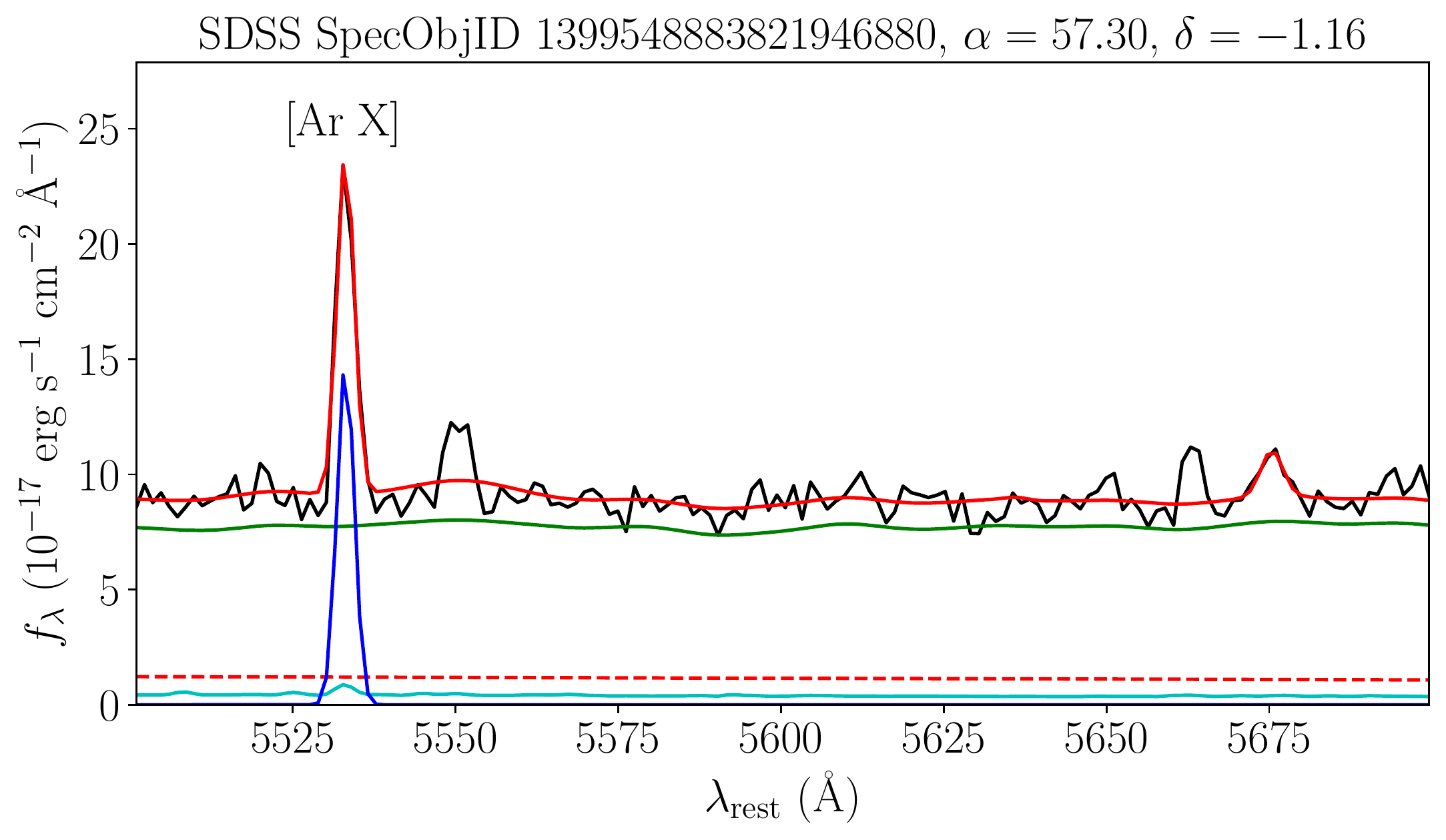}
    \includegraphics[width=.48\textwidth]{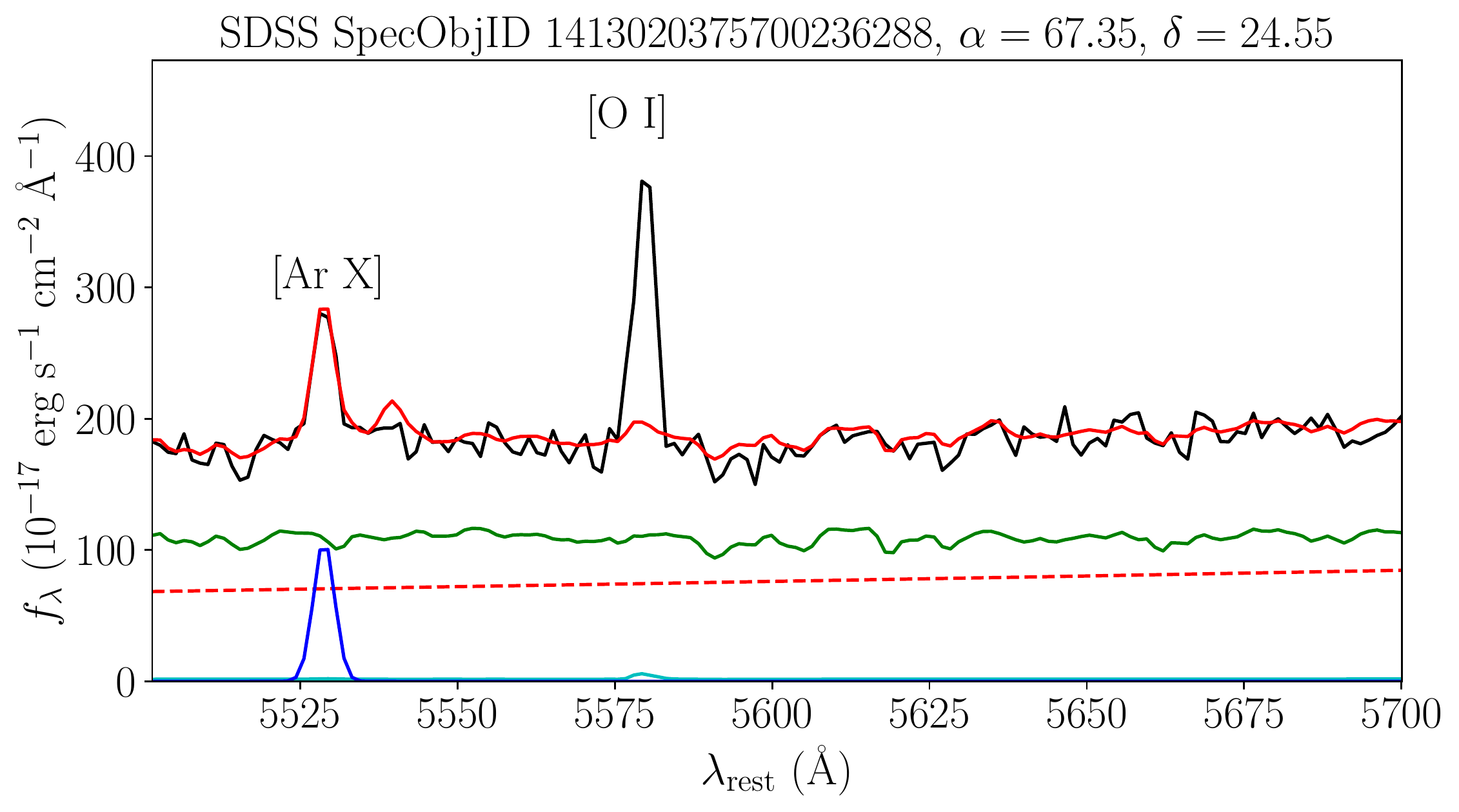}
    \includegraphics[width=.48\textwidth]{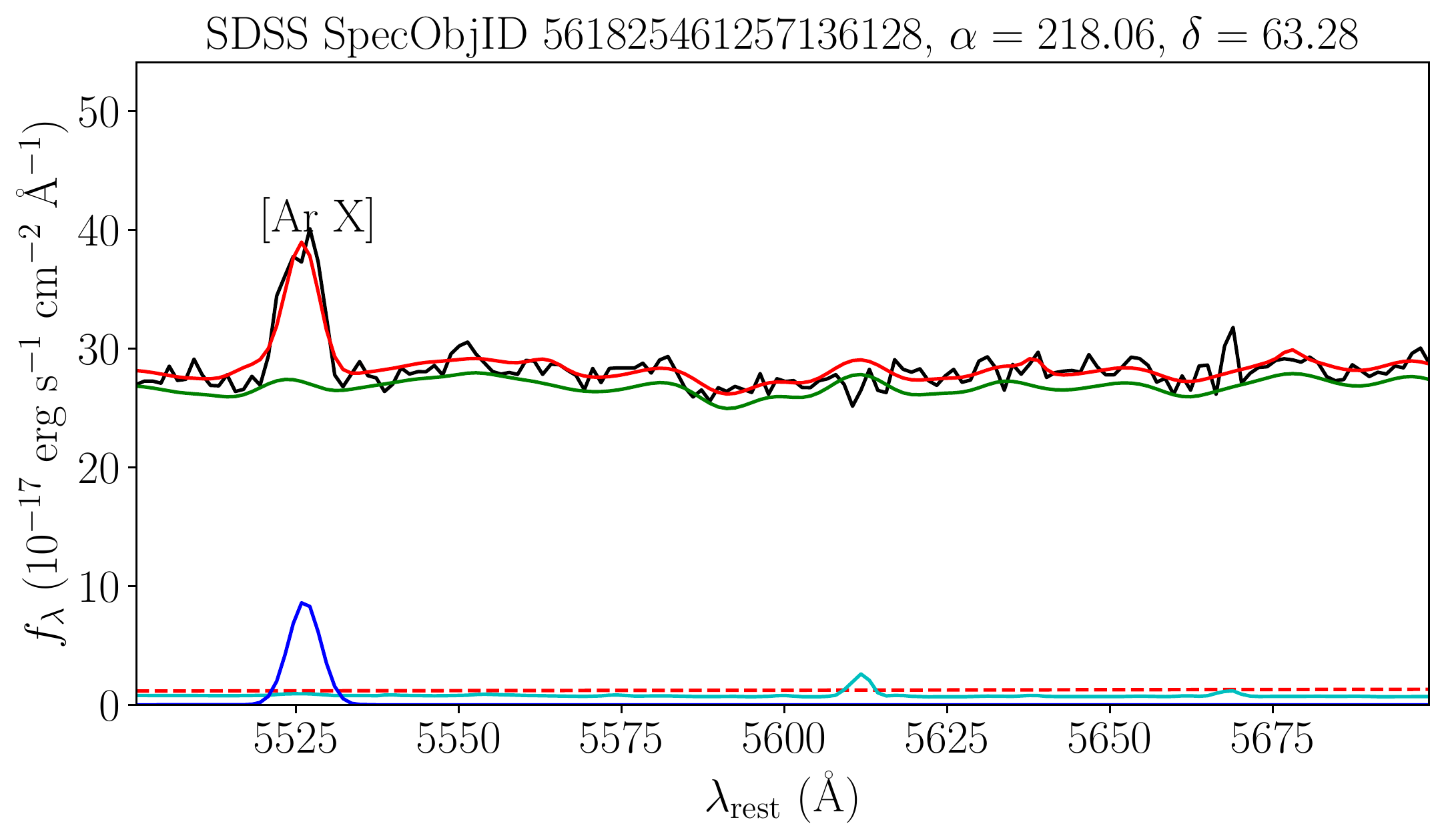}
    \includegraphics[width=.48\textwidth]{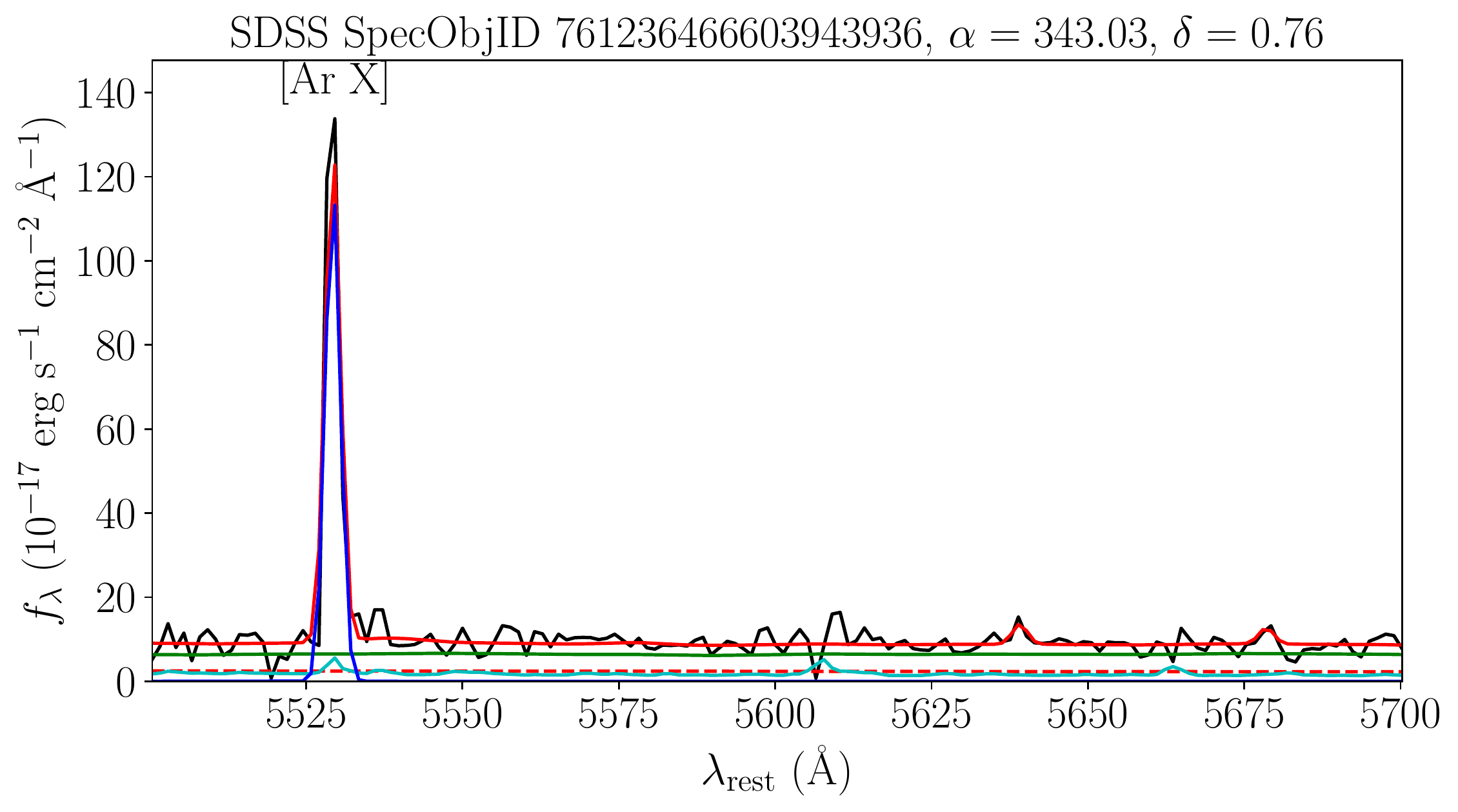}
    \includegraphics[width=.48\textwidth]{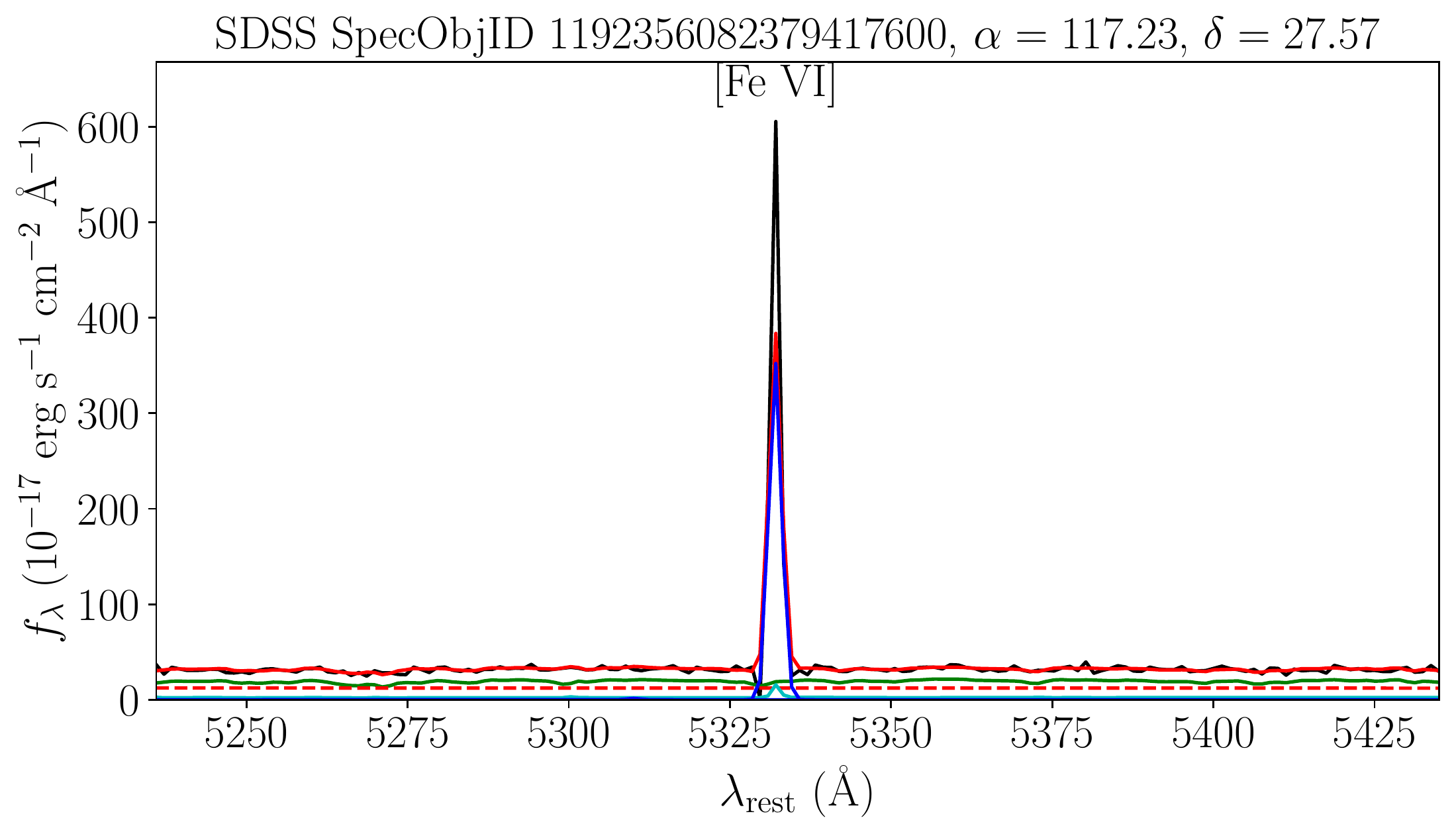}
    \includegraphics[width=.48\textwidth]{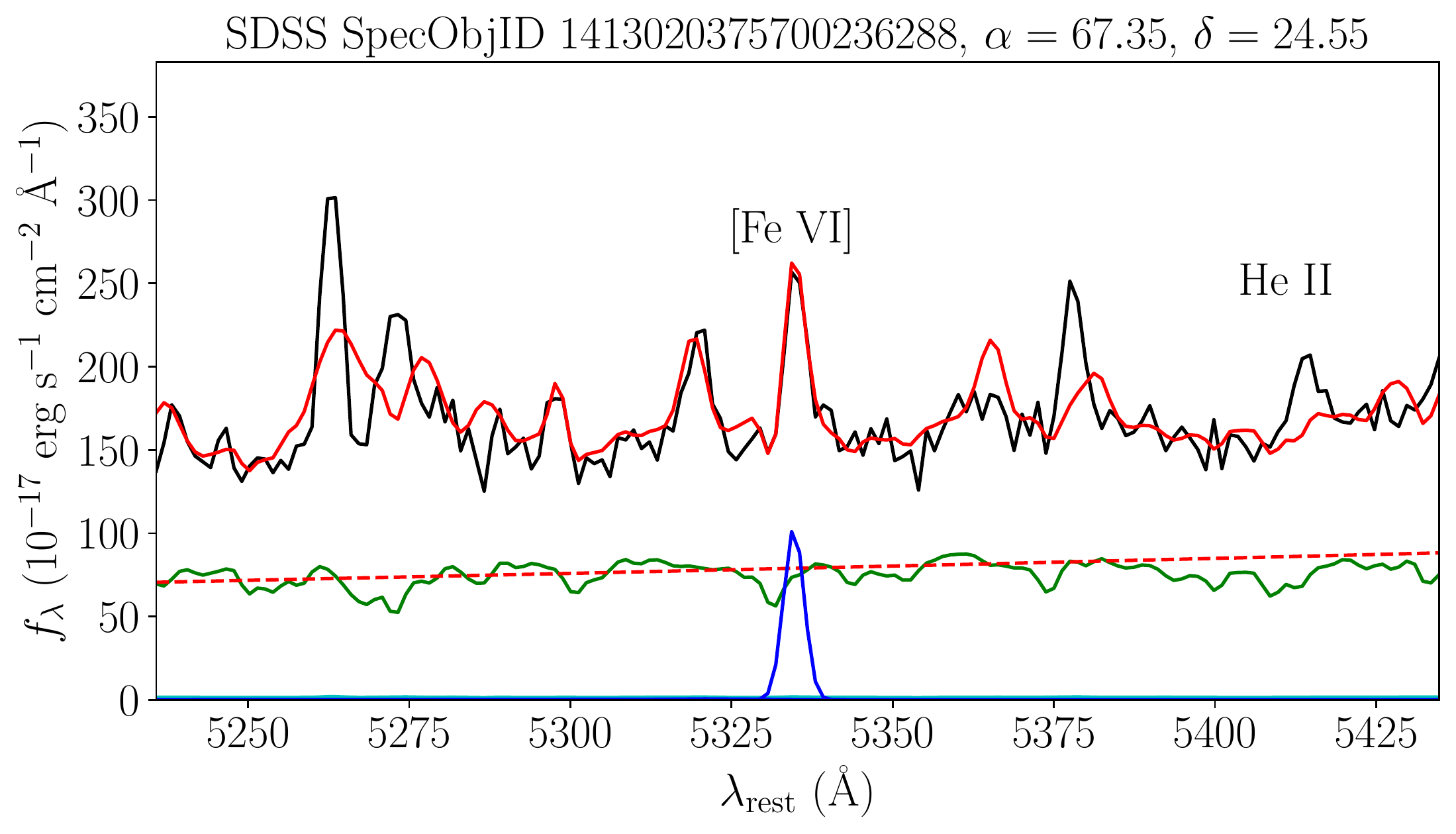}
    \includegraphics[width=.48\textwidth]{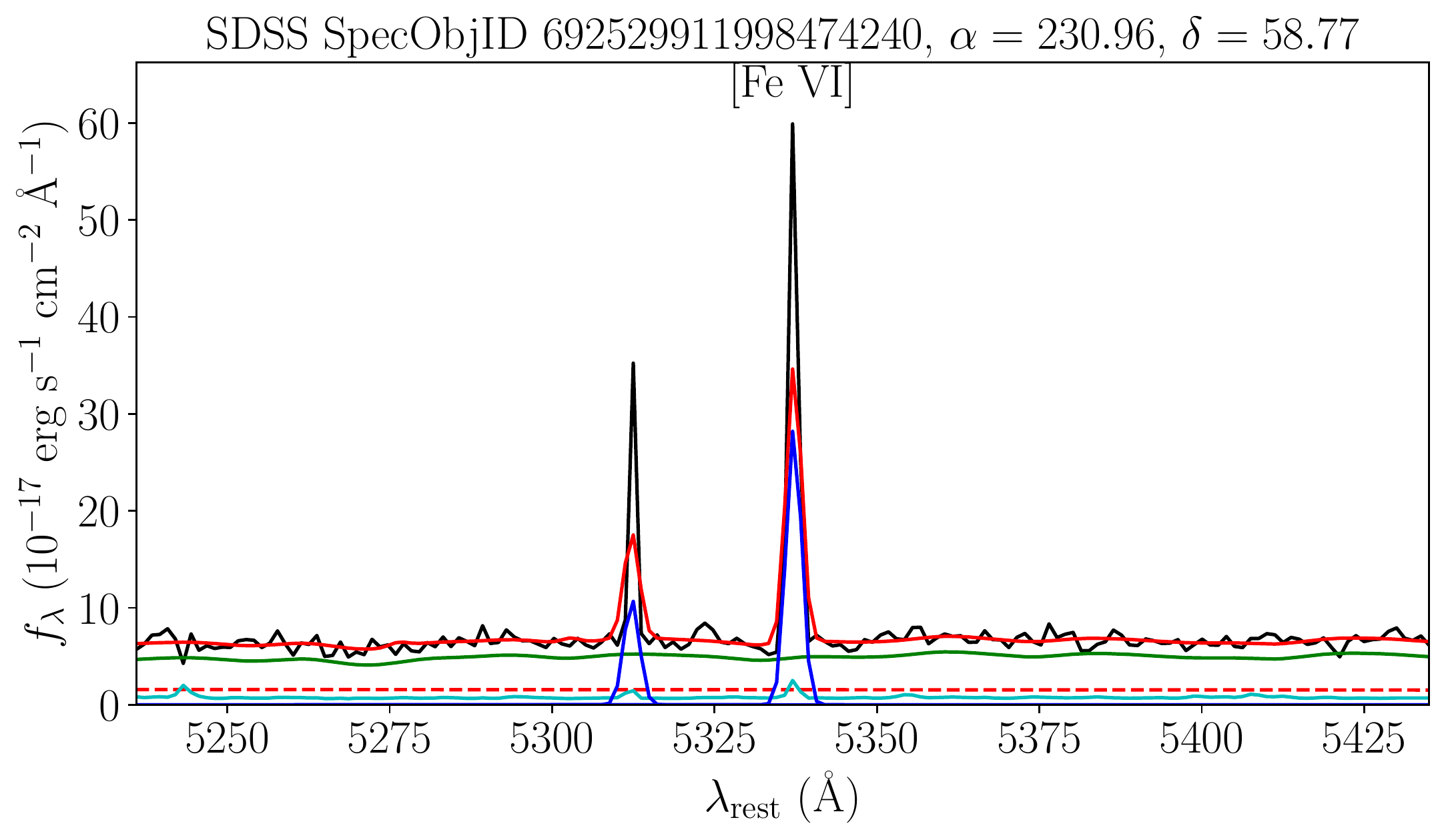}
    \includegraphics[width=.48\textwidth]{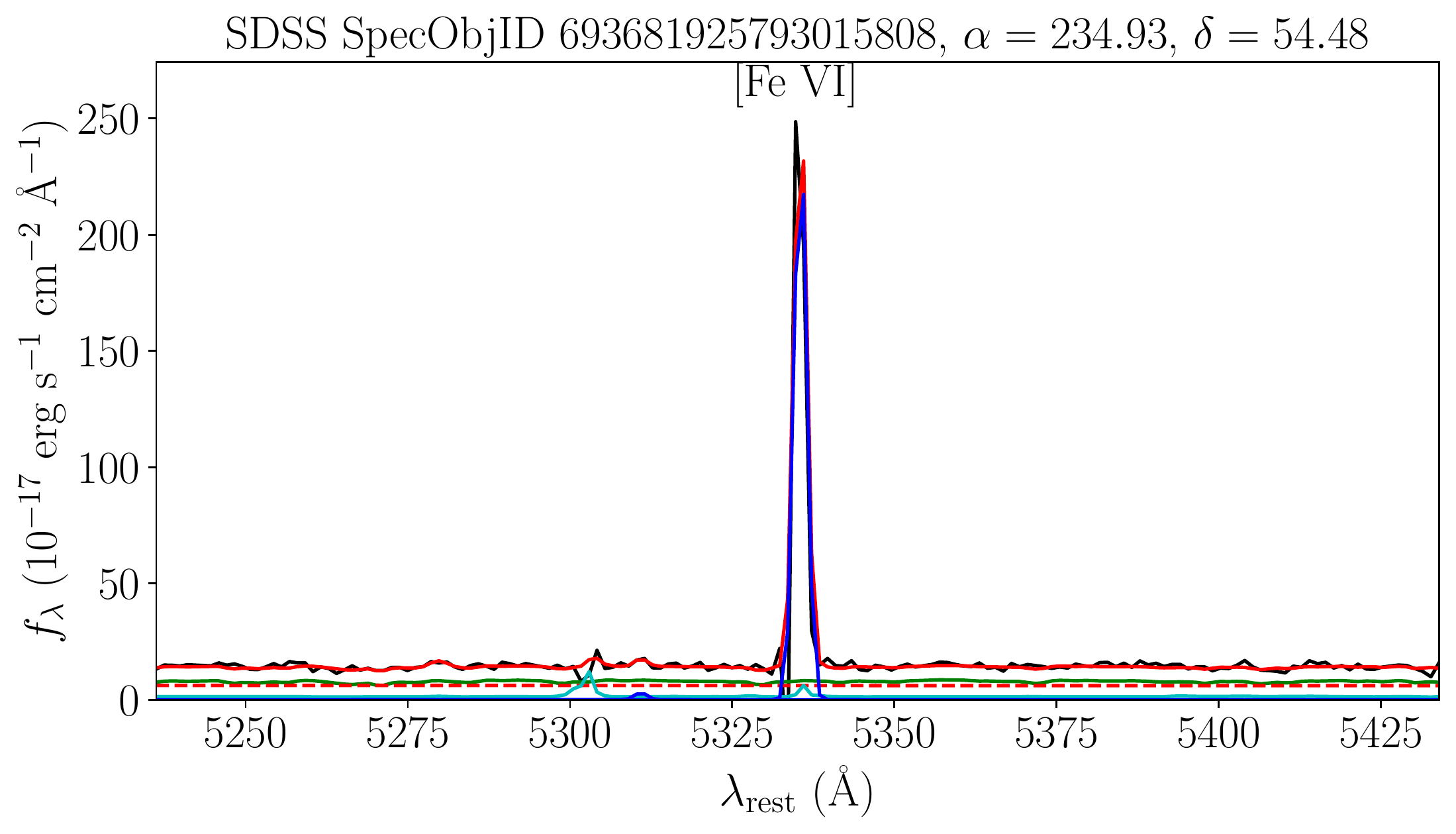}
    \caption{A selection of [\ion{Ar}{10}] $\lambda$5533 and [\ion{Fe}{6}] $\lambda$5335 detections from our subsample. The raw flux, corrected for redshift and galactic extinction, is plotted in black.  The BADASS model and each of its components (emission lines, host galaxy, AGN power law) are plotted in colors indicated by the legend.  Each coronal line is labeled, and the spectra's coordinates and SDSS Spec Object ID are given in the plot titles.}    \label{fig:misc_1}
\end{figure*}

\begin{figure*}
    \centering
    \includegraphics[width=.48\textwidth]{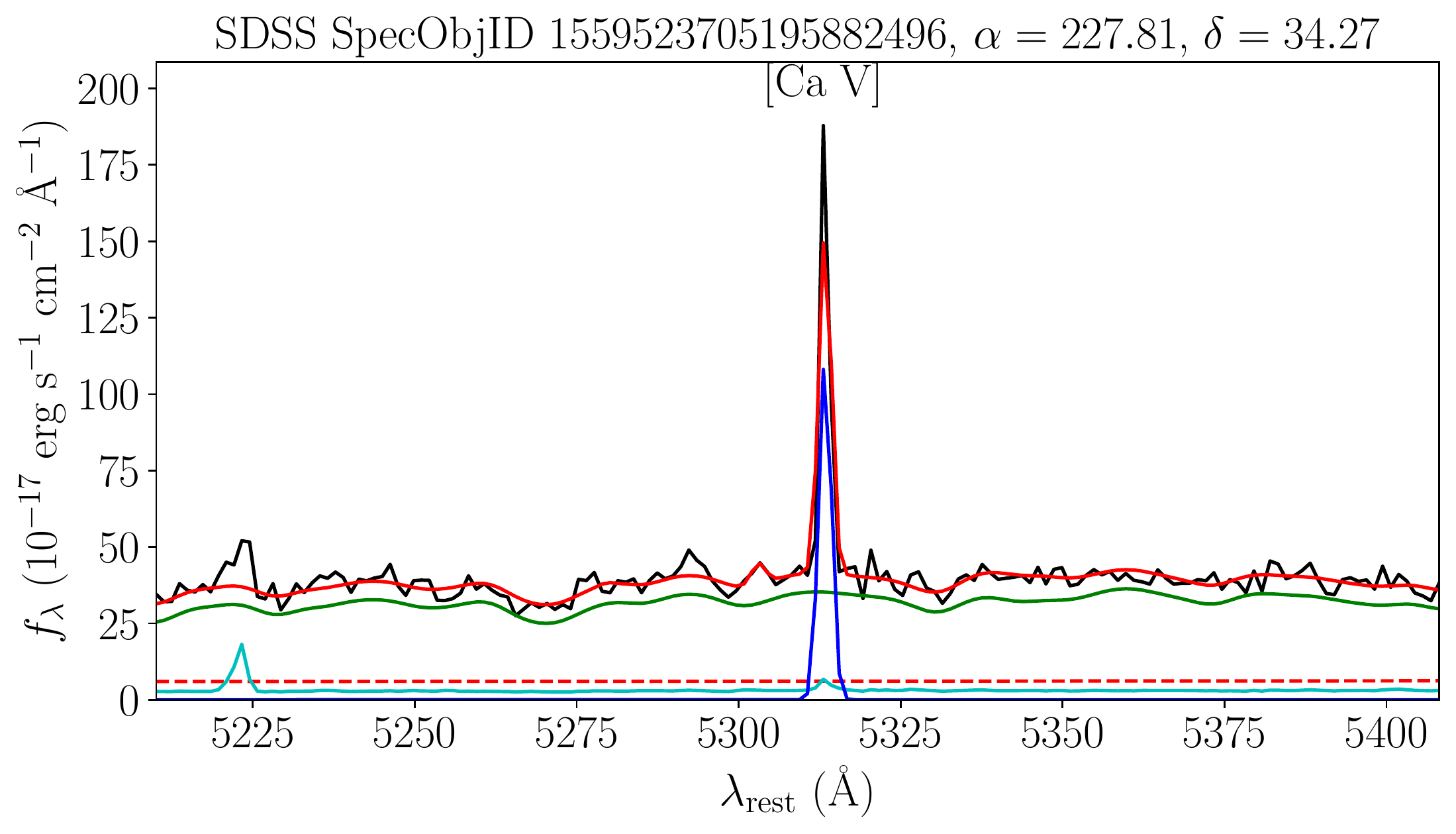}
    \includegraphics[width=.48\textwidth]{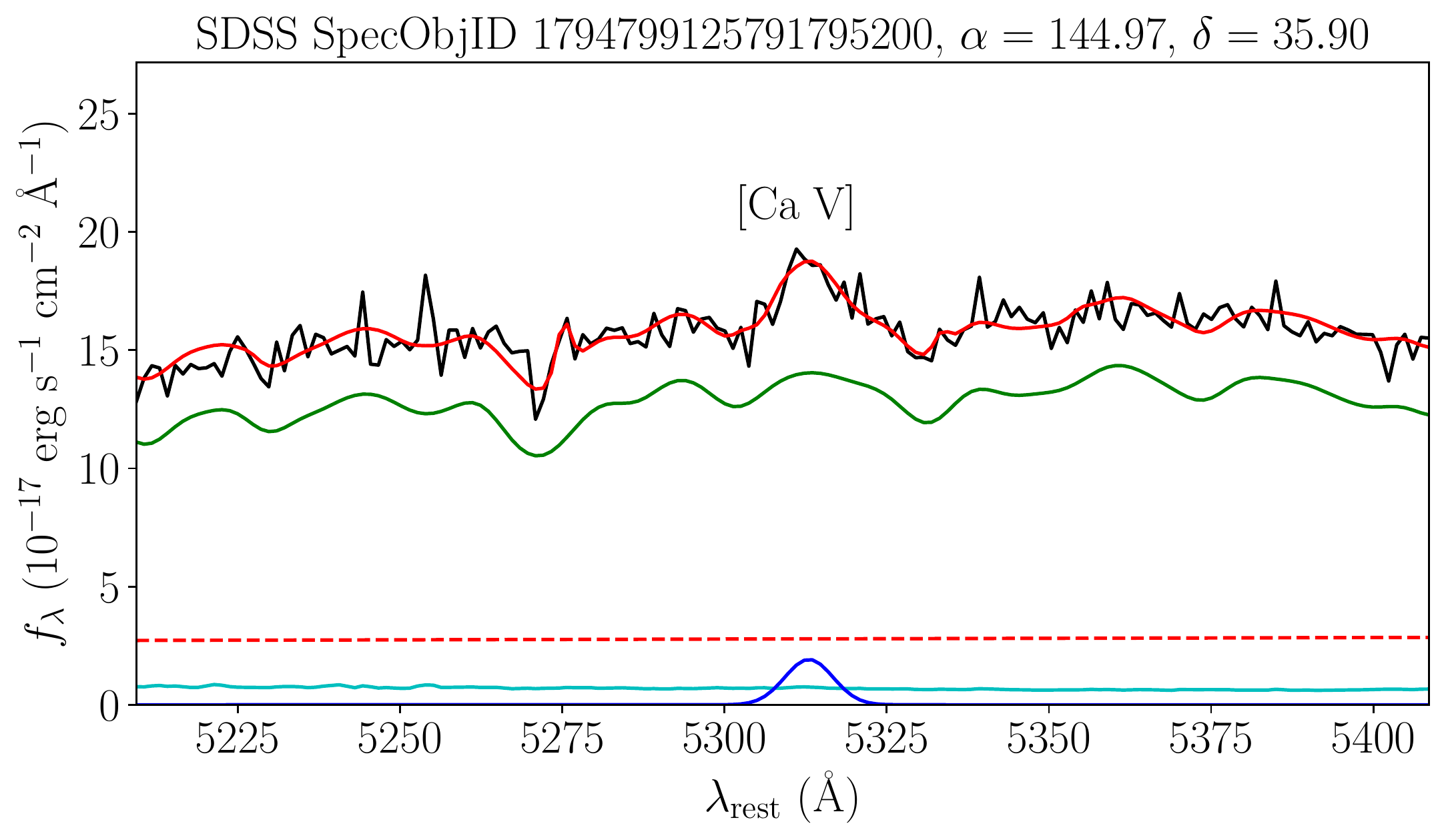}
    \includegraphics[width=.48\textwidth]{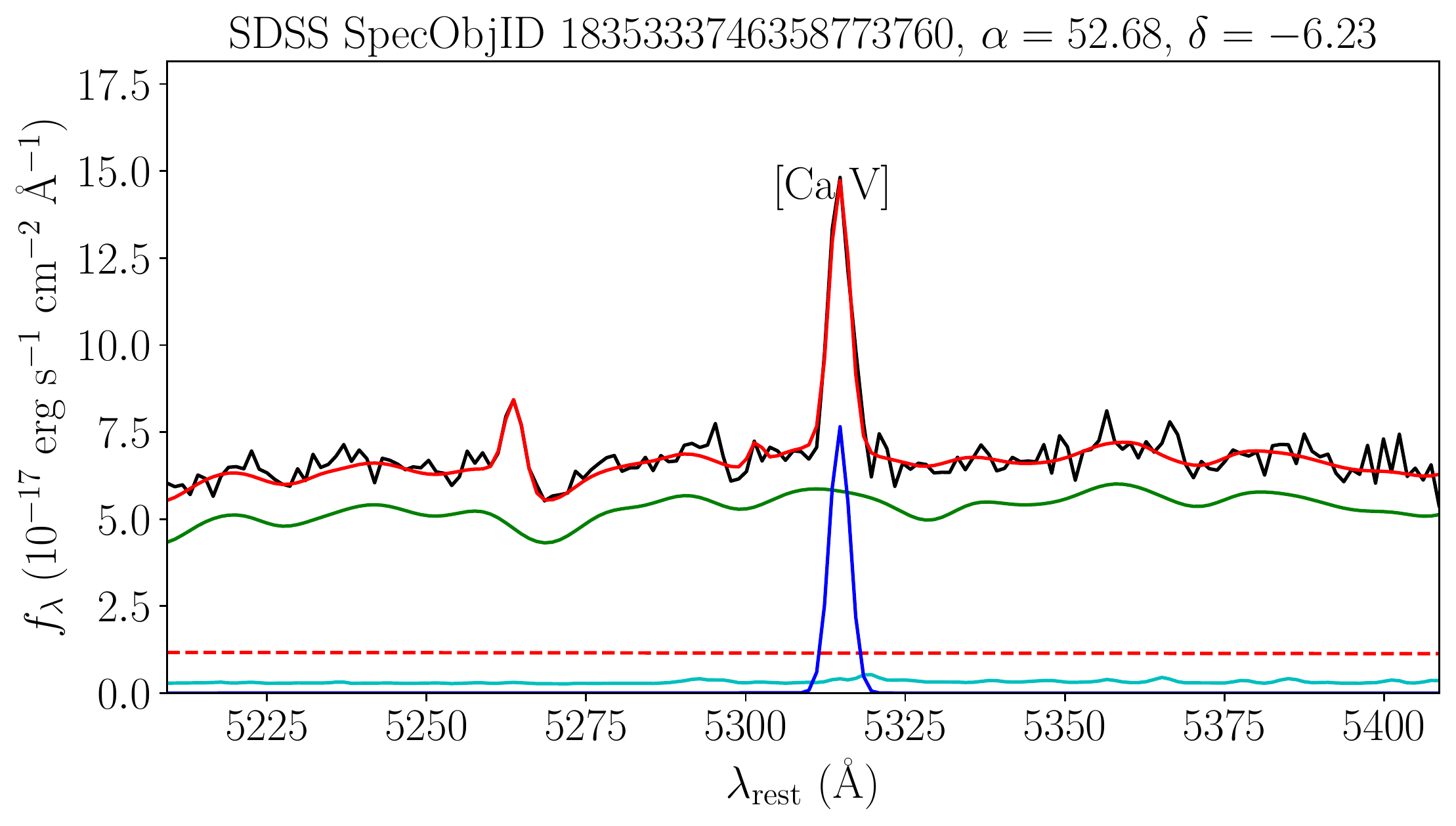}
    \includegraphics[width=.48\textwidth]{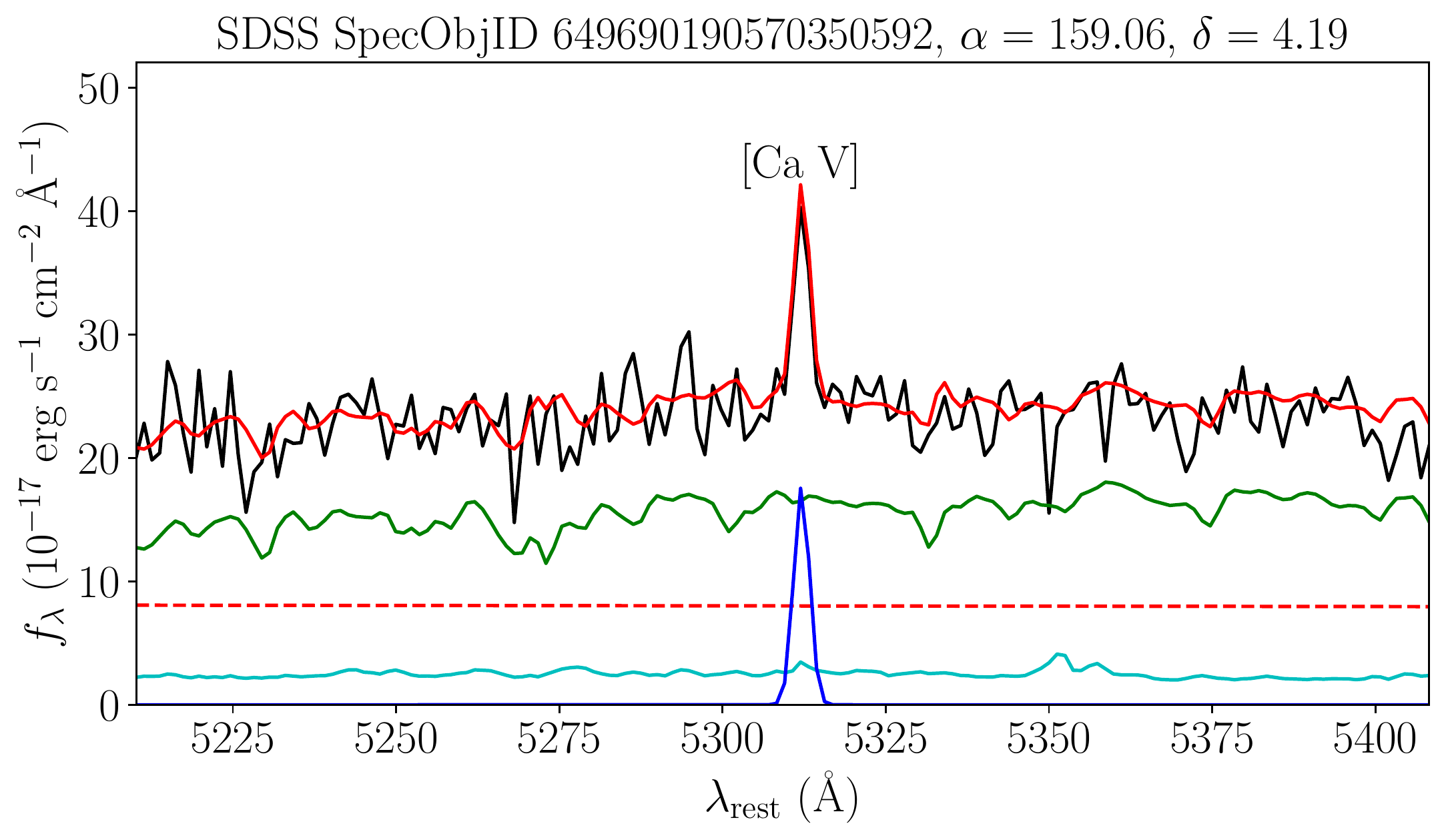}
    \includegraphics[width=.48\textwidth]{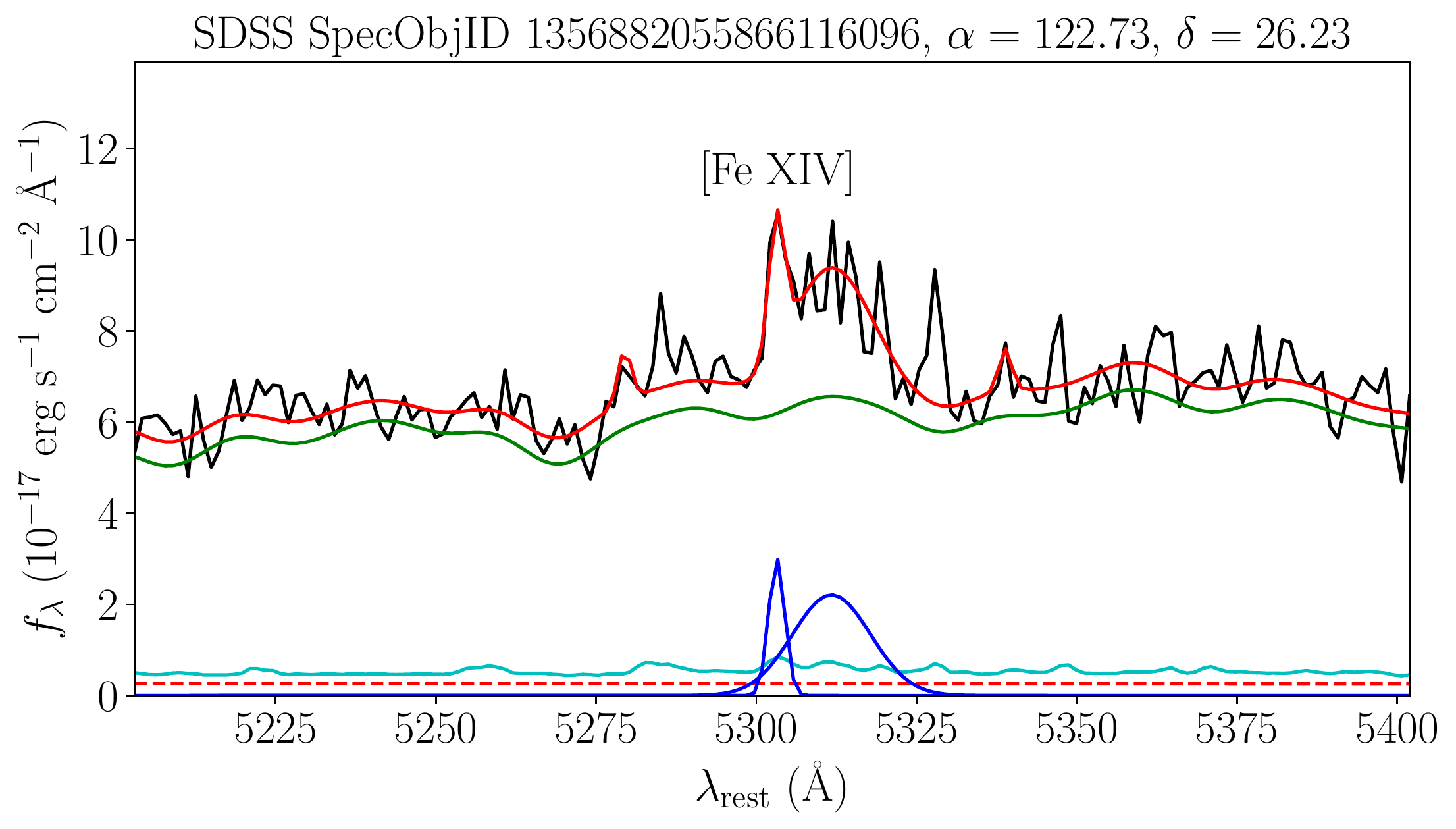}
    \includegraphics[width=.48\textwidth]{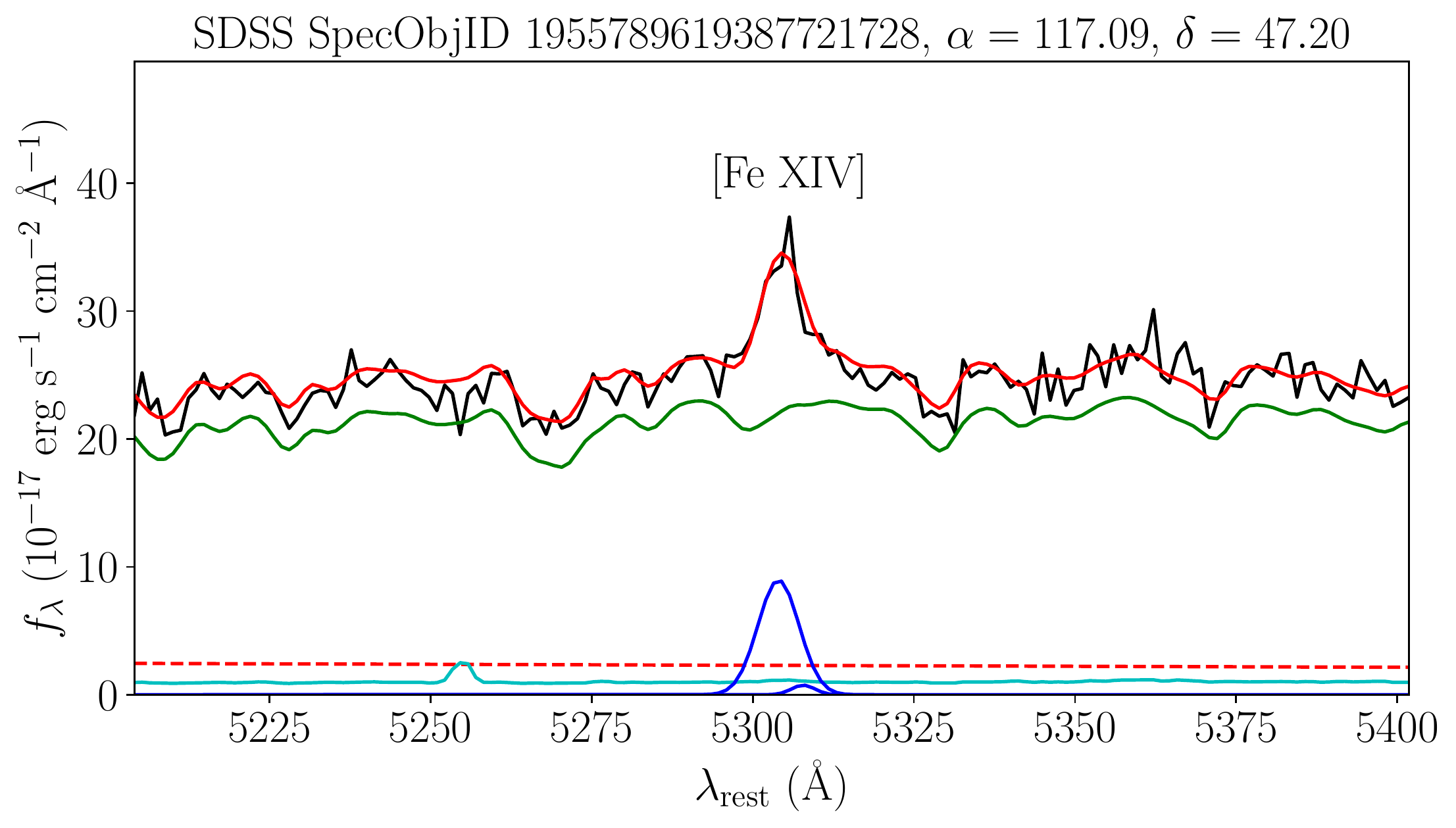}
    \includegraphics[width=.48\textwidth]{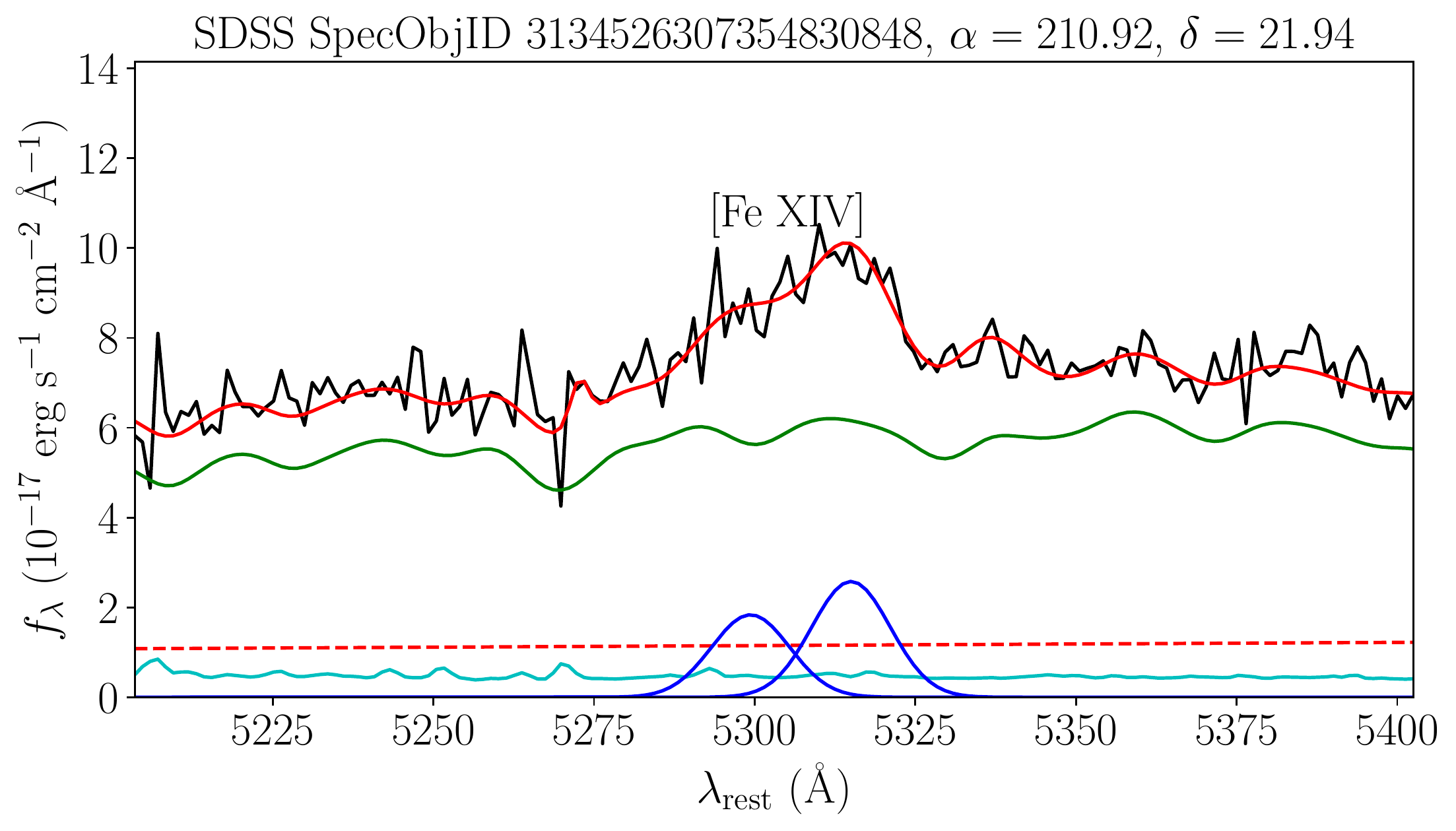}
    \includegraphics[width=.48\textwidth]{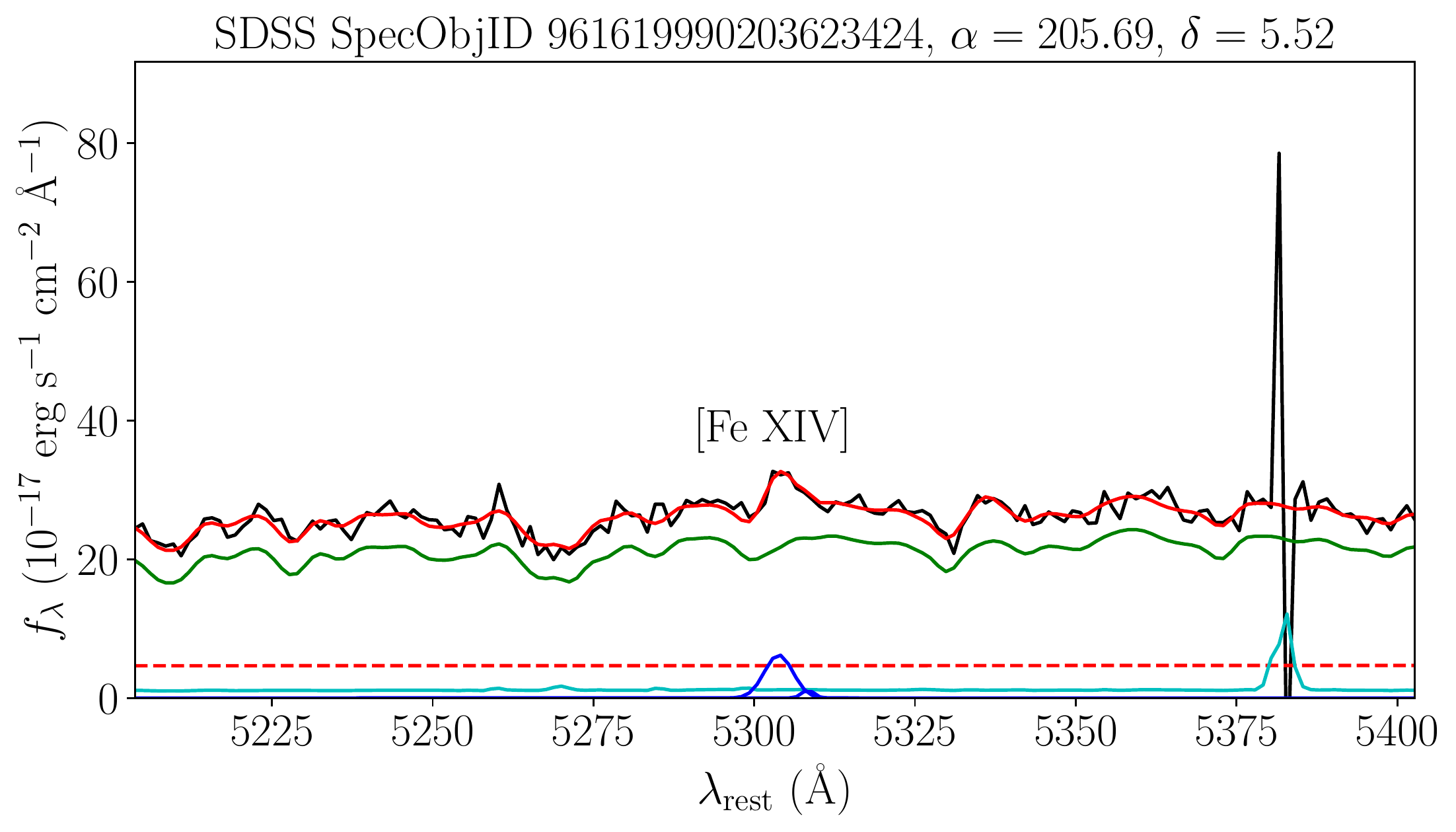}
    \caption{A selection of [\ion{Ca}{5}] $\lambda$5309 and [\ion{Fe}{14}] $\lambda$5303 detections from our subsample. The raw flux, corrected for redshift and galactic extinction, is plotted in black.  The BADASS model and each of its components (emission lines, host galaxy, AGN power law) are plotted in colors indicated by the legend.  Each coronal line is labeled, and the spectra's coordinates and SDSS Spec Object ID are given in the plot titles.}    
    \label{fig:misc_2}
\end{figure*}

\begin{figure*}
    \centering
    \includegraphics[width=.48\textwidth]{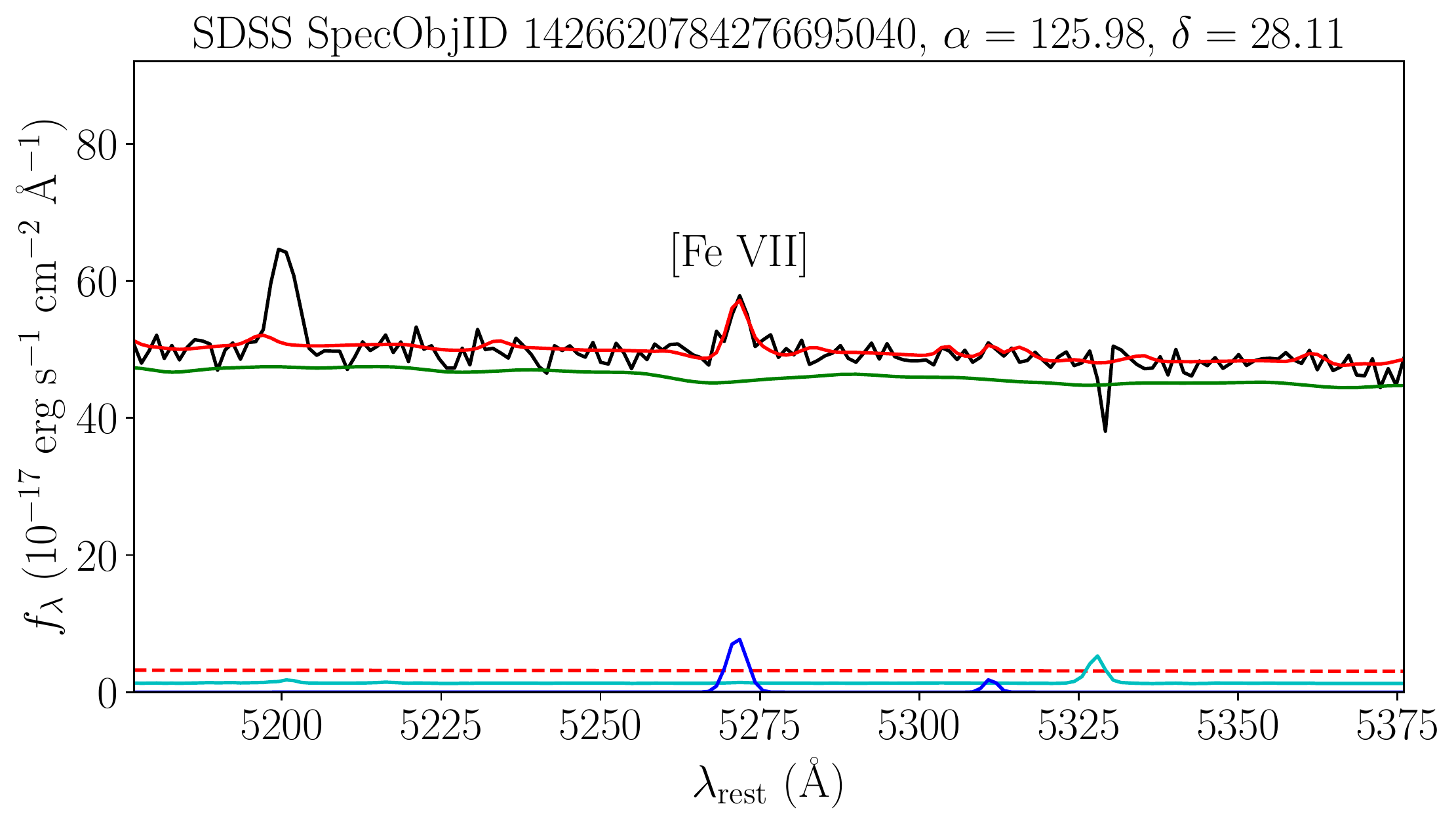}
    \includegraphics[width=.48\textwidth]{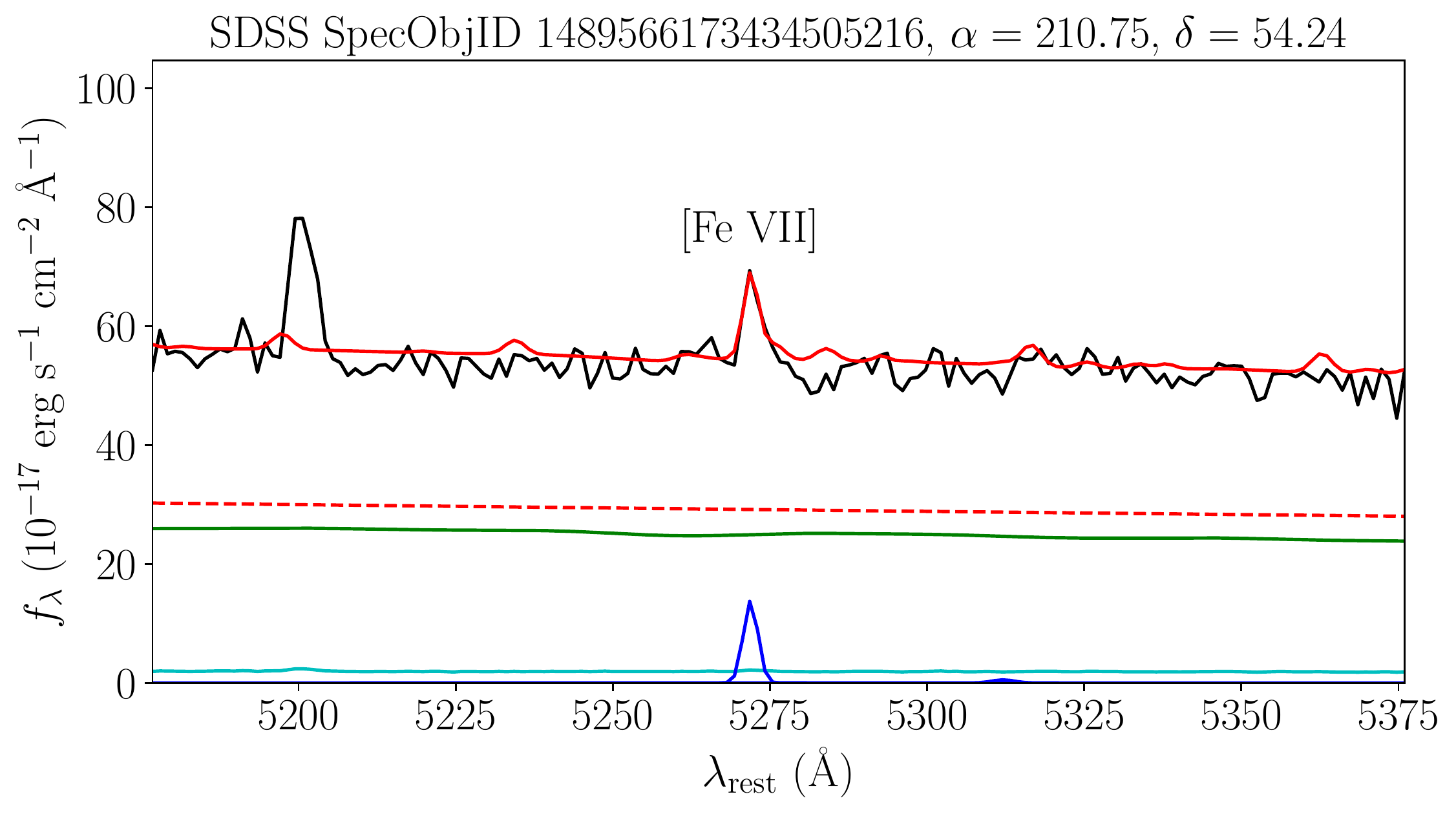}
    \includegraphics[width=.48\textwidth]{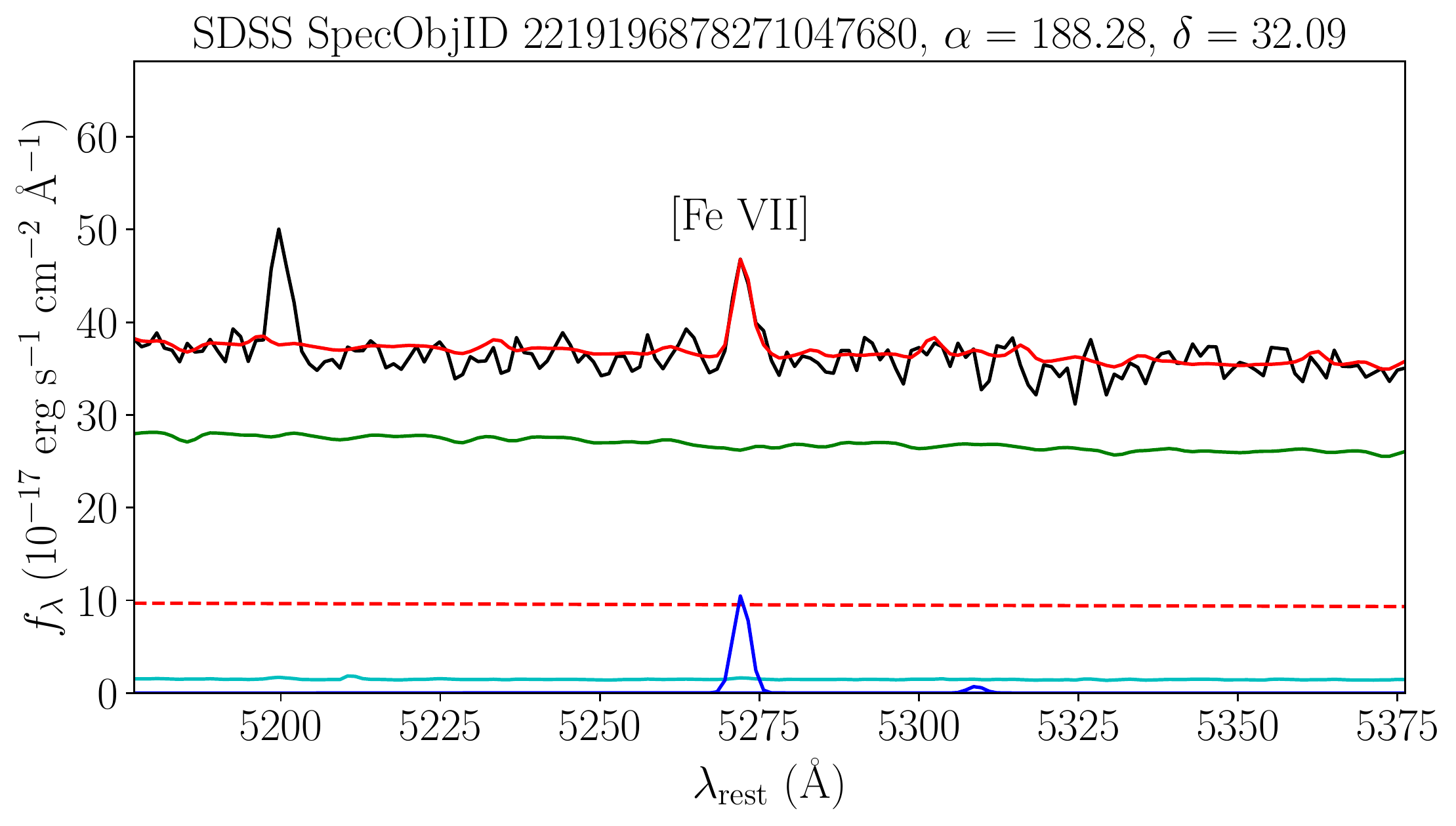}
    \includegraphics[width=.48\textwidth]{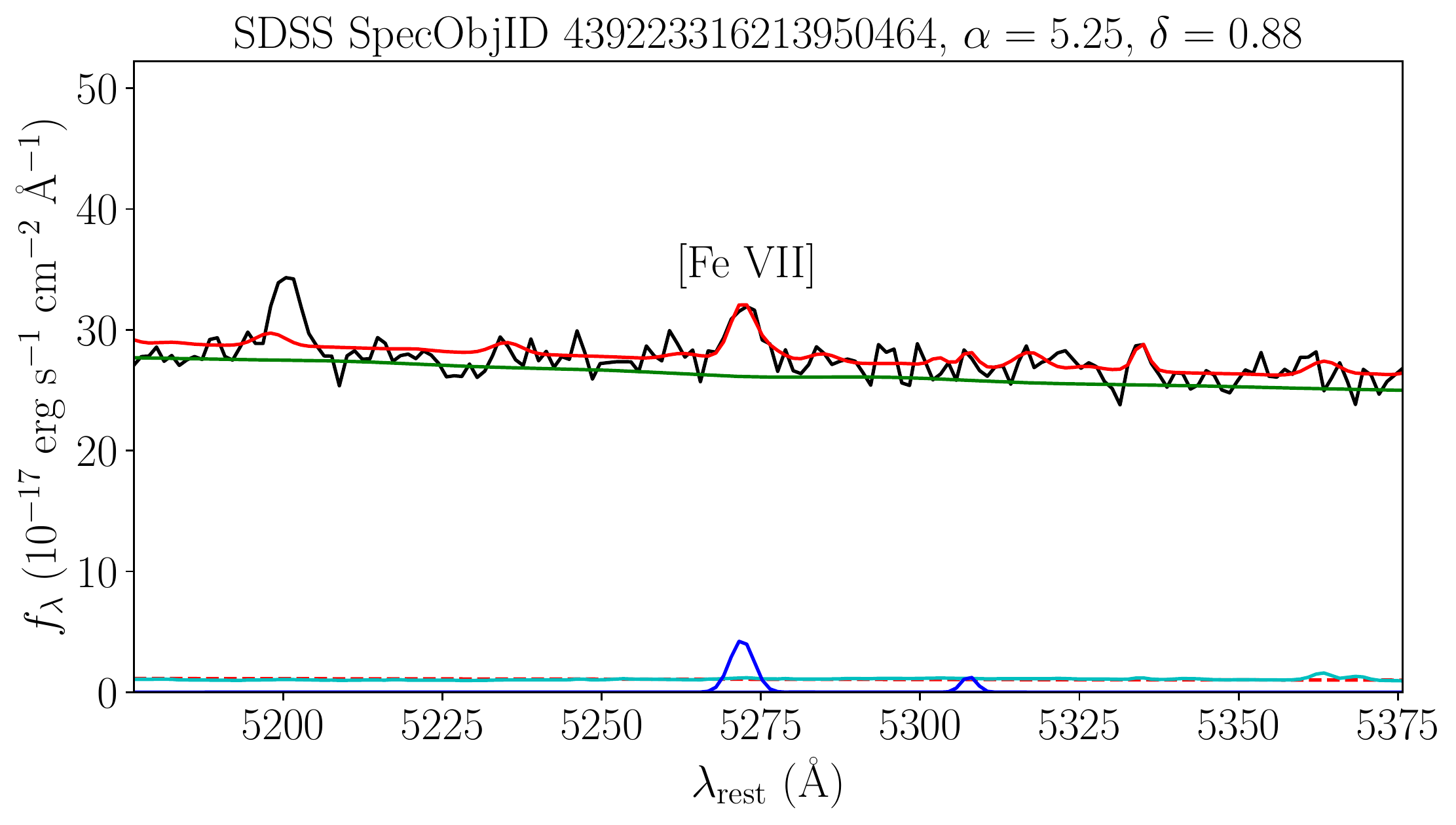}
    \includegraphics[width=.48\textwidth]{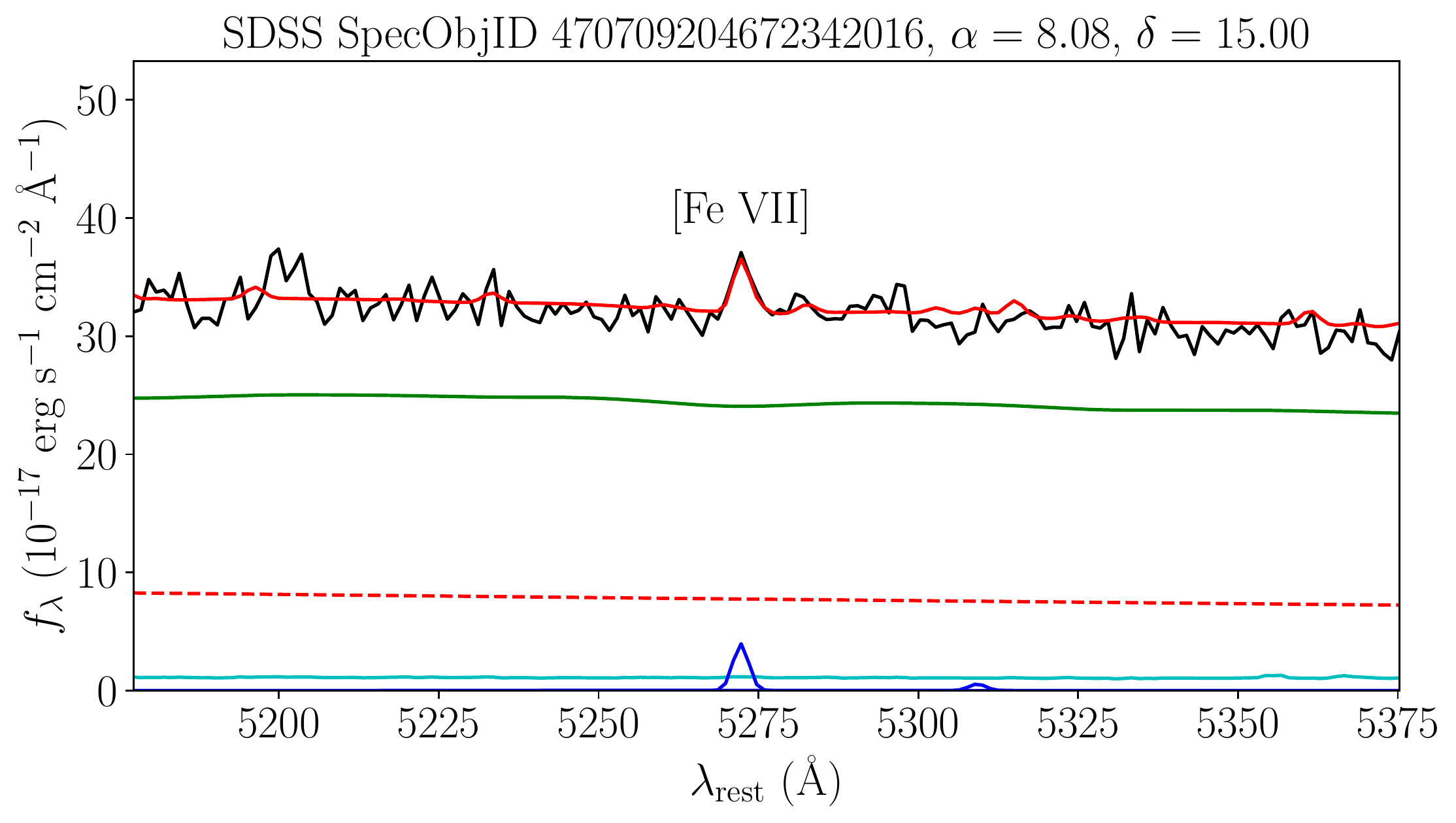}
    \includegraphics[width=.48\textwidth]{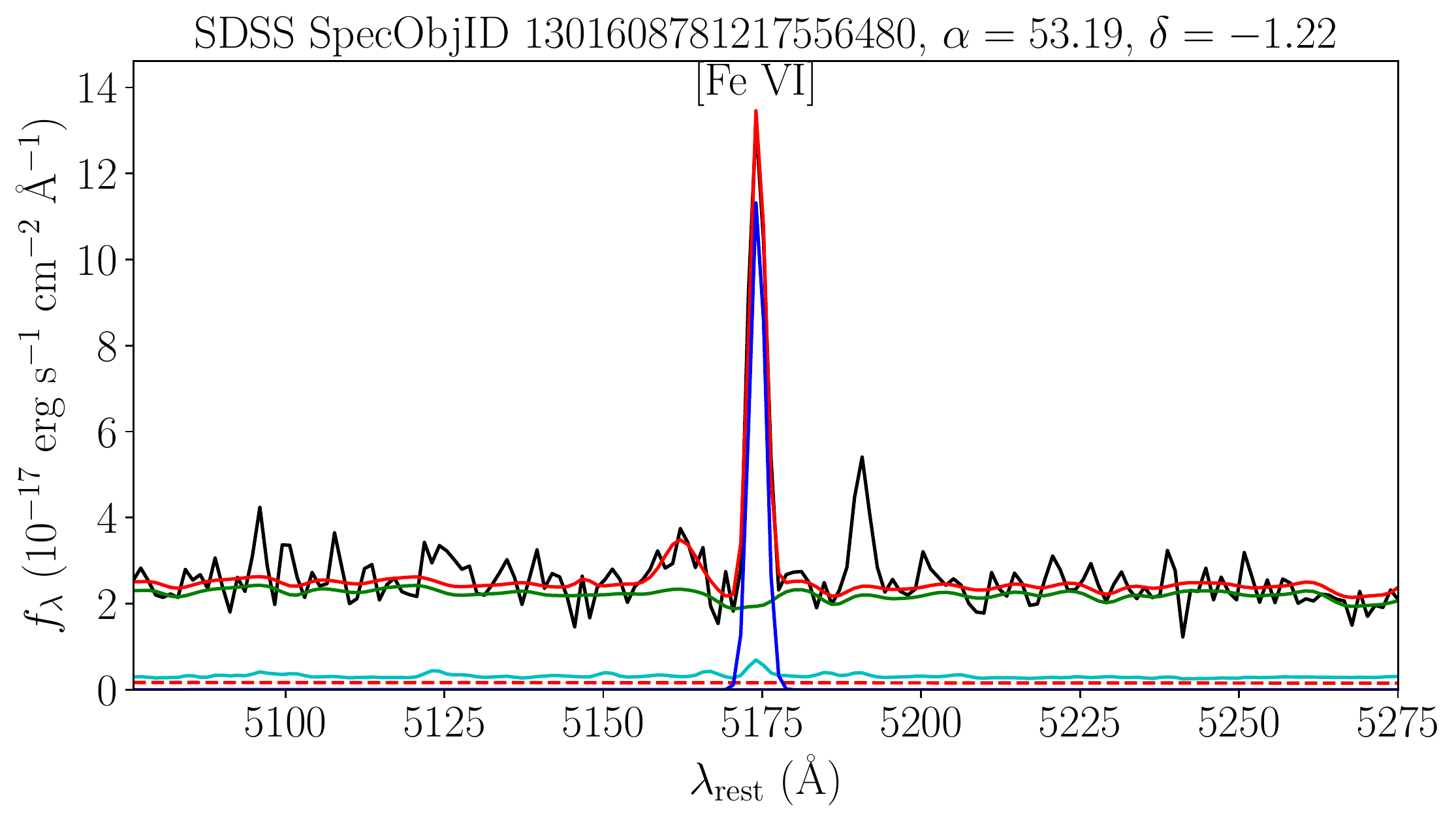}
    \includegraphics[width=.48\textwidth]{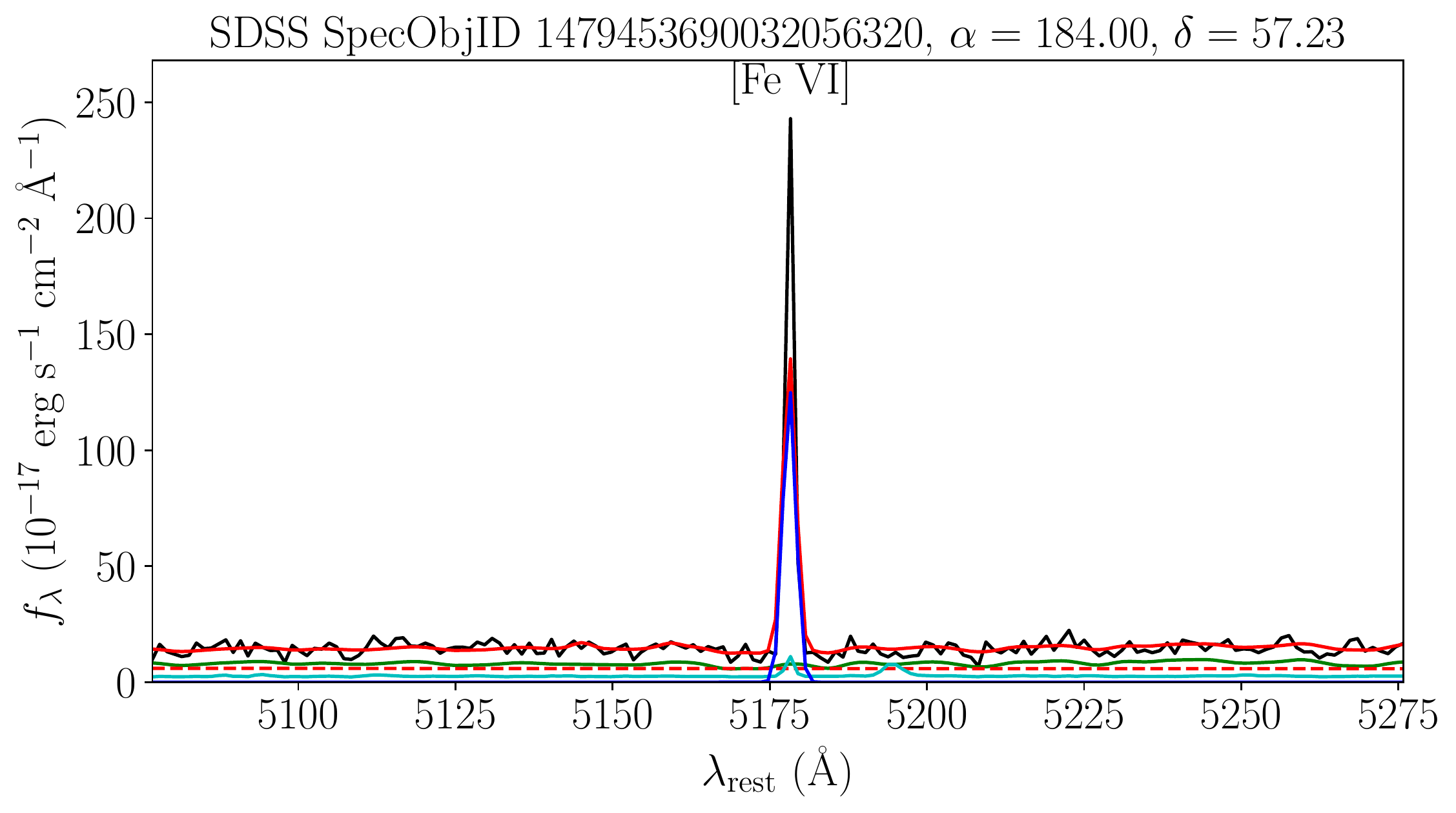}
    \includegraphics[width=.48\textwidth]{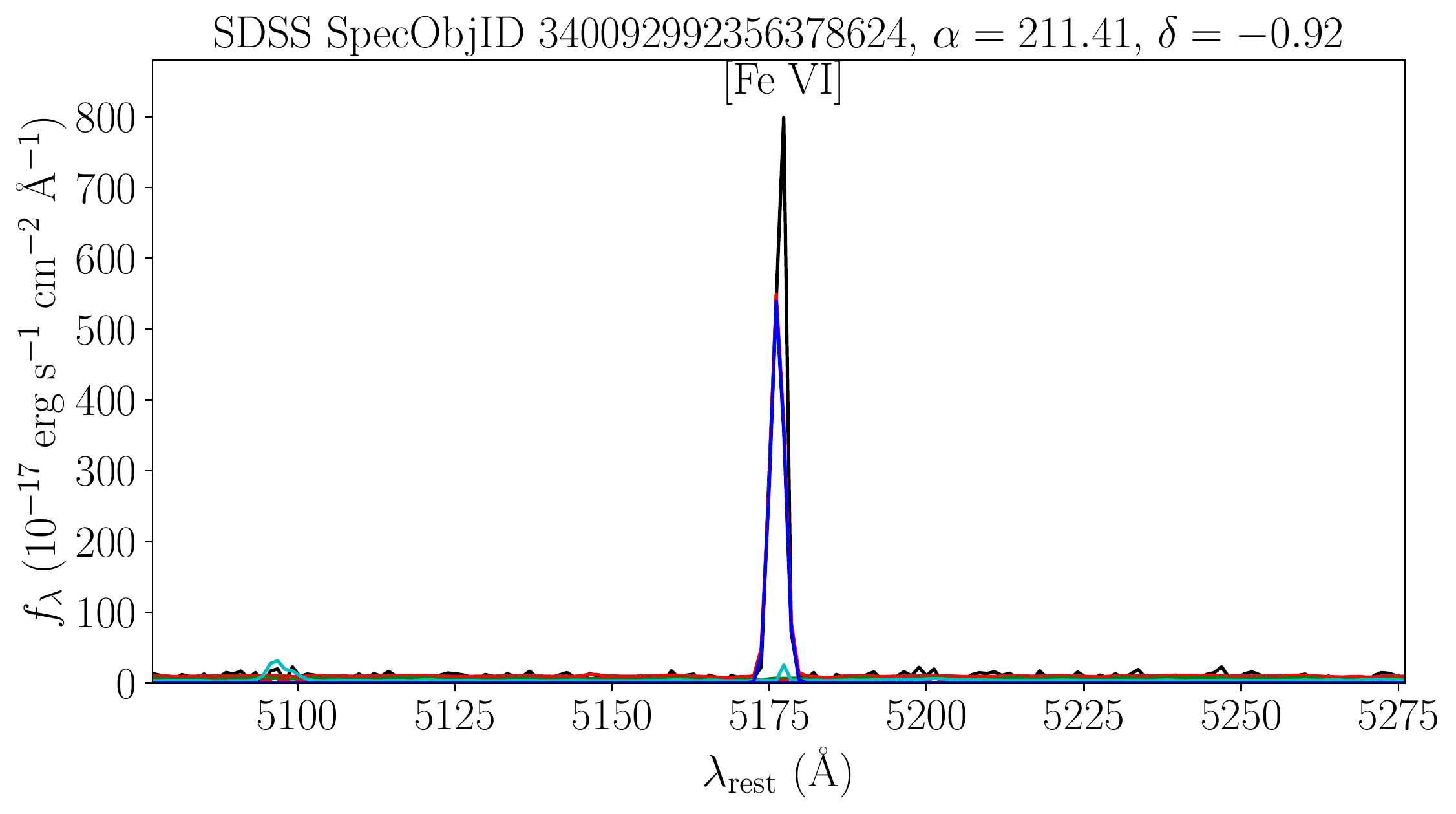}
    \caption{A selection of [\ion{Fe}{7}] $\lambda$5276 and [\ion{Fe}{6}] $\lambda$5176 detections from our subsample. The raw flux, corrected for redshift and galactic extinction, is plotted in black.  The BADASS model and each of its components (emission lines, host galaxy, AGN power law) are plotted in colors indicated by the legend.  Each coronal line is labeled, and the spectra's coordinates and SDSS Spec Object ID are given in the plot titles.}    
    \label{fig:misc_3}
\end{figure*}

\begin{figure*}
    \centering
    \includegraphics[width=.48\textwidth]{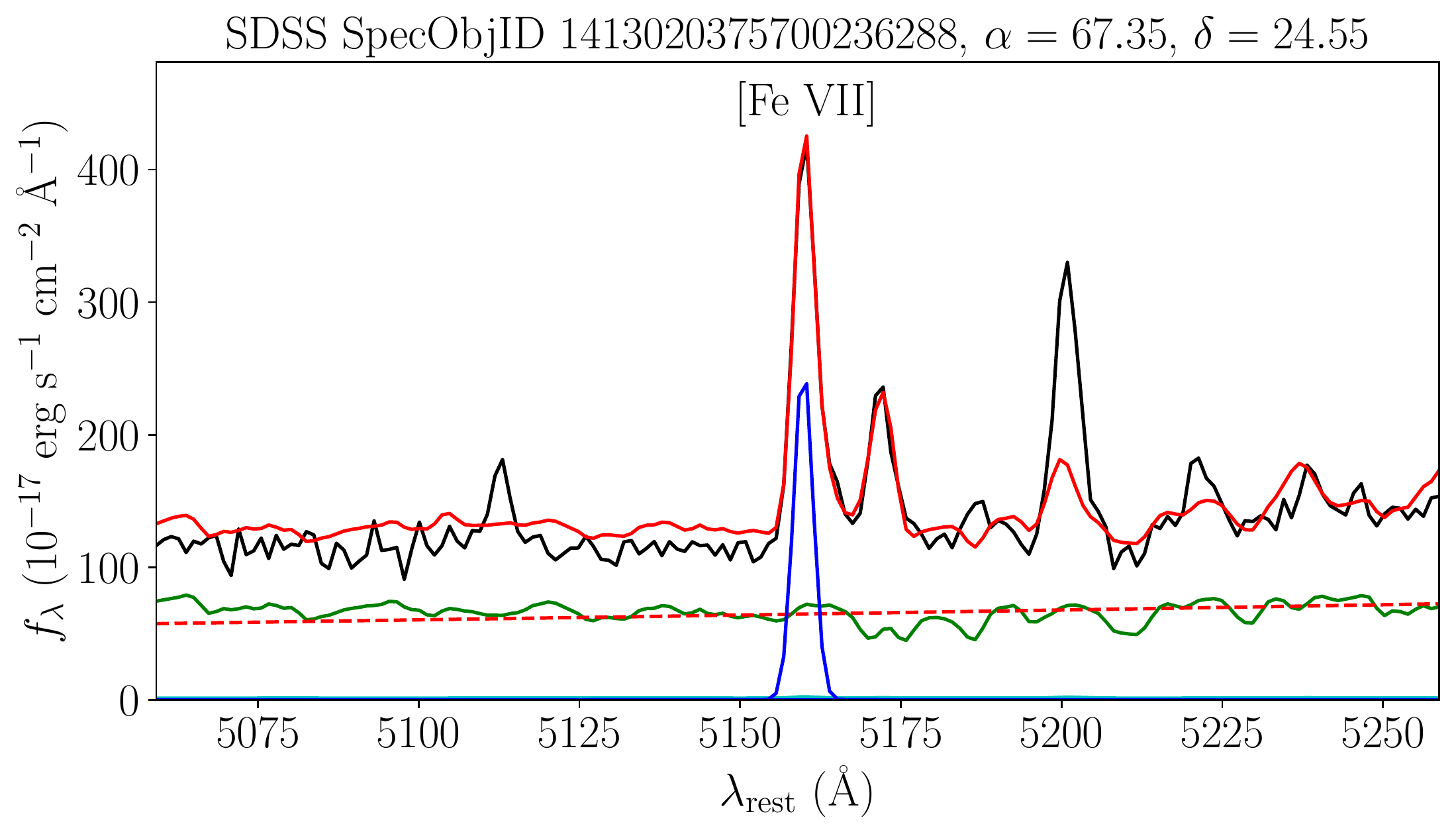}
    \includegraphics[width=.48\textwidth]{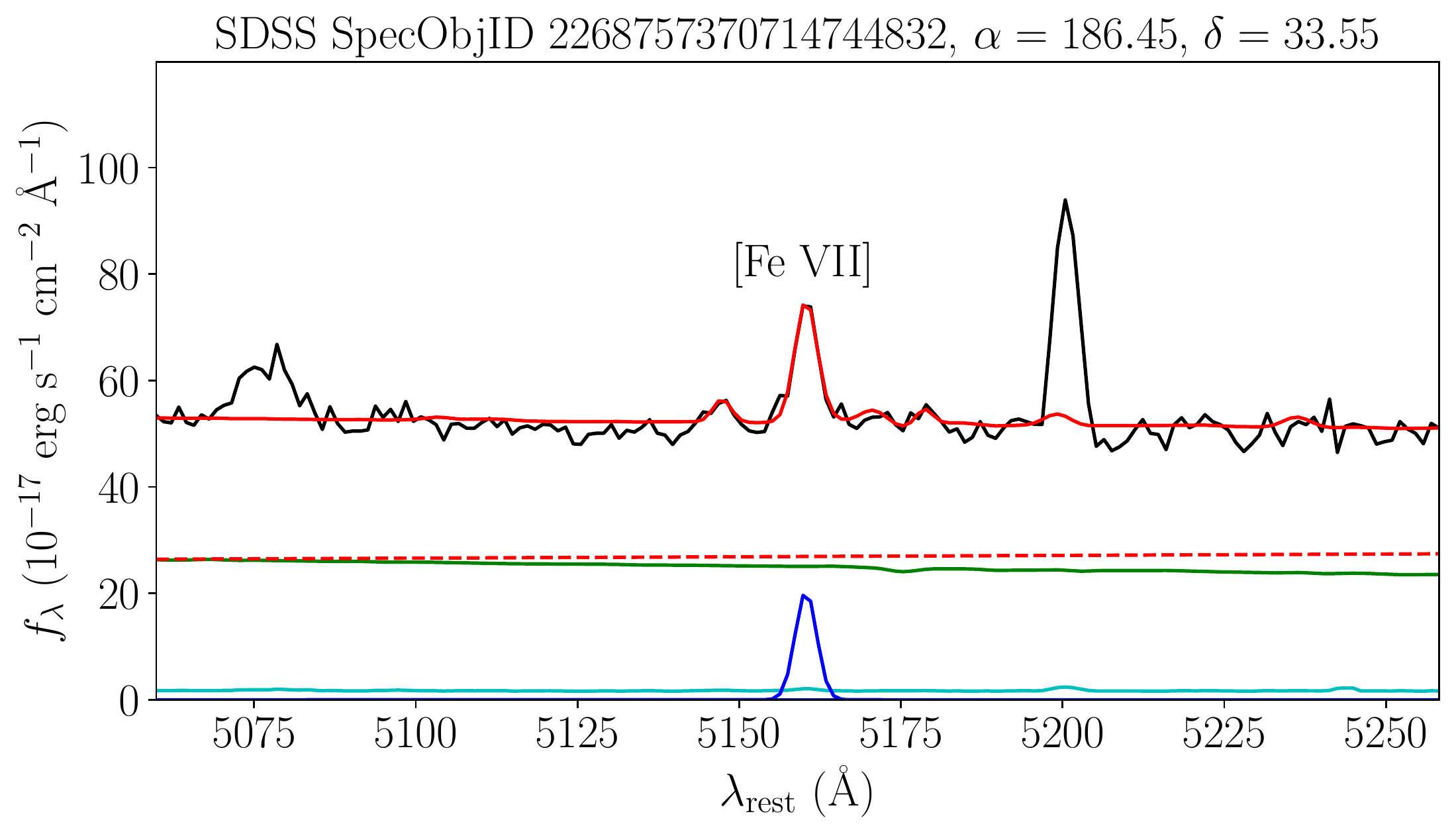}
    \includegraphics[width=.48\textwidth]{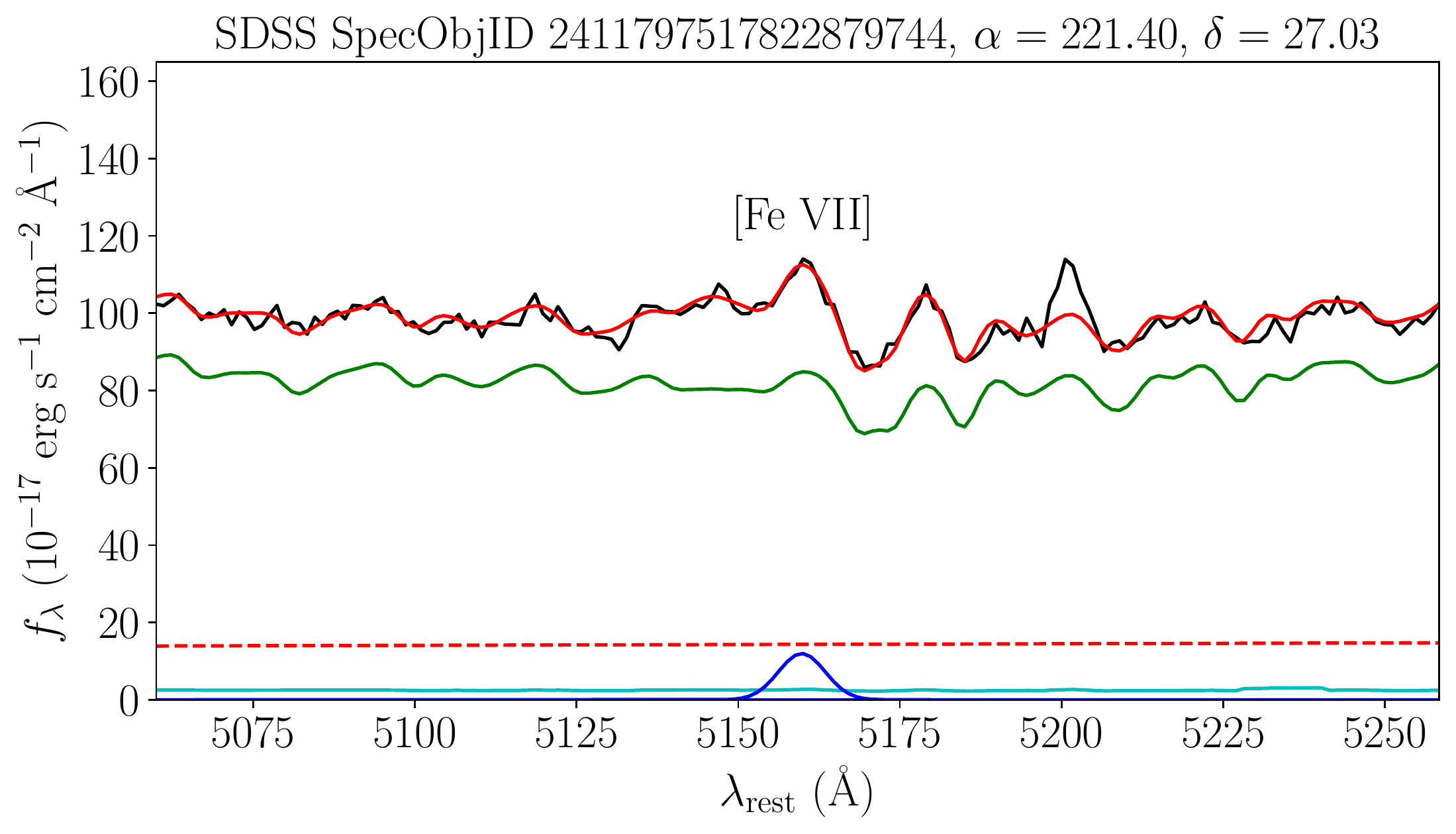}
    \includegraphics[width=.48\textwidth]{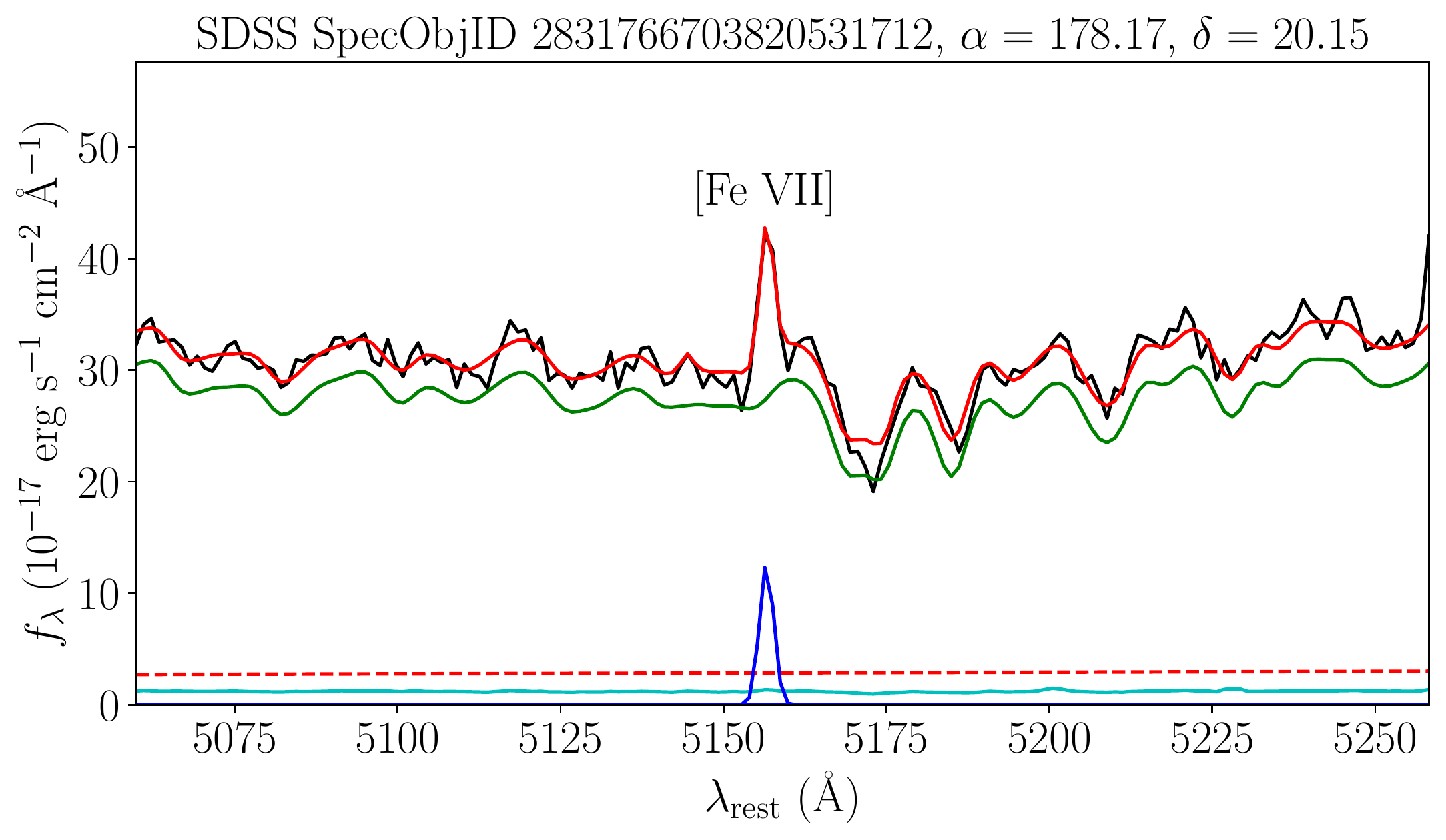}
    \includegraphics[width=.48\textwidth]{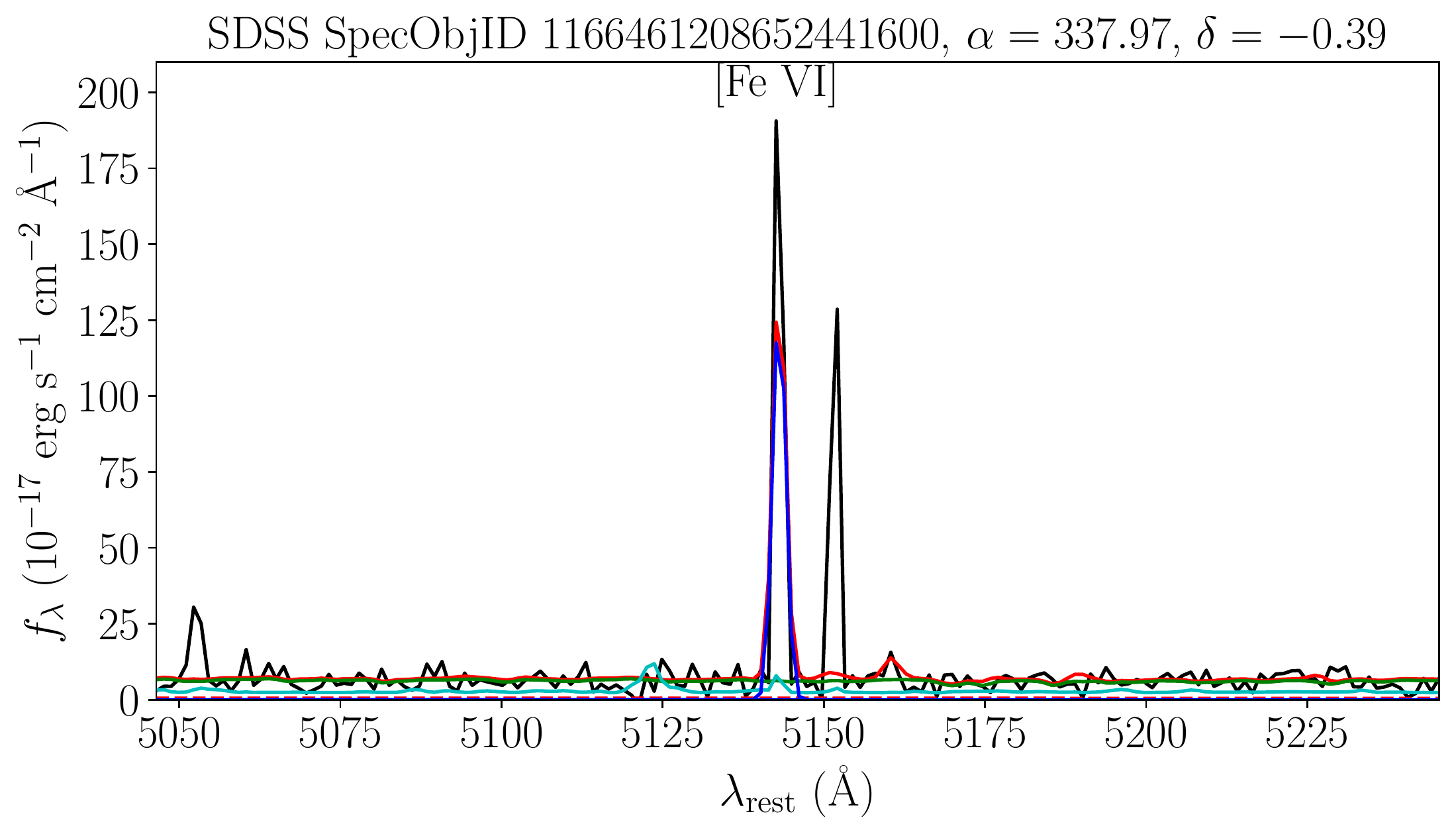}
    \includegraphics[width=.48\textwidth]{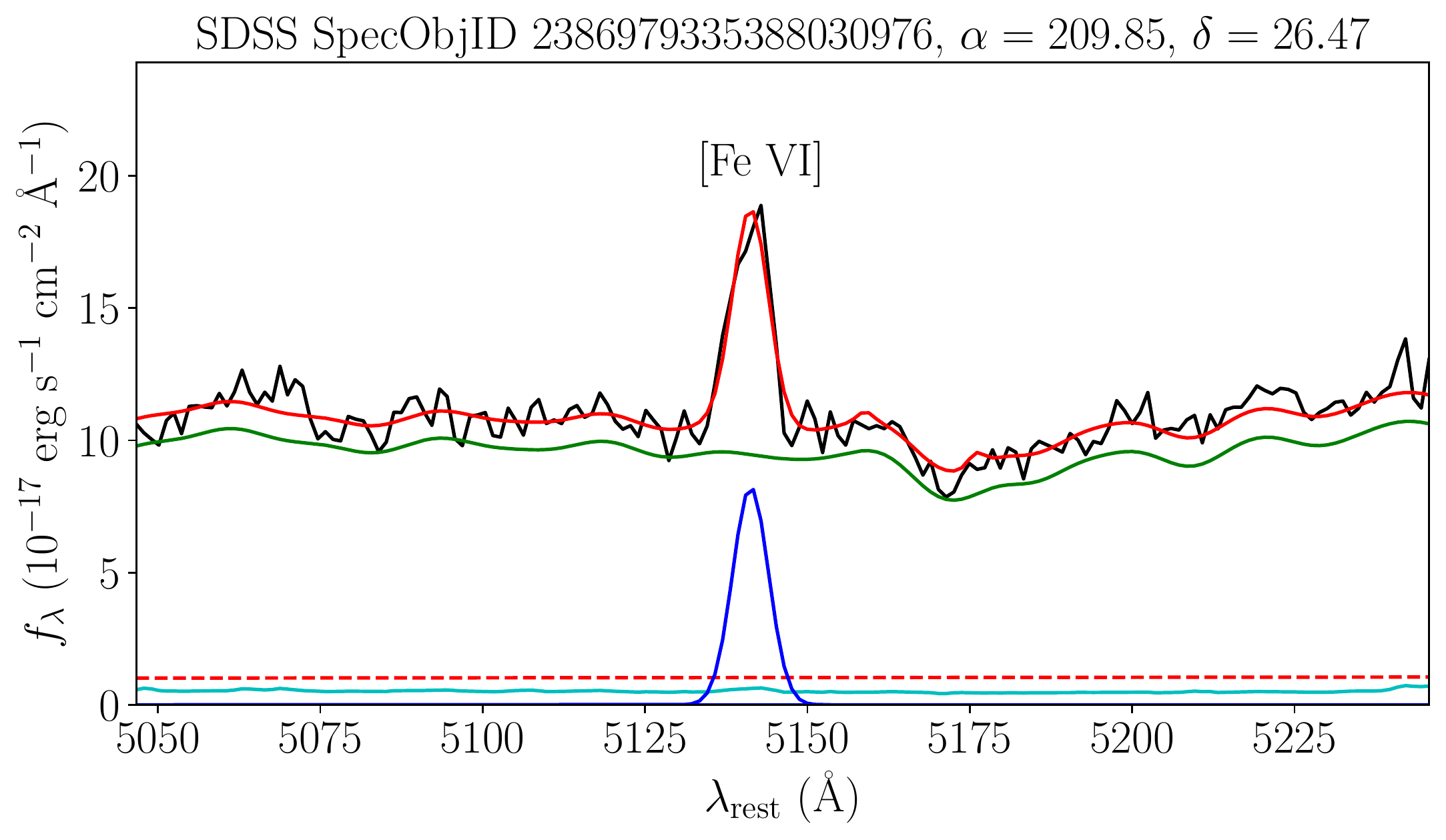}
    \includegraphics[width=.48\textwidth]{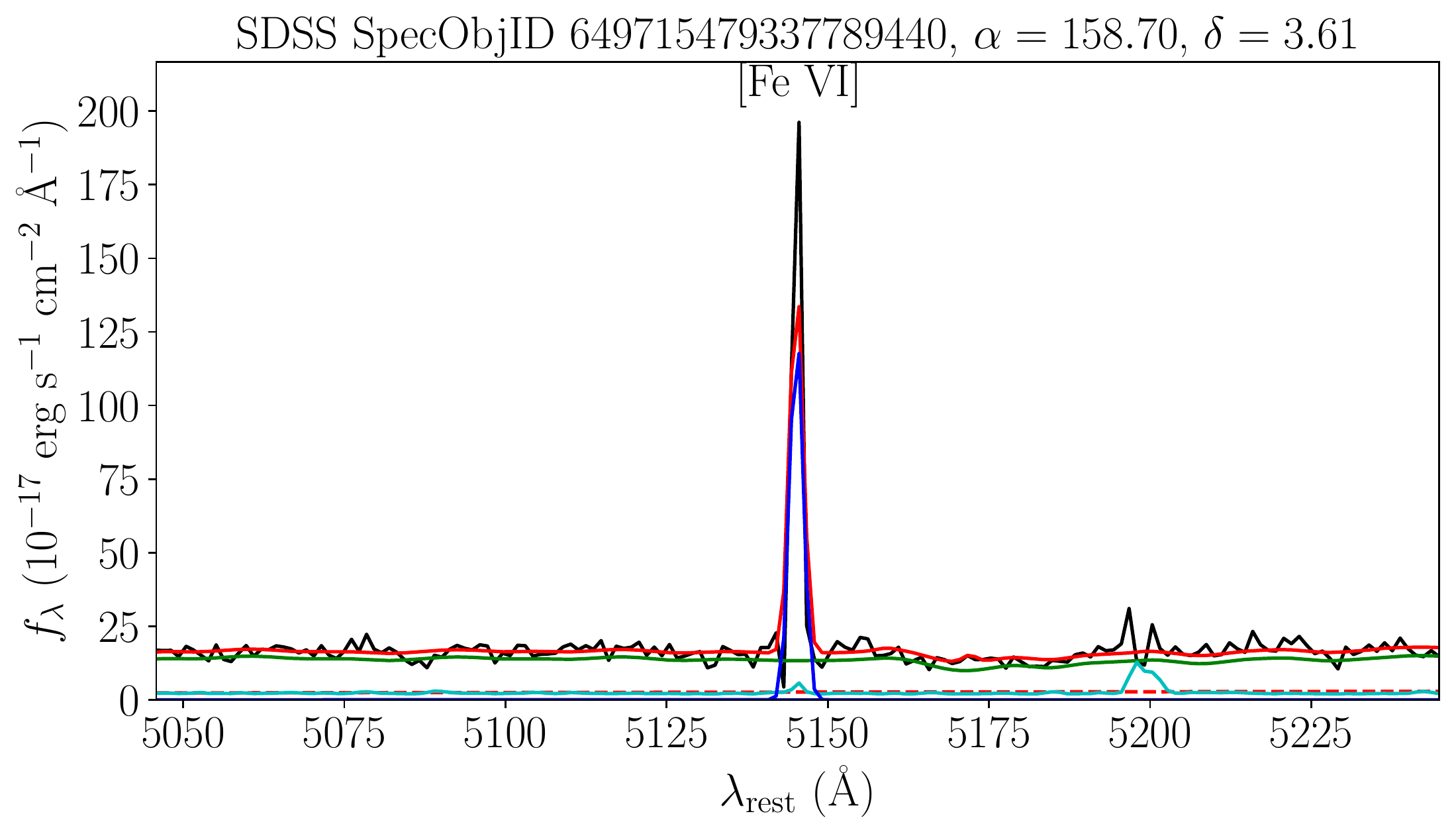}
    \includegraphics[width=.48\textwidth]{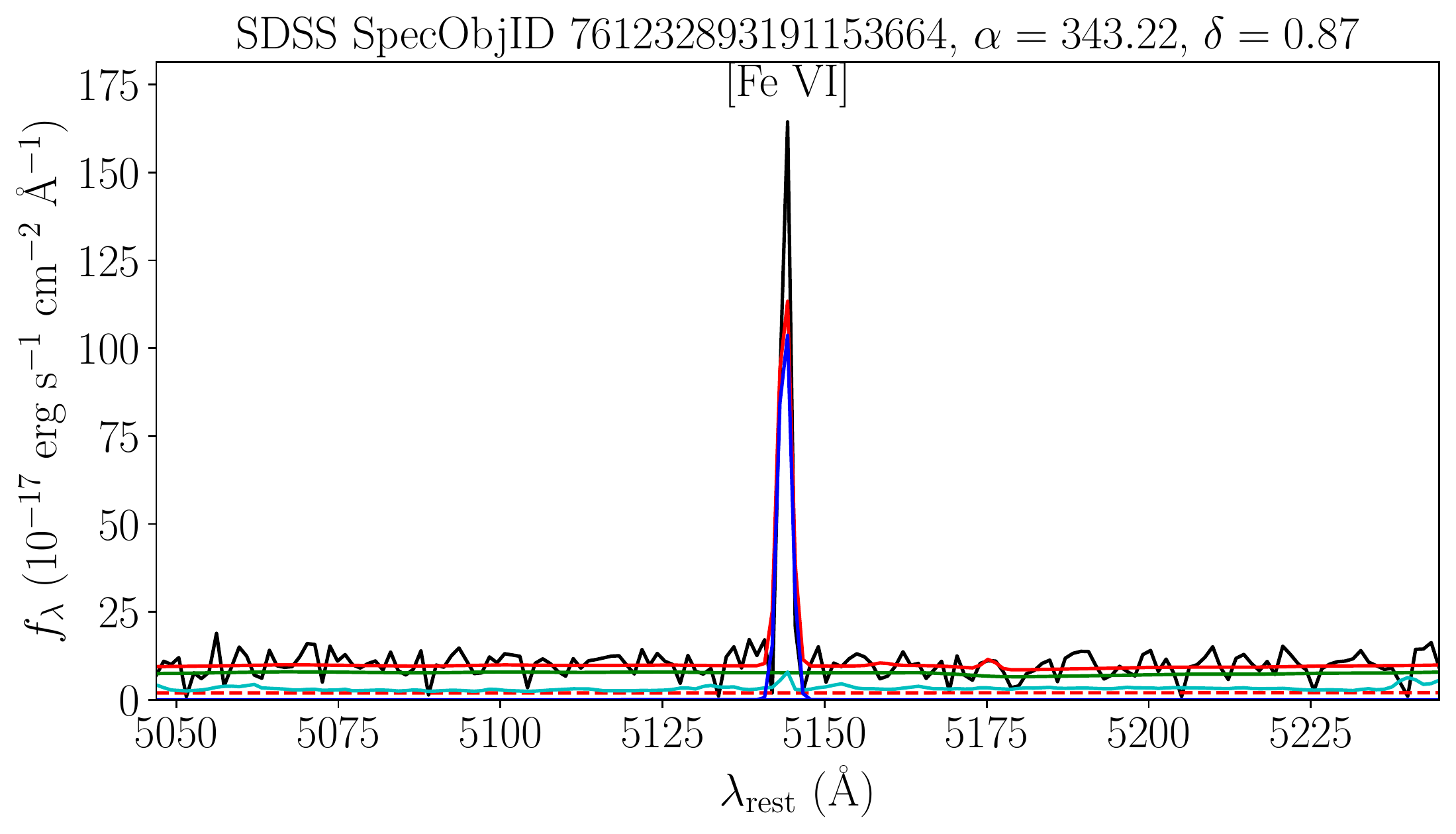}
    \caption{A selection of [\ion{Fe}{7}] $\lambda$5159 and [\ion{Fe}{6}] $\lambda$5146 detections from our subsample. The raw flux, corrected for redshift and galactic extinction, is plotted in black.  The BADASS model and each of its components (emission lines, host galaxy, AGN power law) are plotted in colors indicated by the legend.  Each coronal line is labeled, and the spectra's coordinates and SDSS Spec Object ID are given in the plot titles.}    
    \label{fig:misc_4}
\end{figure*}

\begin{figure*}
    \centering
    \includegraphics[width=.48\textwidth]{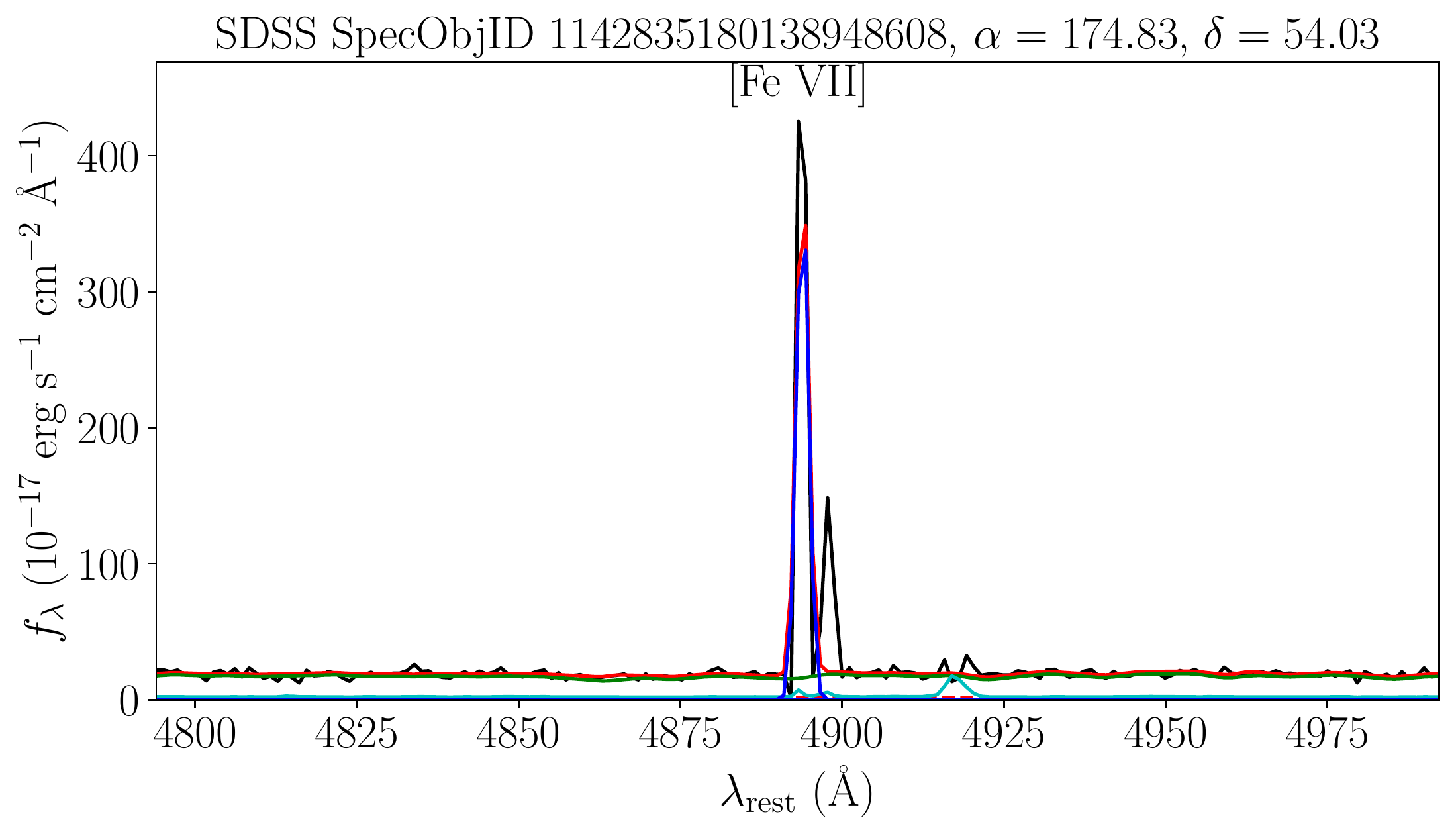}
    \includegraphics[width=.48\textwidth]{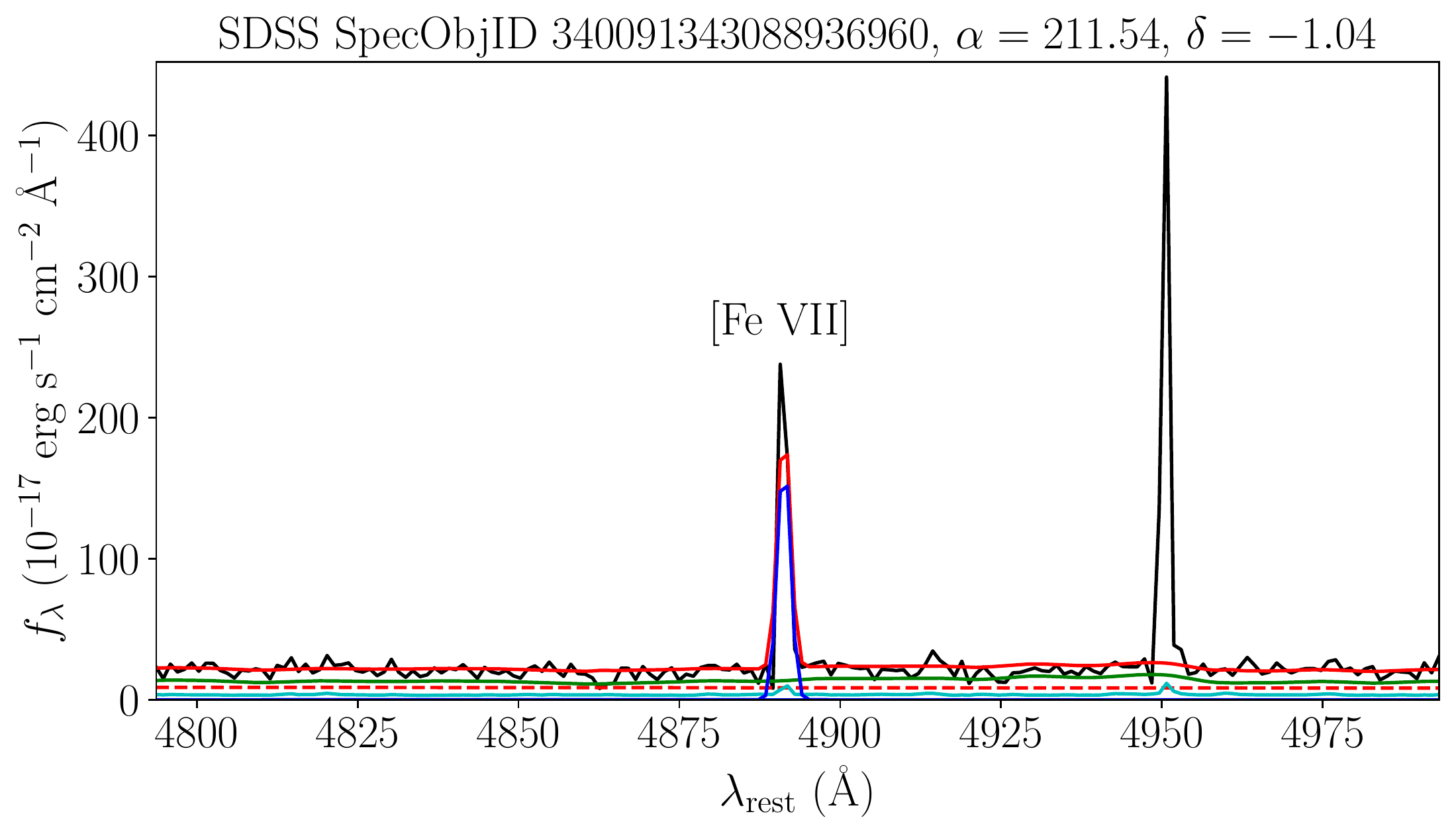}
    \includegraphics[width=.48\textwidth]{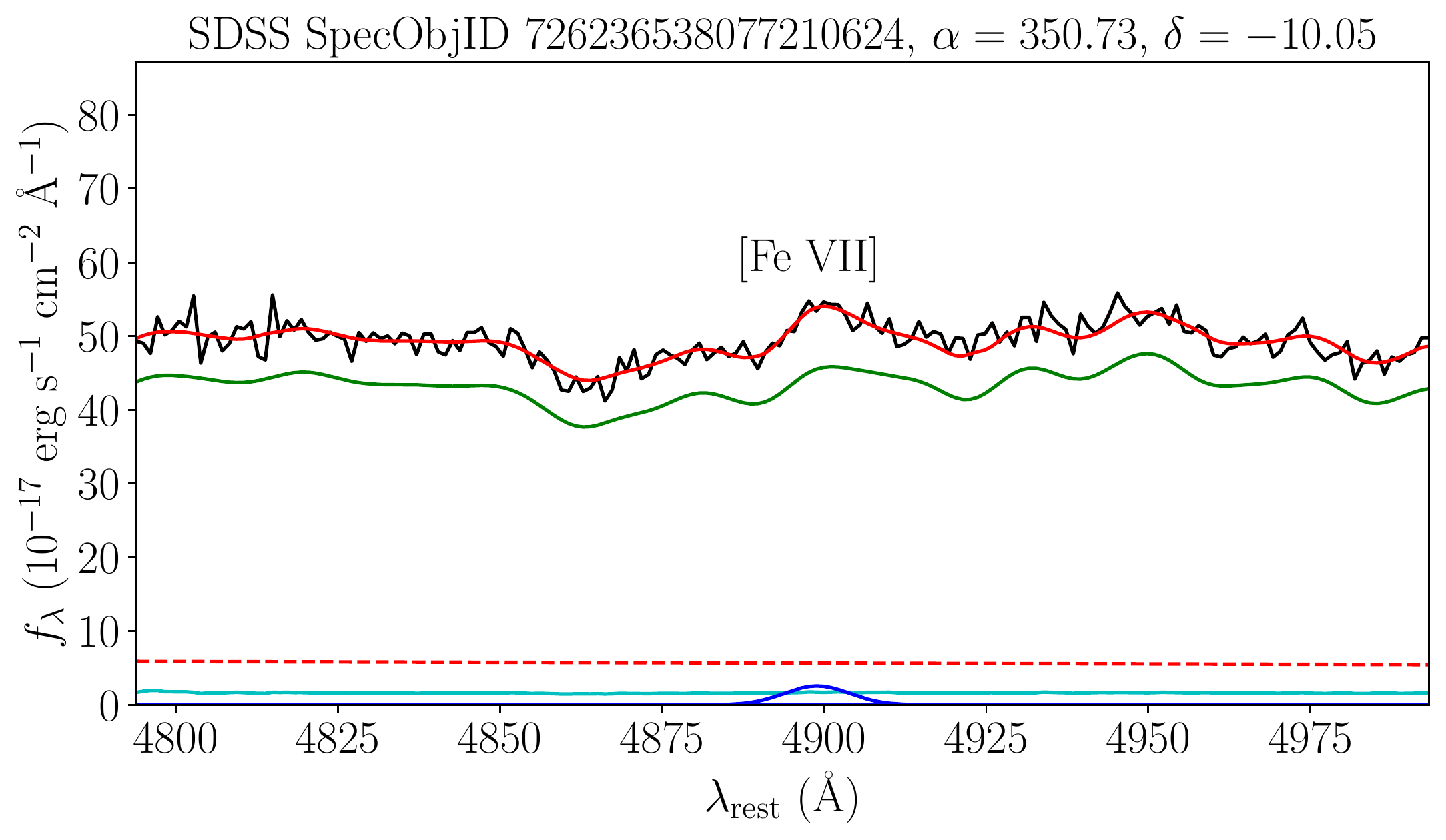}
    \includegraphics[width=.48\textwidth]{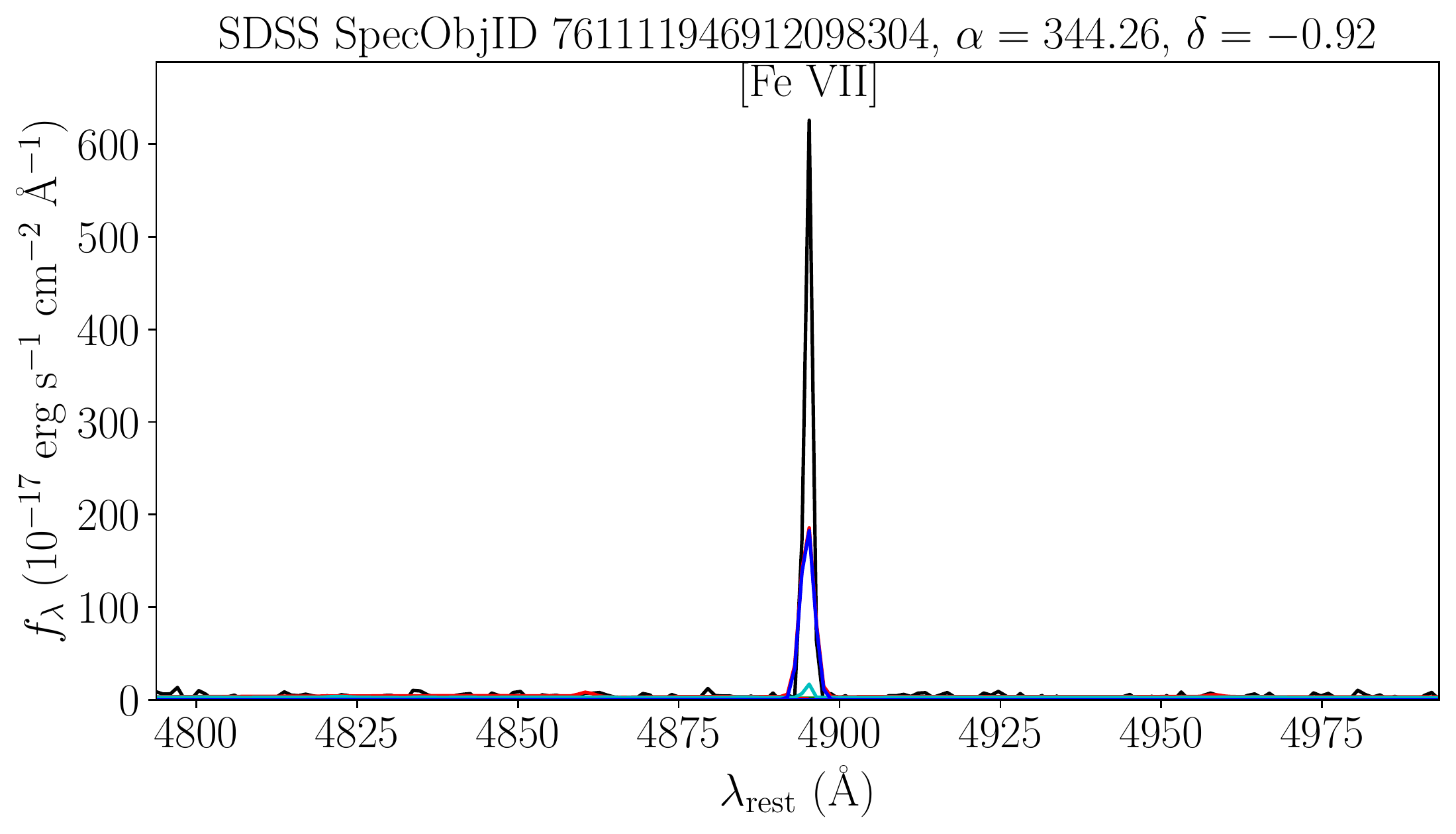}
    \includegraphics[width=.48\textwidth]{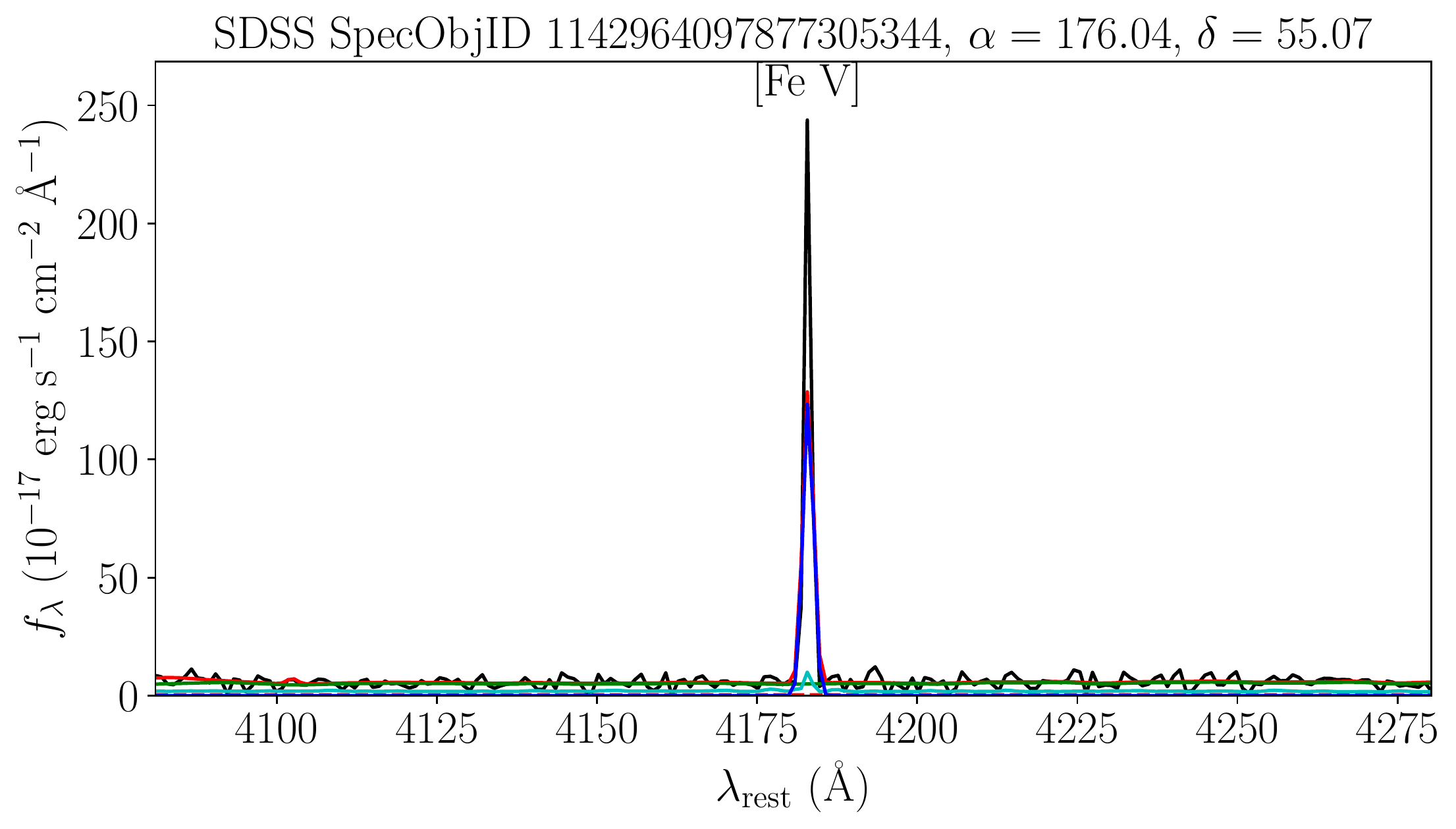}
    \includegraphics[width=.48\textwidth]{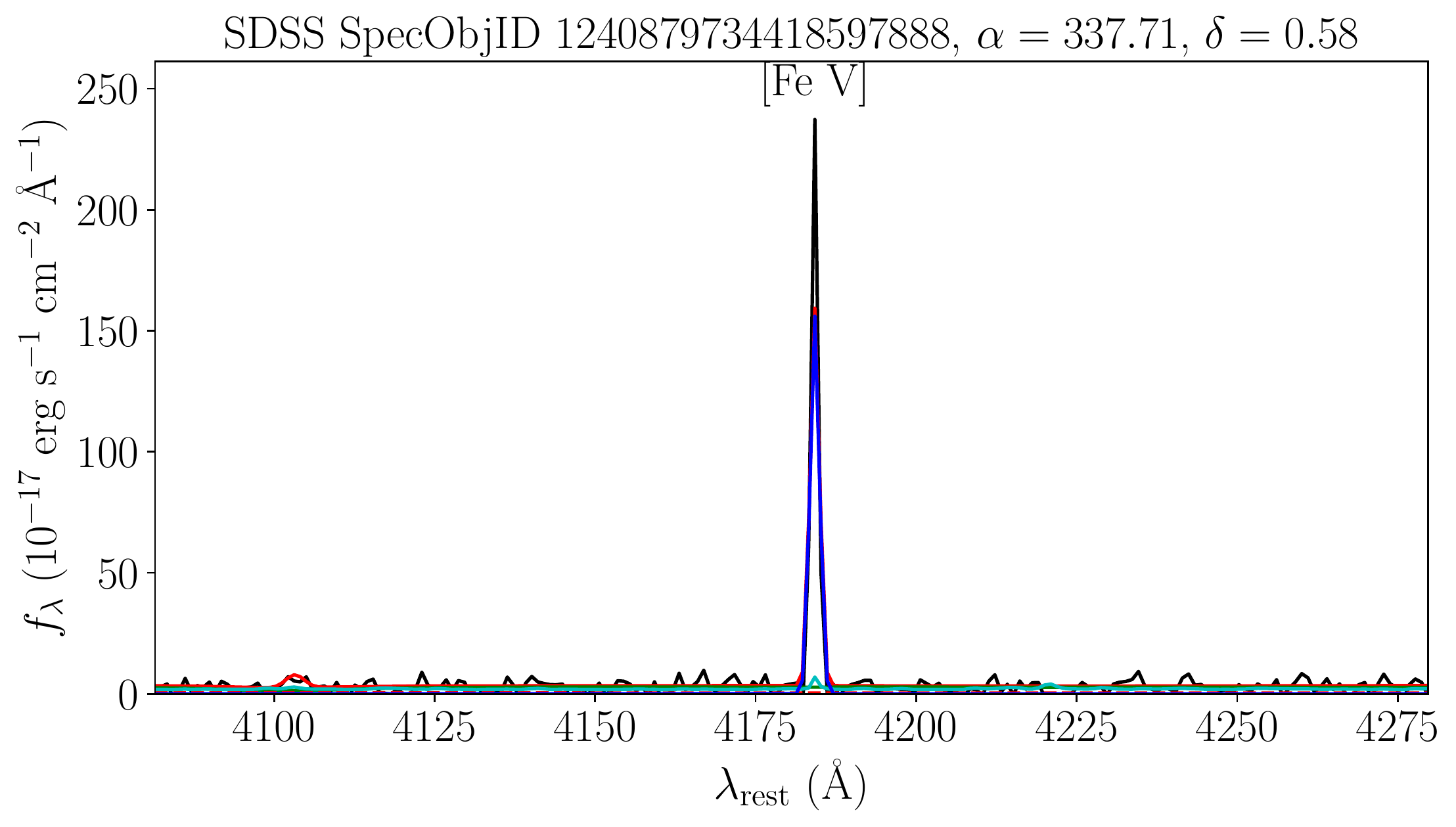}
    \includegraphics[width=.48\textwidth]{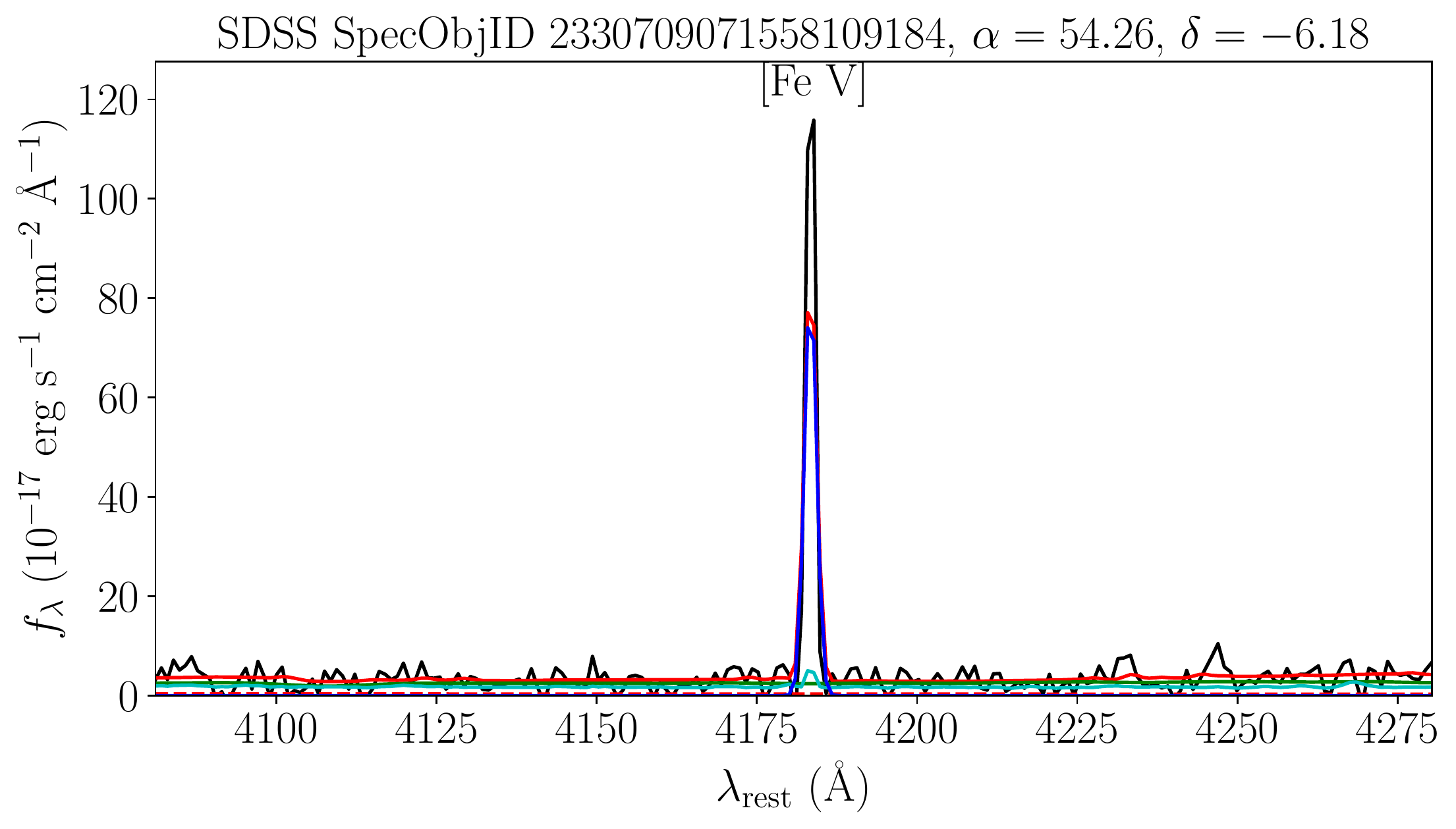}
    \includegraphics[width=.48\textwidth]{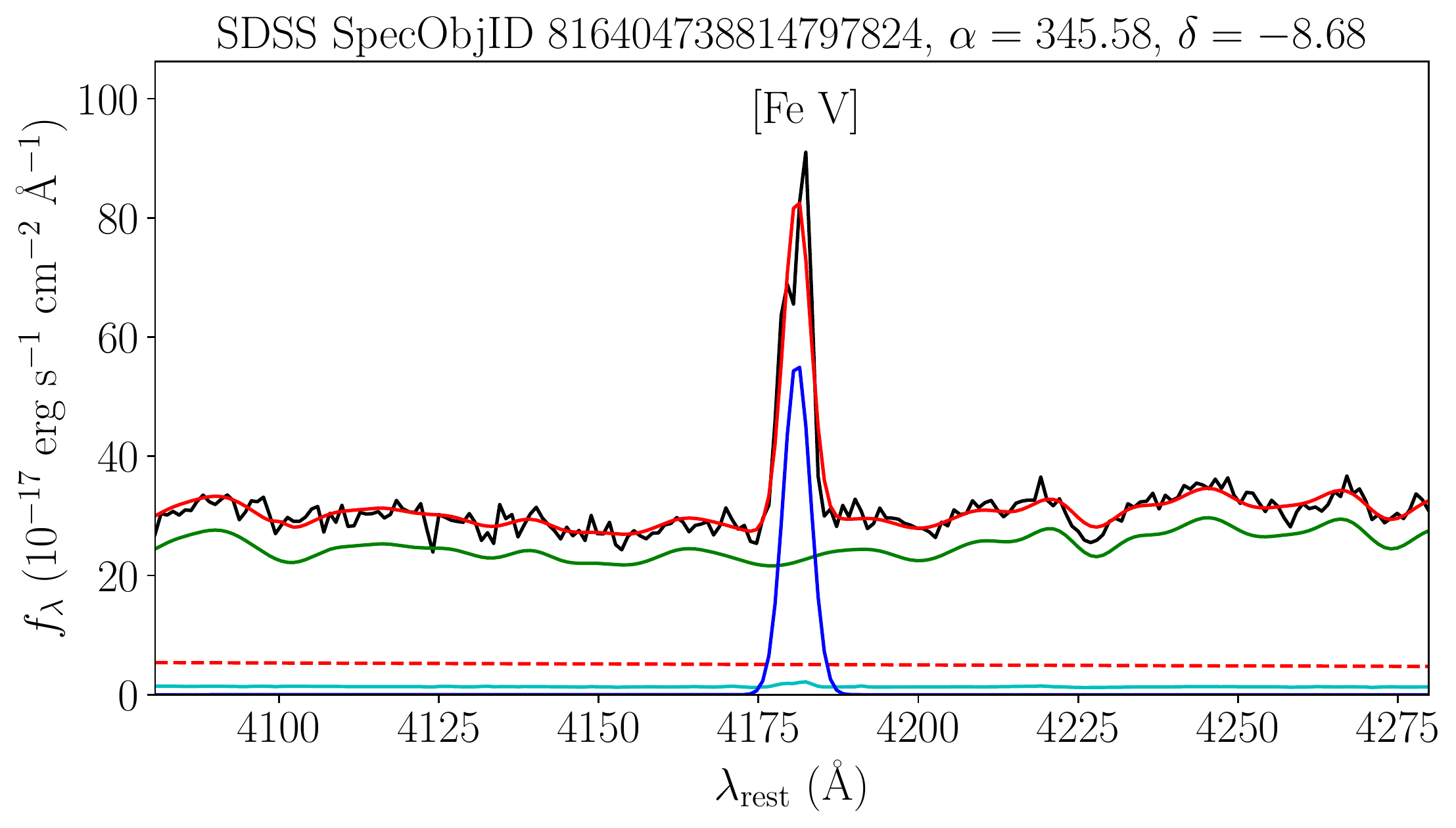}
    \caption{A selection of [\ion{Fe}{7}] $\lambda$4893 and [\ion{Fe}{5}] $\lambda$4181 detections from our subsample. The raw flux, corrected for redshift and galactic extinction, is plotted in black.  The BADASS model and each of its components (emission lines, host galaxy, AGN power law) are plotted in colors indicated by the legend.  Each coronal line is labeled, and the spectra's coordinates and SDSS Spec Object ID are given in the plot titles.}    
    \label{fig:misc_5}
\end{figure*}

\begin{figure*}
    \centering
    \includegraphics[width=.48\textwidth]{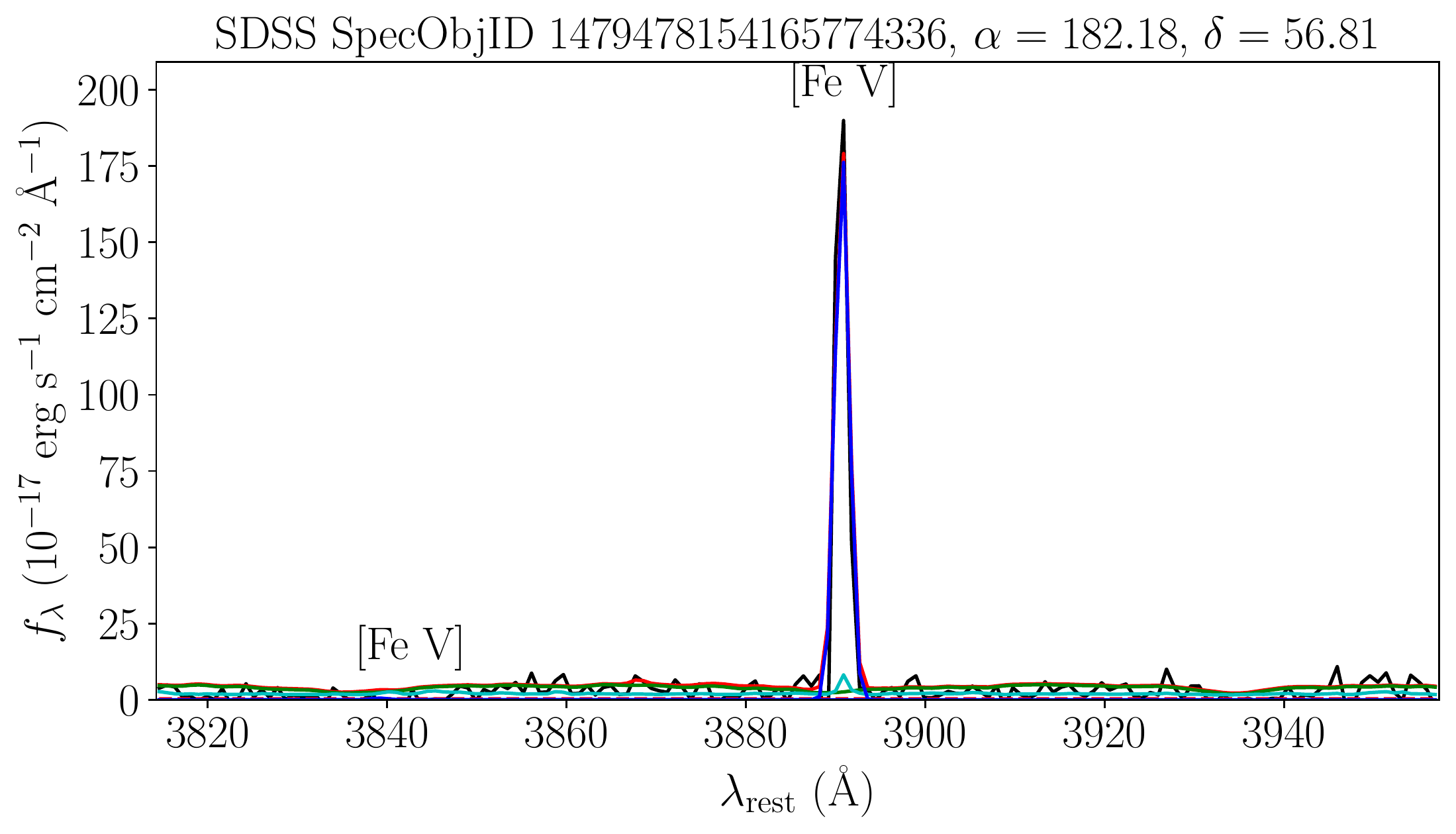}
    \includegraphics[width=.48\textwidth]{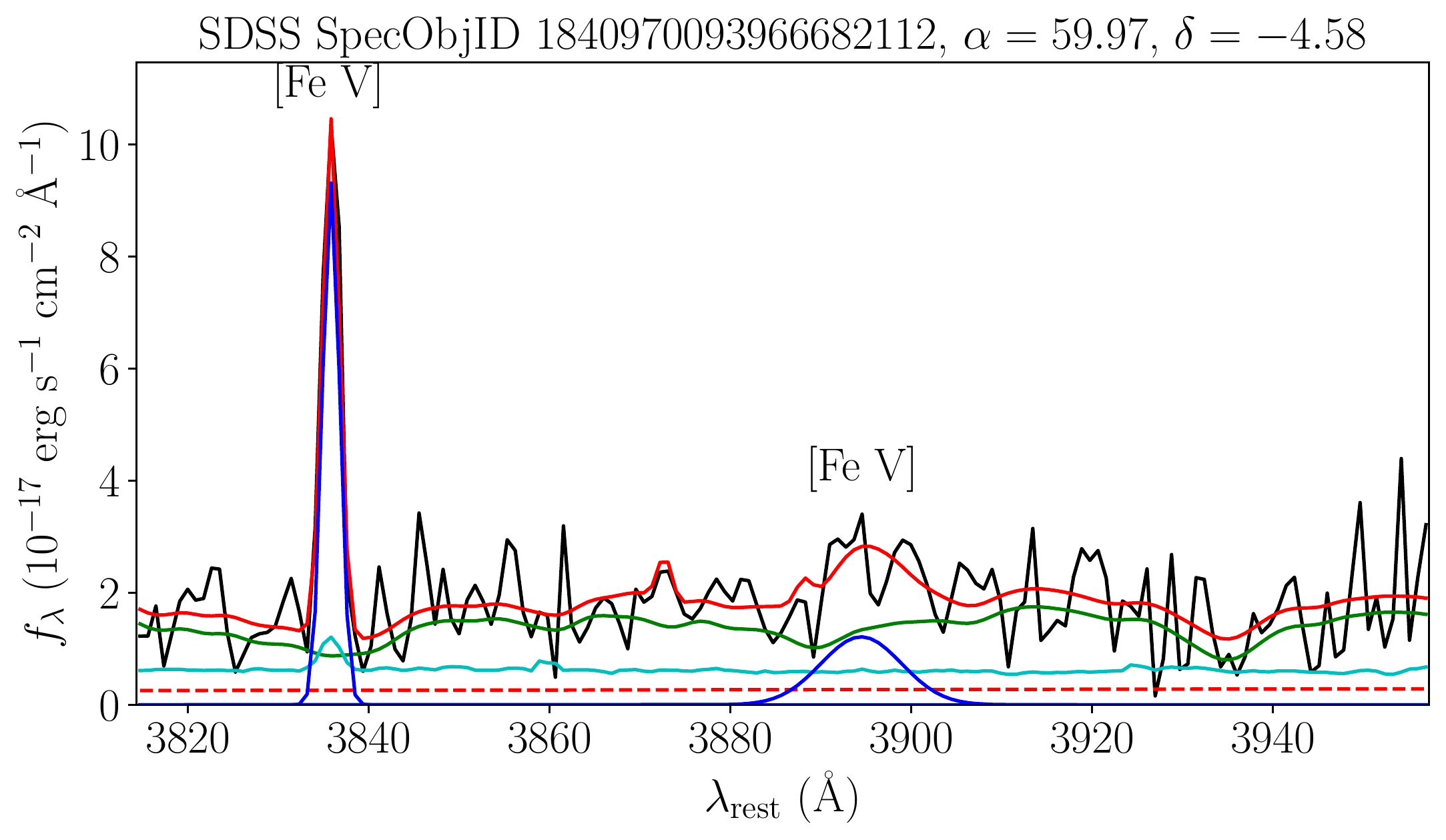}
    \includegraphics[width=.48\textwidth]{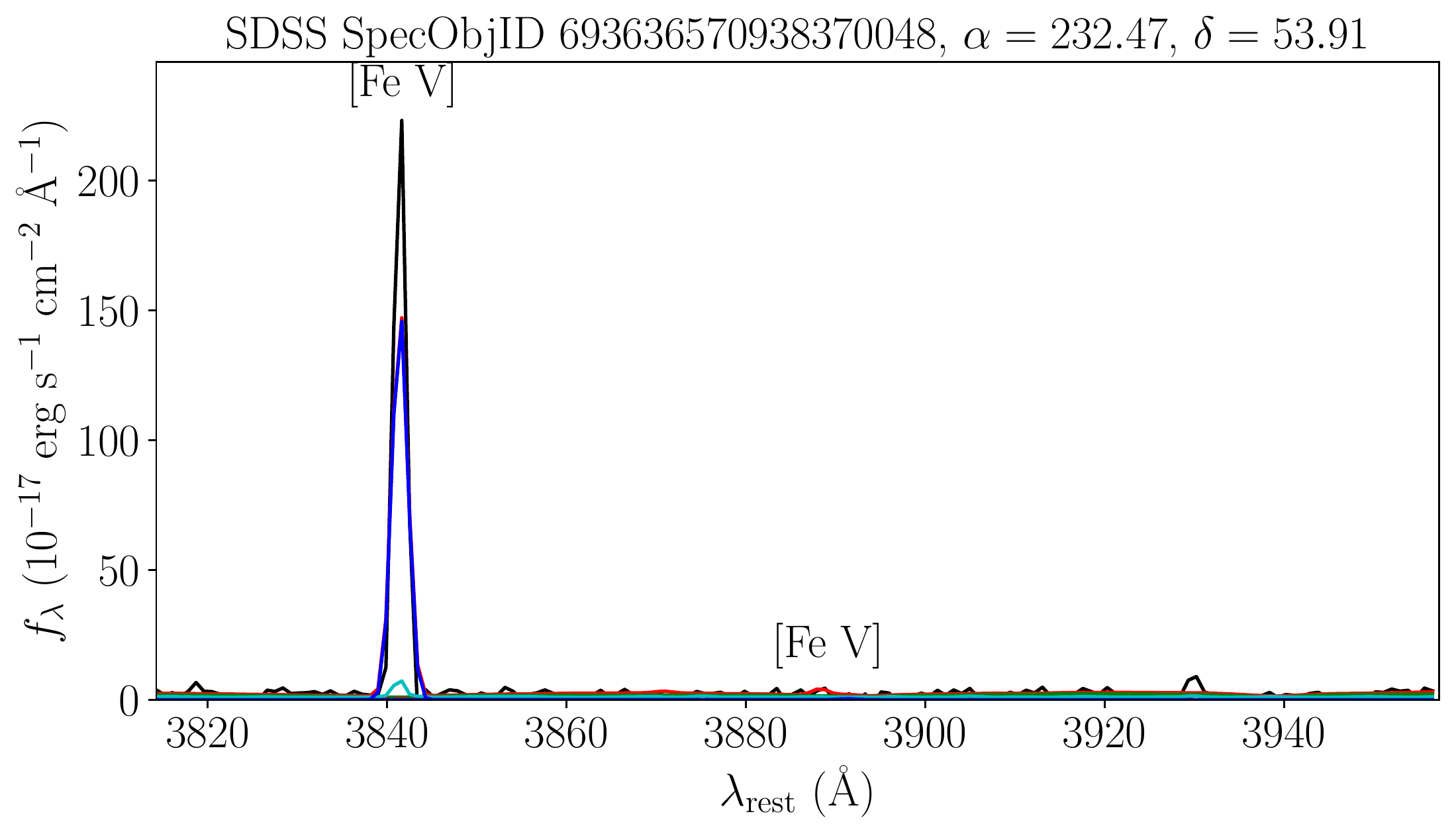}
    \includegraphics[width=.48\textwidth]{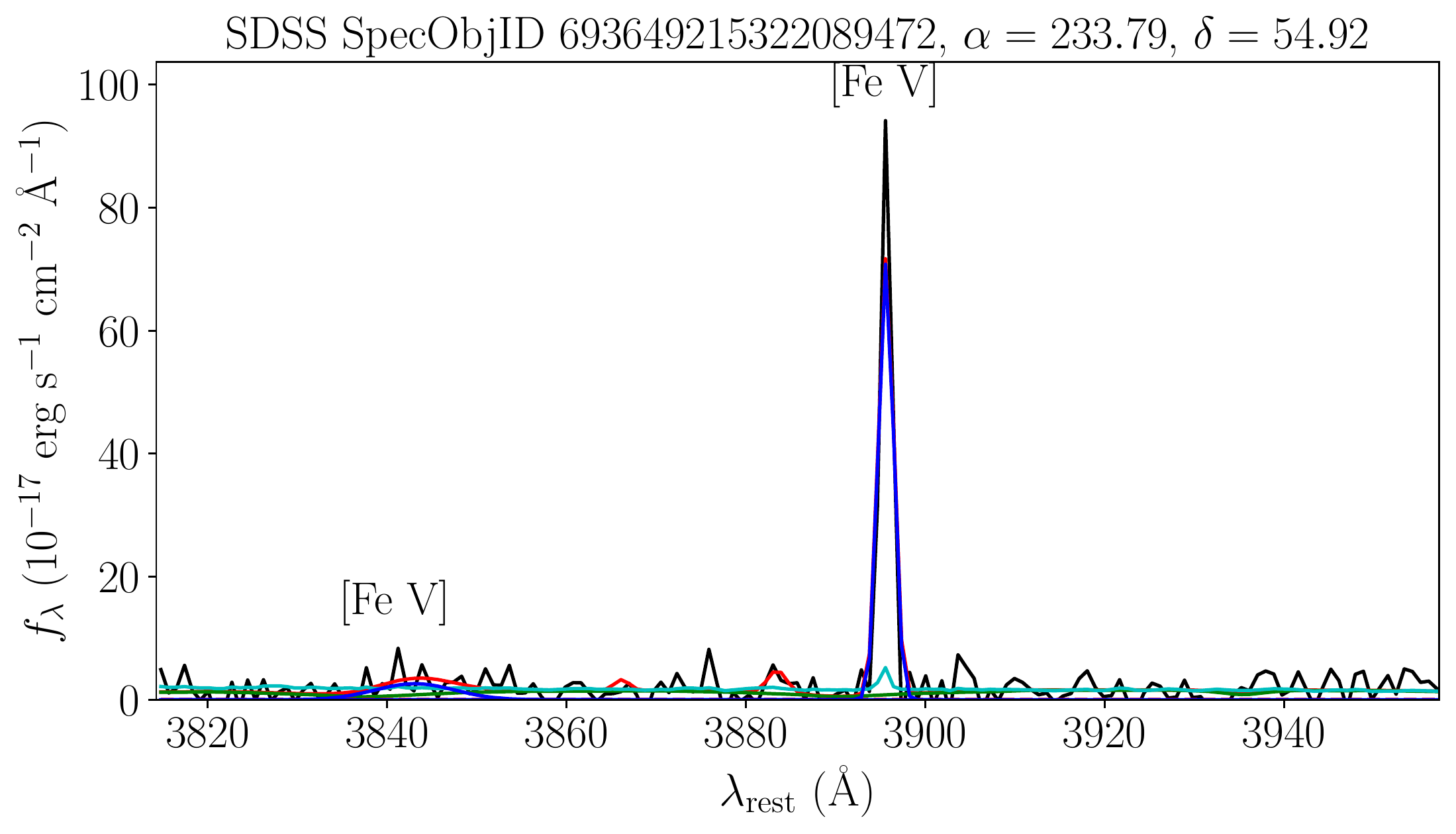}
    \includegraphics[width=.48\textwidth]{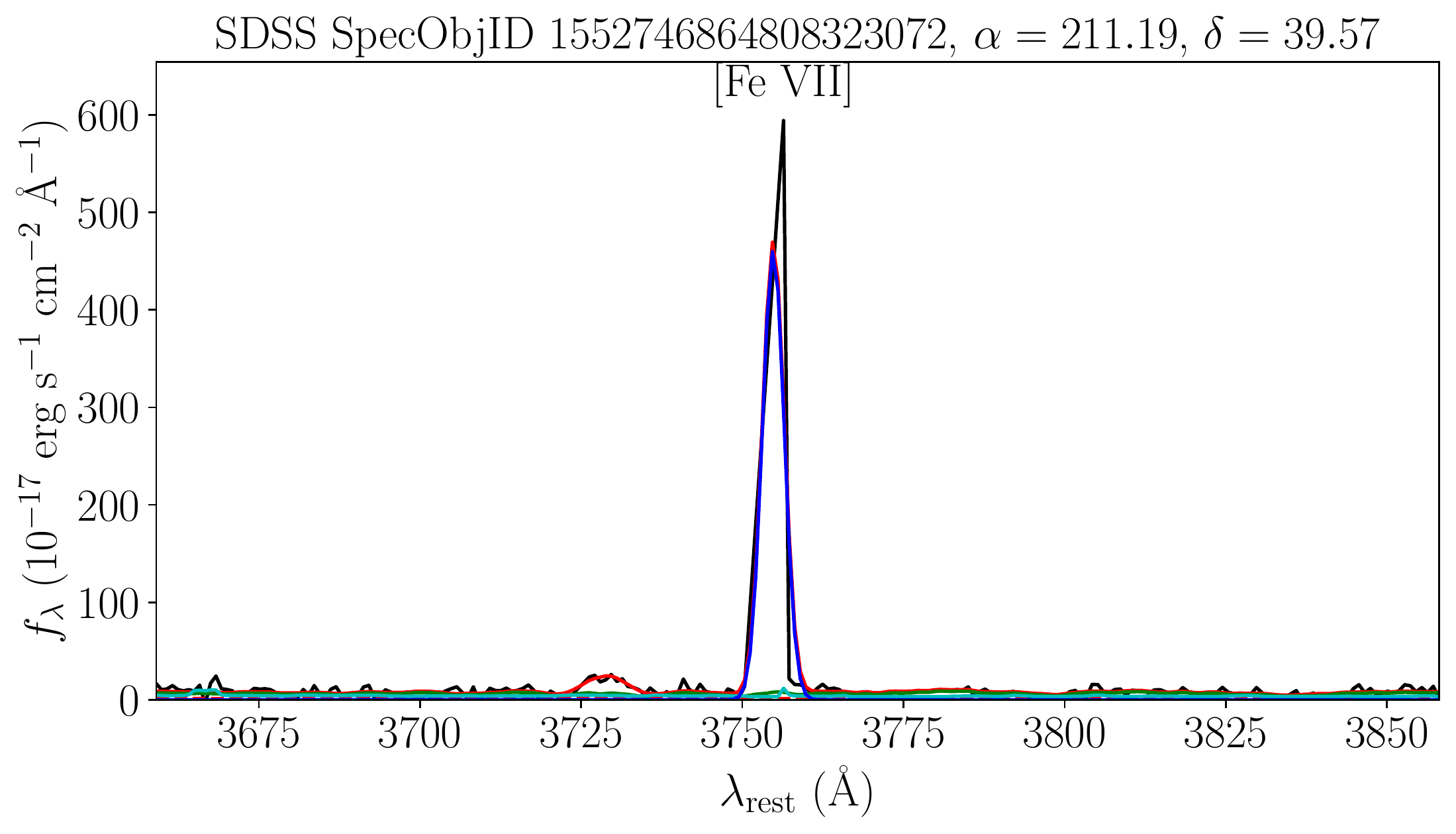}
    \includegraphics[width=.48\textwidth]{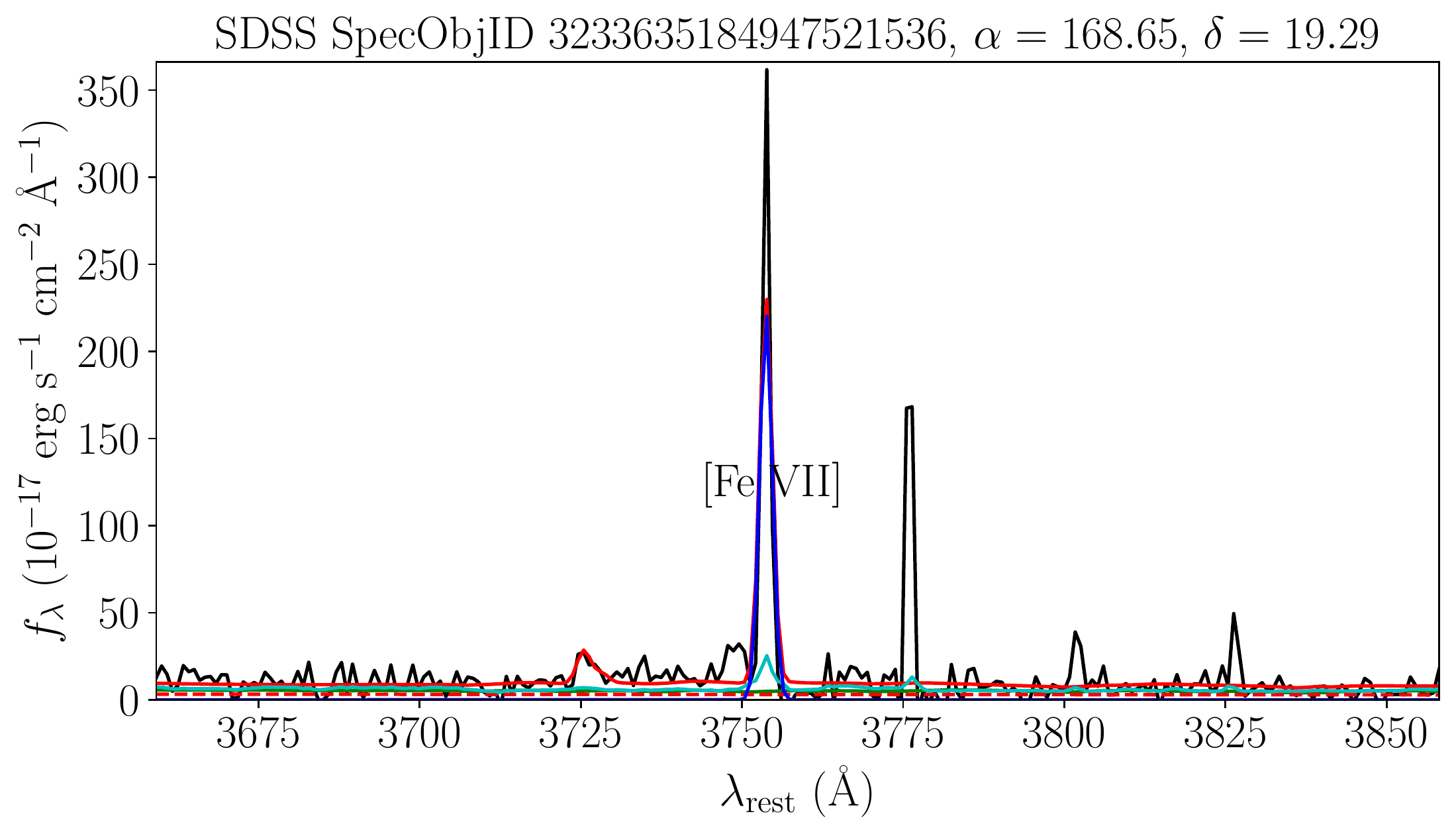}
    \includegraphics[width=.48\textwidth]{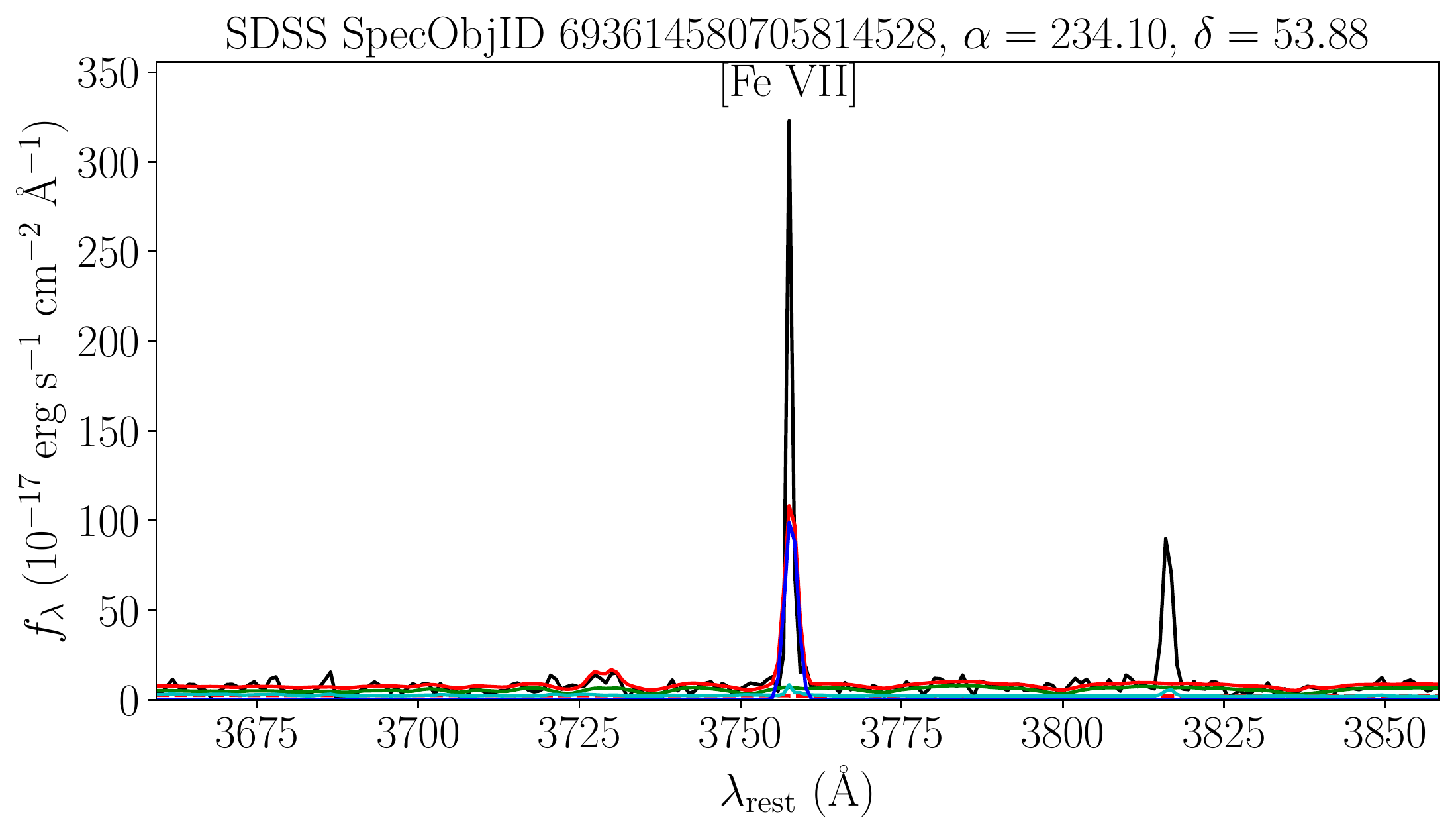}
    \includegraphics[width=.48\textwidth]{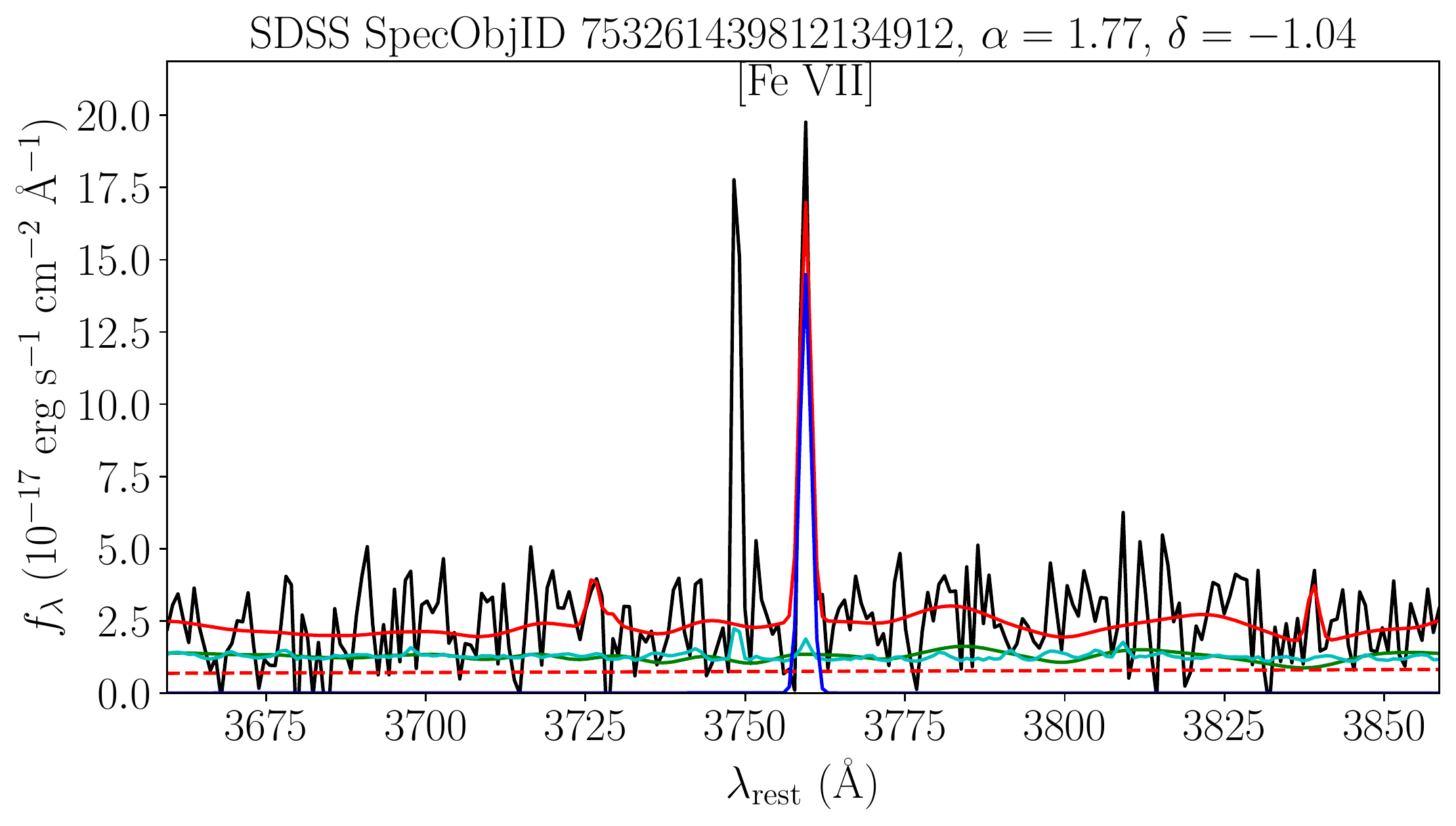}
    \caption{A selection of [\ion{Fe}{5}] $\lambda\lambda$3839,3891 and [\ion{Fe}{7}] $\lambda$3759 detections from our subsample. The raw flux, corrected for redshift and galactic extinction, is plotted in black.  The BADASS model and each of its components (emission lines, host galaxy, AGN power law) are plotted in colors indicated by the legend.  Each coronal line is labeled, and the spectra's coordinates and SDSS Spec Object ID are given in the plot titles.}  
    \label{fig:misc_6}
\end{figure*}

\begin{figure*}
    \centering
    \includegraphics[width=.48\textwidth]{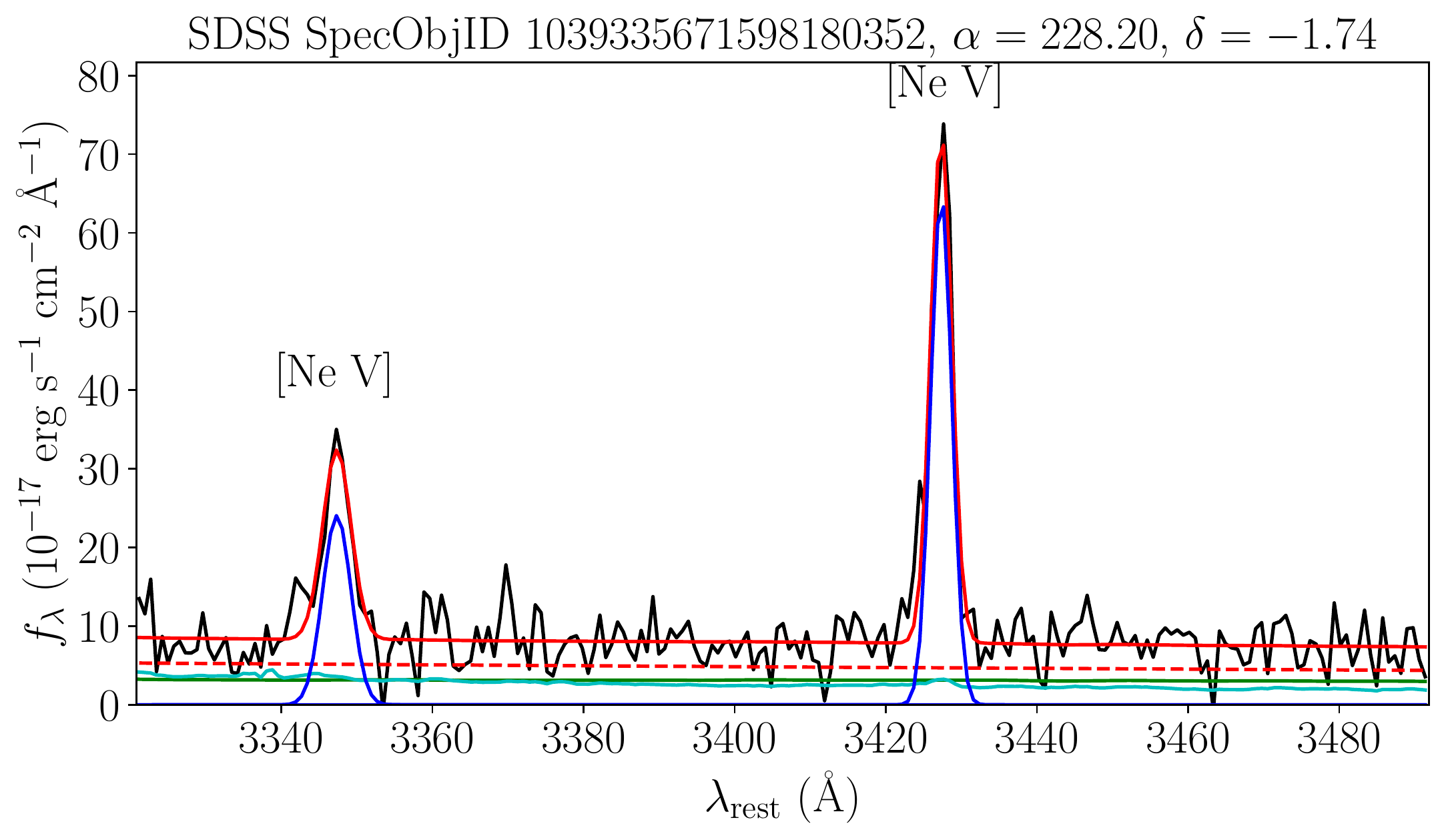}
    \includegraphics[width=.48\textwidth]{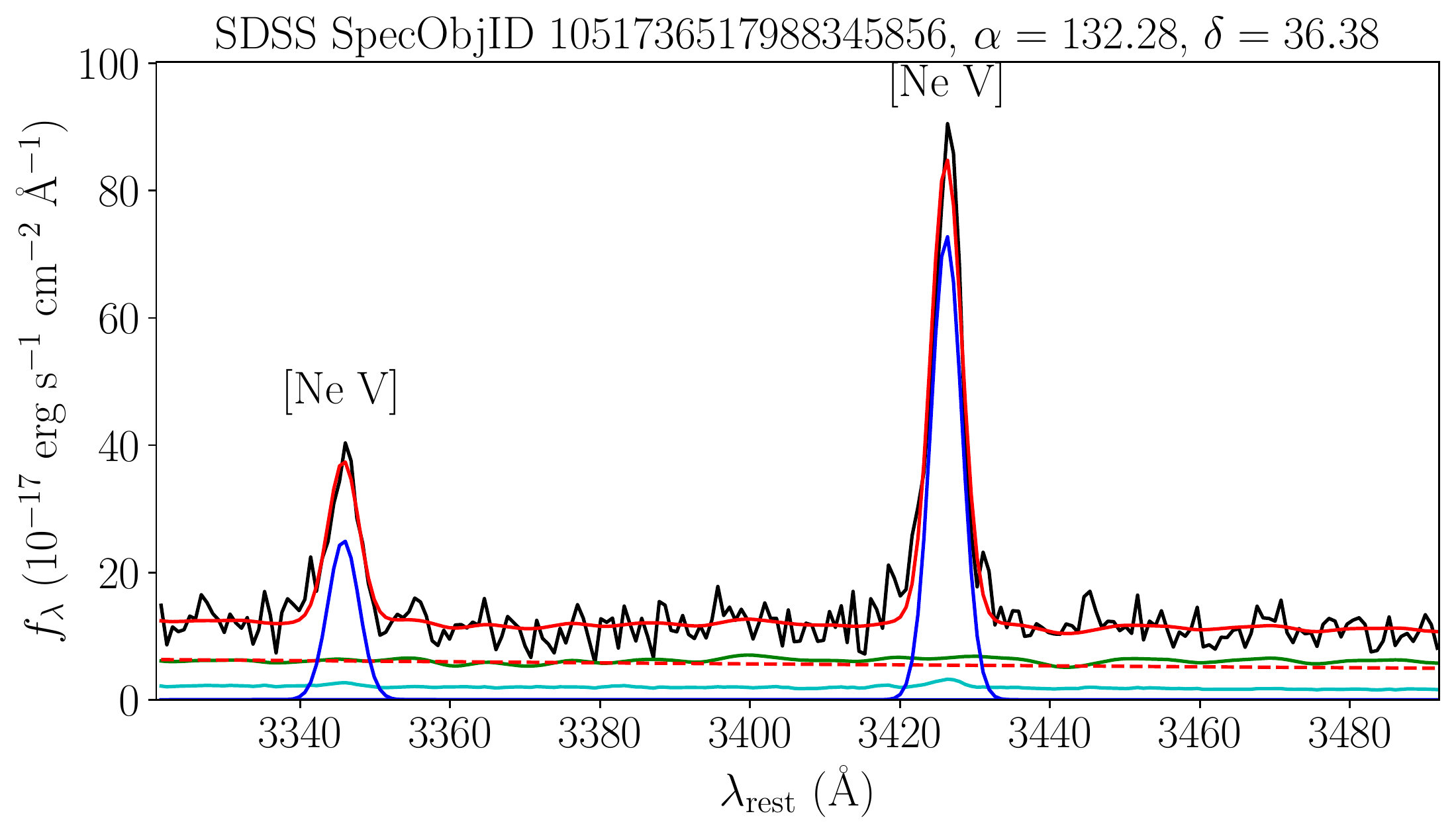}
    \includegraphics[width=.48\textwidth]{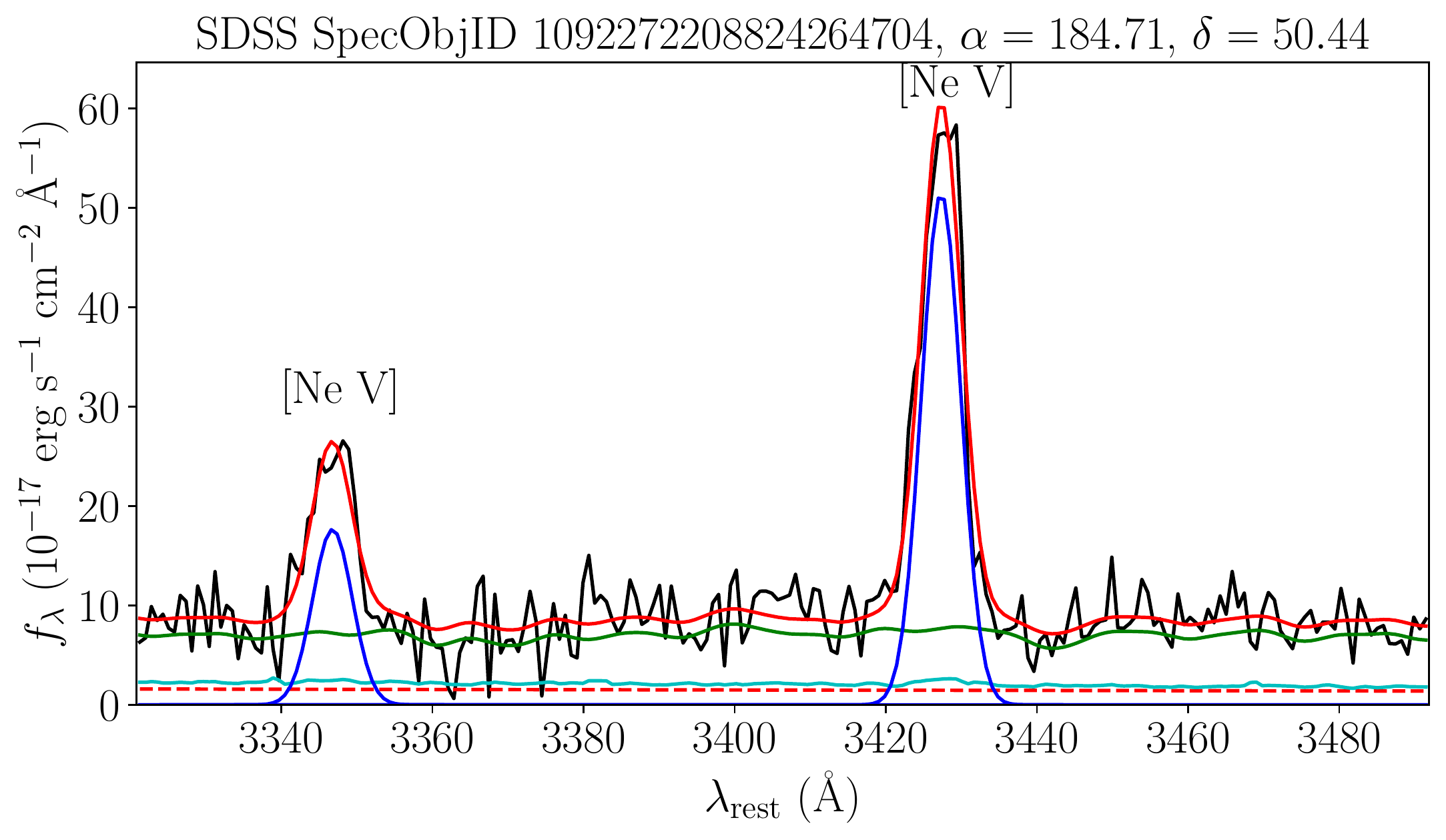}
    \includegraphics[width=.48\textwidth]{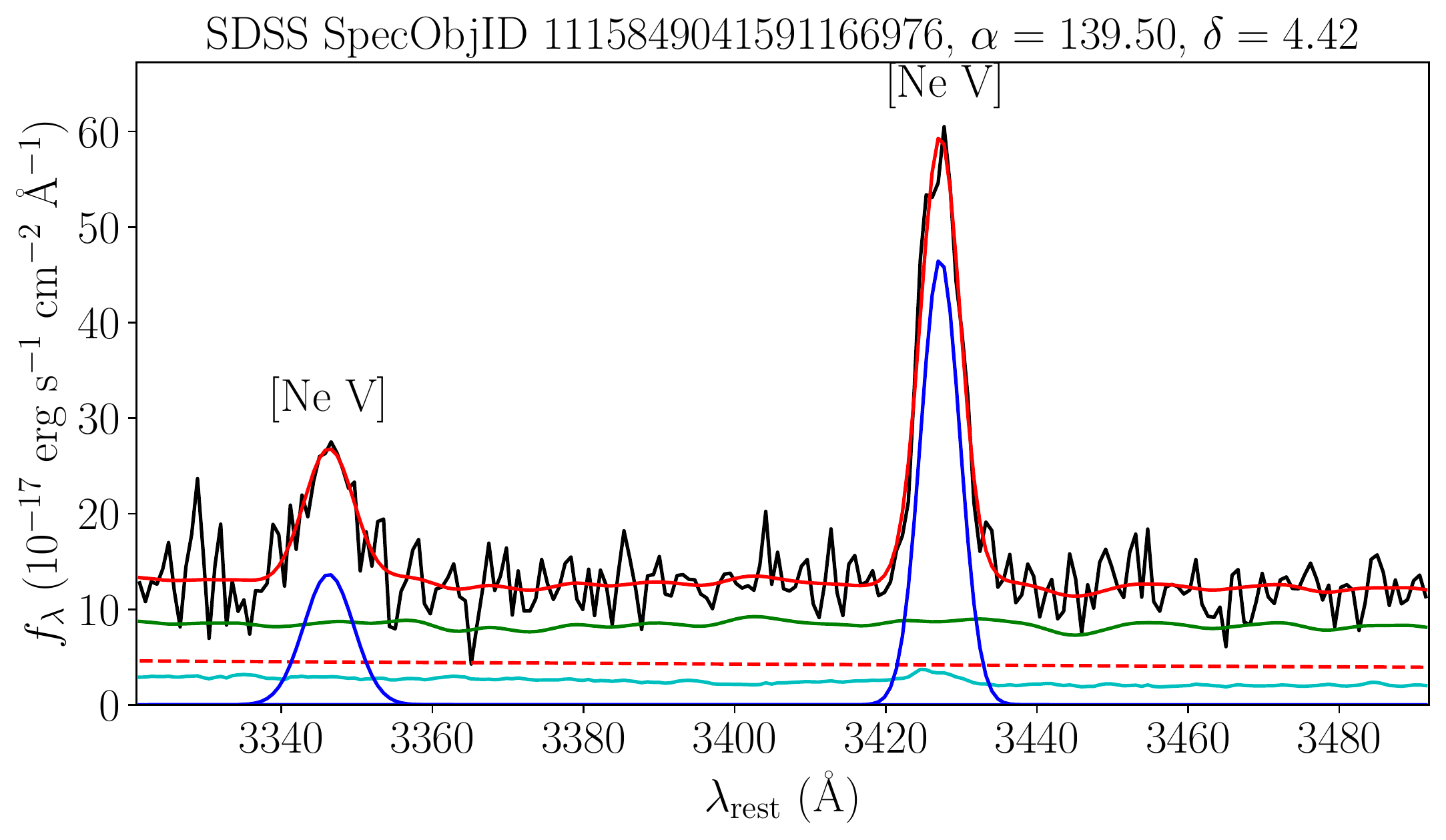}
    \includegraphics[width=.48\textwidth]{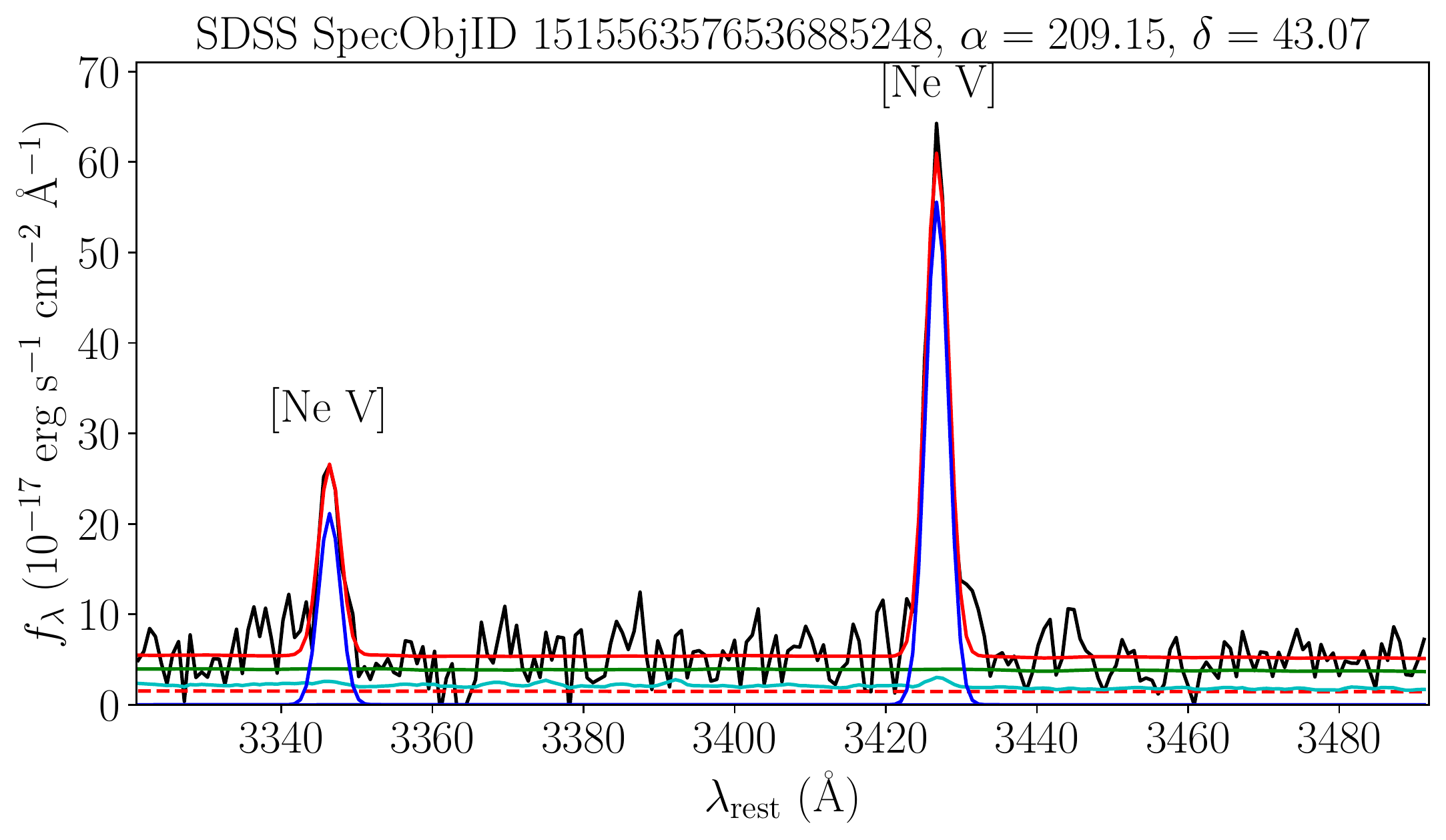}
    \includegraphics[width=.48\textwidth]{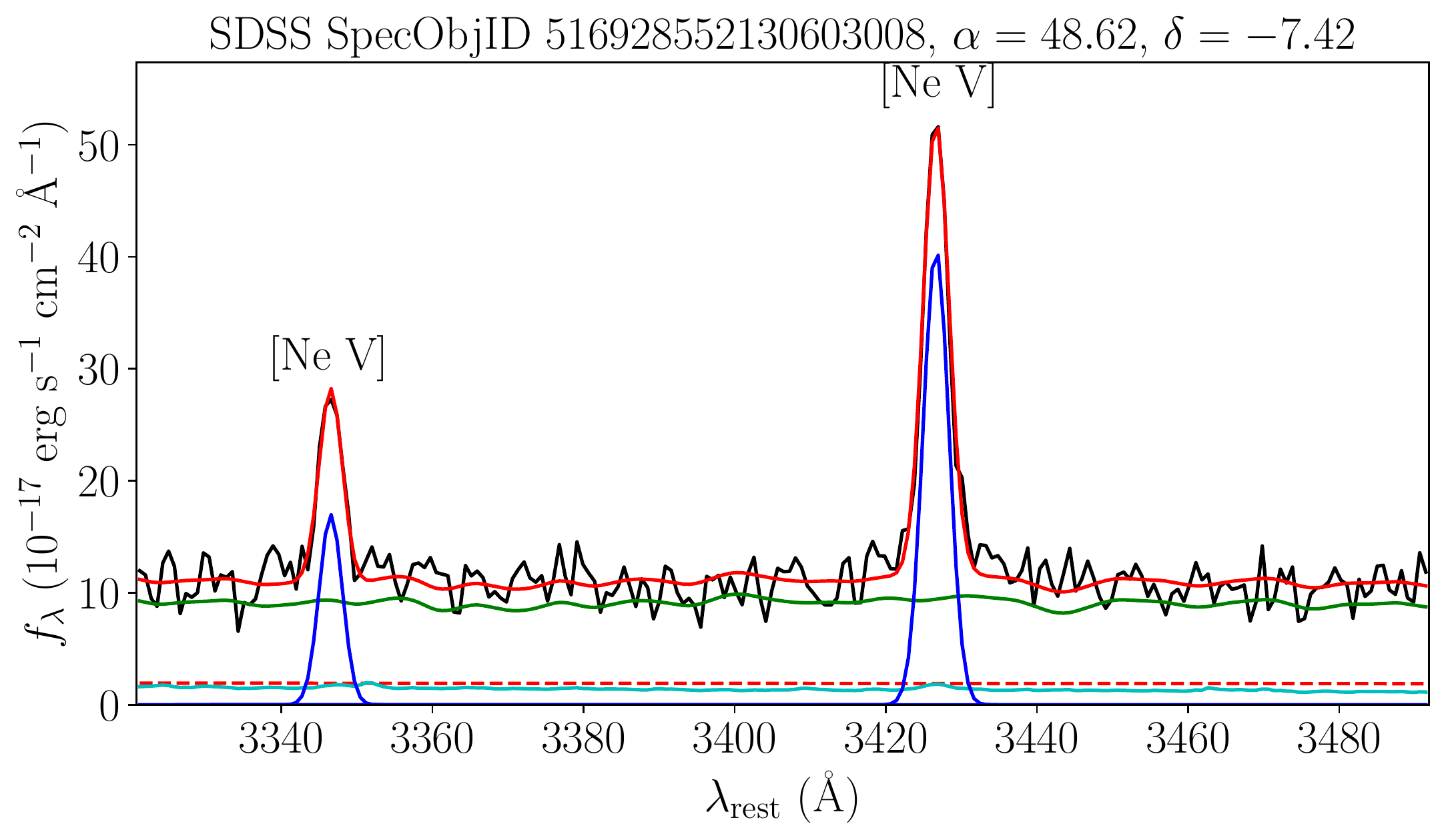}
    \includegraphics[width=.48\textwidth]{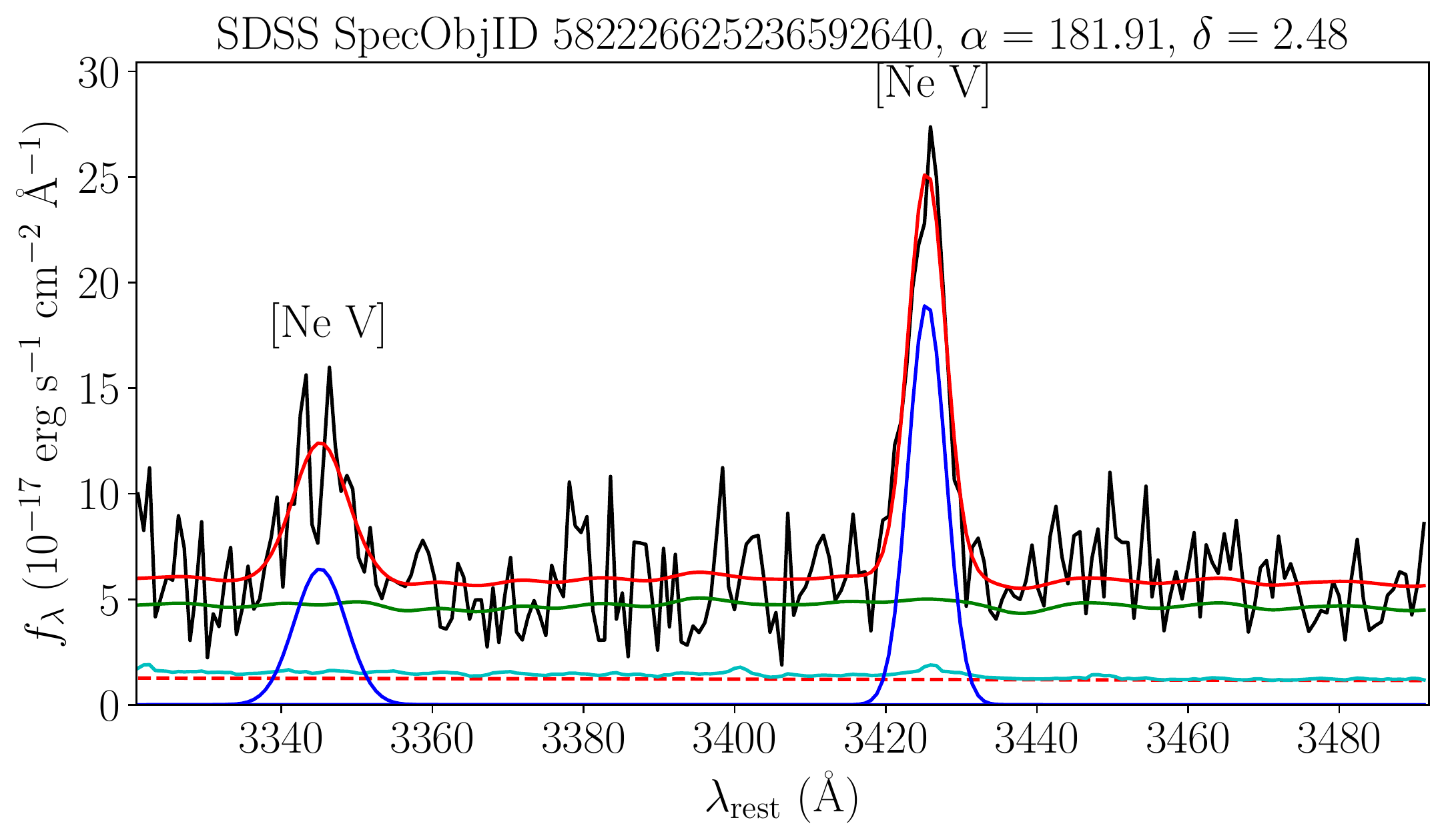}
    \includegraphics[width=.48\textwidth]{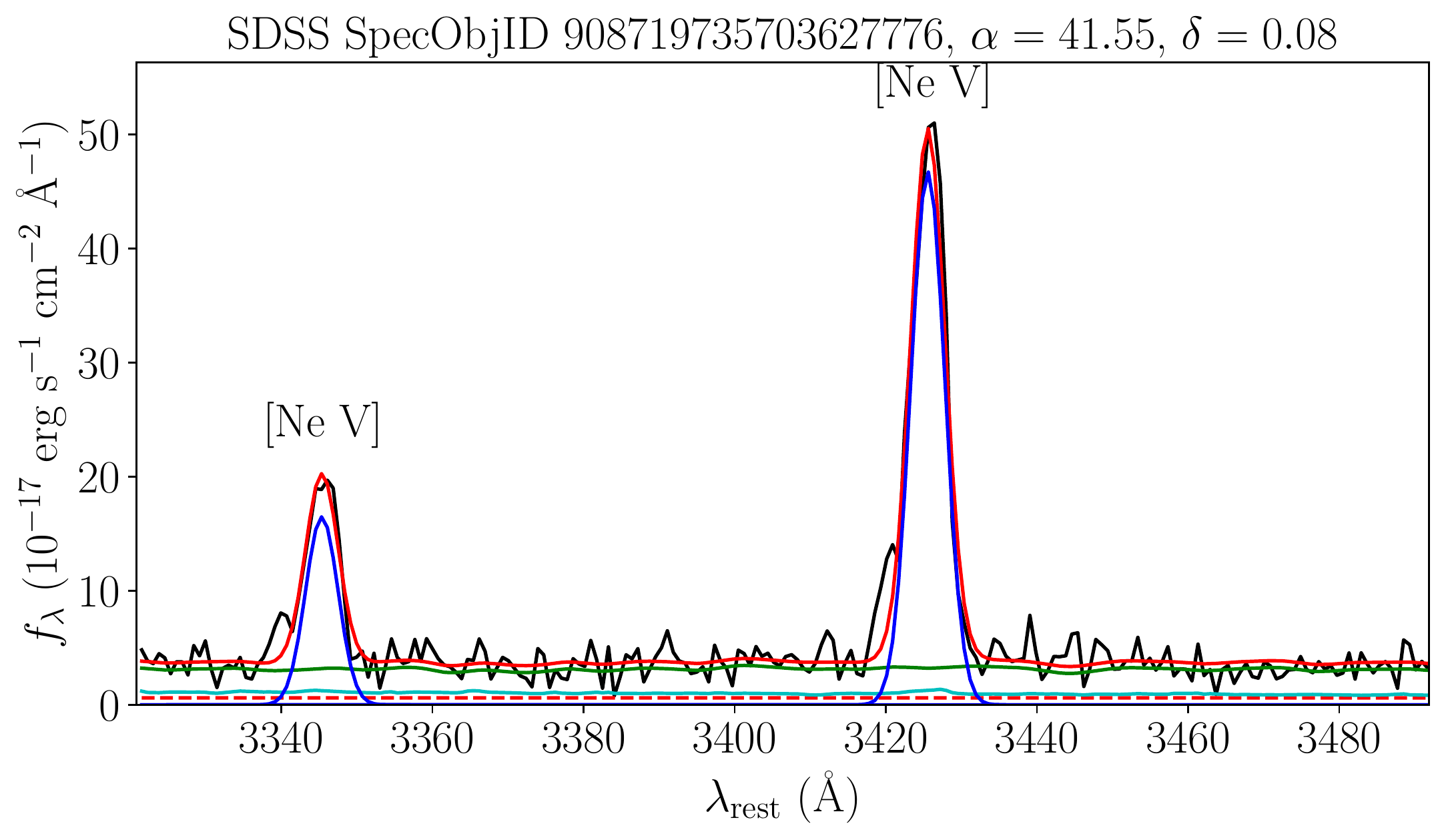}
    \caption{A selection of [\ion{Ne}{5}] $\lambda\lambda$3346,3426 detections from our subsample. The raw flux, corrected for redshift and galactic extinction, is plotted in black.  The BADASS model and each of its components (emission lines, host galaxy, AGN power law) are plotted in colors indicated by the legend.  Each coronal line is labeled, and the spectra's coordinates and SDSS Spec Object ID are given in the plot titles.}  
    \label{fig:nev_3426}
\end{figure*}

Figures \ref{fig:fexi_7892}--\ref{fig:nev_3426} show selected examples of coronal line emitting spectra for each of the 20 lines that we searched for with CLASS. \rev{These examples have been selected out of the sample as the most visible and convincing detections for each line, i.e. those with the largest fluxes and widths and the smallest velocity shifts.}  In Table \ref{tab:column_descriptions} we define the names, units, data types, and sources, and give a brief description of each column in the CLASS survey catalog file.

\startlongtable
\begin{deluxetable*}{lp{7cm}llc}
\label{tab:column_descriptions}
\tablecaption{Column descriptions for the CLASS catalog CSV file.}
\tablehead{\colhead{Column Name} & \colhead{Description} & \colhead{Units} & \colhead{Data Type} & \colhead{Source}}
\decimals
\startdata
SPECOBJID & SDSS Spec Object ID & & 64-bit integer & 2 \\
PLATE & SDSS Plate Number & & integer & 2 \\
MJD & SDSS Modified Julian Date of observation & MJD & integer & 2 \\
FIBERID & SDSS Fiber ID & & integer & 2 \\
ALLWISE\_SOURCEID & AllWISE Source ID & & string & 4 \\
ALLWISE\_COADDID & AllWISE Coadd ID & & string & 4 \\
RA & Right ascension & degrees & float & 2 \\
DEC & Declination & degrees & float & 2 \\
Z & Cosmological redshift & & float & 3 \\
Z\_ERR & Error in cosmological redshift & & float & 3 \\
N\_CL\_DETECT & Number of coronal lines detected in the spectrum & & integer & 1\\
LGM\_TOT\_P50 & Stellar mass median estimate from model photometry & $\log(M/M_\odot)$ & float & 3 \\
SFR\_TOT\_P50 & Star formation rate median estimate from model photometry & $\log($SFR$/M_\odot {\rm yr}^{-1})$ & float & 3 \\
V\_DISP & Stellar velocity dispersion & km s$^{-1}$ & float & 3 \\
V\_DISP\_ERR & Error in stellar velocity dispersion & km s$^{-1}$ & float & 3 \\
D4000 & 4000 \AA\ break from \citet{1983ApJ...273..105B} & & float & 3 \\
D4000\_ERR & Error in the 4000 \AA\ break from \citet{1983ApJ...273..105B} & & float & 3 \\
NII\_HALPHA & The flux ratio of [\ion{N}{2}] and H$\alpha$ & $\log($[\ion{N}{2}]$/$H$\alpha)$ & float & 3 \\
OIII\_HBETA & The flux ratio of [\ion{O}{3}] and H$\beta$ & $\log($[\ion{O}{3}]$/$H$\beta)$ & float & 3 \\
W1MPRO & WISE 1 profile-fit magnitude & mag & float & 4 \\
W2MPRO & WISE 2 profile-fit magnitude & mag & float & 4 \\
W3MPRO & WISE 3 profile-fit magnitude & mag & float & 4 \\
W4MPRO & WISE 4 profile-fit magnitude & mag & float & 4 \\
W1SIGMPRO & Error in WISE 1 profile-fit magnitude & mag & float & 4 \\
W2SIGMPRO & Error in WISE 2 profile-fit magnitude & mag & float & 4 \\
W3SIGMPRO & Error in WISE 3 profile-fit magnitude & mag & float & 4 \\
W4SIGMPRO & Error in WISE 4 profile-fit magnitude & mag & float & 4 \\
W12 & WISE 1 $-$ WISE 2 color & mag & float & 4 \\
W23 & WISE 2 $-$ WISE 3 color & mag & float & 4 \\
W12\_ERR & Error in WISE 1 $-$ WISE 2 color & mag & float & 4 \\
W23\_ERR & Error in WISE 2 $-$ WISE 3 color & mag & float & 4 \\
K01AGN & \citet{2001ApJ...556..121K} BPT AGN classification &  & boolean & 3 \\
K03AGN & \citet{2003MNRAS.346.1055K} BPT AGN classification & & boolean & 3 \\
JARRETTAGN & \citet{2011ApJ...735..112J} WISE AGN classification & & boolean & 4 \\
V\_OFF\_BALMER & Velocity offset of the Balmer lines & km s$^{-1}$ & float & 3 \\
V\_OFF\_BALMER\_ERR & Error in velocity offset of the Balmer lines & km s$^{-1}$ & float & 3 \\
V\_OFF\_FORBIDDEN & Velocity offset of forbidden lines & km s$^{-1}$ & float & 3 \\
V\_OFF\_FORBIDDEN\_ERR & Error in velocity offset of forbidden lines & km s$^{-1}$ & float & 3 \\
SERSIC\_N & S\'ersic index from $r$-band fit & & float & 5 \\
SERSIC\_TH50 & S\'ersic half-light radius along major axis & arcsec & float & 5 \\
ELPETRO\_TH50\_R & Elliptical Petrosian half-light radius in $r$-band & arcsec & float & 5 \\
U\_MAG & SDSS $u'$ magnitude & mag & float & 5 \\
G\_MAG & SDSS $g'$ magnitude & mag & float & 5 \\
R\_MAG & SDSS $r'$ magnitude & mag & float & 5 \\
I\_MAG & SDSS $i'$ magnitude & mag & float & 5 \\
Z\_MAG & SDSS $z'$ magnitude & mag & float & 5 \\
<Coronal line>\_LOGL & Luminosity of the coronal line, \rev{if the line is detected} & $\log(L/$erg s$^{-1})$ & float & 1 \\
<Coronal line>\_LOGL\_ERR & \rev{The error in the luminosity of the coronal line, given by the MCMC posteriors. Readers are encouraged to use caution when utilizing these errors in $S/N$ calculations, as the pure MCMC errors may underestimate errors from systematics.} & dex & float & 1 \\
<Coronal line>\_LOGL\_THRESH & \rev{The luminosity threshold of the spectrum at the line, calculated as described in ${\S}$\ref{sect:results}. A value is not reported if there is a lack of spectral coverage or bad pixels at the location of the line.} & $\log(L/$erg s$^{-1})$ & float & 1 \\
<Coronal line>\_FWHM & FWHM of the coronal line & km s$^{-1}$ & float & 1 \\
<Coronal line>\_FWHM\_ERR & Error in the FWHM of the coronal line & km s$^{-1}$ & float & 1 \\
<Coronal line>\_VOFF & Velocity offset of the coronal line & km s$^{-1}$ & float & 1 \\
<Coronal line>\_VOFF\_ERR & Error in the velocity offset of the coronal line & km s$^{-1}$ & float & 1 \\
<Coronal line>\_EQW & Equivalent width of the coronal line & \AA\ & float & 1 \\
<Coronal line>\_EQW\_ERR & Error in the equivalent width of the coronal line & \AA\ & float & 1 \\
<Coronal line>\_F\_FRAC & $\mathcal{F}$ metric of the coronal line & & float & 1 \\
<Coronal line>\_NPIX & Number of continuous pixels above 3$\sigma$ for the coronal line & & integer & 1 \\
<Coronal line>\_SKY\_FLAG & Sky line proximity warning for the coronal line. True if within 20 \AA\ of 5578.5, 5894.6, 6301.7, or 7246.0 in the observed frame. & & boolean & 1 \\
<Coronal line>\_NN\_CONF & Neural network confidence level that the coronal line is present. & & float & 1 \\
<Other line>\_CONT & Continuum luminosity at the line center from 200-pixel median smoothing of the line-subtracted spectrum & $\log(L/$erg s$^{-1})$ & float & 3 \\
<Other line>\_CONT\_ERR & Error in the continuum luminosity & $\log(L/$erg s$^{-1})$ & float & 3 \\
<Other line>\_EQW & Equivalent width of the emission line, accounting for stellar absorption & \AA\ & float & 3  \\
<Other line>\_EQW\_ERR & Error in the equivalent width of the emission line & \AA\ & float & 3 \\
<Other line>\_REQW & Equivalent width of the emission line, not accounting for stellar absorption & \AA\ & float & 3  \\
<Other line>\_REQW\_ERR & Error in the equivalent width of the emission line & \AA\ & float & 3 \\
<Other line>\_LOGL & Luminosity of the emission line from a Gaussian fit & $\log(L/$erg s$^{-1})$ & float & 3 \\
<Other line>\_LOGL\_ERR & Error in the luminosity of the emission line & $\log(L/$erg s$^{-1})$ & float & 3 \\
<Other line>\_INST\_RES & Instrumental resolution at the line center & km s$^{-1}$ & float & 3 \\
<Other line>\_CHISQ & Reduced $\chi^2$ of the line fit used in the EQW measurements & & float & 3 \\
\enddata
\begin{tablenotes}
    \item[0] Source references:
    \item[1] [1] this work
    \item[2] [2] SDSS
    \item[3] [3] MPA/JHU
    \item[4] [4] AllWISE
    \item[5] [5] NSA
\end{tablenotes}
\end{deluxetable*}
~


\end{document}